\documentclass[longauth]{aa_gaia} 

\usepackage{amsmath}
\usepackage{lscape}
\usepackage{capt-of}
\usepackage{graphicx}
\usepackage{placeins}
\usepackage[utf8]{inputenc}
\usepackage{filecontents}
\usepackage{tablefootnote}
\usepackage{txfonts}
\usepackage{natbib}
\usepackage{arydshln}
\newcommand{\Gaia}{{\it Gaia }}
\newcommand{\sm}{{\rm M}_\odot\, }
\renewcommand{\deg}{{\rm o}}
\newcommand{\masyr}{mas~yr$^{-1}$}
\newcommand{\kms}{km~s$^{-1}$}
\usepackage{color}

\begin{document}

\title{\Gaia Data Release 2: Kinematics of globular clusters and dwarf galaxies around the Milky Way}

\titlerunning{\Gaia DR2: Globular clusters and dwarf galaxies}

\author{
{\it Gaia} Collaboration
\and A.        ~Helmi                         \inst{\ref{inst:0001}}
\and F.        ~van Leeuwen                   \inst{\ref{inst:0002}}
\and P.J.      ~McMillan                      \inst{\ref{inst:0003}}
\and D.        ~Massari                       \inst{\ref{inst:0001}}
\and T.        ~Antoja                        \inst{\ref{inst:0005},\ref{inst:0006}}
\and A.C.      ~Robin                         \inst{\ref{inst:0007}}
\and L.        ~Lindegren                     \inst{\ref{inst:0003}}
\and U.        ~Bastian                       \inst{\ref{inst:0009}}
\and F.        ~Arenou                        \inst{\ref{inst:0010}}
\and C.        ~Babusiaux                     \inst{\ref{inst:0010},\ref{inst:0012}}
\and M.        ~Biermann                      \inst{\ref{inst:0009}}
\and M.A.      ~Breddels                      \inst{\ref{inst:0001}}
\and D.        ~Hobbs                         \inst{\ref{inst:0003}}
\and C.        ~Jordi                         \inst{\ref{inst:0006}}
\and E.        ~Pancino                       \inst{\ref{inst:0017},\ref{inst:0018}}
\and C.        ~Reyl\'{e}                     \inst{\ref{inst:0007}}
\and J.        ~Veljanoski                    \inst{\ref{inst:0001}}
\and A.G.A.    ~Brown                         \inst{\ref{inst:0021}}
\and A.        ~Vallenari                     \inst{\ref{inst:0022}}
\and T.        ~Prusti                        \inst{\ref{inst:0005}}
\and J.H.J.    ~de Bruijne                    \inst{\ref{inst:0005}}
\and C.A.L.    ~Bailer-Jones                  \inst{\ref{inst:0025}}
\and D.W.      ~Evans                         \inst{\ref{inst:0002}}
\and L.        ~Eyer                          \inst{\ref{inst:0027}}
\and F.        ~Jansen                        \inst{\ref{inst:0028}}
\and S.A.      ~Klioner                       \inst{\ref{inst:0029}}
\and U.        ~Lammers                       \inst{\ref{inst:0030}}
\and X.        ~Luri                          \inst{\ref{inst:0006}}
\and F.        ~Mignard                       \inst{\ref{inst:0032}}
\and C.        ~Panem                         \inst{\ref{inst:0033}}
\and D.        ~Pourbaix                      \inst{\ref{inst:0034},\ref{inst:0035}}
\and S.        ~Randich                       \inst{\ref{inst:0017}}
\and P.        ~Sartoretti                    \inst{\ref{inst:0010}}
\and H.I.      ~Siddiqui                      \inst{\ref{inst:0038}}
\and C.        ~Soubiran                      \inst{\ref{inst:0039}}
\and N.A.      ~Walton                        \inst{\ref{inst:0002}}
\and M.        ~Cropper                       \inst{\ref{inst:0041}}
\and R.        ~Drimmel                       \inst{\ref{inst:0042}}
\and D.        ~Katz                          \inst{\ref{inst:0010}}
\and M.G.      ~Lattanzi                      \inst{\ref{inst:0042}}
\and J.        ~Bakker                        \inst{\ref{inst:0030}}
\and C.        ~Cacciari                      \inst{\ref{inst:0046}}
\and J.        ~Casta\~{n}eda                 \inst{\ref{inst:0006}}
\and L.        ~Chaoul                        \inst{\ref{inst:0033}}
\and N.        ~Cheek                         \inst{\ref{inst:0049}}
\and F.        ~De Angeli                     \inst{\ref{inst:0002}}
\and C.        ~Fabricius                     \inst{\ref{inst:0006}}
\and R.        ~Guerra                        \inst{\ref{inst:0030}}
\and B.        ~Holl                          \inst{\ref{inst:0027}}
\and E.        ~Masana                        \inst{\ref{inst:0006}}
\and R.        ~Messineo                      \inst{\ref{inst:0055}}
\and N.        ~Mowlavi                       \inst{\ref{inst:0027}}
\and K.        ~Nienartowicz                  \inst{\ref{inst:0057}}
\and P.        ~Panuzzo                       \inst{\ref{inst:0010}}
\and J.        ~Portell                       \inst{\ref{inst:0006}}
\and M.        ~Riello                        \inst{\ref{inst:0002}}
\and G.M.      ~Seabroke                      \inst{\ref{inst:0041}}
\and P.        ~Tanga                         \inst{\ref{inst:0032}}
\and F.        ~Th\'{e}venin                  \inst{\ref{inst:0032}}
\and G.        ~Gracia-Abril                  \inst{\ref{inst:0064},\ref{inst:0009}}
\and G.        ~Comoretto                     \inst{\ref{inst:0038}}
\and M.        ~Garcia-Reinaldos              \inst{\ref{inst:0030}}
\and D.        ~Teyssier                      \inst{\ref{inst:0038}}
\and M.        ~Altmann                       \inst{\ref{inst:0009},\ref{inst:0070}}
\and R.        ~Andrae                        \inst{\ref{inst:0025}}
\and M.        ~Audard                        \inst{\ref{inst:0027}}
\and I.        ~Bellas-Velidis                \inst{\ref{inst:0073}}
\and K.        ~Benson                        \inst{\ref{inst:0041}}
\and J.        ~Berthier                      \inst{\ref{inst:0075}}
\and R.        ~Blomme                        \inst{\ref{inst:0076}}
\and P.        ~Burgess                       \inst{\ref{inst:0002}}
\and G.        ~Busso                         \inst{\ref{inst:0002}}
\and B.        ~Carry                         \inst{\ref{inst:0032},\ref{inst:0075}}
\and A.        ~Cellino                       \inst{\ref{inst:0042}}
\and G.        ~Clementini                    \inst{\ref{inst:0046}}
\and M.        ~Clotet                        \inst{\ref{inst:0006}}
\and O.        ~Creevey                       \inst{\ref{inst:0032},\ref{inst:0085}}
\and M.        ~Davidson                      \inst{\ref{inst:0086}}
\and J.        ~De Ridder                     \inst{\ref{inst:0087}}
\and L.        ~Delchambre                    \inst{\ref{inst:0088}}
\and A.        ~Dell'Oro                      \inst{\ref{inst:0017}}
\and C.        ~Ducourant                     \inst{\ref{inst:0039}}
\and J.        ~Fern\'{a}ndez-Hern\'{a}ndez   \inst{\ref{inst:0091}}
\and M.        ~Fouesneau                     \inst{\ref{inst:0025}}
\and Y.        ~Fr\'{e}mat                    \inst{\ref{inst:0076}}
\and L.        ~Galluccio                     \inst{\ref{inst:0032}}
\and M.        ~Garc\'{i}a-Torres             \inst{\ref{inst:0095}}
\and J.        ~Gonz\'{a}lez-N\'{u}\~{n}ez    \inst{\ref{inst:0049},\ref{inst:0097}}
\and J.J.      ~Gonz\'{a}lez-Vidal            \inst{\ref{inst:0006}}
\and E.        ~Gosset                        \inst{\ref{inst:0088},\ref{inst:0035}}
\and L.P.      ~Guy                           \inst{\ref{inst:0057},\ref{inst:0102}}
\and J.-L.     ~Halbwachs                     \inst{\ref{inst:0103}}
\and N.C.      ~Hambly                        \inst{\ref{inst:0086}}
\and D.L.      ~Harrison                      \inst{\ref{inst:0002},\ref{inst:0106}}
\and J.        ~Hern\'{a}ndez                 \inst{\ref{inst:0030}}
\and D.        ~Hestroffer                    \inst{\ref{inst:0075}}
\and S.T.      ~Hodgkin                       \inst{\ref{inst:0002}}
\and A.        ~Hutton                        \inst{\ref{inst:0110}}
\and G.        ~Jasniewicz                    \inst{\ref{inst:0111}}
\and A.        ~Jean-Antoine-Piccolo          \inst{\ref{inst:0033}}
\and S.        ~Jordan                        \inst{\ref{inst:0009}}
\and A.J.      ~Korn                          \inst{\ref{inst:0114}}
\and A.        ~Krone-Martins                 \inst{\ref{inst:0115}}
\and A.C.      ~Lanzafame                     \inst{\ref{inst:0116},\ref{inst:0117}}
\and T.        ~Lebzelter                     \inst{\ref{inst:0118}}
\and W.        ~L\"{ o}ffler                  \inst{\ref{inst:0009}}
\and M.        ~Manteiga                      \inst{\ref{inst:0120},\ref{inst:0121}}
\and P.M.      ~Marrese                       \inst{\ref{inst:0122},\ref{inst:0018}}
\and J.M.      ~Mart\'{i}n-Fleitas            \inst{\ref{inst:0110}}
\and A.        ~Moitinho                      \inst{\ref{inst:0115}}
\and A.        ~Mora                          \inst{\ref{inst:0110}}
\and K.        ~Muinonen                      \inst{\ref{inst:0127},\ref{inst:0128}}
\and J.        ~Osinde                        \inst{\ref{inst:0129}}
\and T.        ~Pauwels                       \inst{\ref{inst:0076}}
\and J.-M.     ~Petit                         \inst{\ref{inst:0007}}
\and A.        ~Recio-Blanco                  \inst{\ref{inst:0032}}
\and P.J.      ~Richards                      \inst{\ref{inst:0133}}
\and L.        ~Rimoldini                     \inst{\ref{inst:0057}}
\and L.M.      ~Sarro                         \inst{\ref{inst:0135}}
\and C.        ~Siopis                        \inst{\ref{inst:0034}}
\and M.        ~Smith                         \inst{\ref{inst:0041}}
\and A.        ~Sozzetti                      \inst{\ref{inst:0042}}
\and M.        ~S\"{ u}veges                  \inst{\ref{inst:0025}}
\and J.        ~Torra                         \inst{\ref{inst:0006}}
\and W.        ~van Reeven                    \inst{\ref{inst:0110}}
\and U.        ~Abbas                         \inst{\ref{inst:0042}}
\and A.        ~Abreu Aramburu                \inst{\ref{inst:0143}}
\and S.        ~Accart                        \inst{\ref{inst:0144}}
\and C.        ~Aerts                         \inst{\ref{inst:0087},\ref{inst:0146}}
\and G.        ~Altavilla                     \inst{\ref{inst:0122},\ref{inst:0018},\ref{inst:0046}}
\and M.A.      ~\'{A}lvarez                   \inst{\ref{inst:0120}}
\and R.        ~Alvarez                       \inst{\ref{inst:0030}}
\and J.        ~Alves                         \inst{\ref{inst:0118}}
\and R.I.      ~Anderson                      \inst{\ref{inst:0153},\ref{inst:0027}}
\and A.H.      ~Andrei                        \inst{\ref{inst:0155},\ref{inst:0156},\ref{inst:0070}}
\and E.        ~Anglada Varela                \inst{\ref{inst:0091}}
\and E.        ~Antiche                       \inst{\ref{inst:0006}}
\and B.        ~Arcay                         \inst{\ref{inst:0120}}
\and T.L.      ~Astraatmadja                  \inst{\ref{inst:0025},\ref{inst:0162}}
\and N.        ~Bach                          \inst{\ref{inst:0110}}
\and S.G.      ~Baker                         \inst{\ref{inst:0041}}
\and L.        ~Balaguer-N\'{u}\~{n}ez        \inst{\ref{inst:0006}}
\and P.        ~Balm                          \inst{\ref{inst:0038}}
\and C.        ~Barache                       \inst{\ref{inst:0070}}
\and C.        ~Barata                        \inst{\ref{inst:0115}}
\and D.        ~Barbato                       \inst{\ref{inst:0169},\ref{inst:0042}}
\and F.        ~Barblan                       \inst{\ref{inst:0027}}
\and P.S.      ~Barklem                       \inst{\ref{inst:0114}}
\and D.        ~Barrado                       \inst{\ref{inst:0173}}
\and M.        ~Barros                        \inst{\ref{inst:0115}}
\and M.A.      ~Barstow                       \inst{\ref{inst:0175}}
\and S.        ~Bartholom\'{e} Mu\~{n}oz      \inst{\ref{inst:0006}}
\and J.-L.     ~Bassilana                     \inst{\ref{inst:0144}}
\and U.        ~Becciani                      \inst{\ref{inst:0117}}
\and M.        ~Bellazzini                    \inst{\ref{inst:0046}}
\and A.        ~Berihuete                     \inst{\ref{inst:0180}}
\and S.        ~Bertone                       \inst{\ref{inst:0042},\ref{inst:0070},\ref{inst:0183}}
\and L.        ~Bianchi                       \inst{\ref{inst:0184}}
\and O.        ~Bienaym\'{e}                  \inst{\ref{inst:0103}}
\and S.        ~Blanco-Cuaresma               \inst{\ref{inst:0027},\ref{inst:0039},\ref{inst:0188}}
\and T.        ~Boch                          \inst{\ref{inst:0103}}
\and C.        ~Boeche                        \inst{\ref{inst:0022}}
\and A.        ~Bombrun                       \inst{\ref{inst:0191}}
\and R.        ~Borrachero                    \inst{\ref{inst:0006}}
\and D.        ~Bossini                       \inst{\ref{inst:0022}}
\and S.        ~Bouquillon                    \inst{\ref{inst:0070}}
\and G.        ~Bourda                        \inst{\ref{inst:0039}}
\and A.        ~Bragaglia                     \inst{\ref{inst:0046}}
\and L.        ~Bramante                      \inst{\ref{inst:0055}}
\and A.        ~Bressan                       \inst{\ref{inst:0198}}
\and N.        ~Brouillet                     \inst{\ref{inst:0039}}
\and T.        ~Br\"{ u}semeister             \inst{\ref{inst:0009}}
\and E.        ~Brugaletta                    \inst{\ref{inst:0117}}
\and B.        ~Bucciarelli                   \inst{\ref{inst:0042}}
\and A.        ~Burlacu                       \inst{\ref{inst:0033}}
\and D.        ~Busonero                      \inst{\ref{inst:0042}}
\and A.G.      ~Butkevich                     \inst{\ref{inst:0029}}
\and R.        ~Buzzi                         \inst{\ref{inst:0042}}
\and E.        ~Caffau                        \inst{\ref{inst:0010}}
\and R.        ~Cancelliere                   \inst{\ref{inst:0208}}
\and G.        ~Cannizzaro                    \inst{\ref{inst:0209},\ref{inst:0146}}
\and T.        ~Cantat-Gaudin                 \inst{\ref{inst:0022},\ref{inst:0006}}
\and R.        ~Carballo                      \inst{\ref{inst:0213}}
\and T.        ~Carlucci                      \inst{\ref{inst:0070}}
\and J.M.      ~Carrasco                      \inst{\ref{inst:0006}}
\and L.        ~Casamiquela                   \inst{\ref{inst:0006}}
\and M.        ~Castellani                    \inst{\ref{inst:0122}}
\and A.        ~Castro-Ginard                 \inst{\ref{inst:0006}}
\and P.        ~Charlot                       \inst{\ref{inst:0039}}
\and L.        ~Chemin                        \inst{\ref{inst:0220}}
\and A.        ~Chiavassa                     \inst{\ref{inst:0032}}
\and G.        ~Cocozza                       \inst{\ref{inst:0046}}
\and G.        ~Costigan                      \inst{\ref{inst:0021}}
\and S.        ~Cowell                        \inst{\ref{inst:0002}}
\and F.        ~Crifo                         \inst{\ref{inst:0010}}
\and M.        ~Crosta                        \inst{\ref{inst:0042}}
\and C.        ~Crowley                       \inst{\ref{inst:0191}}
\and J.        ~Cuypers$^\dagger$             \inst{\ref{inst:0076}}
\and C.        ~Dafonte                       \inst{\ref{inst:0120}}
\and Y.        ~Damerdji                      \inst{\ref{inst:0088},\ref{inst:0231}}
\and A.        ~Dapergolas                    \inst{\ref{inst:0073}}
\and P.        ~David                         \inst{\ref{inst:0075}}
\and M.        ~David                         \inst{\ref{inst:0234}}
\and P.        ~de Laverny                    \inst{\ref{inst:0032}}
\and F.        ~De Luise                      \inst{\ref{inst:0236}}
\and R.        ~De March                      \inst{\ref{inst:0055}}
\and D.        ~de Martino                    \inst{\ref{inst:0238}}
\and R.        ~de Souza                      \inst{\ref{inst:0239}}
\and A.        ~de Torres                     \inst{\ref{inst:0191}}
\and J.        ~Debosscher                    \inst{\ref{inst:0087}}
\and E.        ~del Pozo                      \inst{\ref{inst:0110}}
\and M.        ~Delbo                         \inst{\ref{inst:0032}}
\and A.        ~Delgado                       \inst{\ref{inst:0002}}
\and H.E.      ~Delgado                       \inst{\ref{inst:0135}}
\and P.        ~Di Matteo                     \inst{\ref{inst:0010}}
\and S.        ~Diakite                       \inst{\ref{inst:0007}}
\and C.        ~Diener                        \inst{\ref{inst:0002}}
\and E.        ~Distefano                     \inst{\ref{inst:0117}}
\and C.        ~Dolding                       \inst{\ref{inst:0041}}
\and P.        ~Drazinos                      \inst{\ref{inst:0251}}
\and J.        ~Dur\'{a}n                     \inst{\ref{inst:0129}}
\and B.        ~Edvardsson                    \inst{\ref{inst:0114}}
\and H.        ~Enke                          \inst{\ref{inst:0254}}
\and K.        ~Eriksson                      \inst{\ref{inst:0114}}
\and P.        ~Esquej                        \inst{\ref{inst:0256}}
\and G.        ~Eynard Bontemps               \inst{\ref{inst:0033}}
\and C.        ~Fabre                         \inst{\ref{inst:0258}}
\and M.        ~Fabrizio                      \inst{\ref{inst:0122},\ref{inst:0018}}
\and S.        ~Faigler                       \inst{\ref{inst:0261}}
\and A.J.      ~Falc\~{a}o                    \inst{\ref{inst:0262}}
\and M.        ~Farr\`{a}s Casas              \inst{\ref{inst:0006}}
\and L.        ~Federici                      \inst{\ref{inst:0046}}
\and G.        ~Fedorets                      \inst{\ref{inst:0127}}
\and P.        ~Fernique                      \inst{\ref{inst:0103}}
\and F.        ~Figueras                      \inst{\ref{inst:0006}}
\and F.        ~Filippi                       \inst{\ref{inst:0055}}
\and K.        ~Findeisen                     \inst{\ref{inst:0010}}
\and A.        ~Fonti                         \inst{\ref{inst:0055}}
\and E.        ~Fraile                        \inst{\ref{inst:0256}}
\and M.        ~Fraser                        \inst{\ref{inst:0002},\ref{inst:0273}}
\and B.        ~Fr\'{e}zouls                  \inst{\ref{inst:0033}}
\and M.        ~Gai                           \inst{\ref{inst:0042}}
\and S.        ~Galleti                       \inst{\ref{inst:0046}}
\and D.        ~Garabato                      \inst{\ref{inst:0120}}
\and F.        ~Garc\'{i}a-Sedano             \inst{\ref{inst:0135}}
\and A.        ~Garofalo                      \inst{\ref{inst:0279},\ref{inst:0046}}
\and N.        ~Garralda                      \inst{\ref{inst:0006}}
\and A.        ~Gavel                         \inst{\ref{inst:0114}}
\and P.        ~Gavras                        \inst{\ref{inst:0010},\ref{inst:0073},\ref{inst:0251}}
\and J.        ~Gerssen                       \inst{\ref{inst:0254}}
\and R.        ~Geyer                         \inst{\ref{inst:0029}}
\and P.        ~Giacobbe                      \inst{\ref{inst:0042}}
\and G.        ~Gilmore                       \inst{\ref{inst:0002}}
\and S.        ~Girona                        \inst{\ref{inst:0290}}
\and G.        ~Giuffrida                     \inst{\ref{inst:0018},\ref{inst:0122}}
\and F.        ~Glass                         \inst{\ref{inst:0027}}
\and M.        ~Gomes                         \inst{\ref{inst:0115}}
\and M.        ~Granvik                       \inst{\ref{inst:0127},\ref{inst:0296}}
\and A.        ~Gueguen                       \inst{\ref{inst:0010},\ref{inst:0298}}
\and A.        ~Guerrier                      \inst{\ref{inst:0144}}
\and J.        ~Guiraud                       \inst{\ref{inst:0033}}
\and R.        ~Guti\'{e}rrez-S\'{a}nchez     \inst{\ref{inst:0038}}
\and R.        ~Haigron                       \inst{\ref{inst:0010}}
\and D.        ~Hatzidimitriou                \inst{\ref{inst:0251},\ref{inst:0073}}
\and M.        ~Hauser                        \inst{\ref{inst:0009},\ref{inst:0025}}
\and M.        ~Haywood                       \inst{\ref{inst:0010}}
\and U.        ~Heiter                        \inst{\ref{inst:0114}}
\and J.        ~Heu                           \inst{\ref{inst:0010}}
\and T.        ~Hilger                        \inst{\ref{inst:0029}}
\and W.        ~Hofmann                       \inst{\ref{inst:0009}}
\and G.        ~Holland                       \inst{\ref{inst:0002}}
\and H.E.      ~Huckle                        \inst{\ref{inst:0041}}
\and A.        ~Hypki                         \inst{\ref{inst:0021},\ref{inst:0315}}
\and V.        ~Icardi                        \inst{\ref{inst:0055}}
\and K.        ~Jan{\ss}en                    \inst{\ref{inst:0254}}
\and G.        ~Jevardat de Fombelle          \inst{\ref{inst:0057}}
\and P.G.      ~Jonker                        \inst{\ref{inst:0209},\ref{inst:0146}}
\and \'{A}.L.  ~Juh\'{a}sz                    \inst{\ref{inst:0321},\ref{inst:0322}}
\and F.        ~Julbe                         \inst{\ref{inst:0006}}
\and A.        ~Karampelas                    \inst{\ref{inst:0251},\ref{inst:0325}}
\and A.        ~Kewley                        \inst{\ref{inst:0002}}
\and J.        ~Klar                          \inst{\ref{inst:0254}}
\and A.        ~Kochoska                      \inst{\ref{inst:0328},\ref{inst:0329}}
\and R.        ~Kohley                        \inst{\ref{inst:0030}}
\and K.        ~Kolenberg                     \inst{\ref{inst:0331},\ref{inst:0087},\ref{inst:0188}}
\and M.        ~Kontizas                      \inst{\ref{inst:0251}}
\and E.        ~Kontizas                      \inst{\ref{inst:0073}}
\and S.E.      ~Koposov                       \inst{\ref{inst:0002},\ref{inst:0337}}
\and G.        ~Kordopatis                    \inst{\ref{inst:0032}}
\and Z.        ~Kostrzewa-Rutkowska           \inst{\ref{inst:0209},\ref{inst:0146}}
\and P.        ~Koubsky                       \inst{\ref{inst:0341}}
\and S.        ~Lambert                       \inst{\ref{inst:0070}}
\and A.F.      ~Lanza                         \inst{\ref{inst:0117}}
\and Y.        ~Lasne                         \inst{\ref{inst:0144}}
\and J.-B.     ~Lavigne                       \inst{\ref{inst:0144}}
\and Y.        ~Le Fustec                     \inst{\ref{inst:0346}}
\and C.        ~Le Poncin-Lafitte             \inst{\ref{inst:0070}}
\and Y.        ~Lebreton                      \inst{\ref{inst:0010},\ref{inst:0349}}
\and S.        ~Leccia                        \inst{\ref{inst:0238}}
\and N.        ~Leclerc                       \inst{\ref{inst:0010}}
\and I.        ~Lecoeur-Taibi                 \inst{\ref{inst:0057}}
\and H.        ~Lenhardt                      \inst{\ref{inst:0009}}
\and F.        ~Leroux                        \inst{\ref{inst:0144}}
\and S.        ~Liao                          \inst{\ref{inst:0042},\ref{inst:0356},\ref{inst:0357}}
\and E.        ~Licata                        \inst{\ref{inst:0184}}
\and H.E.P.    ~Lindstr{\o}m                  \inst{\ref{inst:0359},\ref{inst:0360}}
\and T.A.      ~Lister                        \inst{\ref{inst:0361}}
\and E.        ~Livanou                       \inst{\ref{inst:0251}}
\and A.        ~Lobel                         \inst{\ref{inst:0076}}
\and M.        ~L\'{o}pez                     \inst{\ref{inst:0173}}
\and S.        ~Managau                       \inst{\ref{inst:0144}}
\and R.G.      ~Mann                          \inst{\ref{inst:0086}}
\and G.        ~Mantelet                      \inst{\ref{inst:0009}}
\and O.        ~Marchal                       \inst{\ref{inst:0010}}
\and J.M.      ~Marchant                      \inst{\ref{inst:0369}}
\and M.        ~Marconi                       \inst{\ref{inst:0238}}
\and S.        ~Marinoni                      \inst{\ref{inst:0122},\ref{inst:0018}}
\and G.        ~Marschalk\'{o}                \inst{\ref{inst:0321},\ref{inst:0374}}
\and D.J.      ~Marshall                      \inst{\ref{inst:0375}}
\and M.        ~Martino                       \inst{\ref{inst:0055}}
\and G.        ~Marton                        \inst{\ref{inst:0321}}
\and N.        ~Mary                          \inst{\ref{inst:0144}}
\and G.        ~Matijevi\v{c}                 \inst{\ref{inst:0254}}
\and T.        ~Mazeh                         \inst{\ref{inst:0261}}
\and S.        ~Messina                       \inst{\ref{inst:0117}}
\and D.        ~Michalik                      \inst{\ref{inst:0003}}
\and N.R.      ~Millar                        \inst{\ref{inst:0002}}
\and D.        ~Molina                        \inst{\ref{inst:0006}}
\and R.        ~Molinaro                      \inst{\ref{inst:0238}}
\and L.        ~Moln\'{a}r                    \inst{\ref{inst:0321}}
\and P.        ~Montegriffo                   \inst{\ref{inst:0046}}
\and R.        ~Mor                           \inst{\ref{inst:0006}}
\and R.        ~Morbidelli                    \inst{\ref{inst:0042}}
\and T.        ~Morel                         \inst{\ref{inst:0088}}
\and D.        ~Morris                        \inst{\ref{inst:0086}}
\and A.F.      ~Mulone                        \inst{\ref{inst:0055}}
\and T.        ~Muraveva                      \inst{\ref{inst:0046}}
\and I.        ~Musella                       \inst{\ref{inst:0238}}
\and G.        ~Nelemans                      \inst{\ref{inst:0146},\ref{inst:0087}}
\and L.        ~Nicastro                      \inst{\ref{inst:0046}}
\and L.        ~Noval                         \inst{\ref{inst:0144}}
\and W.        ~O'Mullane                     \inst{\ref{inst:0030},\ref{inst:0102}}
\and C.        ~Ord\'{e}novic                 \inst{\ref{inst:0032}}
\and D.        ~Ord\'{o}\~{n}ez-Blanco        \inst{\ref{inst:0057}}
\and P.        ~Osborne                       \inst{\ref{inst:0002}}
\and C.        ~Pagani                        \inst{\ref{inst:0175}}
\and I.        ~Pagano                        \inst{\ref{inst:0117}}
\and F.        ~Pailler                       \inst{\ref{inst:0033}}
\and H.        ~Palacin                       \inst{\ref{inst:0144}}
\and L.        ~Palaversa                     \inst{\ref{inst:0002},\ref{inst:0027}}
\and A.        ~Panahi                        \inst{\ref{inst:0261}}
\and M.        ~Pawlak                        \inst{\ref{inst:0411},\ref{inst:0412}}
\and A.M.      ~Piersimoni                    \inst{\ref{inst:0236}}
\and F.-X.     ~Pineau                        \inst{\ref{inst:0103}}
\and E.        ~Plachy                        \inst{\ref{inst:0321}}
\and G.        ~Plum                          \inst{\ref{inst:0010}}
\and E.        ~Poggio                        \inst{\ref{inst:0169},\ref{inst:0042}}
\and E.        ~Poujoulet                     \inst{\ref{inst:0419}}
\and A.        ~Pr\v{s}a                      \inst{\ref{inst:0329}}
\and L.        ~Pulone                        \inst{\ref{inst:0122}}
\and E.        ~Racero                        \inst{\ref{inst:0049}}
\and S.        ~Ragaini                       \inst{\ref{inst:0046}}
\and N.        ~Rambaux                       \inst{\ref{inst:0075}}
\and M.        ~Ramos-Lerate                  \inst{\ref{inst:0425}}
\and S.        ~Regibo                        \inst{\ref{inst:0087}}
\and F.        ~Riclet                        \inst{\ref{inst:0033}}
\and V.        ~Ripepi                        \inst{\ref{inst:0238}}
\and A.        ~Riva                          \inst{\ref{inst:0042}}
\and A.        ~Rivard                        \inst{\ref{inst:0144}}
\and G.        ~Rixon                         \inst{\ref{inst:0002}}
\and T.        ~Roegiers                      \inst{\ref{inst:0432}}
\and M.        ~Roelens                       \inst{\ref{inst:0027}}
\and M.        ~Romero-G\'{o}mez              \inst{\ref{inst:0006}}
\and N.        ~Rowell                        \inst{\ref{inst:0086}}
\and F.        ~Royer                         \inst{\ref{inst:0010}}
\and L.        ~Ruiz-Dern                     \inst{\ref{inst:0010}}
\and G.        ~Sadowski                      \inst{\ref{inst:0034}}
\and T.        ~Sagrist\`{a} Sell\'{e}s       \inst{\ref{inst:0009}}
\and J.        ~Sahlmann                      \inst{\ref{inst:0030},\ref{inst:0441}}
\and J.        ~Salgado                       \inst{\ref{inst:0442}}
\and E.        ~Salguero                      \inst{\ref{inst:0091}}
\and N.        ~Sanna                         \inst{\ref{inst:0017}}
\and T.        ~Santana-Ros                   \inst{\ref{inst:0315}}
\and M.        ~Sarasso                       \inst{\ref{inst:0042}}
\and H.        ~Savietto                      \inst{\ref{inst:0447}}
\and M.        ~Schultheis                    \inst{\ref{inst:0032}}
\and E.        ~Sciacca                       \inst{\ref{inst:0117}}
\and M.        ~Segol                         \inst{\ref{inst:0450}}
\and J.C.      ~Segovia                       \inst{\ref{inst:0049}}
\and D.        ~S\'{e}gransan                 \inst{\ref{inst:0027}}
\and I-C.      ~Shih                          \inst{\ref{inst:0010}}
\and L.        ~Siltala                       \inst{\ref{inst:0127},\ref{inst:0455}}
\and A.F.      ~Silva                         \inst{\ref{inst:0115}}
\and R.L.      ~Smart                         \inst{\ref{inst:0042}}
\and K.W.      ~Smith                         \inst{\ref{inst:0025}}
\and E.        ~Solano                        \inst{\ref{inst:0173},\ref{inst:0460}}
\and F.        ~Solitro                       \inst{\ref{inst:0055}}
\and R.        ~Sordo                         \inst{\ref{inst:0022}}
\and S.        ~Soria Nieto                   \inst{\ref{inst:0006}}
\and J.        ~Souchay                       \inst{\ref{inst:0070}}
\and A.        ~Spagna                        \inst{\ref{inst:0042}}
\and F.        ~Spoto                         \inst{\ref{inst:0032},\ref{inst:0075}}
\and U.        ~Stampa                        \inst{\ref{inst:0009}}
\and I.A.      ~Steele                        \inst{\ref{inst:0369}}
\and H.        ~Steidelm\"{ u}ller            \inst{\ref{inst:0029}}
\and C.A.      ~Stephenson                    \inst{\ref{inst:0038}}
\and H.        ~Stoev                         \inst{\ref{inst:0472}}
\and F.F.      ~Suess                         \inst{\ref{inst:0002}}
\and J.        ~Surdej                        \inst{\ref{inst:0088}}
\and L.        ~Szabados                      \inst{\ref{inst:0321}}
\and E.        ~Szegedi-Elek                  \inst{\ref{inst:0321}}
\and D.        ~Tapiador                      \inst{\ref{inst:0477},\ref{inst:0478}}
\and F.        ~Taris                         \inst{\ref{inst:0070}}
\and G.        ~Tauran                        \inst{\ref{inst:0144}}
\and M.B.      ~Taylor                        \inst{\ref{inst:0481}}
\and R.        ~Teixeira                      \inst{\ref{inst:0239}}
\and D.        ~Terrett                       \inst{\ref{inst:0133}}
\and P.        ~Teyssandier                   \inst{\ref{inst:0070}}
\and W.        ~Thuillot                      \inst{\ref{inst:0075}}
\and A.        ~Titarenko                     \inst{\ref{inst:0032}}
\and F.        ~Torra Clotet                  \inst{\ref{inst:0487}}
\and C.        ~Turon                         \inst{\ref{inst:0010}}
\and A.        ~Ulla                          \inst{\ref{inst:0489}}
\and E.        ~Utrilla                       \inst{\ref{inst:0110}}
\and S.        ~Uzzi                          \inst{\ref{inst:0055}}
\and M.        ~Vaillant                      \inst{\ref{inst:0144}}
\and G.        ~Valentini                     \inst{\ref{inst:0236}}
\and V.        ~Valette                       \inst{\ref{inst:0033}}
\and A.        ~van Elteren                   \inst{\ref{inst:0021}}
\and E.        ~Van Hemelryck                 \inst{\ref{inst:0076}}
\and M.        ~van Leeuwen                   \inst{\ref{inst:0002}}
\and M.        ~Vaschetto                     \inst{\ref{inst:0055}}
\and A.        ~Vecchiato                     \inst{\ref{inst:0042}}
\and Y.        ~Viala                         \inst{\ref{inst:0010}}
\and D.        ~Vicente                       \inst{\ref{inst:0290}}
\and S.        ~Vogt                          \inst{\ref{inst:0432}}
\and C.        ~von Essen                     \inst{\ref{inst:0503}}
\and H.        ~Voss                          \inst{\ref{inst:0006}}
\and V.        ~Votruba                       \inst{\ref{inst:0341}}
\and S.        ~Voutsinas                     \inst{\ref{inst:0086}}
\and G.        ~Walmsley                      \inst{\ref{inst:0033}}
\and M.        ~Weiler                        \inst{\ref{inst:0006}}
\and O.        ~Wertz                         \inst{\ref{inst:0509}}
\and T.        ~Wevers                        \inst{\ref{inst:0002},\ref{inst:0146}}
\and \L{}.     ~Wyrzykowski                   \inst{\ref{inst:0002},\ref{inst:0411}}
\and A.        ~Yoldas                        \inst{\ref{inst:0002}}
\and M.        ~\v{Z}erjal                    \inst{\ref{inst:0328},\ref{inst:0516}}
\and H.        ~Ziaeepour                     \inst{\ref{inst:0007}}
\and J.        ~Zorec                         \inst{\ref{inst:0518}}
\and S.        ~Zschocke                      \inst{\ref{inst:0029}}
\and S.        ~Zucker                        \inst{\ref{inst:0520}}
\and C.        ~Zurbach                       \inst{\ref{inst:0111}}
\and T.        ~Zwitter                       \inst{\ref{inst:0328}}
}
\institute{
     Kapteyn Astronomical Institute, University of Groningen, Landleven 12, 9747 AD Groningen, The Netherlands\relax                                                                                         \label{inst:0001}
\and Institute of Astronomy, University of Cambridge, Madingley Road, Cambridge CB3 0HA, United Kingdom\relax                                                                                                \label{inst:0002}
\and Lund Observatory, Department of Astronomy and Theoretical Physics, Lund University, Box 43, 22100 Lund, Sweden\relax                                                                                    \label{inst:0003}
\and Science Support Office, Directorate of Science, European Space Research and Technology Centre (ESA/ESTEC), Keplerlaan 1, 2201AZ, Noordwijk, The Netherlands\relax                                       \label{inst:0005}
\and Institut de Ci\`{e}ncies del Cosmos, Universitat  de  Barcelona  (IEEC-UB), Mart\'{i} i  Franqu\`{e}s  1, 08028 Barcelona, Spain\relax                                                                  \label{inst:0006}
\and Institut UTINAM UMR6213, CNRS, OSU THETA Franche-Comt\'{e} Bourgogne, Universit\'{e} Bourgogne Franche-Comt\'{e}, 25000 Besan\c{c}on, France\relax                                                      \label{inst:0007}
\and Astronomisches Rechen-Institut, Zentrum f\"{ u}r Astronomie der Universit\"{ a}t Heidelberg, M\"{ o}nchhofstr. 12-14, 69120 Heidelberg, Germany\relax                                                   \label{inst:0009}
\and GEPI, Observatoire de Paris, Universit\'{e} PSL, CNRS, 5 Place Jules Janssen, 92190 Meudon, France\relax                                                                                                \label{inst:0010}
\and Univ. Grenoble Alpes, CNRS, IPAG, 38000 Grenoble, France\relax                                                                                                                                          \label{inst:0012}
\and INAF - Osservatorio Astrofisico di Arcetri, Largo Enrico Fermi 5, 50125 Firenze, Italy\relax                                                                                                            \label{inst:0017}
\and Space Science Data Center - ASI, Via del Politecnico SNC, 00133 Roma, Italy\relax                                                                                                                       \label{inst:0018}
\and Leiden Observatory, Leiden University, Niels Bohrweg 2, 2333 CA Leiden, The Netherlands\relax                                                                                                           \label{inst:0021}
\and INAF - Osservatorio astronomico di Padova, Vicolo Osservatorio 5, 35122 Padova, Italy\relax                                                                                                             \label{inst:0022}
\and Max Planck Institute for Astronomy, K\"{ o}nigstuhl 17, 69117 Heidelberg, Germany\relax                                                                                                                 \label{inst:0025}
\and Department of Astronomy, University of Geneva, Chemin des Maillettes 51, 1290 Versoix, Switzerland\relax                                                                                                \label{inst:0027}
\and Mission Operations Division, Operations Department, Directorate of Science, European Space Research and Technology Centre (ESA/ESTEC), Keplerlaan 1, 2201 AZ, Noordwijk, The Netherlands\relax          \label{inst:0028}
\and Lohrmann Observatory, Technische Universit\"{ a}t Dresden, Mommsenstra{\ss}e 13, 01062 Dresden, Germany\relax                                                                                           \label{inst:0029}
\and European Space Astronomy Centre (ESA/ESAC), Camino bajo del Castillo, s/n, Urbanizacion Villafranca del Castillo, Villanueva de la Ca\~{n}ada, 28692 Madrid, Spain\relax                                \label{inst:0030}
\and Universit\'{e} C\^{o}te d'Azur, Observatoire de la C\^{o}te d'Azur, CNRS, Laboratoire Lagrange, Bd de l'Observatoire, CS 34229, 06304 Nice Cedex 4, France\relax                                        \label{inst:0032}
\and CNES Centre Spatial de Toulouse, 18 avenue Edouard Belin, 31401 Toulouse Cedex 9, France\relax                                                                                                          \label{inst:0033}
\and Institut d'Astronomie et d'Astrophysique, Universit\'{e} Libre de Bruxelles CP 226, Boulevard du Triomphe, 1050 Brussels, Belgium\relax                                                                 \label{inst:0034}
\and F.R.S.-FNRS, Rue d'Egmont 5, 1000 Brussels, Belgium\relax                                                                                                                                               \label{inst:0035}
\and Telespazio Vega UK Ltd for ESA/ESAC, Camino bajo del Castillo, s/n, Urbanizacion Villafranca del Castillo, Villanueva de la Ca\~{n}ada, 28692 Madrid, Spain\relax                                       \label{inst:0038}
\and Laboratoire d'astrophysique de Bordeaux, Univ. Bordeaux, CNRS, B18N, all{\'e}e Geoffroy Saint-Hilaire, 33615 Pessac, France\relax                                                                       \label{inst:0039}
\and Mullard Space Science Laboratory, University College London, Holmbury St Mary, Dorking, Surrey RH5 6NT, United Kingdom\relax                                                                            \label{inst:0041}
\and INAF - Osservatorio Astrofisico di Torino, via Osservatorio 20, 10025 Pino Torinese (TO), Italy\relax                                                                                                   \label{inst:0042}
\and INAF - Osservatorio di Astrofisica e Scienza dello Spazio di Bologna, via Piero Gobetti 93/3, 40129 Bologna, Italy\relax                                                                                \label{inst:0046}
\and Serco Gesti\'{o}n de Negocios for ESA/ESAC, Camino bajo del Castillo, s/n, Urbanizacion Villafranca del Castillo, Villanueva de la Ca\~{n}ada, 28692 Madrid, Spain\relax                                \label{inst:0049}
\and ALTEC S.p.a, Corso Marche, 79,10146 Torino, Italy\relax                                                                                                                                                 \label{inst:0055}
\and Department of Astronomy, University of Geneva, Chemin d'Ecogia 16, 1290 Versoix, Switzerland\relax                                                                                                      \label{inst:0057}
\and Gaia DPAC Project Office, ESAC, Camino bajo del Castillo, s/n, Urbanizacion Villafranca del Castillo, Villanueva de la Ca\~{n}ada, 28692 Madrid, Spain\relax                                            \label{inst:0064}
\and SYRTE, Observatoire de Paris, Universit\'{e} PSL, CNRS,  Sorbonne Universit\'{e}, LNE, 61 avenue de l’Observatoire 75014 Paris, France\relax                                                          \label{inst:0070}
\and National Observatory of Athens, I. Metaxa and Vas. Pavlou, Palaia Penteli, 15236 Athens, Greece\relax                                                                                                   \label{inst:0073}
\and IMCCE, Observatoire de Paris, Universit\'{e} PSL, CNRS,  Sorbonne Universit\'{e}, Univ. Lille, 77 av. Denfert-Rochereau, 75014 Paris, France\relax                                                      \label{inst:0075}
\and Royal Observatory of Belgium, Ringlaan 3, 1180 Brussels, Belgium\relax                                                                                                                                  \label{inst:0076}
\and Institut d'Astrophysique Spatiale, Universit\'{e} Paris XI, UMR 8617, CNRS, B\^{a}timent 121, 91405, Orsay Cedex, France\relax                                                                          \label{inst:0085}
\and Institute for Astronomy, University of Edinburgh, Royal Observatory, Blackford Hill, Edinburgh EH9 3HJ, United Kingdom\relax                                                                            \label{inst:0086}
\and Instituut voor Sterrenkunde, KU Leuven, Celestijnenlaan 200D, 3001 Leuven, Belgium\relax                                                                                                                \label{inst:0087}
\and Institut d'Astrophysique et de G\'{e}ophysique, Universit\'{e} de Li\`{e}ge, 19c, All\'{e}e du 6 Ao\^{u}t, B-4000 Li\`{e}ge, Belgium\relax                                                              \label{inst:0088}
\and ATG Europe for ESA/ESAC, Camino bajo del Castillo, s/n, Urbanizacion Villafranca del Castillo, Villanueva de la Ca\~{n}ada, 28692 Madrid, Spain\relax                                                   \label{inst:0091}
\and \'{A}rea de Lenguajes y Sistemas Inform\'{a}ticos, Universidad Pablo de Olavide, Ctra. de Utrera, km 1. 41013, Sevilla, Spain\relax                                                                     \label{inst:0095}
\and ETSE Telecomunicaci\'{o}n, Universidade de Vigo, Campus Lagoas-Marcosende, 36310 Vigo, Galicia, Spain\relax                                                                                             \label{inst:0097}
\and Large Synoptic Survey Telescope, 950 N. Cherry Avenue, Tucson, AZ 85719, USA\relax                                                                                                                      \label{inst:0102}
\and Observatoire Astronomique de Strasbourg, Universit\'{e} de Strasbourg, CNRS, UMR 7550, 11 rue de l'Universit\'{e}, 67000 Strasbourg, France\relax                                                       \label{inst:0103}
\and Kavli Institute for Cosmology, University of Cambridge, Madingley Road, Cambride CB3 0HA, United Kingdom\relax                                                                                          \label{inst:0106}
\and Aurora Technology for ESA/ESAC, Camino bajo del Castillo, s/n, Urbanizacion Villafranca del Castillo, Villanueva de la Ca\~{n}ada, 28692 Madrid, Spain\relax                                            \label{inst:0110}
\and Laboratoire Univers et Particules de Montpellier, Universit\'{e} Montpellier, Place Eug\`{e}ne Bataillon, CC72, 34095 Montpellier Cedex 05, France\relax                                                \label{inst:0111}
\and Department of Physics and Astronomy, Division of Astronomy and Space Physics, Uppsala University, Box 516, 75120 Uppsala, Sweden\relax                                                                  \label{inst:0114}
\and CENTRA, Universidade de Lisboa, FCUL, Campo Grande, Edif. C8, 1749-016 Lisboa, Portugal\relax                                                                                                           \label{inst:0115}
\and Universit\`{a} di Catania, Dipartimento di Fisica e Astronomia, Sezione Astrofisica, Via S. Sofia 78, 95123 Catania, Italy\relax                                                                        \label{inst:0116}
\and INAF - Osservatorio Astrofisico di Catania, via S. Sofia 78, 95123 Catania, Italy\relax                                                                                                                 \label{inst:0117}
\and University of Vienna, Department of Astrophysics, T\"{ u}rkenschanzstra{\ss}e 17, A1180 Vienna, Austria\relax                                                                                           \label{inst:0118}
\and CITIC – Department of Computer Science, University of A Coru\~{n}a, Campus de Elvi\~{n}a S/N, 15071, A Coru\~{n}a, Spain\relax                                                                        \label{inst:0120}
\and CITIC – Astronomy and Astrophysics, University of A Coru\~{n}a, Campus de Elvi\~{n}a S/N, 15071, A Coru\~{n}a, Spain\relax                                                                            \label{inst:0121}
\and INAF - Osservatorio Astronomico di Roma, Via di Frascati 33, 00078 Monte Porzio Catone (Roma), Italy\relax                                                                                              \label{inst:0122}
\and University of Helsinki, Department of Physics, P.O. Box 64, 00014 Helsinki, Finland\relax                                                                                                               \label{inst:0127}
\and Finnish Geospatial Research Institute FGI, Geodeetinrinne 2, 02430 Masala, Finland\relax                                                                                                                \label{inst:0128}
\and Isdefe for ESA/ESAC, Camino bajo del Castillo, s/n, Urbanizacion Villafranca del Castillo, Villanueva de la Ca\~{n}ada, 28692 Madrid, Spain\relax                                                       \label{inst:0129}
\and STFC, Rutherford Appleton Laboratory, Harwell, Didcot, OX11 0QX, United Kingdom\relax                                                                                                                   \label{inst:0133}
\and Dpto. de Inteligencia Artificial, UNED, c/ Juan del Rosal 16, 28040 Madrid, Spain\relax                                                                                                                 \label{inst:0135}
\and Elecnor Deimos Space for ESA/ESAC, Camino bajo del Castillo, s/n, Urbanizacion Villafranca del Castillo, Villanueva de la Ca\~{n}ada, 28692 Madrid, Spain\relax                                         \label{inst:0143}
\and Thales Services for CNES Centre Spatial de Toulouse, 18 avenue Edouard Belin, 31401 Toulouse Cedex 9, France\relax                                                                                      \label{inst:0144}
\and Department of Astrophysics/IMAPP, Radboud University, P.O.Box 9010, 6500 GL Nijmegen, The Netherlands\relax                                                                                             \label{inst:0146}
\and European Southern Observatory, Karl-Schwarzschild-Str. 2, 85748 Garching, Germany\relax                                                                                                                 \label{inst:0153}
\and ON/MCTI-BR, Rua Gal. Jos\'{e} Cristino 77, Rio de Janeiro, CEP 20921-400, RJ,  Brazil\relax                                                                                                             \label{inst:0155}
\and OV/UFRJ-BR, Ladeira Pedro Ant\^{o}nio 43, Rio de Janeiro, CEP 20080-090, RJ, Brazil\relax                                                                                                               \label{inst:0156}
\and Department of Terrestrial Magnetism, Carnegie Institution for Science, 5241 Broad Branch Road, NW, Washington, DC 20015-1305, USA\relax                                                                 \label{inst:0162}
\and Universit\`{a} di Torino, Dipartimento di Fisica, via Pietro Giuria 1, 10125 Torino, Italy\relax                                                                                                        \label{inst:0169}
\and Departamento de Astrof\'{i}sica, Centro de Astrobiolog\'{i}a (CSIC-INTA), ESA-ESAC. Camino Bajo del Castillo s/n. 28692 Villanueva de la Ca\~{n}ada, Madrid, Spain\relax                                \label{inst:0173}
\and Leicester Institute of Space and Earth Observation and Department of Physics and Astronomy, University of Leicester, University Road, Leicester LE1 7RH, United Kingdom\relax                           \label{inst:0175}
\and Departamento de Estad\'{i}stica, Universidad de C\'{a}diz, Calle Rep\'{u}blica \'{A}rabe Saharawi s/n. 11510, Puerto Real, C\'{a}diz, Spain\relax                                                       \label{inst:0180}
\and Astronomical Institute Bern University, Sidlerstrasse 5, 3012 Bern, Switzerland (present address)\relax                                                                                                 \label{inst:0183}
\and EURIX S.r.l., Corso Vittorio Emanuele II 61, 10128, Torino, Italy\relax                                                                                                                                 \label{inst:0184}
\and Harvard-Smithsonian Center for Astrophysics, 60 Garden Street, Cambridge MA 02138, USA\relax                                                                                                            \label{inst:0188}
\and HE Space Operations BV for ESA/ESAC, Camino bajo del Castillo, s/n, Urbanizacion Villafranca del Castillo, Villanueva de la Ca\~{n}ada, 28692 Madrid, Spain\relax                                       \label{inst:0191}
\and SISSA - Scuola Internazionale Superiore di Studi Avanzati, via Bonomea 265, 34136 Trieste, Italy\relax                                                                                                  \label{inst:0198}
\and University of Turin, Department of Computer Sciences, Corso Svizzera 185, 10149 Torino, Italy\relax                                                                                                     \label{inst:0208}
\and SRON, Netherlands Institute for Space Research, Sorbonnelaan 2, 3584CA, Utrecht, The Netherlands\relax                                                                                                  \label{inst:0209}
\and Dpto. de Matem\'{a}tica Aplicada y Ciencias de la Computaci\'{o}n, Univ. de Cantabria, ETS Ingenieros de Caminos, Canales y Puertos, Avda. de los Castros s/n, 39005 Santander, Spain\relax             \label{inst:0213}
\and Unidad de Astronom\'ia, Universidad de Antofagasta, Avenida Angamos 601, Antofagasta 1270300, Chile\relax                                                                                               \label{inst:0220}
\and CRAAG - Centre de Recherche en Astronomie, Astrophysique et G\'{e}ophysique, Route de l'Observatoire Bp 63 Bouzareah 16340 Algiers, Algeria\relax                                                       \label{inst:0231}
\and University of Antwerp, Onderzoeksgroep Toegepaste Wiskunde, Middelheimlaan 1, 2020 Antwerp, Belgium\relax                                                                                               \label{inst:0234}
\and INAF - Osservatorio Astronomico d'Abruzzo, Via Mentore Maggini, 64100 Teramo, Italy\relax                                                                                                               \label{inst:0236}
\and INAF - Osservatorio Astronomico di Capodimonte, Via Moiariello 16, 80131, Napoli, Italy\relax                                                                                                           \label{inst:0238}
\and Instituto de Astronomia, Geof\`{i}sica e Ci\^{e}ncias Atmosf\'{e}ricas, Universidade de S\~{a}o Paulo, Rua do Mat\~{a}o, 1226, Cidade Universitaria, 05508-900 S\~{a}o Paulo, SP, Brazil\relax          \label{inst:0239}
\and Department of Astrophysics, Astronomy and Mechanics, National and Kapodistrian University of Athens, Panepistimiopolis, Zografos, 15783 Athens, Greece\relax                                            \label{inst:0251}
\and Leibniz Institute for Astrophysics Potsdam (AIP), An der Sternwarte 16, 14482 Potsdam, Germany\relax                                                                                                    \label{inst:0254}
\and RHEA for ESA/ESAC, Camino bajo del Castillo, s/n, Urbanizacion Villafranca del Castillo, Villanueva de la Ca\~{n}ada, 28692 Madrid, Spain\relax                                                         \label{inst:0256}
\and ATOS for CNES Centre Spatial de Toulouse, 18 avenue Edouard Belin, 31401 Toulouse Cedex 9, France\relax                                                                                                 \label{inst:0258}
\and School of Physics and Astronomy, Tel Aviv University, Tel Aviv 6997801, Israel\relax                                                                                                                    \label{inst:0261}
\and UNINOVA - CTS, Campus FCT-UNL, Monte da Caparica, 2829-516 Caparica, Portugal\relax                                                                                                                     \label{inst:0262}
\and School of Physics, O'Brien Centre for Science North, University College Dublin, Belfield, Dublin 4, Ireland\relax                                                                                       \label{inst:0273}
\and Dipartimento di Fisica e Astronomia, Universit\`{a} di Bologna, Via Piero Gobetti 93/2, 40129 Bologna, Italy\relax                                                                                      \label{inst:0279}
\and Barcelona Supercomputing Center - Centro Nacional de Supercomputaci\'{o}n, c/ Jordi Girona 29, Ed. Nexus II, 08034 Barcelona, Spain\relax                                                               \label{inst:0290}
\and Department of Computer Science, Electrical and Space Engineering, Lule\aa{} University of Technology, Box 848, S-981 28 Kiruna, Sweden\relax                                                            \label{inst:0296}
\and Max Planck Institute for Extraterrestrial Physics, High Energy Group, Gie{\ss}enbachstra{\ss}e, 85741 Garching, Germany\relax                                                                           \label{inst:0298}
\and Astronomical Observatory Institute, Faculty of Physics, Adam Mickiewicz University, S{\l}oneczna 36, 60-286 Pozna{\'n}, Poland\relax                                                                    \label{inst:0315}
\and Konkoly Observatory, Research Centre for Astronomy and Earth Sciences, Hungarian Academy of Sciences, Konkoly Thege Mikl\'{o}s \'{u}t 15-17, 1121 Budapest, Hungary\relax                               \label{inst:0321}
\and E\"{ o}tv\"{ o}s Lor\'and University, Egyetem t\'{e}r 1-3, H-1053 Budapest, Hungary\relax                                                                                                               \label{inst:0322}
\and American Community Schools of Athens, 129 Aghias Paraskevis Ave. \& Kazantzaki Street, Halandri, 15234 Athens, Greece\relax                                                                             \label{inst:0325}
\and Faculty of Mathematics and Physics, University of Ljubljana, Jadranska ulica 19, 1000 Ljubljana, Slovenia\relax                                                                                         \label{inst:0328}
\and Villanova University, Department of Astrophysics and Planetary Science, 800 E Lancaster Avenue, Villanova PA 19085, USA\relax                                                                           \label{inst:0329}
\and Physics Department, University of Antwerp, Groenenborgerlaan 171, 2020 Antwerp, Belgium\relax                                                                                                           \label{inst:0331}
\and McWilliams Center for Cosmology, Department of Physics, Carnegie Mellon University, 5000 Forbes Avenue, Pittsburgh, PA 15213, USA\relax                                                                 \label{inst:0337}
\and Astronomical Institute, Academy of Sciences of the Czech Republic, Fri\v{c}ova 298, 25165 Ond\v{r}ejov, Czech Republic\relax                                                                            \label{inst:0341}
\and Telespazio for CNES Centre Spatial de Toulouse, 18 avenue Edouard Belin, 31401 Toulouse Cedex 9, France\relax                                                                                           \label{inst:0346}
\and Institut de Physique de Rennes, Universit{\'e} de Rennes 1, 35042 Rennes, France\relax                                                                                                                  \label{inst:0349}
\and Shanghai Astronomical Observatory, Chinese Academy of Sciences, 80 Nandan Rd, 200030 Shanghai, China\relax                                                                                              \label{inst:0356}
\and School of Astronomy and Space Science, University of Chinese Academy of Sciences, Beijing 100049, China\relax                                                                                           \label{inst:0357}
\and Niels Bohr Institute, University of Copenhagen, Juliane Maries Vej 30, 2100 Copenhagen {\O}, Denmark\relax                                                                                              \label{inst:0359}
\and DXC Technology, Retortvej 8, 2500 Valby, Denmark\relax                                                                                                                                                  \label{inst:0360}
\and Las Cumbres Observatory, 6740 Cortona Drive Suite 102, Goleta, CA 93117, USA\relax                                                                                                                      \label{inst:0361}
\and Astrophysics Research Institute, Liverpool John Moores University, 146 Brownlow Hill, Liverpool L3 5RF, United Kingdom\relax                                                                            \label{inst:0369}
\and Baja Observatory of University of Szeged, Szegedi \'{u}t III/70, 6500 Baja, Hungary\relax                                                                                                               \label{inst:0374}
\and Laboratoire AIM, IRFU/Service d'Astrophysique - CEA/DSM - CNRS - Universit\'{e} Paris Diderot, B\^{a}t 709, CEA-Saclay, 91191 Gif-sur-Yvette Cedex, France\relax                                        \label{inst:0375}
\and Warsaw University Observatory, Al. Ujazdowskie 4, 00-478 Warszawa, Poland\relax                                                                                                                         \label{inst:0411}
\and Institute of Theoretical Physics, Faculty of Mathematics and Physics, Charles University in Prague, Czech Republic\relax                                                                                \label{inst:0412}
\and AKKA for CNES Centre Spatial de Toulouse, 18 avenue Edouard Belin, 31401 Toulouse Cedex 9, France\relax                                                                                                 \label{inst:0419}
\and Vitrociset Belgium for ESA/ESAC, Camino bajo del Castillo, s/n, Urbanizacion Villafranca del Castillo, Villanueva de la Ca\~{n}ada, 28692 Madrid, Spain\relax                                           \label{inst:0425}
\and HE Space Operations BV for ESA/ESTEC, Keplerlaan 1, 2201AZ, Noordwijk, The Netherlands\relax                                                                                                            \label{inst:0432}
\and Space Telescope Science Institute, 3700 San Martin Drive, Baltimore, MD 21218, USA\relax                                                                                                                \label{inst:0441}
\and QUASAR Science Resources for ESA/ESAC, Camino bajo del Castillo, s/n, Urbanizacion Villafranca del Castillo, Villanueva de la Ca\~{n}ada, 28692 Madrid, Spain\relax                                     \label{inst:0442}
\and Fork Research, Rua do Cruzado Osberno, Lt. 1, 9 esq., Lisboa, Portugal\relax                                                                                                                            \label{inst:0447}
\and APAVE SUDEUROPE SAS for CNES Centre Spatial de Toulouse, 18 avenue Edouard Belin, 31401 Toulouse Cedex 9, France\relax                                                                                  \label{inst:0450}
\and Nordic Optical Telescope, Rambla Jos\'{e} Ana Fern\'{a}ndez P\'{e}rez 7, 38711 Bre\~{n}a Baja, Spain\relax                                                                                              \label{inst:0455}
\and Spanish Virtual Observatory\relax                                                                                                                                                                       \label{inst:0460}
\and Fundaci\'{o}n Galileo Galilei - INAF, Rambla Jos\'{e} Ana Fern\'{a}ndez P\'{e}rez 7, 38712 Bre\~{n}a Baja, Santa Cruz de Tenerife, Spain\relax                                                          \label{inst:0472}
\and INSA for ESA/ESAC, Camino bajo del Castillo, s/n, Urbanizacion Villafranca del Castillo, Villanueva de la Ca\~{n}ada, 28692 Madrid, Spain\relax                                                         \label{inst:0477}
\and Dpto. Arquitectura de Computadores y Autom\'{a}tica, Facultad de Inform\'{a}tica, Universidad Complutense de Madrid, C/ Prof. Jos\'{e} Garc\'{i}a Santesmases s/n, 28040 Madrid, Spain\relax            \label{inst:0478}
\and H H Wills Physics Laboratory, University of Bristol, Tyndall Avenue, Bristol BS8 1TL, United Kingdom\relax                                                                                              \label{inst:0481}
\and Institut d'Estudis Espacials de Catalunya (IEEC), Gran Capita 2-4, 08034 Barcelona, Spain\relax                                                                                                         \label{inst:0487}
\and Applied Physics Department, Universidade de Vigo, 36310 Vigo, Spain\relax                                                                                                                               \label{inst:0489}
\and Stellar Astrophysics Centre, Aarhus University, Department of Physics and Astronomy, 120 Ny Munkegade, Building 1520, DK-8000 Aarhus C, Denmark\relax                                                   \label{inst:0503}
\and Argelander-Institut f\"{ ur} Astronomie, Universit\"{ a}t Bonn,  Auf dem H\"{ u}gel 71, 53121 Bonn, Germany\relax                                                                                       \label{inst:0509}
\and Research School of Astronomy and Astrophysics, Australian National University, Canberra, ACT 2611 Australia\relax                                                                                       \label{inst:0516}
\and Sorbonne Universit\'{e}s, UPMC Univ. Paris 6 et CNRS, UMR 7095, Institut d'Astrophysique de Paris, 98 bis bd. Arago, 75014 Paris, France\relax                                                          \label{inst:0518}
\and Department of Geosciences, Tel Aviv University, Tel Aviv 6997801, Israel\relax                                                                                                                          \label{inst:0520}
}

  \date{}


  \abstract
  {}
  {The goal of this paper is to demonstrate the outstanding quality of the second data release of the \Gaia mission and its power for constraining many different aspects of the dynamics of the satellites of the Milky Way. We focus here on determining the proper motions of 75 Galactic globular clusters, nine dwarf spheroidal galaxies, one ultra-faint system, and the Large and Small Magellanic Clouds.}
  {Using data extracted from the \Gaia archive, we derived the proper
    motions and parallaxes for these systems, as well as their
    uncertainties. We demonstrate that the errors, statistical and
    systematic, are relatively well understood. We integrated the
    orbits of these objects in three different Galactic potentials, and
    characterised their properties. We present the derived proper motions, space velocities, and characteristic orbital parameters in
    various tables to facilitate their use by the astronomical
    community.}
  {Our limited and straightforward analyses have allowed us for
    example to $(i)$ determine absolute and very precise proper
    motions for globular clusters; $(ii)$ detect clear rotation
    signatures in the proper motions of at least five globular clusters;
    $(iii)$ show that the satellites of the Milky Way are all on high-inclination orbits, but that they do not share a single plane of
    motion; $(iv)$ derive a lower limit for the mass of the Milky Way
    of $9.1^{+6.2}_{-2.6} \times 10^{11}\,\sm$ based on the assumption
    that the Leo~I dwarf spheroidal is bound; $(v)$ derive a rotation curve for the Large Magellanic Cloud
    based solely on proper motions that is competitive with
    line-of-sight velocity curves, now using many orders of magnitude
    more sources; and $(vi)$ unveil the dynamical effect of the bar on
    the motions of stars in the Large Magellanic Cloud.}
   {All these results
    highlight the incredible power of the \Gaia astrometric mission, and in particular of its second data release.}


   \maketitle
\authorrunning{{\gaia} Collaboration}

%

\section{Introduction}

The possibility of determining for the first time the absolute proper
motions of stars in the satellites of the Milky Way opens up a whole
new window for understanding their dynamics, origin, and evolution, as
well as that of the Milky Way itself. The data presented in the Second
\Gaia Data Release \citep[hereafter DR2,][]{DR2-DPACP-36} allows us to
achieve this goal. In this paper we study the proper motions (PM
hereafter) of stars in a large sample of globular clusters, in the
classical dwarf spheroidal galaxies and one ultra-faint system, and in
the Large and Small Magellanic Clouds (LMC and SMC hereafter).

A plethora of interesting science questions can be
addressed with this dataset. In this Introduction, we do not aim to be
fully comprehensive, but we mention a few topics to set the
context, to highlight the power of the unprecedentedly accurate absolute
PM measurements, and also to fan curiosity in the community for exploring
this outstanding dataset themselves.

Proper motion studies of satellite systems, such as the globular
clusters and dwarf galaxies of the Milky Way, have a long history,
starting from the use of photographic plates that were sometimes taken with a
time baseline longer than 100 years (see \citet{1997A&ARv...8....1M}
and \citet{2000A&A...360..472V} for interesting and thorough
historical reviews on the determination of PM of stars in globular
clusters). More recently, the space missions \textsc{Hipparcos} and the Hubble
Space Telescope (HST), and of course the \Gaia mission in its first
data release \citep{2016A&A...595A...2G}, have demonstrated the
enormous power of space-based astrometry. \textsc{Hipparcos} data
\citep{1997A&A...323L..49P} have been used for many purposes, and in
particular, for studying the dynamics of nearby open clusters
\citep[e.g.][]{1999A&A...341L..71V,2009A&A...497..209V}, and although
\textsc{Hipparcos} did not observe stars in globular clusters, it provided
an absolute reference frame that was used to derive the
orbits of 15 globular clusters from photographic plates, for
example \citep[][]{1997NewA....2..477O}. On the other hand, the
HST has carried
out several large (legacy) surveys \citep[e.g.][]{2017AJ....153...19S}
that have allowed studies of the dynamics of globular clusters and of
the Milky Way satellites, and it has even constrained the motions of our
largest neighbouring galaxy M31 \citep{2012ApJ...753....7S}. In all
these cases, relative astrometry is done using background quasars and
distant galaxies to define a reference frame, and typically, a time
baseline of 5 -- 10 years is used. This has been a highly successful
approach, and has, for example, allowed researchers to develop the
idea that the Magellanic Clouds may be on their first infall
\citep{2006ApJ...638..772K,2007ApJ...668..949B}, to place constraints on
the mass of the Milky Way from its most distant satellite Leo~I
\citep{2013ApJ...768..140B}, and also to argue in support of the
conjecture that dwarf galaxy satellites may lie on a vast polar plane
based on the first constraints on their orbits
\citep{2013MNRAS.435.2116P}.

This brief overview gives a flavour of the palette of scientific
results that can be derived from accurate PM information of the
satellites of the Milky Way. In combination with knowledge of the
line-of-sight velocities, PM can be used to derive orbits for these
systems. This is interesting for very many reasons, some of which
we highlight below.

The orbits of globular clusters can shed light on their formation and
evolution, for example, which may have formed
in situ and which could be accreted
\citep{1978ApJ...225..357S,2004MNRAS.355..504M,2017MNRAS.465.3622R}. Furthermore,
knowledge of the orbits helps understanding the effect of tides and
the interplay with internal processes, such as evaporation, mass
segregation, and two-body relaxation
\citep{1994AJ....108.1292D,2003MNRAS.340..227B}.  Based on the orbits it
is also possible to aid the search for extra-tidal stars and
streamers, which are very useful for constraining the gravitational
potential of the Milky Way because of the coldness of such streams
\citep{2015ApJ...803...80K}.

In the case of the dwarf galaxy satellites of the Milky Way, knowledge
of the orbits also has multiple implications that range from the scale
of the formation of the smallest galaxies in the Universe to
constraints and challenges to the cosmological model. By determining
the orbits of dwarf galaxies, we can establish the effect of the
environment on their evolution, including star formation and chemical
enrichment histories \citep{2009ARA&A..47..371T}, and also 
the effect of ram pressure stripping, and we can place constraints on the hot
gaseous halo of the Milky Way \citep{2011ApJ...732...17N}.  The
structure of these small galaxies may also have been strongly affected
by tidal interactions with the Milky Way, and to quantify the
importance of this process, knowledge of the orbits is imperative
\citep{2011ApJ...726...98K}. Furthermore, such knowledge also allows
to establish whether there is internal rotation and its amplitude
\citep{2008ApJ...681L..13B}, which is relevant for understanding the
formation path of dwarf spheroidal (dSph) galaxies. For the ultra-faint galaxies, whose
nature is debated, PM are also useful to identify
interlopers, which is particularly important for establishing whether these
systems are (on the verge of being) disrupted or embedded in a dark
matter halo.

The orbits of the Milky Way satellites (both globular clusters and dwarf
galaxies) also provide information on the Milky Way
itself, such as its dynamical mass
\citep[e.g.][]{1999MNRAS.310..645W}.  It is likely that the internal
dynamics of the Milky Way have also been affected by the gravitational
influence of, in particular, the Sagittarius dwarf
\citep{2013MNRAS.429..159G} and the LMC
\citep{2012MNRAS.422.1957B,2015ApJ...802..128G}, and improved
knowledge of the orbits of these objects will allow us to understand
what their effect has been.  On the other hand, orbits also allow us
to gain insight into how a galaxy acquires its satellite
population. For example, it has been argued that the satellites lie
preferentially on streams \citep{1995MNRAS.275..429L}, on a thin plane
\citep{2005A&A...431..517K}, or that they have fallen in groups
\citep{2008MNRAS.385.1365L}, of which the LMC/SMC and their recently
discovered satellites are direct proof
\citep{2015ApJ...807...50B,2015ApJ...805..130K}. The \Gaia DR2 data
will allow us to establish how real and important these associations
are, and also whether the orbits found are consistent with the
expectations from the concordance cosmological model.

In this paper we analyse 75 globular clusters in our Galaxy, and
we demonstrate that the \Gaia DR2 PM measurements for
these clusters are of outstanding quality, with the formal and
systematic uncertainties being effectively negligible. In comparison
to previous efforts
\citep[e.g.][]{2003AJ....125.1373D,2007AJ....134..195C,2010AJ....140.1282C,2013AJ....146...33C},
the errors are reduced by nearly two orders of magnitude. This dramatic improvement will also enable detailed studies of
the internal dynamics that could shed light onto how these objects
formed and their evolutionary path \citep{2012A&ARv..20...50G}. Some
of the questions that might be addressed include whether globular clusters have formed in mini-halos or are fully
devoid of dark matter \citep{2013MNRAS.428.3648I}. Do they host
intermediate mass black holes \citep{2017MNRAS.464.2174B}?  Are there
dynamical differences between the different populations known to be
present in many globular clusters
\citep{2012A&A...538A..18B,2015ApJ...810L..13B,2013MNRAS.429.1913V}?
Has the formation process and evolution for in situ clusters been the same
as for those that have been accreted? Have these processes left an
imprint on the internal phase-space distribution of their stars? How
many clusters show rotation, and what is the link to how they have
formed \citep{2015MNRAS.450.1164H}? Many of the globular clusters are also 
being targeted by radial velocity surveys
\citep[e.g.][]{2015A&A...573A.115L,2017arXiv171007257K}, and the
combination of \Gaia DR2 with such datasets will be extremely
powerful.

We also study the Magellanic Clouds, the nine
classical dSph, and include the UFD Bootes~I as an example of what can
be achieved with \Gaia DR2 data.  Even though the dwarf galaxies are on
average farther away, their mean PMs can be very well
determined using \Gaia DR2, and they are still above the systematic
level. Although for many objects, the uncertainties are comparable to
those achievable using the HST, the advantage of having a full view of
these galaxies and of the PMs being in an absolute
reference frame cannot be over-emphasised. For the dSph, establishing their internal dynamics using this dataset is not yet feasible, however, although perhaps the combination of \Gaia and
the HST will allow to make progress before the end of the \Gaia mission
\citep[as recently demonstrated by][]{2017arXiv171108945M}.  For the
Magellanic Clouds, \Gaia DR2 gives a clearer, more detailed view of the
internal dynamics than has ever been possible before, with measured
PMs for millions of sources.

The paper is structured as follows. The main part introduces the DR2
data, methods, and analysis, including orbit integrations, and details are given in the appendix. The appendix
also contains tables with the measured PM for the objects we
studied, as
well as a list of the orbital parameters we derived. More specifically, in
Sec.~\ref{sec:data-method} of the main paper we present the \Gaia DR2
data, with emphasis on the astrometry, the selection procedures, and
the methods. Sec.~\ref{sec:data-method-gcdw} focuses on deriving the proper motions of the globular clusters and
dSph, and in Sec.~\ref{sec:data-method-MCs} we describe the procedures
that are tailored for the LMC and SMC. We then present the various
analyses of the datasets that we have carried out, and which allow us
to show the superb quality of the data. Sec.~\ref{sec:gc} concentrates
on the globular clusters, Sec.~\ref{sec:dw} on the dSph, and
Sec.~\ref{sec:LMCSMC} on the Magellanic Clouds. In
Sec.~\ref{sec:orbits} we determine the orbits of the satellites using
different Galactic potentials, a showcase of the fantastic
possibilities that \Gaia DR2 offers for studies of the dynamics and
origin of the satellites of the Milky Way. In
Sec.~\ref{sec:discussion} we discuss our findings, provide an example
of the use of DR2 astrometry to find tidal debris, present a summary
of what lies beyond a straightforward analysis of the data such as
that presented here, and also what will need to wait for later \Gaia
data releases (i.e. the limitations of the \Gaia DR2
dataset). We present our conclusions in Sec.~\ref{sec:concl}.


\section{Data and methods}
\label{sec:data-method}

The data we used are the second \Gaia data release as
described in \cite{DR2-DPACP-36}. Further details on its validation
may be found in \cite{DR2-DPACP-39}.  The procedures to derive
  the \Gaia astrometric solution (also known as AGIS) are described in
  detail in \citet{2016A&A...595A...4L} and \citet{DR2-DPACP-51}. We recall that the astrometric
  parameters are absolute in the sense that they do not rely on an
  external reference frame.

\subsection{Globular clusters and dwarf galaxies}
\label{sec:data-method-gcdw}
\begin{figure}
\centering
\includegraphics[width=8.5cm]{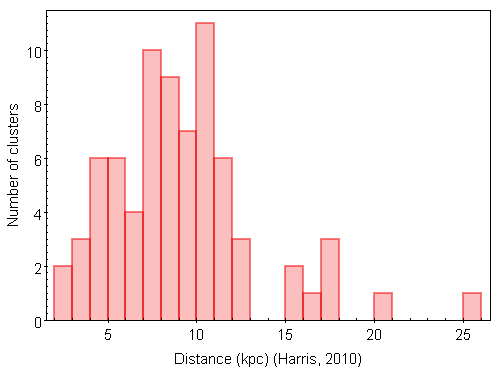}
\includegraphics[width=8.5cm]{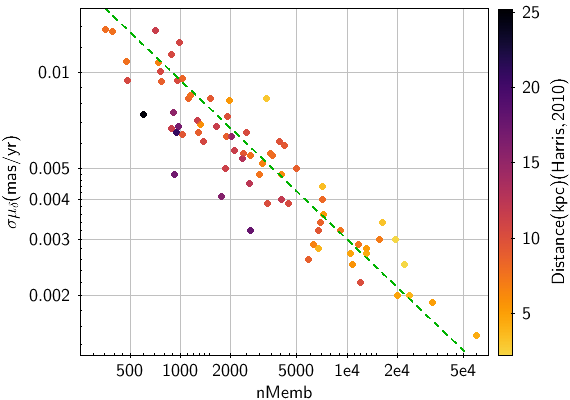}
\caption{{\it Top}: Distance distribution of the 75 globular clusters included in the present study. {\it Bottom}: Standard uncertainties on the PM in declination as a function of number of cluster members $nMemb$ used in the solution. The diagonal line represents a fit to the relation $\sigma\mu_\delta = a/\sqrt{nMemb}$ where we find $a = 0.3$~[mas/yr]. A similar dependence on the number of members is found for the parallax uncertainty (with $a = 0.15$~[mas]) and for $\sigma\mu_{\alpha*}$  (where $a = 0.25$~[mas/yr]).}
\label{fig:globhistdist}
\end{figure}
\begin{figure*}
\centering
\includegraphics[width=18cm,trim={0 1cm 0 2cm},clip]{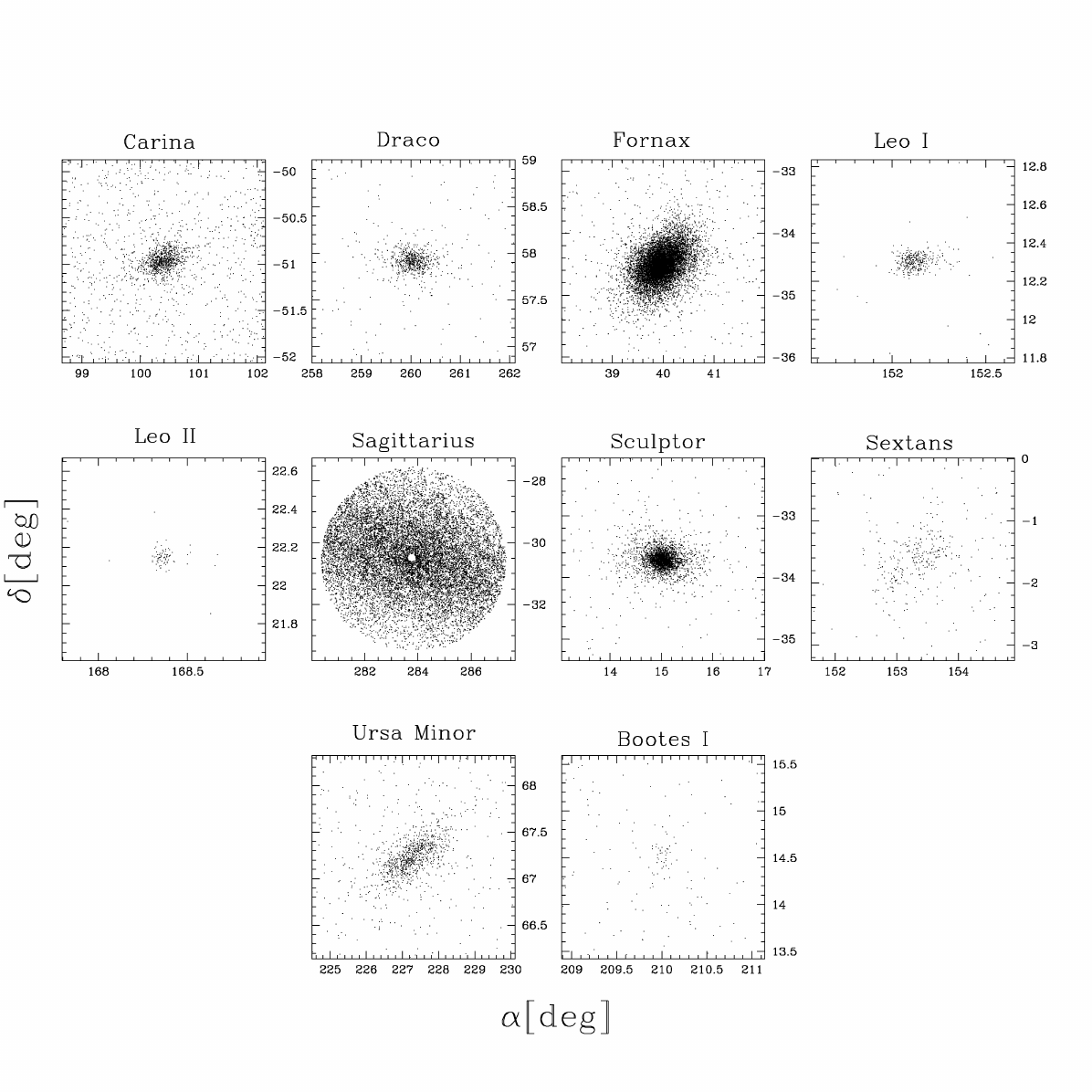}
\caption{Field-of-view towards the dSph galaxies (the nine classical and one UFD)
  in our sample. The stars shown correspond to members according to
  the photometric selection (on the RGB and BHB) and the astrometric procedure (within
  2$\sigma$ from the mean PM of the object). The striping apparent in Sagittarius and Sextans is driven in part by the scanning law. The hole in the centre of Sagittarius corresponds to the location of the globular cluster NGC 6715 (M54).}
\label{fig:fov_dwarfs}
\end{figure*}

The sample of globular clusters analysed in this paper includes half
of the whole population of globular clusters in the Milky Way. We
focus mostly on the clusters that are located within a distance limit of 12
to 13 kpc to achieve a reasonable compromise on the number of stars
with reliable astrometric solutions.  It is important to bear in mind
that the astrometric solutions for stars in areas of high stellar
density, such as the cores of the clusters, are more likely to be
disturbed by image blending and onboard image selection. This
plays a significant role when observing more distant clusters and
affects the fainter stars in particular \citep[see
e.g.][]{2017MNRAS.467..412P}.  Our selection also takes into account
the ability of distinguishing (in PM and parallax space) the cluster
stars from those in the field, both as a function of distance from the
cluster centre and of magnitude. Furthermore, clusters at low galactic
latitude have also generally been avoided to escape confusion with
field stars. The top panel of Figure~\ref{fig:globhistdist} shows a
histogram of the distance distribution for the 75 globular clusters. The bottom panel exemplifies how the
standard uncertainties on the cluster PM in declination
vary as a function of the number of cluster members used\footnote{Note
  the tendency for more distant clusters to show smaller uncertainties
  at a fixed number of members. This is driven by the fact that for
  more distant clusters, only the brighter and less populated part of
  the luminosity function is effectively sampled, and this implies a
  lower crowding impact \citep{2017MNRAS.467..412P}.}.

As Fig.~\ref{fig:fov_dwarfs} shows, we also studied the
classical dSph and one ultra-faint dwarf (UFD) galaxy, Bootes I. UFD
galaxies are intrinsically very faint, as their name indicates, and this implies that
there are very few stars on the red giant branch (RGB), and depending on the distance to
the system, there may be even fewer because of the somewhat bright faint
magnitude limit of \Gaia ($G$ = 21). Bootes~I is the best UFD
case for \Gaia DR2, because its RGB is relatively well populated (at
least in comparison with other UFDs), and it is relatively near
\citep[at 60 kpc,][]{Belok06}. These conditions allow us to apply a homogeneous selection and
analysis procedure to all the dwarfs in our sample, which we find highly desirable at this
point. With external knowledge of radial velocity members, for
instance, it
might be possible to derive the PM for more UFDs, but the
Bootes~I case already illustrates the problems to be faced with \Gaia
DR2 data for this type of system. 
\begin{figure*}
\centering
\includegraphics[width=18cm,trim={0 0.5cm 0 2cm},clip]{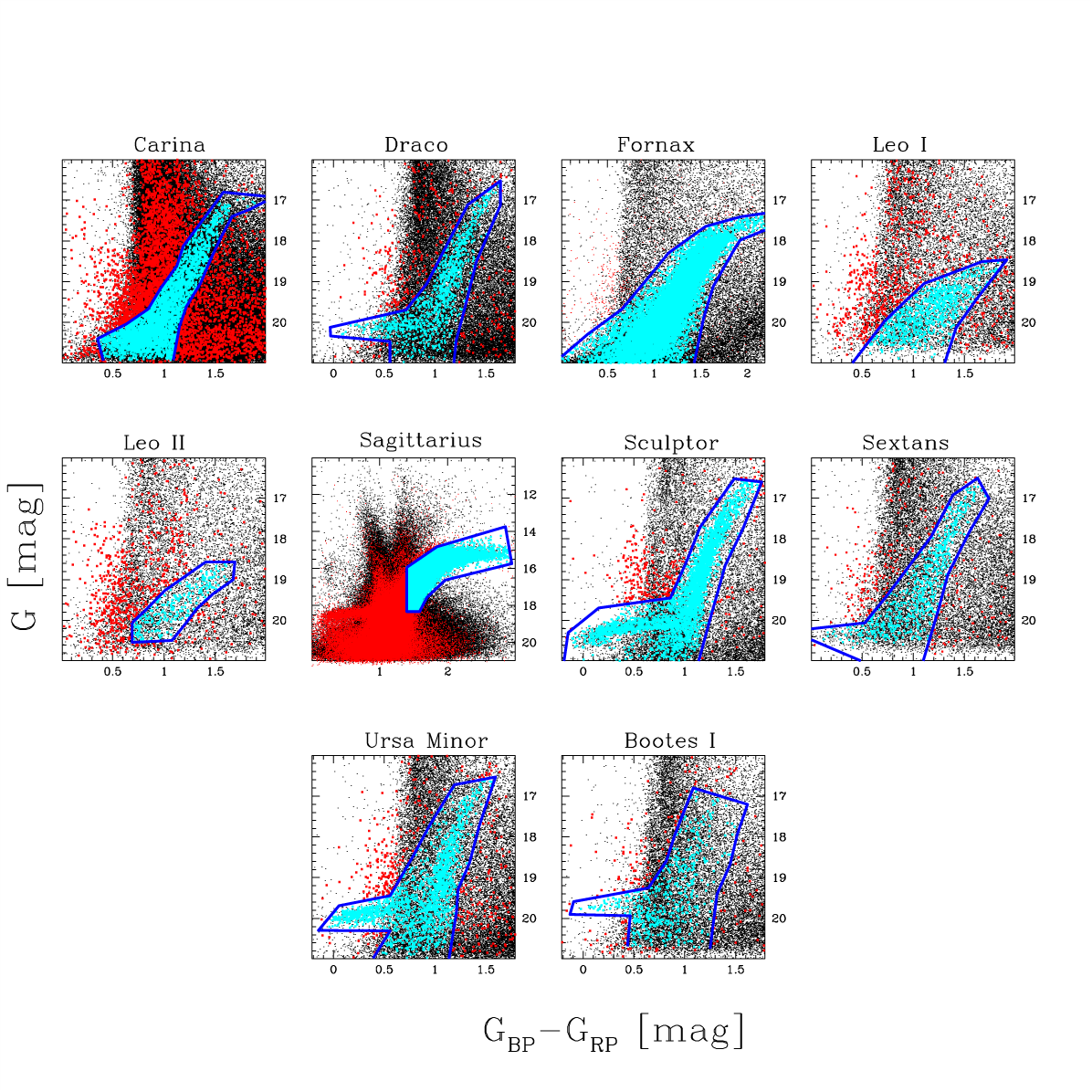}
\caption{Colour-magnitude of the stars in the field of view towards
  the dSph galaxies (the eight classical and one UFD) in our sample. The blue lines mark our (relatively tight)
  pre-selection of tentative members (on the RGB and BHB) that is fed
  to the pipeline to derive mean PMs. The coloured points
  indicate stars within 3$\sigma$ of our determination of the mean PM of the
  object. This means that cyan points satisfy both the PM and CMD selections.}
\label{fig:cmd_dwarfs}
\end{figure*}

The selection procedure, which is described in detail in
Appendix~\ref{app:sequence}, starts with the extraction of data for
each object from the GACS archive.  The archive provides us with the
astrometric parameters, their standard uncertainties and error
correlations, the photometric data with standard uncertainties (flux
values and fluxes converted into magnitudes), various statistics
relating to the astrometric and photometric solutions, and radial
velocities where available for our analysis. Depending on the nature
of the object analysed, we set different magnitude limits. For the
dwarf galaxies, we first considered stars with $G <21$. For the
globular clusters, the limit was generally set at $G = 20$, but in a
few cases, we took a brighter value to limit the contamination by
field stars.  This was necessary for clusters at low
galactic latitude in particular.

The \Gaia sky coverage can locally show strong variations that can affect the
selection of members with good astrometric solutions
\citep{DR2-DPACP-36,DR2-DPACP-39}. In addition, for many of the
globular clusters, the central core is often poorly resolved. These
conditions are reflected in the standard uncertainties of the derived
parameters, but are unlikely to cause a systematic bias in the
results. The most strongly affected cluster $\omega$~Cen (see
Fig.~\ref{fig:maps}) still shows good astrometric data for very many stars.  In the case of the dSph galaxies, the most affected
object is Sextans, as can be seen from Fig.~\ref{fig:fov_dwarfs}. The
inhomogeneous distribution of sources is related to the number of
independent scans in the field of view towards the dwarf. To determine
the astrometric parameters reliably, a sufficiently high number of
truly independent scans is necessary. This is measured by the
parameter {\tt visibility-periods-used,} which has to reach a value
greater than 5 for a five-parameter solution for an object (i.e. including the
PMs and parallax) to be considered reliable
\citep{DR2-DPACP-51}, otherwise, only its position on the sky is
determined. There are other instrumental effects that affect the
astrometric parameters, and these are discussed elsewhere in the
paper and in \cite{DR2-DPACP-51}.

The first step in our procedure to derive the motions of the
satellites is to focus on an area of the sky, centred on the assumed
centre of the object of interest and with an assumed maximum
radius. For the dwarf galaxies, these radii were fixed at 2~deg, except
for the Sagittarius dwarf, for which we took 3~deg (we also excluded stars
within one tidal radius of its nuclear globular cluster M54).  For the
globular clusters, we interactively explored the data using the TopCat
software \citep{2005ASPC..347...29T}, and then made a pre-selection of
members based on the concentration of the PMs, followed by
a cutoff in parallax, as well as on inspection of the colour-magnitude
diagram (for more details, see Appendix \ref{app:sequence}).

Because of their low stellar density contrast and the consequently
higher
number of contaminants (non-member stars) in the field of view, we applied additional selection criteria for the dSph galaxies in order to
obtain a more robust estimate of the mean PMs.  First, we
only considered stars within $1.5\times$ the tidal radius ($r_{t}$) of
each dwarf \citep[taken from][for Bootes]{IH1995,bootes}, except for
Sagittarius, where we considered all the stars in the 3 deg radius
field of view.  Then for all dSph, we also performed a cut in relative
parallax error to remove foreground sources, as nearby stars will have
relatively good parallaxes, especially in comparison to the stars in
the dwarf galaxies. The relative error we used is $0<
\sigma_\varpi/\varpi < 0.5$ (which is equivalent to $\varpi - 2 \sigma_\varpi > 0$), and corresponds to removing stars within
roughly 5~kpc from the Sun.  Finally, we used the distribution of
sources in the colour-magnitude diagram (CMD) to isolate the giant branch (RGB and HB), as shown
in Fig.~\ref{fig:cmd_dwarfs} with the blue lines. In the case of the
Sagittarius dwarf, we used a slightly different selection and
focused on the reddest part of the RGB. The reason for this is the very
large foreground, which overlaps substantially with the bluer portions
of the Sagittarius RGB.

The astrometric solution to derive the PMs and parallaxes
for the globular clusters and the dwarf galaxies follows the
procedures described in \citet[][see also Appendix \ref{app:sequence}]{2009A&A...497..209V,
  2017A&A...601A..19G}.  A joint solution for the PM and
parallax is obtained that takes into account the full error
correlation matrix as evaluated for each contributing star:
\begin{equation}
\mathcal{N} = \mathcal{N}_a + \mathcal{N}_v + \mathcal{N}_d.
\label{equ:noisematrix}
\end{equation}
The three main contributions to the noise matrix in
Eq.~(\ref{equ:noisematrix}) come from the astrometric solution
$\mathcal{N}_a$, the estimated contributions from the internal
velocity dispersion on the PM dispersion $\mathcal{N}_v$,
and the dispersion of the parallaxes from the depth of the cluster
$\mathcal{N}_d$, respectively. For the dwarf galaxies, the second and
third of these contributions could be ignored, as even the
brightest stars in these systems still have standard uncertainties on
the astrometric parameters that are relatively large in
comparison\footnote{We chose to set the intrinsic dispersion to
    the characteristic 10~\kms ~value found for the dwarfs from radial
    velocity data.  However, we have tested different input values and
  found the results on the mean PM to be robust.}. Although
for most globular clusters the velocity dispersion shows a clear
gradient with respect to distance from the cluster centre (see
Fig.~\ref{fig:pmrtdisp}), we did not take it into account. This would
have required a detailed investigation of the actual distribution of
the PMs as a function of radial distance, which is beyond
the scope of the present paper. The internal velocity dispersion as
implemented is an average over the cluster.

The procedure we used to determine the astrometric parameters is
iterative and requires a first guess for the parallax and PMs. For the globular clusters, this first guess was obtained
using the TopCat software \citep{2005ASPC..347...29T}, as described
above. While iterating, several diagnostics are produced, and in
particular, we plot the surface density as a function of distance from
the centre of the cluster. Such a diagram often shows that the maximum
radius initially considered in the data extraction step can be
extended farther out (i.e. the background density has not yet been
reached). In that case, we retrieved more data from the GACS archive
using an increased radius and the latest values found for the PM and parallax. We then repeated the procedure, now with the
starting guesses being those given by the latest astrometric
solution. This process was repeated until it was clear that the maximum
radius had been reached.

\begin{figure}[t]
\centering
\includegraphics[width=8.5cm]{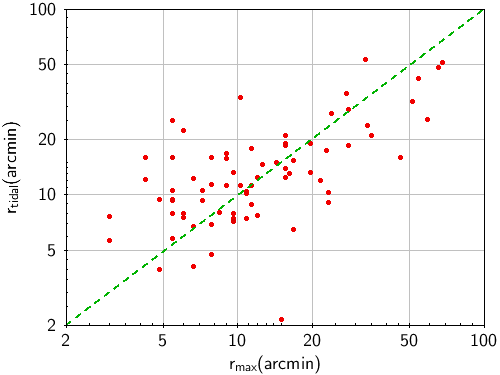}
\caption{Comparison between the tidal radii $r_t$ (according to
  Harris10) of the 75 globular clusters in our sample and the maximum
  radii at which we have been able to detect cluster members in the
  present study. The diagonal line represents the one-to-one
  relation.}.
\label{fig:rtcomp}
\end{figure}

The maximum radius for the cluster, that is, the distance from the centre
within which we still detect cluster stars (3$\sigma$ from the mean
PM, where $\sigma$ is the error on the PM derived using
Eq.~\ref{equ:noisematrix}) was compared to the tidal radii $r_t$
extracted from \cite{1996AJ....112.1487H} and its 2010 update
\citep[hereafter Harris10,][]{Harris10}. Fig.~\ref{fig:rtcomp} shows
that for the majority of the clusters, this maximum radius is between
$1/2$ and $2$ times the published estimate of the tidal
radius. Clusters for which the maximum radius was found to be much
smaller than $r_t$ are often affected by a high-density field star
population, making the detection of cluster members problematic. We note
that $r_t$ has typically been estimated by fitting a King profile
to the projected density distribution of stars, and thus does not
necessarily nor always reflect the true extent of a cluster \citep[see
e.g.][]{2010MNRAS.407.2241K}.

In the case of the globular clusters, the contamination by field stars was
checked through the dispersion diagrams (see Fig.~\ref{fig:parpm} for
two examples), in which the distribution of PM and parallax
was plotted against the standard uncertainties of the
measurements, and compared with the expected distributions that
include all noise contributions. A contaminating source, such as the SMC for 47~Tuc (NGC~104), shows as an offset over-density in one or more of these charts, and in that case was removed by applying a $3\sigma$ filter to the residuals in all three observables, that is,\ relative PMs and parallax. 

We also note that the parallax reference value used for the data
extraction was the \Gaia parallax for the cluster. This can differ from
what is considered the best value for the cluster based on the
distance from the literature (see Sec.~\ref{sec:gc} for more details).

In the case of the dwarf galaxies, the iterative procedures are
similar, except that further iterations with the GACS archive are not
necessary given our choices of initial field sizes. We thus worked only
with the data extracted in the first step, as described earlier in this
section.  We have found, however, that we obtained more reliable mean PM
using only stars brighter than a magnitude limit in the range
$19.1<G<20$. This is the faintest magnitude at which the mean value
of the astrometric parameters becomes stable and where the effects of
contaminating field stars and the large measurement uncertainties of
very faint stars are minimised.

\subsection{Magellanic Clouds}
\label{sec:data-method-MCs}

\begin{figure*}[t]
\centering
\includegraphics[width=8cm]{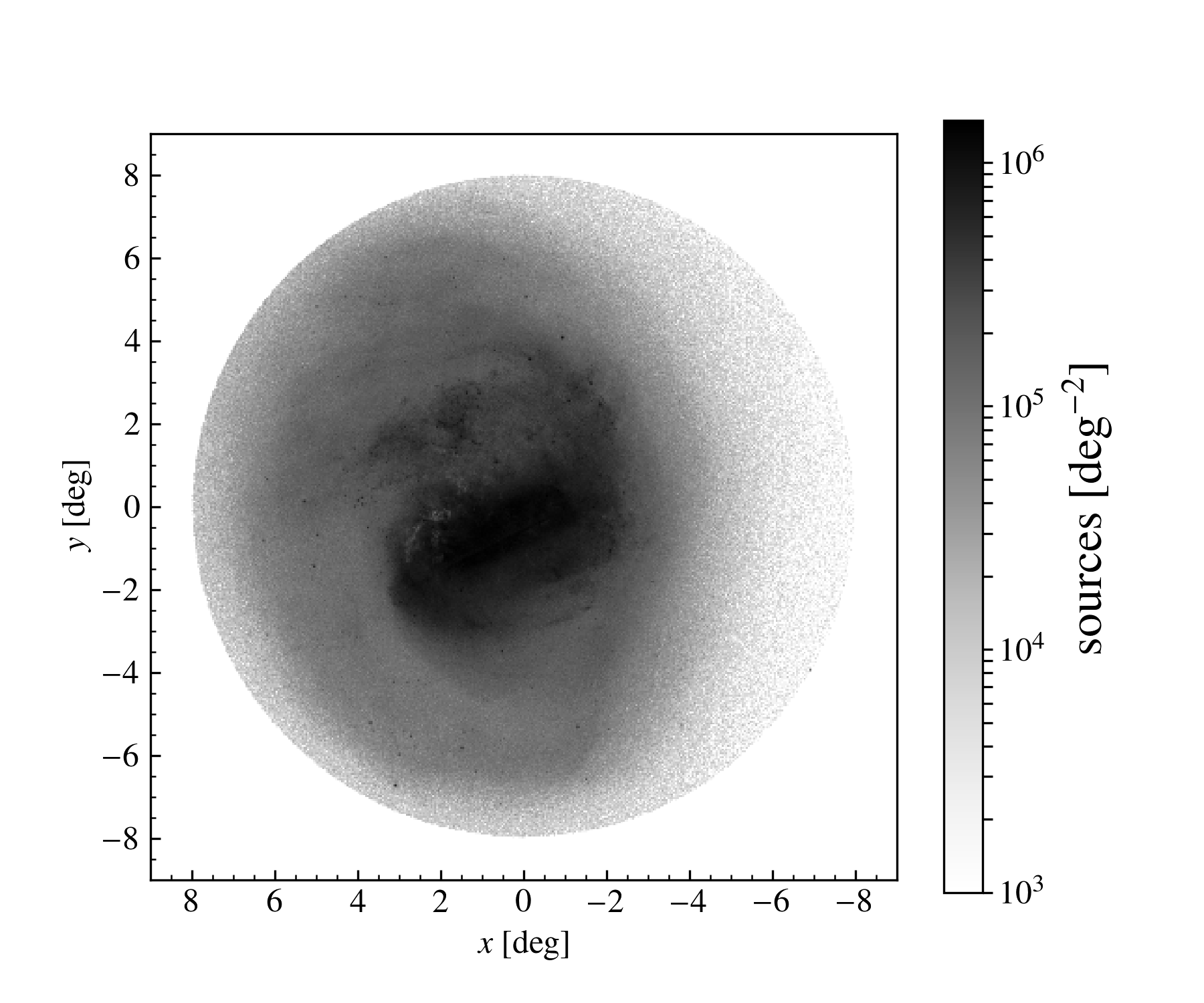}
\includegraphics[width=8cm]{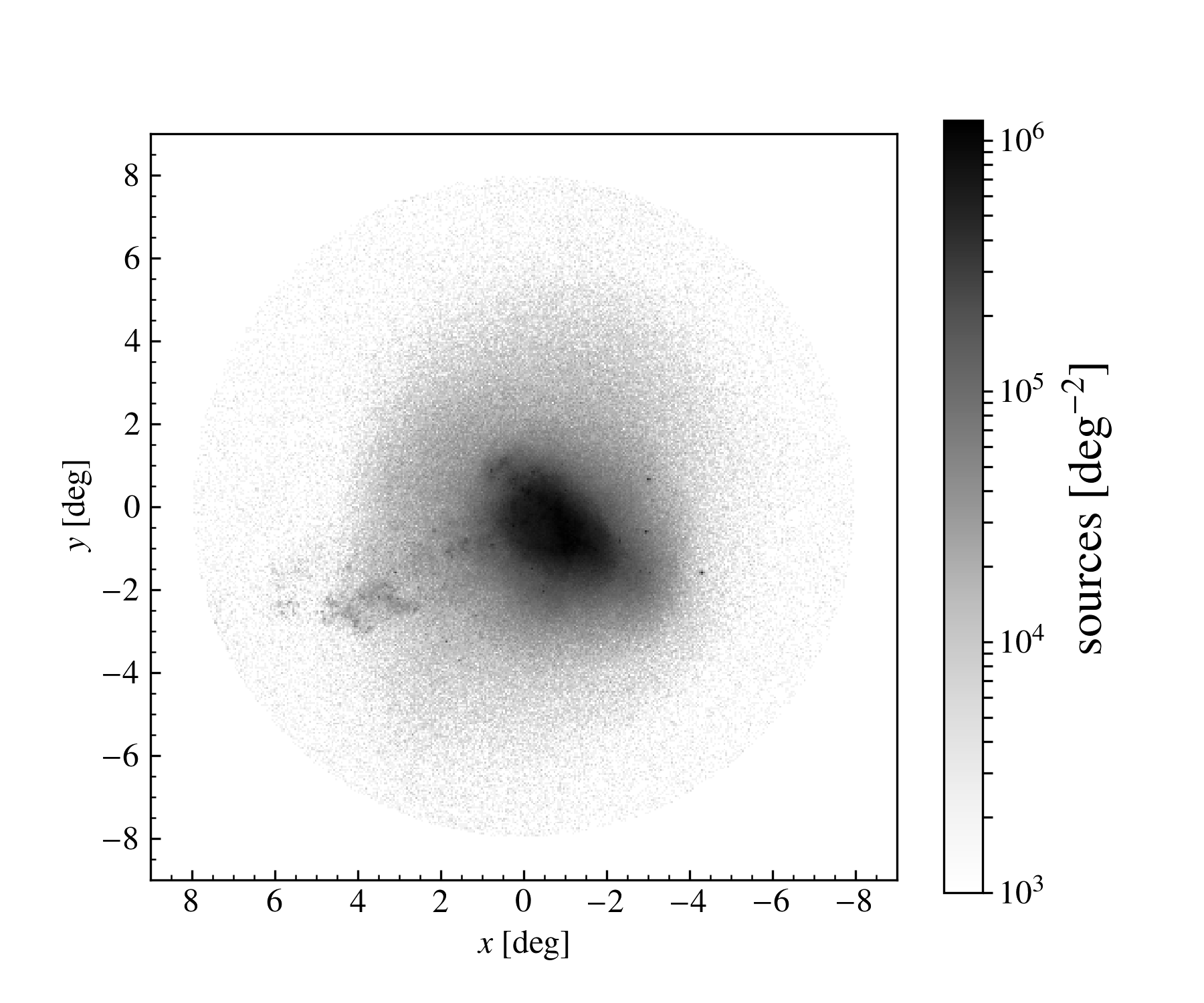}
\caption{Density distribution on the sky of the stars selected as members of the LMC (left) and SMC (right). Positions are shown in the ($x,y$) coordinates described in the text (Eq.~\ref{eq:xy}). In all figures that use this coordinate system, the $x$-axis has been inverted so that it corresponds to the usual inversion of right ascension.}
\label{fig:LMCSMCDensity}
\end{figure*}

The LMC and SMC present a different analytical challenge to the analysis of dwarfs and globular clusters, because they are very extended on the sky and contain two orders of magnitudes more \Gaia sources than any of the dwarfs or clusters analysed.

To simplify our analysis and ensure that the quoted (and plotted) PMs are relatively easy to interpret in terms of internal velocities, it is particularly helpful to define an orthographic projection of the usual celestial coordinates and PMs:
\begin{equation}\label{eq:xy}
\begin{aligned}
x & = \cos\delta\sin(\alpha-\alpha_C)\\
y & = \sin\delta\cos\delta_C-\cos\delta\sin\delta_C\cos(\alpha-\alpha_C) \\
\mu_{x} &= \mu_{\alpha*}\cos(\alpha-\alpha_C)-\mu_\delta\sin\delta\sin(\alpha-\alpha_C)\\
\mu_{y} &= \mu_{\alpha*}\sin\delta_C\sin(\alpha-\alpha_C)\\
& \quad+\mu_\delta\left(\cos\delta\cos\delta_C+\sin\delta\sin\delta_C\cos(\alpha-\alpha_C)\right)
\end{aligned}
.\end{equation}  

The centres of the coordinate systems are chosen to be the dynamical centre of  the H\textsc{i} gas for the LMC and SMC, $(\alpha_{C,{\rm LMC}},\delta_{C,{\rm LMC}}) = (78\fdg 77,-69\fdg 01)$ and  $(\alpha_{C,{\rm SMC}},\delta_{C,{\rm SMC}}) = (16\fdg 26,-72\fdg 42)$ \citep{1992A&A...263...41L,1998ApJ...503..674K,2004ApJ...604..176S}\footnote{Following \cite{2014ApJ...781..121V}, we have taken the LMC centre to be the average of the centres determined by \cite{1998ApJ...503..674K} and \cite{1992A&A...263...41L}.}.  Figure~\ref{fig:LMCSMCDensity} shows the density of stars in this $x,y$-plane for the LMC and SMC, with these centres assumed.

If we approximate each cloud as a thin disc with some bulk motion that rotates about a point with celestial coordinates $(\alpha_C,\delta_C)$ with a constant angular velocity $\omega$ and no other streaming motion, it can be shown (Appendix \ref{app:LMCSMC}) that these coordinates are, to first order, straightforwardly related to the parameters that describe the position and motion of the disc. These approximations are reasonable towards the centre of the LMC, and serve as a first approximation for the SMC. 

It is convenient to define $\vec{n}$ to be the unit vector normal to the disc (such that rotation is positive about $\vec{n}$), with $\vec{z}$ the unit vector from the observer to the reference centre $(\alpha_C,\delta_C)$ at the reference epoch. We then have the mutually orthogonal unit vectors in the plane of the disc $\vec{l}$ =  $\vec{z}\times\vec{n}/|\vec{z}\times\vec{n}|$ and $\vec{m}$ =  $\vec{n}\times\vec{l}$. These have the property that $\vec{l}$ points in the direction of the receding node (the intersection of the disc with the tangent plane of the celestial sphere). 

When we define $v_x$, $v_y$ to be the centre-of-mass motion of the cloud in the $x$ and $y$ directions and $v_z$ to be the same along the line of sight (divided by the distance to the cloud, to put it in the same units) then we have, to first order, 
\begin{equation}\label{eq:xyderivs}
\begin{aligned}
{\partial\mu_{x}}/{\partial x} &\approx av_x - v_z + al_xm_z\omega \\[6pt]
{\partial\mu_{x}}/{\partial y} &\approx  bv_x - n_z\omega + bl_xm_z\omega \\[6pt]
{\partial\mu_{y}}/{\partial x} &\approx  av_y + n_z\omega + al_ym_z\omega \\[6pt]
{\partial\mu_{y}}/{\partial y} &\approx  bv_y - v_z + bl_ym_z\omega
\end{aligned}
\end{equation}
where with inclination $i$ (the angle between the line-of-sight direction to the cloud centre and the rotation axis of the disc, with $i>90\degr$ for retrograde motion)\footnote{`Retrograde' here means negative spin about the line of sight ($\omega_z<0$),
  which means counter-clockwise as seen by the observer. In our notation, the LMC 
  has prograde rotation, that is, positive spin about the line of sight or clockwise as seen
  by the observer. According to some conventions (e.g.\ for binary orbits), this would 
  be regarded as retrograde.}, and 
$\Omega$ the position angle of the receding node, measured from $\vec{y}$ 
towards $\vec{x}$, that is,\ from north towards east, we have the components of $\vec{l},\vec{m}, \text{and }\vec{ n}$ being
\begin{equation}\label{eq:lmn}
\begin{bmatrix} l_x & m_x & n_x \\ l_y & m_y & n_y \\ l_z & m_z & n_z \end{bmatrix} =
\begin{bmatrix} 
\sin\Omega & -\cos i\cos\Omega & \phantom{-}\sin i\cos\Omega \\ 
\cos\Omega & \phantom{-}\cos i\sin\Omega & -\sin i\sin\Omega \\ 
0 & \sin i & \cos i 
\end{bmatrix}
,\end{equation}
and
\begin{equation}\label{e42}
a = \tan i \cos\Omega \, , \quad b = -\tan i\sin\Omega \, .
\end{equation}

This means that simply by finding a linear fit to the PM
as a function of position on the sky, yielding the bulk motion
perpendicular to the line of sight, and four gradients, we have four
equations for four (in principle) free parameters: $v_z$, $i$,
$\Omega,$ and $\omega$. The first, $v_z$, produces a perspective
contraction (or expansion) as the clouds appear to shrink as they move
away from us (or the opposite). The last three describe the
orientation and rotation of the disc, which also leave a signature in
the PMs.

In practice, neither cloud is flat or expected to have perfectly circular streaming motion. The assumption of a constant angular velocity is approximately valid in the central few degrees of the LMC, but this breaks down at larger radii. Nonetheless, these approximations allow us to draw tentative conclusions about the orientation and velocity curve of the Cloud from these gradients
that are simple to measure.

We could take some of the four `free' parameters from other studies, but in practice, we only ever did this for $v_z$. For the LMC, we took the line-of-sight velocity from \citet[$262.2\pm3.4\, \rm{km}\,\rm{s}^{-1}$]{2002AJ....124.2639V}, and the distance from \citet[$50.1\pm2.5\,{\rm kpc,}$]{2001ApJ...553...47F} and for the SMC, we took the line-of-sight velocity from \citet[$145.6\pm0.6\,\rm{km}\,\rm{s}^{-1}$]{2006AJ....131.2514H} , and the distance from \citet[$62.8\pm2.4\,{\rm kpc}$]{2000A&A...359..601C}. This gives us $v_{z,{\rm LMC}} = 1.104 \pm 0.057\, \rm{mas}\,\rm{yr}^{-1}$ and $v_{z,{\rm SMC}} = 0.489 \pm 0.019 \rm{mas}\,\rm{yr}^{-1}$.

To determine the PMs of the Clouds, we selected sources using the following procedure: 
\begin{enumerate}
\item To create a filter, we initially selected stars with $\rho=\sqrt{x^2+y^2}<\sin r_{\rm sel}$ ($ r_{\rm sel} = 5\degr$ for the LMC, $ r_{\rm sel} = 3\degr$ for the SMC) and $\varpi/\sigma_\varpi<10$ (to minimise foreground contamination).  We also selected only stars with $G<19$ in this step to ensure that the spread in PM due to uncertainties is small compared to the difference between the PM of the Cloud and of the bulk of the foreground.
\item We determined the median PM of this sample, and preliminarily filtered on PM by removing any source where $\mu_x$ or $\mu_y$ lies more than four times the robust scatter estimate\footnote{The robust scatter estimate (RSE) is defined in terms of the 10th and 90th percentile values, $P_{10}$ and $P_{90}$ as RSE$=C\times(P_{90}-P_{10})$, where $C = {\left(2\sqrt{2}\,{\rm erf}^{-1}\bigl(4/5\bigr)\right)}^{-1}\approx0.390152$ 
. For a Gaussian distribution, it is equal to the standard deviation. } of that PM component from the median. 
\item We determined the covariance matrix of $\mu_x,\mu_y$ for these stars, $\mbox{\boldmath\textsf{$\sigma$}}$, and used this to define a filter on PM, requiring that  $\mbox{\boldmath$\mu$}^T$$\mbox{\boldmath\textsf{$\sigma$}$^{-1}$}$$\mbox{\boldmath$\mu$} < 9.21$ to correspond to a 99\%\ confidence region.
\item We applied this filter in PM, along with that in $\varpi$, to all stars with $G<20$ within 8 degrees of the assumed centre of LMC or SMC to define our complete sample.
\end{enumerate}

We iterated this procedure twice, first using the expected $\mu_x$, $\mu_y$ given the
quoted $\mu_\alpha*$, $\mu_\delta$. This gave us a median parallax
for the stars in the two Clouds: $-19 \mu {\rm as}$ for the LMC, and $-0.9
\mu {\rm as}$ for the SMC (compared to the expected values of 
$\sim20 \mu {\rm as}$ and $\sim16 \mu {\rm as}$, respectively). 
This is consistent with the offset and variation
reported in other sections of this paper and in \citet{DR2-DPACP-39}. 
We then repeated the procedure using the values of
$\mu_\alpha*$, $\mu_\delta$ implied by the data, conditional on the
source parallax taking this median value (taking into account the
quoted uncertainties and correlations). This procedure left us with 
8~million sources in the LMC and 1.4~million in the SMC.

\section{Analysis: Globular clusters}
\label{sec:gc}

As described earlier, we have analysed 75 globular clusters, for which
the data are presented in Table~\ref{tab:overview}. For each cluster
we have derived the PM and parallax, and where data were
available, the radial velocity.

\subsection{First analysis and comparisons}
\begin{figure}
\centering
\includegraphics[width=8.5cm]{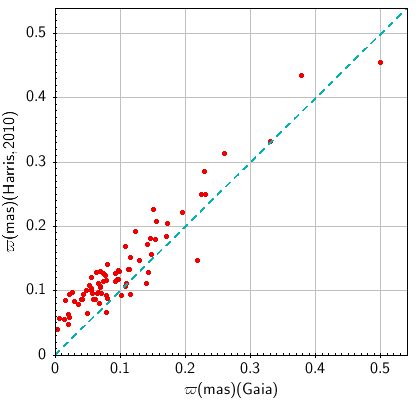}
\caption{Comparison between parallaxes as derived from the \Gaia DR2 data and parallaxes derived from the cluster distances as given in Harris10.}
\label{fig:parcomp}
\end{figure}
\begin{figure}
\centering
\includegraphics[width=8.5cm]{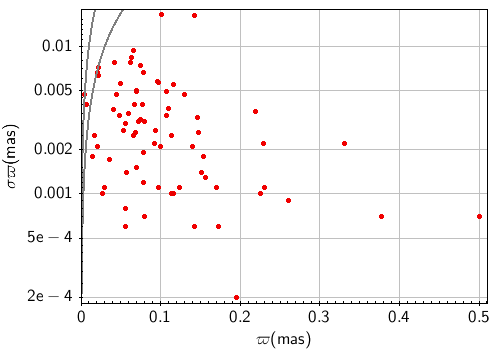} 
\includegraphics[width=8.5cm]{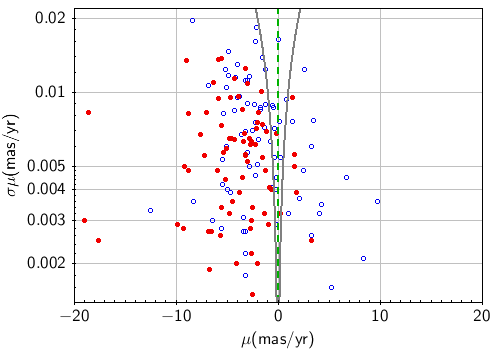}
\caption{{\it Top}: Parallax error against the \Gaia parallax as
  determined from the \Gaia data for 75 globular clusters. The black
  curves are the 1 and 3$\sigma$ limits. The vertical dashed line
  corresponds to 3 times our estimate of the systematic error on the
  parallax. {\it Bottom}: PM errors against the PMs in right ascension  (open circles) and declination (solid circles) for the clusters in our sample.  The curves represent the value of the PM for 100$\sigma_\mu$. The PM measurement has a significance lower than 10$\sigma$ only for NGC 6453, for which $\mu_{\alpha*}/\sigma_{\mu_{\alpha*} }\sim  4$.}
\label{fig:pmdistr}
\end{figure}

Fig.~\ref{fig:parcomp} compares the parallaxes derived from the \Gaia
data to those from the cluster distances given in Harris10. There is a
systematic difference of -0.029 mas (the \Gaia parallaxes being
smaller), originating largely from the \Gaia data, and a calibration
noise level around that relation of 0.025 mas
\citep{DR2-DPACP-39,DR2-DPACP-51}. A small
contribution might also come from the values given by Harris10. However, we
have made a provisional check on these distance estimates
using the \Gaia photometric data by superimposing the HR diagrams for
all the clusters using the distances and reddening values as presented
in Harris10 \citep[see][]{DR2-DPACP-31}. We found that all the clusters
are neatly aligned for the critical elements (mainly the
position of the blue horizontal branch). This indicates that, as a
group, the distance moduli and colour corrections
are confirmed to be in mutual agreement to better than 0.1 magnitude.

The standard uncertainties, which measure the precision rather than
the accuracy, of the cluster-parallax determinations are smaller or
very much smaller than the overall calibration noise level, as shown
in the top panel of Fig.~\ref{fig:pmdistr}. The actual errors
on these parallax determinations are therefore dominated by the
overall \Gaia calibration noise and offset in the parallax values. As
discussed in depth in \cite{DR2-DPACP-51}, these systematic errors are
also apparent in the parallax distribution of quasi-stellar objects
(QSOs, which reveal the
same offset), and as we show in Appendix \ref{app:app-ah}, also in the
parallaxes of stars in the LMC (localised fluctuations) and other
dSph, and are due to the basic angle variation and scanning
law of \textit{Gaia}. It is therefore expected that their amplitude will be
significantly smaller in future \Gaia data releases. For the
time being, and because the parallax uncertainties derived
photometrically are smaller, we use the distances as given by Harris10
in the analyses that follow.

\begin{figure}[t]
\centering
\includegraphics[width=8.5cm]{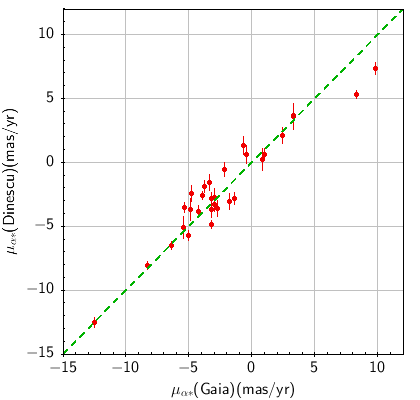}
\includegraphics[width=8.5cm]{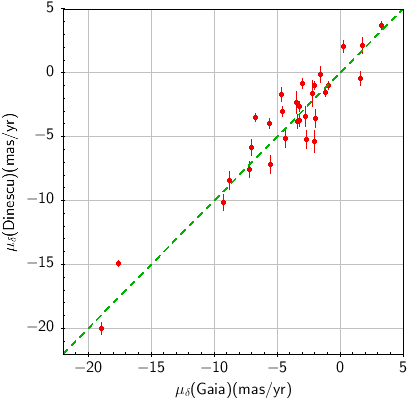}
\caption{Comparison of the \Gaia PMs (in right ascension: top, and declination: bottom) to measurements reported in \cite{1999AJ....117..277D, 2003AJ....125.1373D, 2007AJ....134..195C, 2010AJ....140.1282C, 2013AJ....146...33C} for 31 globular clusters.}
\label{fig:dinescupmcomp}
\end{figure}

The observed PMs are mostly about one to two orders of
magnitude larger than the parallaxes, and thus the measurements are
very robust and significant (see the bottom panel of
Fig.~\ref{fig:pmdistr}). A comparison with a series of studies \citep{1999AJ....117..277D, 2003AJ....125.1373D, 2007AJ....134..195C, 2010AJ....140.1282C, 2013AJ....146...33C} is shown in Fig.~\ref{fig:dinescupmcomp},
and indicates overall good agreement, and most notably that the errors
have been reduced by nearly two orders of magnitude. It remains somewhat uncertain, however, if the same calibration noise level can be
assumed for the PMs as for the parallax (but see
e.g. Sec.~\ref{sec:dwarfs_sys}). Nonetheless, this systematic will be
much smaller than the amplitude of the PMs themselves.

\begin{figure}
\centering
\includegraphics[width=8.5cm]{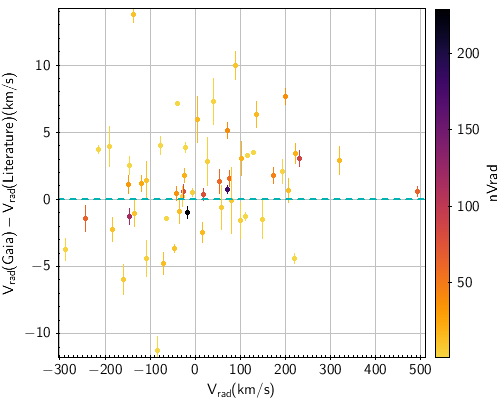}
\includegraphics[width=8.5cm]{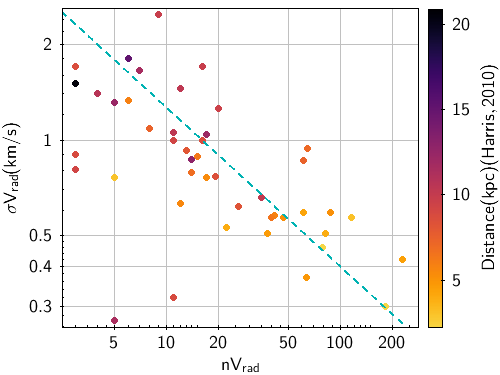}
\caption{{\it Top}: Comparison between the ground-based and \Gaia
  radial velocities for 52 clusters with at least two measurements. {\it
    Bottom}: Standard uncertainties on the mean radial velocities as a
  function of the number of stars contributing to the mean. The
  diagonal line shows the effect of an additional contribution of
  4~km~s$^{-1}$ originating from the internal velocity dispersion.}
\label{fig:radvelcomp}
\end{figure}

Radial velocities as measured by \Gaia \citep{DR2-DPACP-46} are
available for 57 of the 75 clusters, although there
were 3 or more cluster stars with measured radial velocities
for only 46 clusters. 
  While future \Gaia data releases will contain radial velocities for
  more of these sources, this highlights a need for dedicated
  high-precision spectroscopy of these clusters to properly complement
  the \Gaia astrometry. Figure~\ref{fig:radvelcomp} shows a comparison
between ground-based (from Harris10) and \Gaia radial velocity
measurements, indicating a good relation for clusters for which enough
stars have spectroscopic measurements (darker points). The relation
between the number of stars and the standard uncertainty on the mean
cluster velocity indicates an average internal velocity dispersion of
the order of 4~km~s$^{-1}$. This estimate of the intrinsic velocity
dispersions is very similar to what is observed for the PMs.

\begin{figure*}
\centering
\includegraphics[width=17cm]{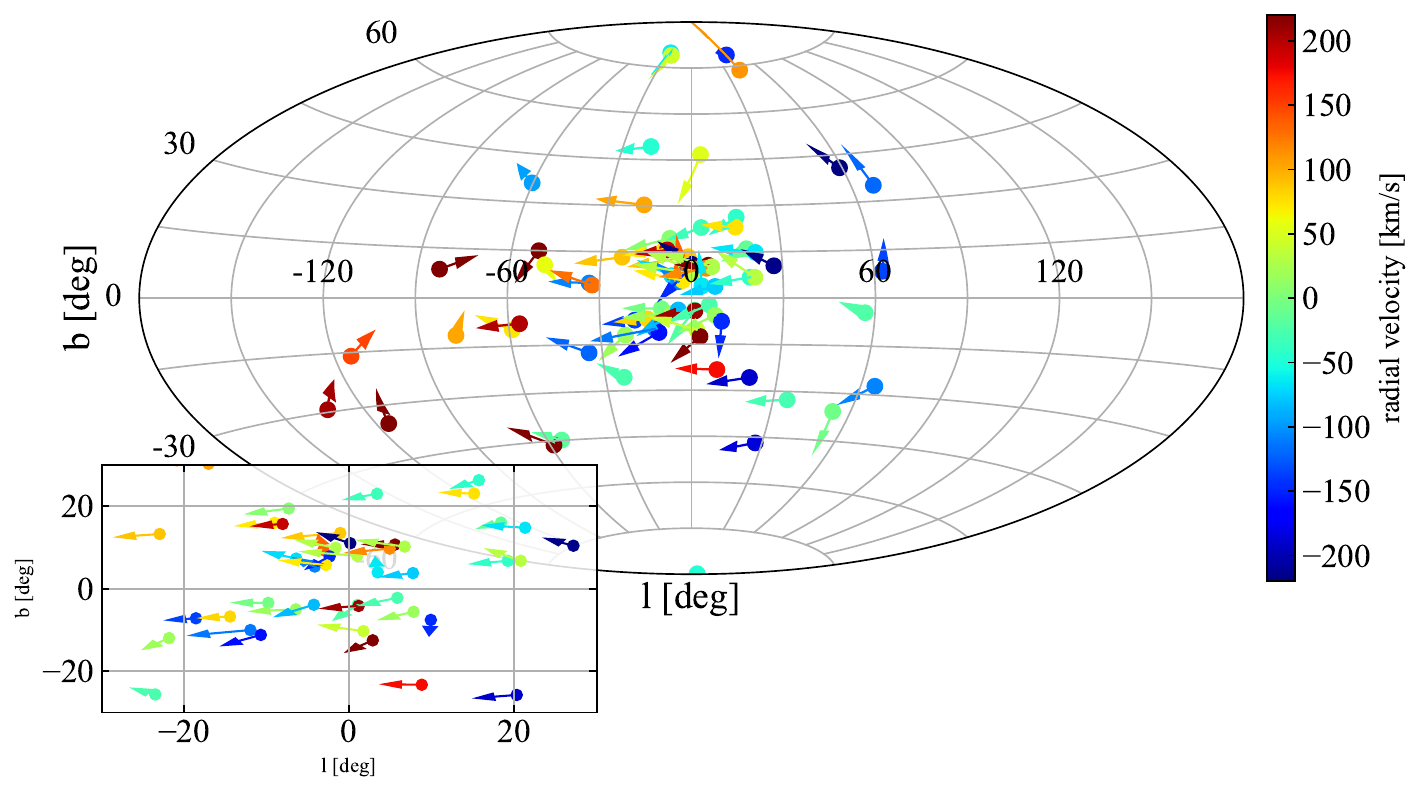}
\caption{Sky distribution of the 75 globular clusters in our sample in Galactic coordinates. Their tangential velocities are denoted by the size and direction of the arrows, while the colours indicate their line-of-sight velocities. The inset shows a zoom-in of the central $60\times60$ deg$^2$.}
\label{fig:skydistrglob}
\end{figure*}

Fig.~\ref{fig:skydistrglob} shows the distribution on the sky of the
globular clusters in our sample, where the arrows indicate the
direction of motion and the colour-coding reflects the amplitude of
the tangential velocities. These were derived using the PMs
listed in Table~\ref{tab:overview} and the distances from
Harris10.

\subsection{Further results from the globular cluster astrometric data}

\begin{figure*}[t]
\centering
\includegraphics[totalheight=4.5cm]{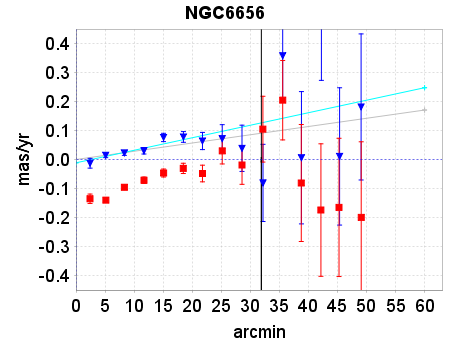}
\includegraphics[totalheight=4.5cm]{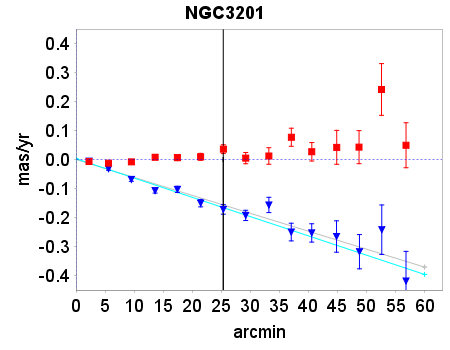}
\includegraphics[totalheight=4.5cm]{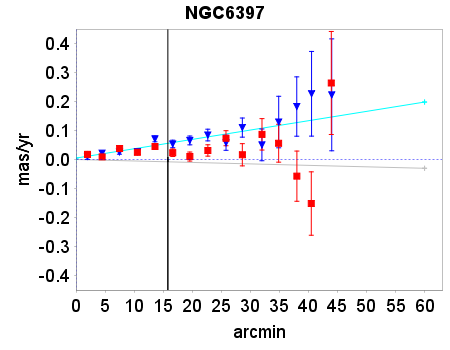}
\caption{PM ``systematics'' in radial (blue triangles) and transverse (red squares) directions, as a function of distance to the cluster centre. The cyan line is a fit to the variation of the mean radial component of the PMs as a function of distance, while the grey line is the expected trend resulting from perspective contraction or expansion. From left to right: NGC~6656 shows a strong rotation signal as well as perspective expansion; NGC~3201 shows no rotation but very strong perspective contraction; and NGC~6397 shows halo expansion associated with core collapse, clearly different from the signal expected from the small perspective contraction. The vertical line indicates the tidal radius of the cluster reported in Harris10.}
\label{fig:globclpm}
\end{figure*}

The outstanding quality of the \Gaia DR2 data together with the absolute reference frame (free
of expansion and rotation) in which the PMs are presented has also allowed us to clearly detect
rotation in 5 of the 75 globular clusters in our sample. For 3 of
these clusters (NGC~104, NGC~5139, and NGC~7078), this was already known
\citep{2013ApJ...772...67B}, but we have also detected rotation in
NGC~5904 and NGC~6656 (see e.g. the left panel of
Fig.~\ref{fig:globclpm}). An indication of rotation can also be
observed in NGC~5272, NGC~6752, and NGC~6809.  Similarly, \Gaia data
allow measuring expansion and contraction in globular clusters. For
example, NGC~3201 (Fig.~\ref{fig:globclpm}, middle) shows very clear
perspective contraction, which is due to its very high radial velocity
and relatively large parallax. From this we may determine the parallax
of this cluster in the same way as this used to be done for the nearby
Hyades open cluster \citep[see][and references
therein]{2009A&A...497..209V}. The \Gaia data as presented here for
the radial velocity and the PMs thus provide a cluster
parallax of $0.221\pm 0.0086$~mas, at about $2\sigma$ from the value
of 0.204~mas given by Harris10. Finally, for NGC~6397, a cluster
considered to have been subject to core collapse, we can still see a
signal of the expanding halo (Fig.~\ref{fig:globclpm}, right), clearly
different from the expected very weak perspective contraction
signal.

Furthermore, we find that our clusters have velocity dispersion
profiles that decline with radius (Fig.~\ref{fig:pmrtdisp}), and that
several clusters show a slight increase in the outskirts, probably as
the result of a halo of more loosely bound stars \citep[as evidenced
also by their spatial extent, see
e.g.][]{2009AJ....138.1570O,2012MNRAS.419...14C,2016ApJ...829..123N,2018MNRAS.473.2881K}. This
increase is found at a distance where contamination by field stars
should not yet be important.

\begin{figure*}
\centering
\includegraphics[width=18cm,trim={0 1cm 0 2cm},clip]{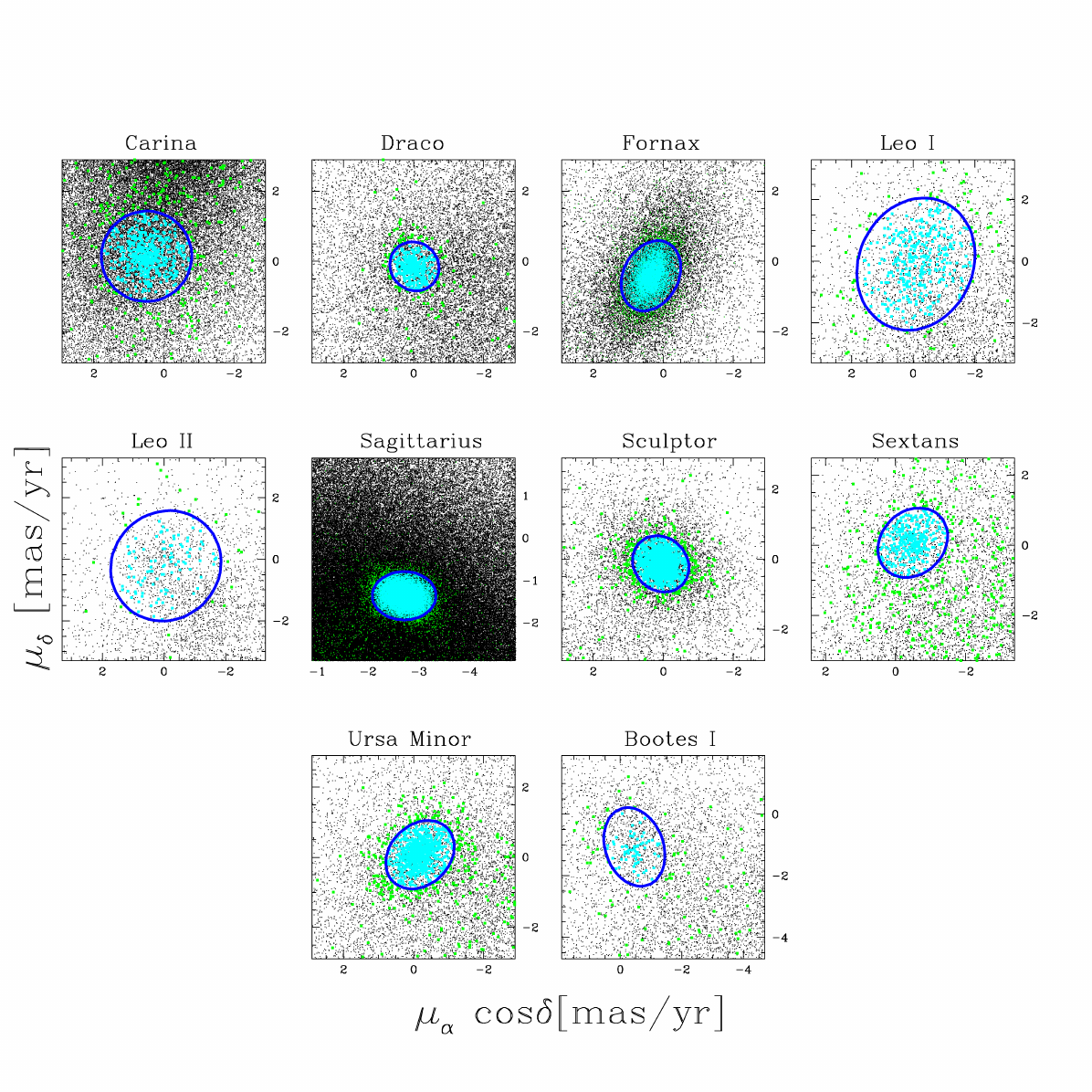}
\caption{PMs of the stars in the field of view towards the
  different dSph galaxies in our sample. Members, defined as stars
  within 3$\sigma$ of the mean measured PM and located in
  the expected region of the CMD, are shown in cyan. The green points
  correspond to stars that also fall in the CMD-selected box, but are
  not within 3$\sigma$ of the systemic PM. Especially for
  Sculptor and Fornax, it is quite clear that there may be more
  members, but very likely, the large errors on the PMs of
  individual stars place them beyond the 3$\sigma$ ellipse.}
\label{fig:VPD_dwarfs}
\end{figure*}

\section{Analysis: Dwarf spheroidal galaxies}
\label{sec:dw}

The procedures described in Section~\ref{sec:data-method-gcdw} allow us to determine
the mean PMs of the dSph in our sample. As discussed
earlier, we focus in this paper on the classical dSph, and have
included in our sample one example of an ultra-faint galaxy, Bootes I.
The resulting mean $\mu_\alpha^*$ and $\mu_\delta$, as well as the
astrometric parameters and uncertainties, are listed in
Table~\ref{tab:PM_dwarfs}.

The efficiency of our selection procedure in removing most of the
foreground contamination becomes clear in
Fig.~\ref{fig:VPD_dwarfs}. Stars surviving the proposed criteria are
shown as cyan dots for each of the dwarfs, and they clearly clump much more strongly in
the diagrams than the 
likely non-members (shown as black points).  In this figure the blue
ellipses indicate the contours corresponding to 3$\sigma$ dispersion
around the mean $\mu_\alpha^*$ and $\mu_\delta$ ($\sigma$ is
computed taking into account the covariances, using the standard error
on the mean $\times \sqrt{N_{*}}$, where $N_*$ is the number of stars used
to measure the mean PMs).

The possibility of selecting members via
their PM that the \Gaia DR2 data provide opens a new window for understanding the structure
and extent of the dSph. In particular, Fig.~\ref{fig:fov_dwarfs} shows that some
of the dwarfs in our sample present spatial asymmetries (e.g. Fornax in the
top right corner, the Sculptor outskirts appear somewhat boxy), while there is an indication of tidal streams in the case of the Carina dSph. A more detailed
analysis of these features is beyond the scope of this paper, but the quality of the{} dataset certainly makes this possible.

\begin{figure*}
\centering
\includegraphics[width=6cm]{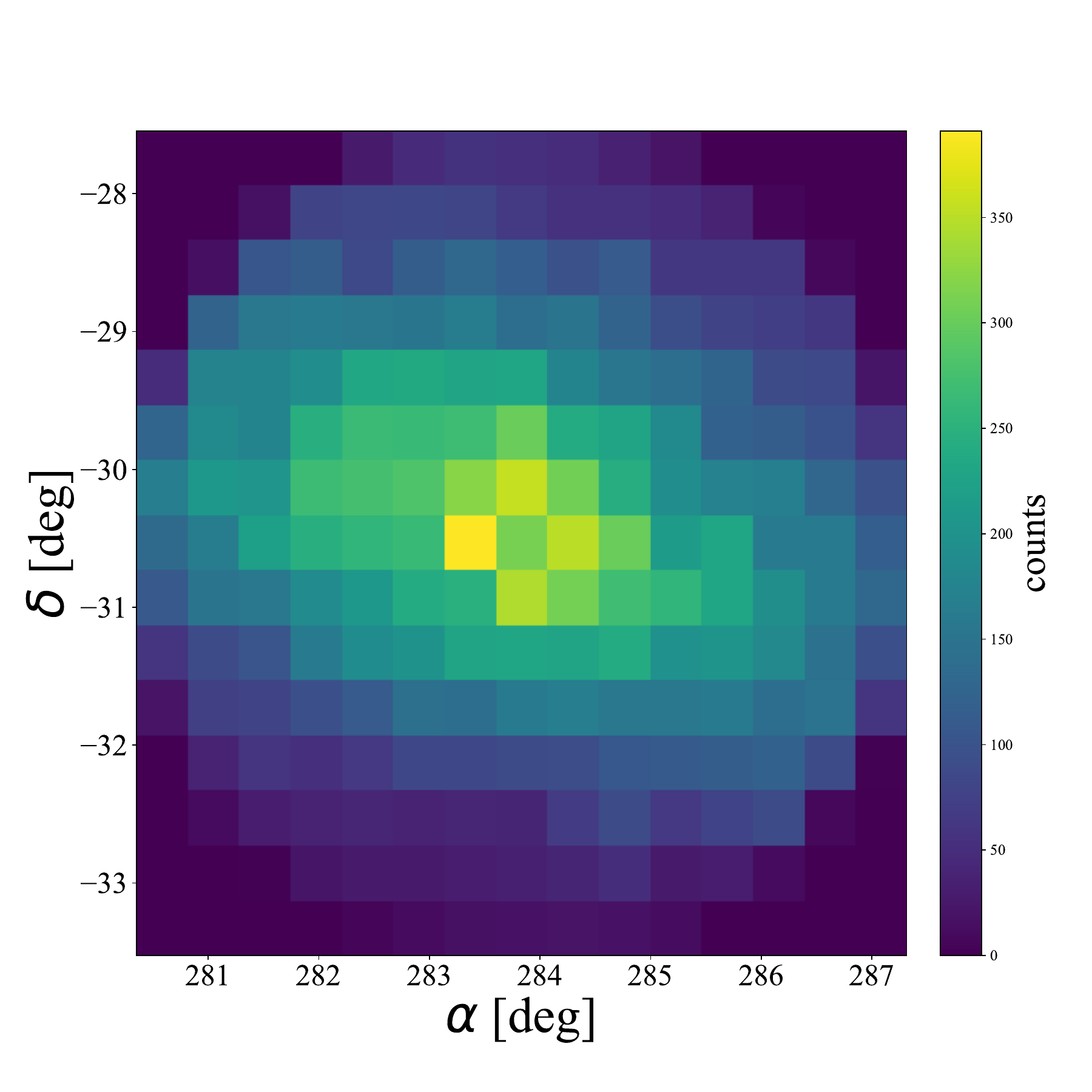}
\includegraphics[width=6cm]{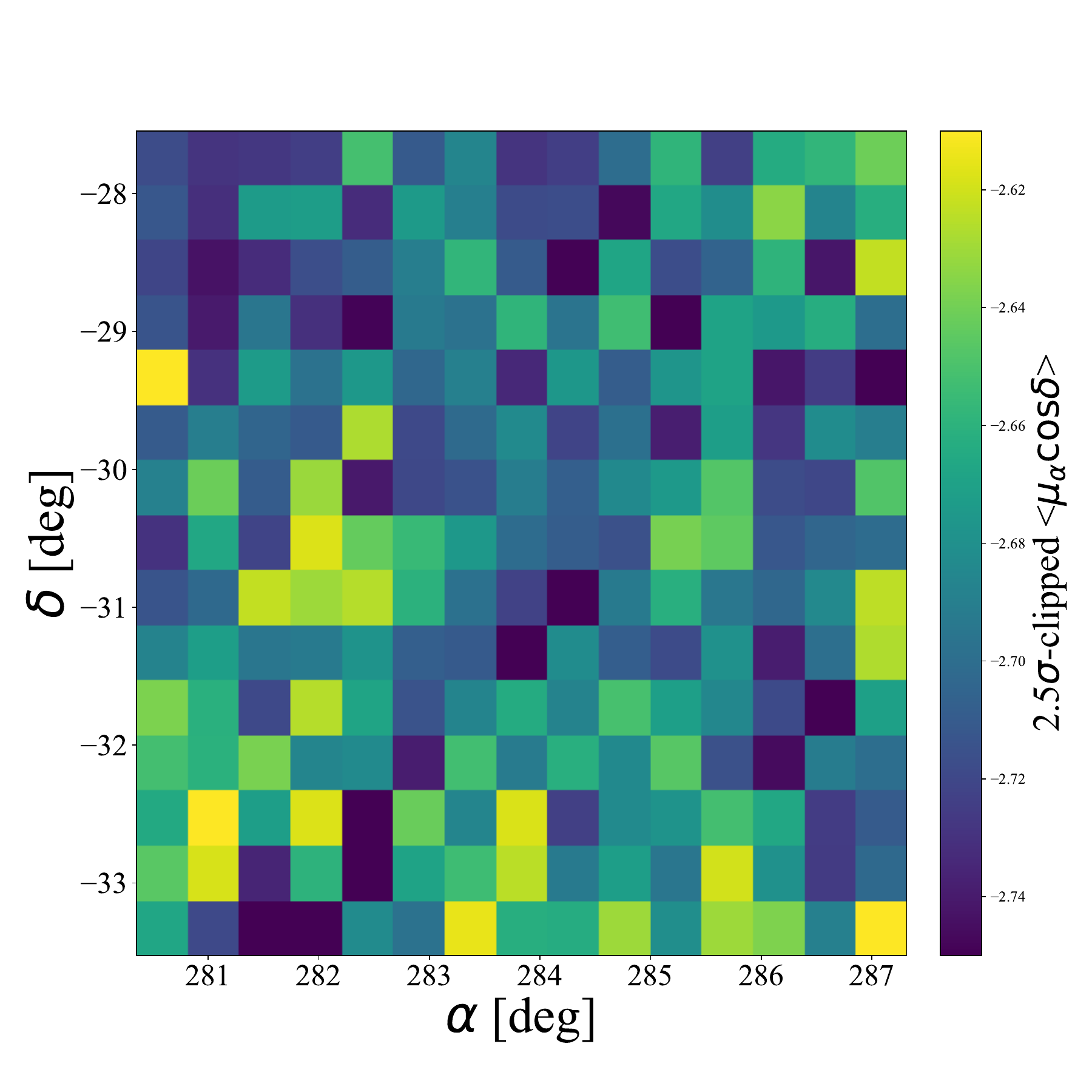}
\includegraphics[width=6cm]{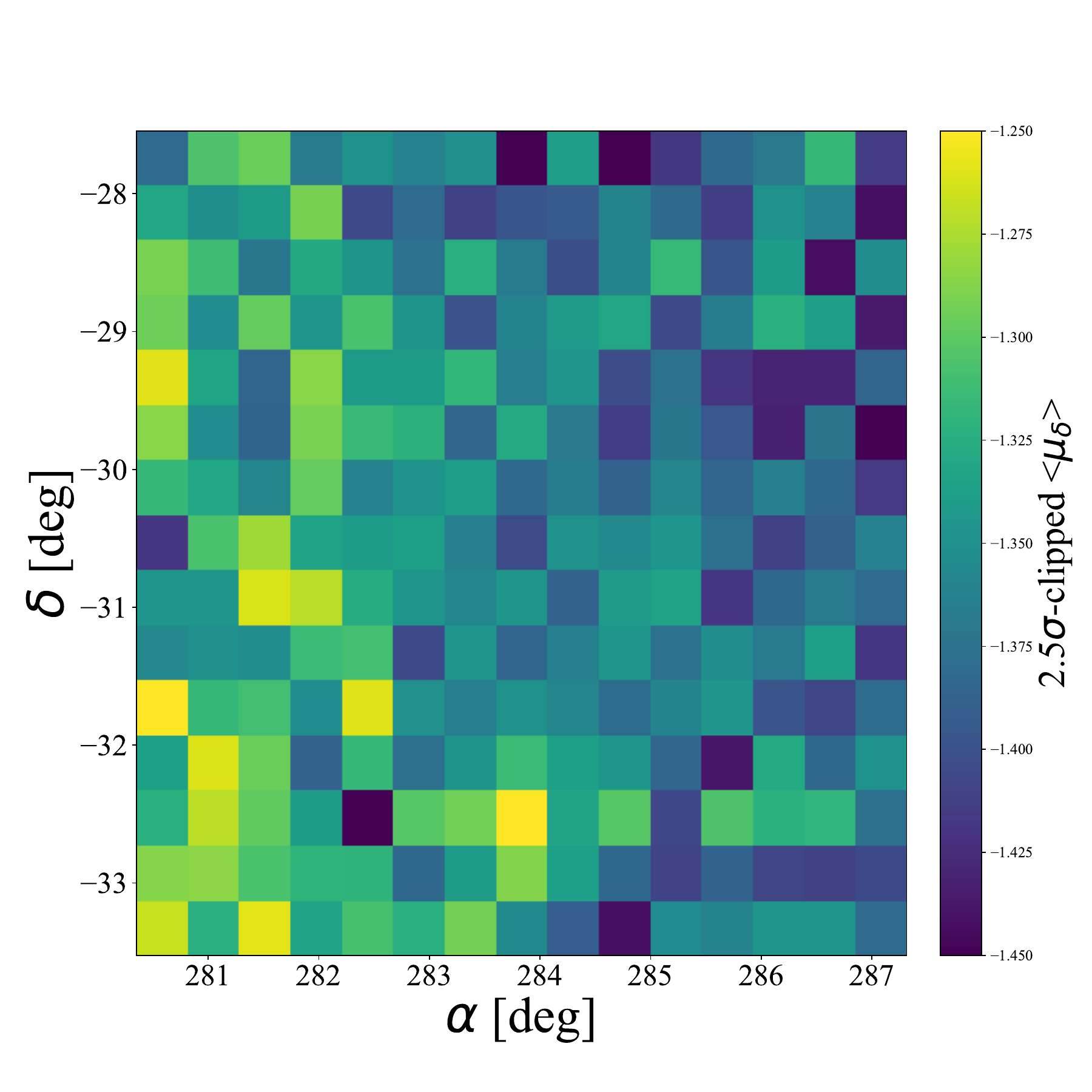}
\caption{Counts in 0.2 deg wide bins on the sky with at least 100 stars for stars in the Sagittarius
dwarf (left) and the average PM in $\mu_{\alpha*}$ (middle) and $\mu_\delta$ (right) for each of these bins.}
\label{fig:Sgr_systematic}
\end{figure*}

\begin{figure}
\centering
\includegraphics[width=4.4cm]{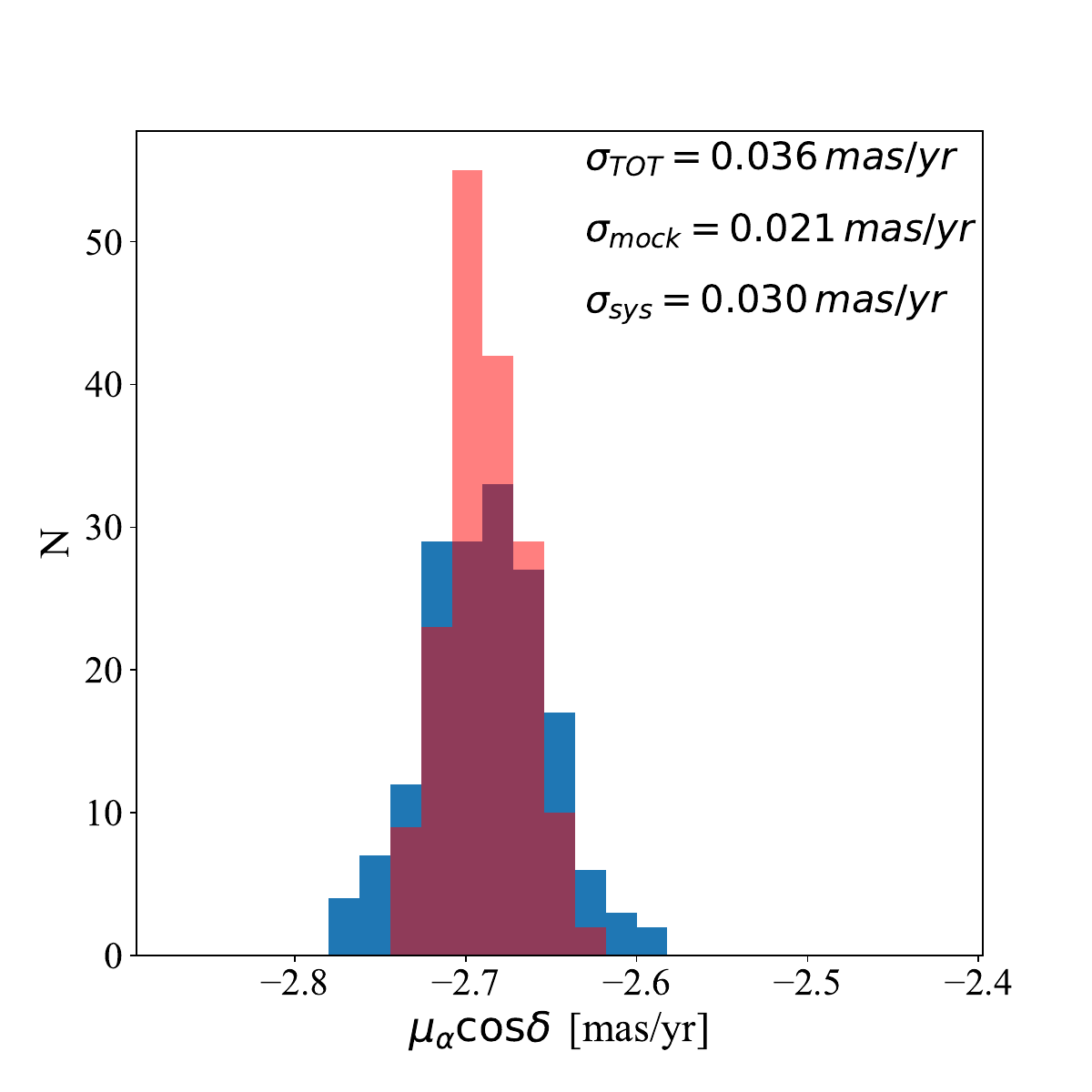}
\includegraphics[width=4.4cm]{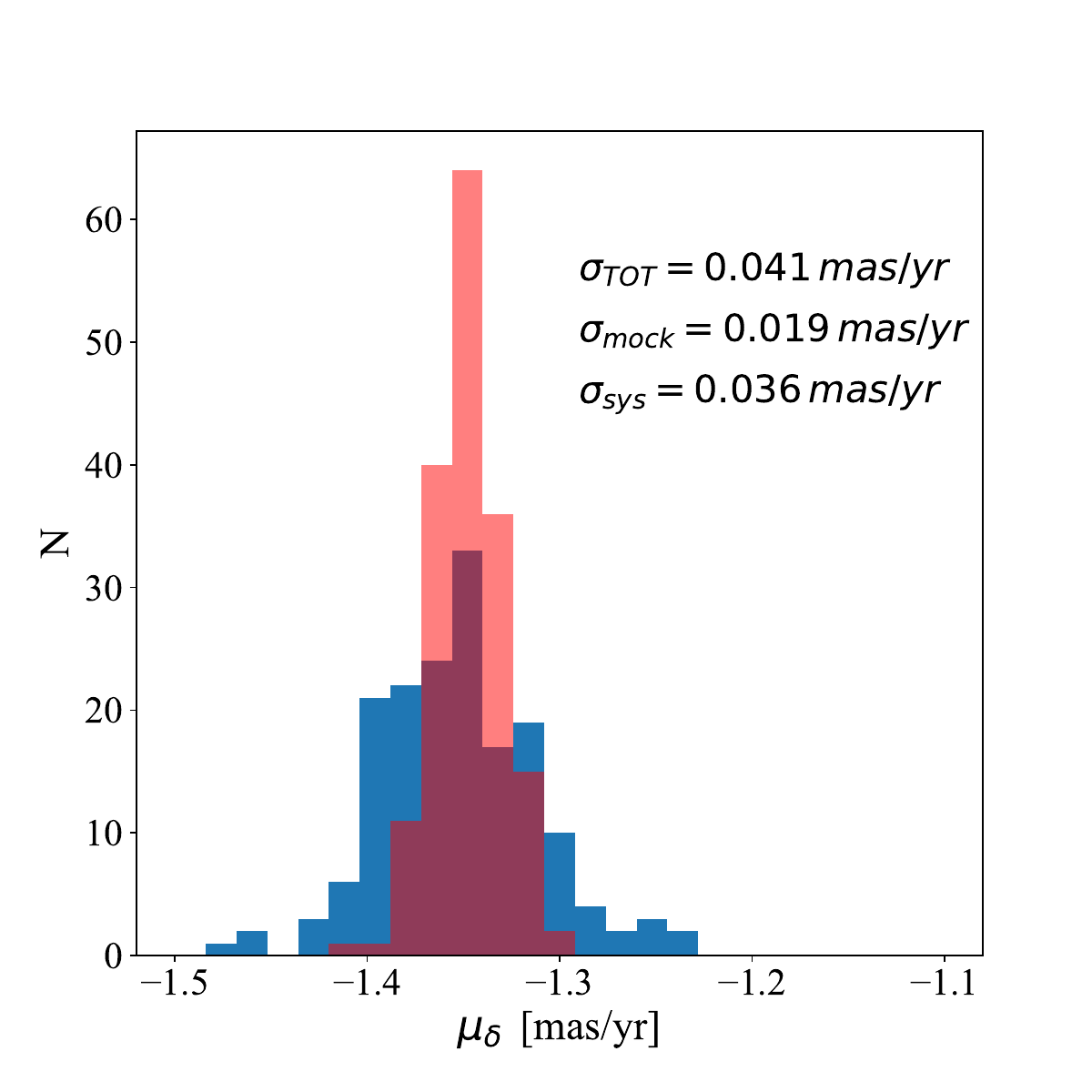}
\caption{Blue histograms show the distribution of average PM in
  the bins shown in Fig.~\ref{fig:Sgr_systematic}, while the red
  histograms correspond to a model with the derived average PM
  convolved with the random errors provided by the astrometric
  solution. The difference shows the amplitude of the systematic on
  the mean value of the PM for a bin of this size.}
\label{fig:Sgr_dispersion_sys}
\end{figure}

\subsection{Systematics, correlations, and dispersion}
\label{sec:dwarfs_sys}

Table~\ref{tab:PM_dwarfs} shows that the average
parallax is negative for several dSph  \footnote{Since the parallax is a measured,
  unconstrained quantity with an associated measurement error,  the probability density function for the observed
  parallax will increasingly cover negative values with increasing
  error (and especially for distant objects whose parallax
  is close to zero). A parallax
  zero-point offset, as found in Gaia DR2, further affects this
  distribution.}. This systematic error, similar to that found for the
globular clusters and the Magellanic Clouds \citep[and the
QSOs][]{DR2-DPACP-51}, is present in the different fields, and its
amplitude varies from object to object (for more details, see
Appendix~\ref{app:app-ah}). The average offset, computed as the
difference between the expected parallax \citep[based on the distances
from][]{McConnachie12}, and the parallax from DR2 for all dSph, is
$-0.056$~mas\footnote{\cite{DR2-DPACP-39} reported a comparable offset that was
  computed using an average over spectroscopically identified member
  stars of all the dSph simultaneously.}, and when Leo I is excluded,
it is $-0.038$~mas (the parallax offset is $-0.21$~mas
for Leo I).

The PM maps shown in Fig.~\ref{fig:VPD_dwarfs} reveal that
the dwarfs are extended in PM space. The main contributor of this
dispersion is not intrinsic but is due to the uncertainties, which
are typically very large. For example, for an object such as Sculptor,
the individual velocity errors for $G \sim 18$ mag stars are of the order of
$80$ \kms (for $G \sim 20$, they are $> 200$~km~s$^{-1}$),
compared to the expected internal dispersion of order of 10
km~s$^{-1}$. Therefore, measuring the intrinsic dispersion for these
systems does not appear to be feasible with the data provided by
DR2. However, it may begin to become feasible with later data
releases, and certainly with an extension of the \Gaia mission.

\begin{figure*}[t]
\centering
\includegraphics[width=16cm,trim={0 0 0 0.5cm},clip]{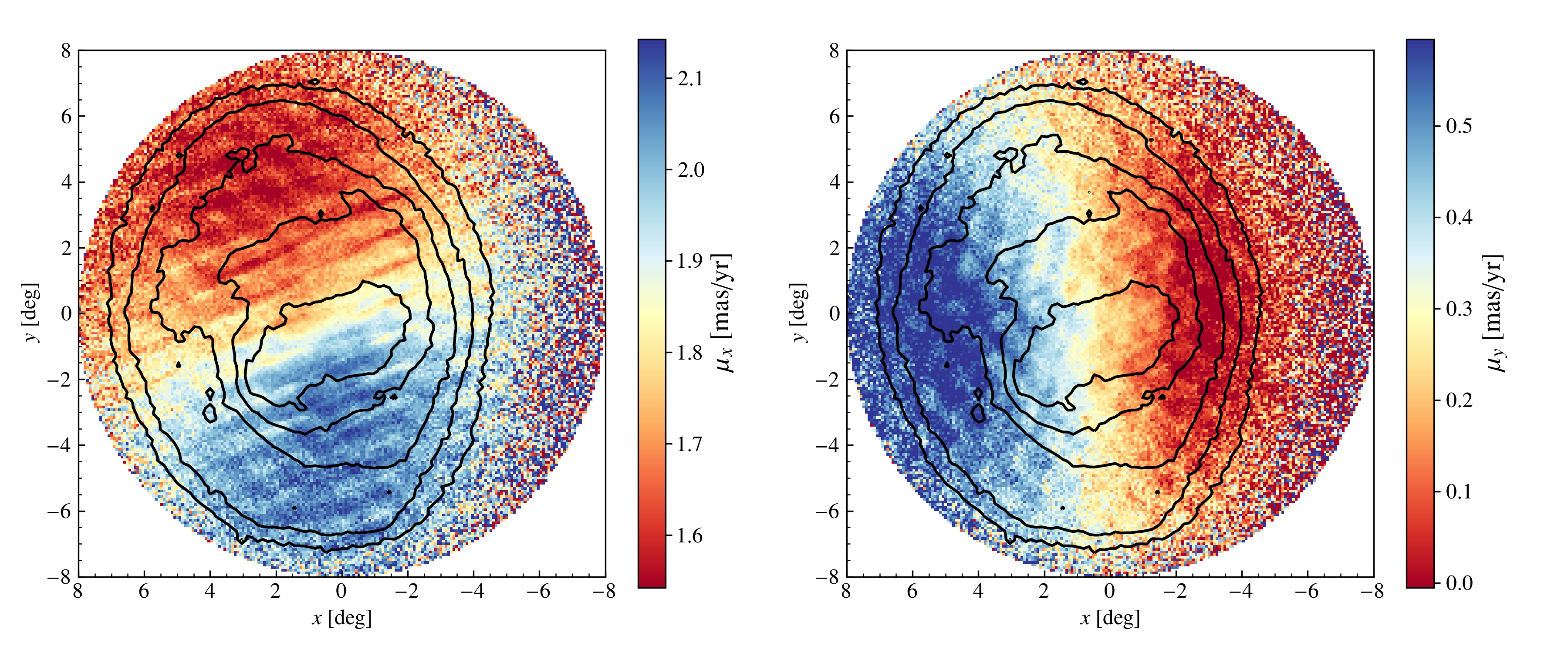}
\includegraphics[width=16cm,trim={0 0 0 0.5cm},clip]{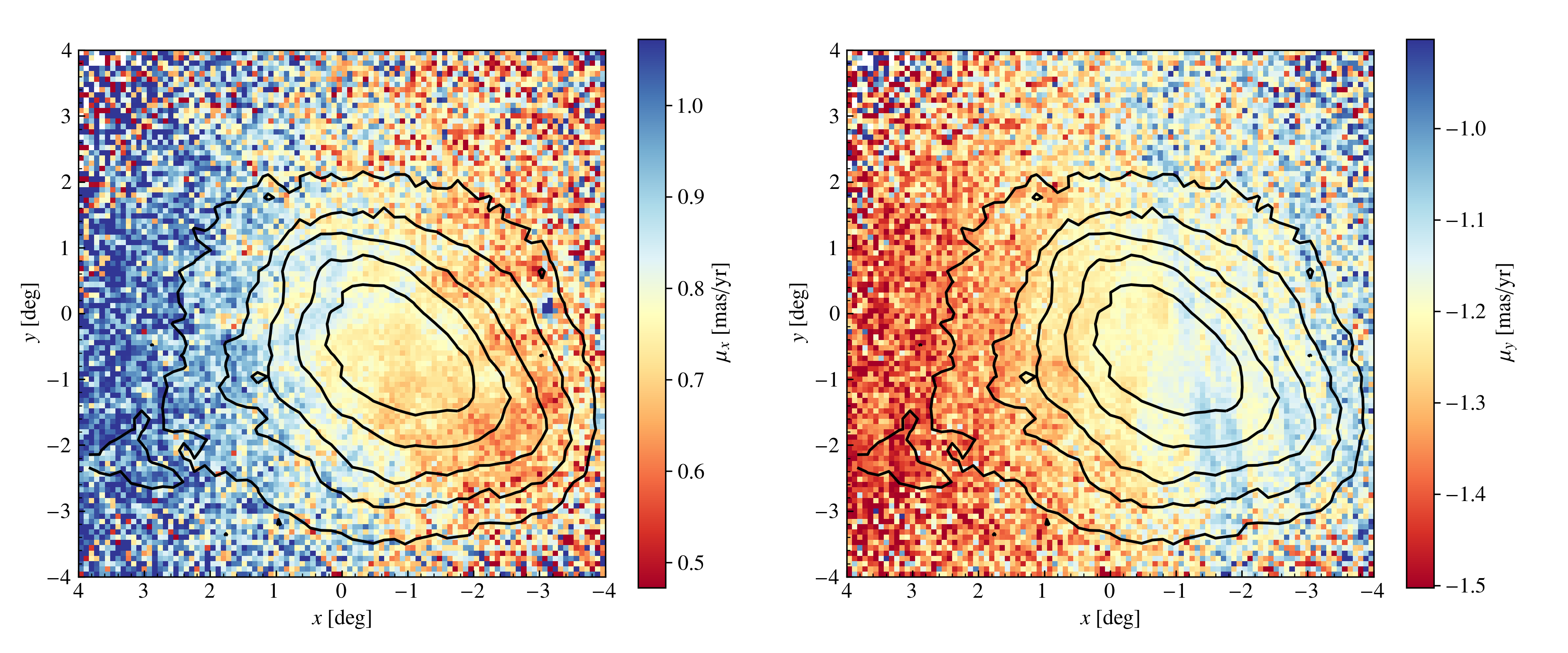}
\caption{PMs of stars in the LMC (upper) and SMC (lower) showing the components $\mu_x$ (left) and $\mu_y$ (right), which are described in the text (Eq.~\ref{eq:xyderivs}). The colour shows the median PM in each pixel (after filtering). The black density contours are logarithmically spaced, such that the outermost contour is at a source density 100 times lower than the highest density. The centre of each colour bar is chosen to be the median PM of all sources. This is not the same as the PM derived for the clouds below, because the sources are not distributed symmetrically around the assumed dynamical centre; the photometric and dynamical  (from the H\textsc{i} disc) centres are offset from one another.  Trends in PM, particularly the trend associated with rotation in the LMC, are clearly visible.  The banding associated with
the {\Gaia} scanning law, and as seen in the parallaxes, are
clearly visible as well. }
\label{fig:LMCSMCPM}
\end{figure*}

The measurements of the PM are also affected by the scans and varying astrometric incompleteness, which introduce
a pattern in the parallax and PM field that is only readily apparent
for sufficiently large objects on the sky \citep[see
also][]{DR2-DPACP-39,DR2-DPACP-51}. This is illustrated for the LMC
and SMC in Figs.~\ref{fig:LMCSMCPar} and \ref{fig:LMCSMCPM}, and it is
also present for example for the Sagittarius dSph, as shown in
Figs.~\ref{fig:fov_dwarfs} and \ref{fig:Sgr_systematic}.

For sufficiently large objects on the sky, the banding pattern is
averaged out, and the mean PM is more robust\footnote{Although there
  may still be a residual effect of $\sim$0.028 mas~yr$^{-1}$ amplitude on scales of 10-20 degrees, as
  reported for the QSOs in \cite{DR2-DPACP-51}.}. The global dispersion is larger than expected just from random errors, however (because of the
offset from bin to bin in the pattern). We quantify this effect in
Fig.~\ref{fig:Sgr_dispersion_sys}, where we show the distribution of
the mean PM in $\alpha$ and $\delta$ computed in bins of
0.2$\times$0.2~deg$^2$ size that contain at least 100 stars\footnote{We
  considered 0.2 deg bins because this is the smallest angular size of a
  dSph in our sample.}, and after 2.5$\sigma$ clipping to remove
outliers. The fact that the mean value changes from bin to bin is at
least partly caused by the finite number of stars in each bin, as well
as by the random errors. We tried to estimate the residual systematic
error by modelling this distribution assuming the mean PM
derived using all member stars, and assuming that the errors are
Gaussian. We drew a new PM from this mean for each
star, assuming
its quoted uncertainty and correlations, and recomputed the mean using
all the stars in the bin (we also assumed an intrinsic velocity
dispersion of 10 km~s$^{-1}$). This is the red histogram in
Fig.~\ref{fig:Sgr_dispersion_sys}. Clearly, the observed distribution
is wider (blue), and the difference between the two can be used to
compute the systematic error as $\sigma_{mock}^2 + \sigma_{sys}^2 =
\sigma_{tot}^2$. We find this to be $\sigma_{sys} \sim 0.030$~\masyr
and $\sim 0.036$ \masyr in right ascension and declination
directions, respectively. These values are of slightly smaller
amplitude than those derived by \cite{DR2-DPACP-51} from a sample of
QSOs.

For systems that are smaller on the sky, and in particular for those
that would fall in a single bin, their PM may be offset by this
much. Objects such as the larger dSph Fornax, Sculptor, etc., are
likely not affected by this systematic (because it averages out), but
for systems such as Leo II, it should be considered. In our
subsequent analyses we thus considered the amplitude of the systematic
uncertainty to be $0.035$~mas~yr$^{-1}$. Table~\ref{tab:PM_dwarfs} shows that in many cases, this systematic error is
larger than the random error on the measurement of the mean PM of a dSph.

Table~\ref{tab:PM_dwarfs} also shows strong
correlations in the different mean astrometric parameters derived for
the dSph in our sample. These correlations vary from object to object
in amplitude and direction (see Appendix \ref{app:app-ah}), and it is
important to take them into account in the derivation of the
orbital parameters, for instance.

\subsection{Comparison to the literature}

We have compared the PM we derived with our selection criteria and
those we would obtain if we were to use only stars identified as members from
publicly available radial velocity catalogues
\citep[from][]{1995AJ....110.2131A,2002MNRAS.330..792K,2006ApJ...649..201M,2006A&A...459..423B,2011MNRAS.411.1013B,2007ApJ...663..960S,2008ApJ...675..201M,2009AJ....137.3100W,2015MNRAS.448.2717W}. We
have found very good agreement \citep[i.e. the estimates differ by less than $1\sigma$,
see][]{DR2-DPACP-39}. The main disadvantage of using external
information is its heterogeneous nature. Furthermore, this information
is not available for all the dSph in our sample, and the
sample of stars with radial velocities for any given dSph is typically smaller
by a factor $\sim 2$ than is found with our selection and analysis,
which are based
exclusively on \Gaia DR2 data, even down to the same magnitude limit.

\begin{figure}
\centering
\includegraphics[width=9cm,trim={0 0 0 1cm},clip]{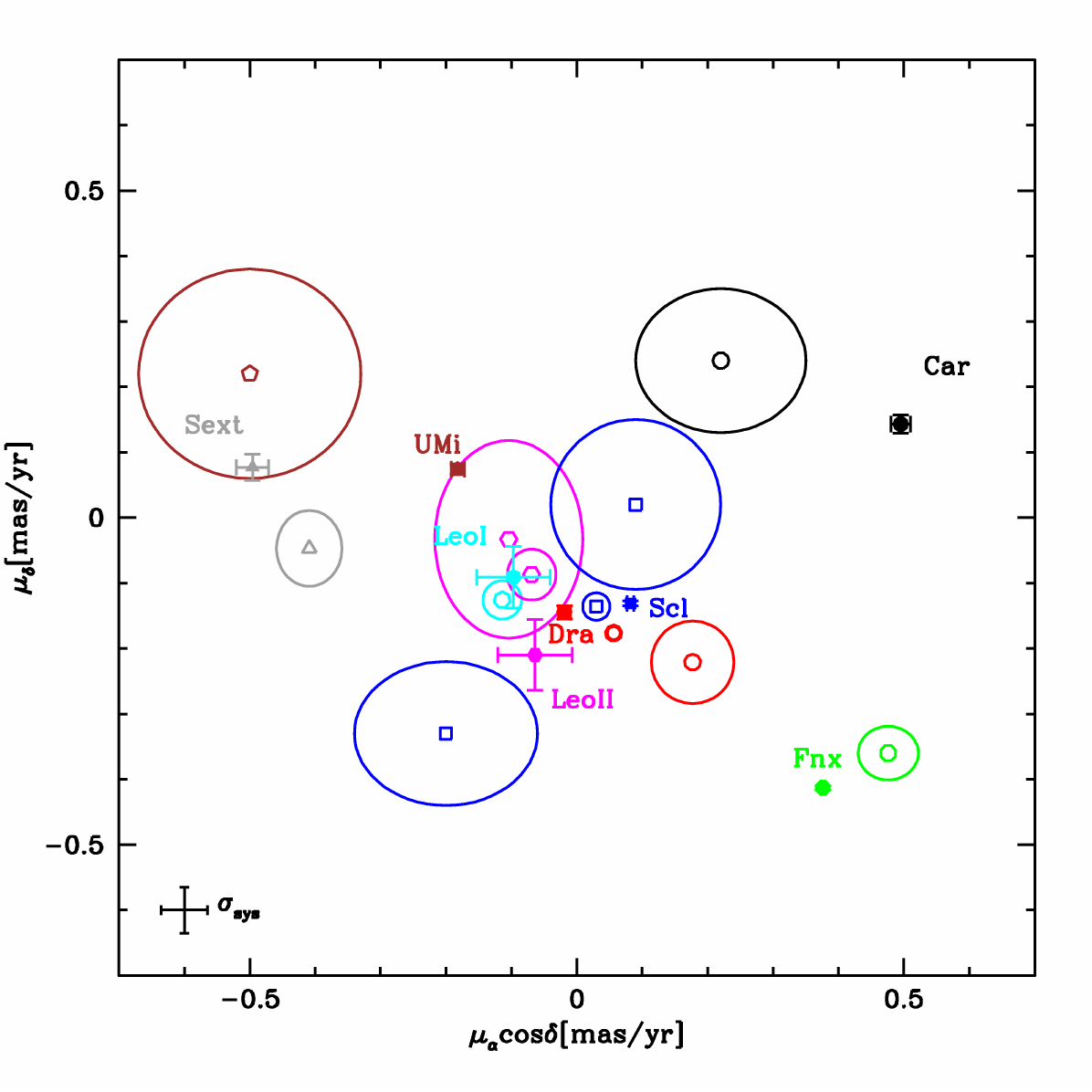}
\caption{Comparison to literature values of the proper motions of the
  dSph. Symbols with the same colour correspond to the same dSph,
  where filled symbols with error bars are those derived in this
  paper, and open symbols surrounded by ellipses correspond to the
  literature values. The error bars have the size of
  $\epsilon_{\mu_\alpha*}$ and $\epsilon_{\mu_\delta}$ , as reported in
  Table~\ref{tab:PM_dwarfs}. The black cross in the bottom left corner
  indicates our estimate of the systematic uncertainty on the PMs.}  
\label{fig:literature-dwarfs} 
\end{figure}

Figure \ref{fig:literature-dwarfs} compares our measurements of the PM of the dSph to
astrometrically derived values reported in the literature. These are
from \cite{2003AJ....126.2346P,2004AJ....128..951P} for Carina; from \cite{2015AJ....149...42P,2017ApJ...849...93S} for
Draco; from
\cite{2005AJ....130...95P} for Ursa Minor; from
\cite{2006AJ....131.1445P,2017ApJ...849...93S,2017arXiv171108945M}
for Sculptor; from
\cite{2007AJ....133..818P} for Fornax; from
\cite{2011ApJ...741..100L,2016AJ....152..166P} for Leo II; from
\cite{2013ApJ...768..139S} for Leo I, and from
\cite{2018MNRAS.473.4064C} for Sextans. Most of these measurements were obtained
from space, except for Sextans, whose measurements are based on Subaru imaging. We
excluded measurements for Sagittarius and Bootes~I for visualisation
purposes (because their PMs are much larger than for the other,
more distant dSph). 

A striking difference between previous estimates of the dSph PM and
those obtained using \Gaia DR2 data is the extent of the error bars,
particularly for the objects for which more than 400 (and up to several
thousand) astrometric members have been identified, such as
Carina, Ursa Minor, Fornax, Sculptor, and Draco. In many cases, our
measured PMs are consistent with the literature values at
the 2$\sigma$ level (given the large error bars of the latter). For
the most recent astrometric measurements with the HST (which
are therefore typically
based on a larger baseline), the values appear to be closer and
consistent with each other (e.g. Leo I), especially when the systematic
uncertainties are taken into account (e.g. Sculptor and Draco). In the case of Sagittarius, we find that the \Gaia DR2 PM is
consistent with that of \cite{2013ApJ...779...81M}, although it is now
much more accurate. We here present the first measurement of
the PM of the UFD Bootes~I.

\section{Analysis: LMC and SMC} 
\label{sec:LMCSMC}

\begin{figure*}
\centering
\includegraphics[width=16cm,trim={0 0 0 0.5cm},clip]{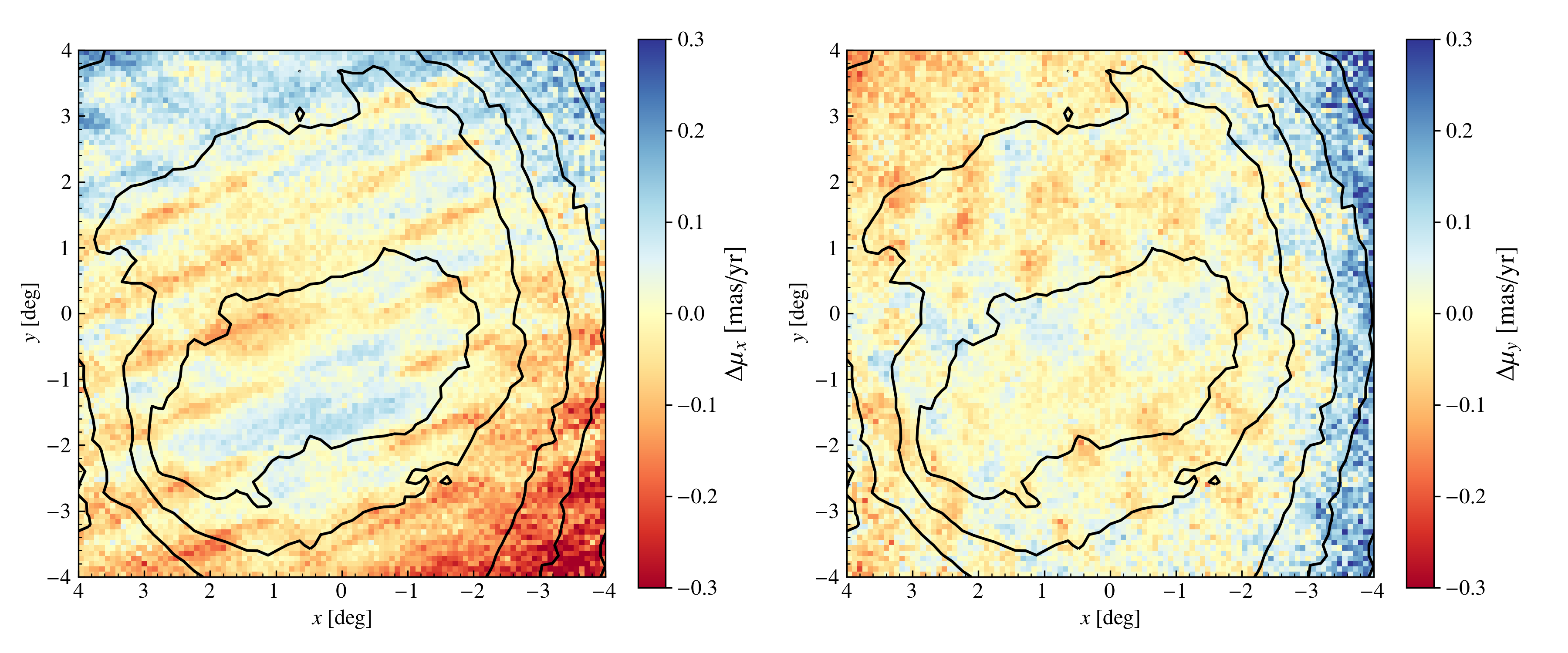}
\includegraphics[width=16cm,trim={0 0 0 0.5cm},clip]{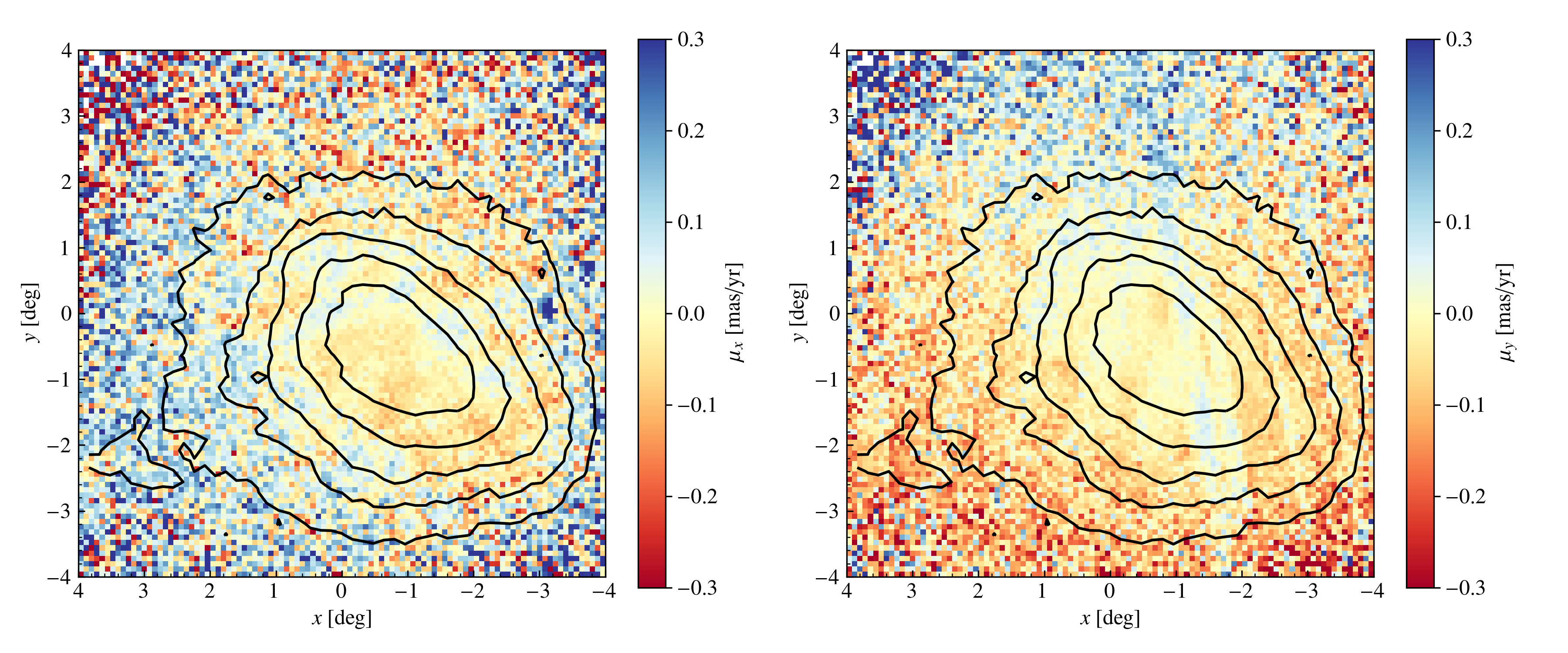}
\caption{Residual PM, after subtraction of a model PM field for the LMC (top) and the SMC (bottom). For both objects, the majority of the variation disappears, and the banding in PM is more clearly visible. Black density contours are spaced in the same way as in Figure~\ref{fig:LMCSMCPM}. For the LMC, the imprint of orbits on the bar can be seen as the bluer area on the lower side of the bar (as it appears in the plot) and the redder area on the upper side of the bar. The model that is subtracted is fit from sources within angular radii $\rho_{\rm max} = 3\degr$ of the centre for the LMC, and $\rho_{\rm max} = 2\degr$ of the centre for the SMC.} 
\label{fig:LMCSMCPMresid}
\end{figure*}

\subsection{Basic analysis}\label{sec:LMCSMCBasic}
In Figure~\ref{fig:LMCSMCPM} we show the median PM, in the $x,y$ coordinate system defined in Section~\ref{sec:data-method-MCs}, of sources that meet our membership criteria for the LMC and SMC, binned by position on the sky. This is a demonstration of the extraordinary precision of the \Gaia PM measurements (see also Figure~\ref{fig:LMCPMarrows}). The rotation signature is clearly visible in the LMC, and trends in the PMs of stars in the SMC are visible as well. Figure~\ref{fig:LMCSMCPM} also serves as a demonstration of the shortcomings of this data release. The banding or striping of the PMs that we discussed above, which
is associated with different scans and has been investigated by \cite{DR2-DPACP-51}, is also clearly visible. The parallaxes show this as well (Figure~\ref{fig:LMCSMCPar}).

We can characterise the trends seen in Figure~\ref{fig:LMCSMCPM}, to first order, by the central values and gradients. We calculated these either directly as a least-squares fit to the data or as a least-squares fit to the median PMs calculated in $0\fdg 04$-by-$0\fdg 04$ bins in $x$,$y$ plane. This latter fit was performed to reflect the fact that the most important errors in this analysis are systematic and depend on position on the sky, but it becomes less appropriate towards larger radii as Poisson noise becomes more important. The differences between the values derived using these two methods give a sense of the scale of the uncertainty associated with the position-dependent systematic errors.

In Tables~\ref{tab:LMCcircles}~and~\ref{tab:SMCcircles} we show the central values and gradients (Eqs.~\ref{eq:xyderivs}) for the PMs. We provide values for all sources within various angular radii (i.e. $\rho =\sin^{-1}(x^2+y^2)^{1/2}<\rho_{\rm max}$) and for annuli.
For both the LMC and SMC, we show the values of $i,\Omega,\omega$ that we found when we performed a least-squares fit to the derived gradients under the assumption that $v_z$ takes the value implied by the known values of the line-of-sight velocity and distance to the Clouds. The results are reasonably consistent with one another (allowing for the fact that we expect $\omega$ to decrease farther out).

For the LMC we also give the implied values of $v_z,i,\Omega,\omega$
when we place no constraint on the line-of-sight velocity. The value of
$v_z$ that this implies is of the order of
$1.3$-$2.0\;\rm{mas}\,\rm{yr}^{-1}$, which is similar to (but somewhat
higher than) the value $1.104 \pm 0.057\;\rm{mas}\,\rm{yr}^{-1}$ that
is expected given the measured line-of-sight velocity and distance of
the LMC. The effect of the line-of-sight velocity of the Clouds is to
produce a perspective shrinking of the Cloud on the sky (a
negative contribution to both ${\partial\mu_{x}}/{\partial x}$ and
${\partial\mu_{y}}/{\partial y}$ in Eq.~\ref{eq:xyderivs}), similar to
the effect seen in NGC6656 or NGC3201 (Figure~\ref{fig:globclpm}). The mismatch between the value derived from \Gaia
astrometry and that found from spectroscopy might be related to an actual
contraction of the LMC disc (similar to the apparent expansion of NGC6397
in Figure~\ref{fig:globclpm}). However, the orientation of the LMC disc
also plays an important role in these values, and the values of $i$
derived when $v_z$ for the LMC is fixed lie closer to those found in
photometric studies, which tend to be in the range $25-40\degr$.

For the SMC, directly inverting Eqs.~\ref{eq:xyderivs} gives line-of-sight velocities that are completely inconsistent with those measured from spectroscopy ($\sim$$\,-0.8$~mas~yr$^{-1}$ as opposed to $0.489 \pm 0.019$~mas~yr$^{-1}$). This may be due to the inadequacy of modelling the SMC as a flat disc, or a real expansion of the SMC (which, again, is degenerate with line-of-sight motion). However, forcing $v_z$ to take the value expected from the measured distance to the SMC and its line-of-sight velocity gives us a model that has a disc inclination $\sim 74\degr$, which is broadly similar to that measured for the Cepheid population \citep[$64\fdg 4\pm0\fdg 7$:][]{2015A&A...573A.135S}.

In Figure~\ref{fig:LMCSMCPMresid} we show the residual PMs after we subtracted off a gradient in PM corresponding to our first-order approximation, with the parameters $v_z,i,\Omega,\omega$ found for sources within angular radii $\rho_{\rm max} = 3\degr$ of the centre for the LMC, and $\rho_{\rm max} = 2\degr$ of the centre for the SMC. This shows the scale of the spatially correlated errors in PM more clearly, and in Figure~\ref{fig:LMC1Dvar} we show the variation in 1D stripes across the LMC (as well as the variation of the parallaxes), to allow an easier quantification. The residuals are comparable to those found in Section~\ref{sec:dwarfs_sys}. The residuals in the centre are rather small, but become larger far from the centre for the LMC, in the opposite sense to the variation from the median shown in Figure~\ref{fig:LMCSMCPM}. This is because of our assumption of constant $\omega$, which breaks down badly at large radii, as the rotation curve becomes flat. 

Figure~\ref{fig:LMCSMCPMresid} shows indications of the impact of the LMC bar on the kinematics of the disc. The residual PM near the upper (as we see it) side of the bar tend to be negative, while those on the lower side tend to be positive. This indicates that the stars are moving at faster-than-circular velocities at these points; this is consistent with stars belonging to the x1 orbit family, which is elongated along the bar \citep{1980A&A....92...33C}.

\subsection{Uncertainties and comparison to the literature}
The uncertainties on the measurement of the centre-of-mass motion of the LMC and SMC using \Gaia data are completely dominated by systematic, rather than random, uncertainties. The estimates of these quantities that we show in Tables~\ref{tab:LMCcircles}~and~\ref{tab:SMCcircles} are consistent to around the $10\,\mu{\rm as}$ level. This is smaller than the systematic uncertainty on PMs  calculated in section~\ref{sec:dw} or that on a large scale (of a few tens of degrees) derived from the PMs of quasars observed by \Gaia \citep{DR2-DPACP-51}, which is $\sim 28\,\mu{\rm as}\,\rm{yr}^{-1}$ in each component. There is no clear choice of the correct values to take from Tables~\ref{tab:LMCcircles}~and~\ref{tab:SMCcircles}. However, we can note that the approximation of constant angular velocity becomes poor for the LMC beyond about $3\degr$, and that the density of stars in the SMC is small beyond a similar radius, so this appears to be a sensible choice. We therefore adopted the values we found using all stars within these radii as our best estimates. These are $(\mu_{\alpha*,0,{\rm LMC}}, \mu_{\delta,0,{\rm LMC}})= (1.850\pm0.030,0.234\pm0.030) \,\rm{mas}\,\rm{yr}^{-1}$ and $(\mu_{\alpha*,0,{\rm SMC}}, \mu_{\delta,0,{\rm SMC}})= (0.797\pm0.030,-1.220\pm0.030)\, \rm{mas}\,\rm{yr}^{-1}$, where our uncertainty was estimated from the $\sim 10\,\mu{\rm as}\,\rm{yr}^{-1}$ variation listed in Tables~\ref{tab:LMCcircles}~and~\ref{tab:SMCcircles} and the $\sim 28\,\mu{\rm as}\,\rm{yr}^{-1}$ large-scale systematic uncertainty.

\cite{2013ApJ...764..161K} give an overview of recent estimates of the PMs of the LMC and SMC, including their own, found using HST three-epoch astrometry. Our estimates are consistent with theirs and with almost all of the values they cite, as well as with the values found by \cite{2016ApJ...832L..23V} using PMs found from the Tycho-\Gaia Astrometric Solution \citep{2016A&A...595A...2G,2016A&A...595A...4L}, which were $(1.872\pm 0.045, 0.224 \pm 0.054)\, \rm{mas}\,\rm{yr}^{-1}$ and $(0.874 \pm 0.066, -1.229 \pm 0.047)\, \rm{mas}\,\rm{yr}^{-1}$ for the LMC and SMC, respectively. 

It is worth noting that the measured centre-of-mass PM is dependent on the chosen (or derived) centre of the Clouds. The centre of the H\textsc{i} gas disc of the LMC, which we assume to be the dynamical centre of the Cloud, is close to, but not exactly the same as, the centre that was derived from HST PMs by \cite{2014ApJ...781..121V}, which was that assumed by \cite{2013ApJ...764..161K} and \cite{2016ApJ...832L..23V}. 

Instead of assuming that the stars' dynamical centres lie at the dynamical centre of the H\textsc{i} gas disc in each case, we could assume that they lie at the photometric centres of the Clouds at  $(\alpha_{C,LMC,{\rm phot}},\delta_{C,LMC,{\rm phot}}) = (81\fdg 28,-69\fdg 78)$ \citep{2001AJ....122.1827V} and  $(\alpha_{C,SMC,{\rm phot}},\delta_{C,SMC,{\rm phot}}) = (12\fdg 80,-73\fdg 15)$ \citep{2000A&A...358L...9C},
respectively. When we do so, we derive mean PMs ($\mu_{\alpha*,0,{\rm phot, LMC}}$, $\mu_{\delta,0,{\rm phot,  LMC}}$)$= (1.890, 0.314)\;\rm{mas}\,\rm{yr}^{-1}$  and $(\mu_{\alpha*,0,{\rm phot, SMC}}, \mu_{\delta,0,{\rm phot,  SMC}}) = (0.685,-1.230)\;\rm{mas}\,\rm{yr}^{-1}$. Therefore, this variation
is stronger than the variation due to the large-scale systematic uncertainties.

Our uncertainties, including systematics, are comparable with those of previous studies. The sheer number of sources spread across the Clouds for which \Gaia provides accurate PMs is extraordinary, and this will allow astronomers using DR2 to make a detailed mapping of the dynamics of the Magellanic Clouds.

\begin{figure}[t]
\centering
\includegraphics[width=9cm,trim={1cm 0.5cm 0 0.5cm},clip]{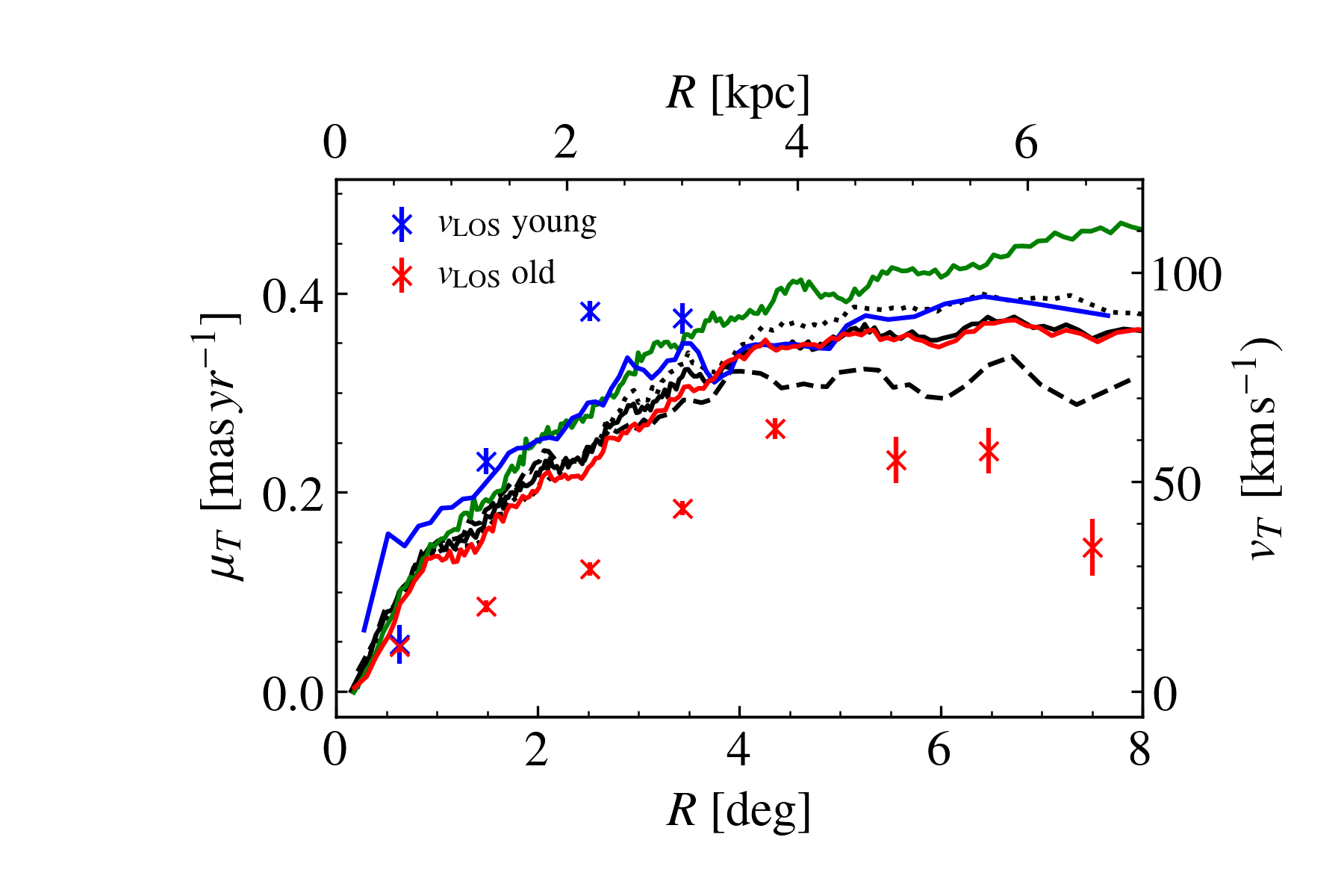}
\includegraphics[width=9cm,trim={1cm 0.5cm 0 0.5cm},clip]{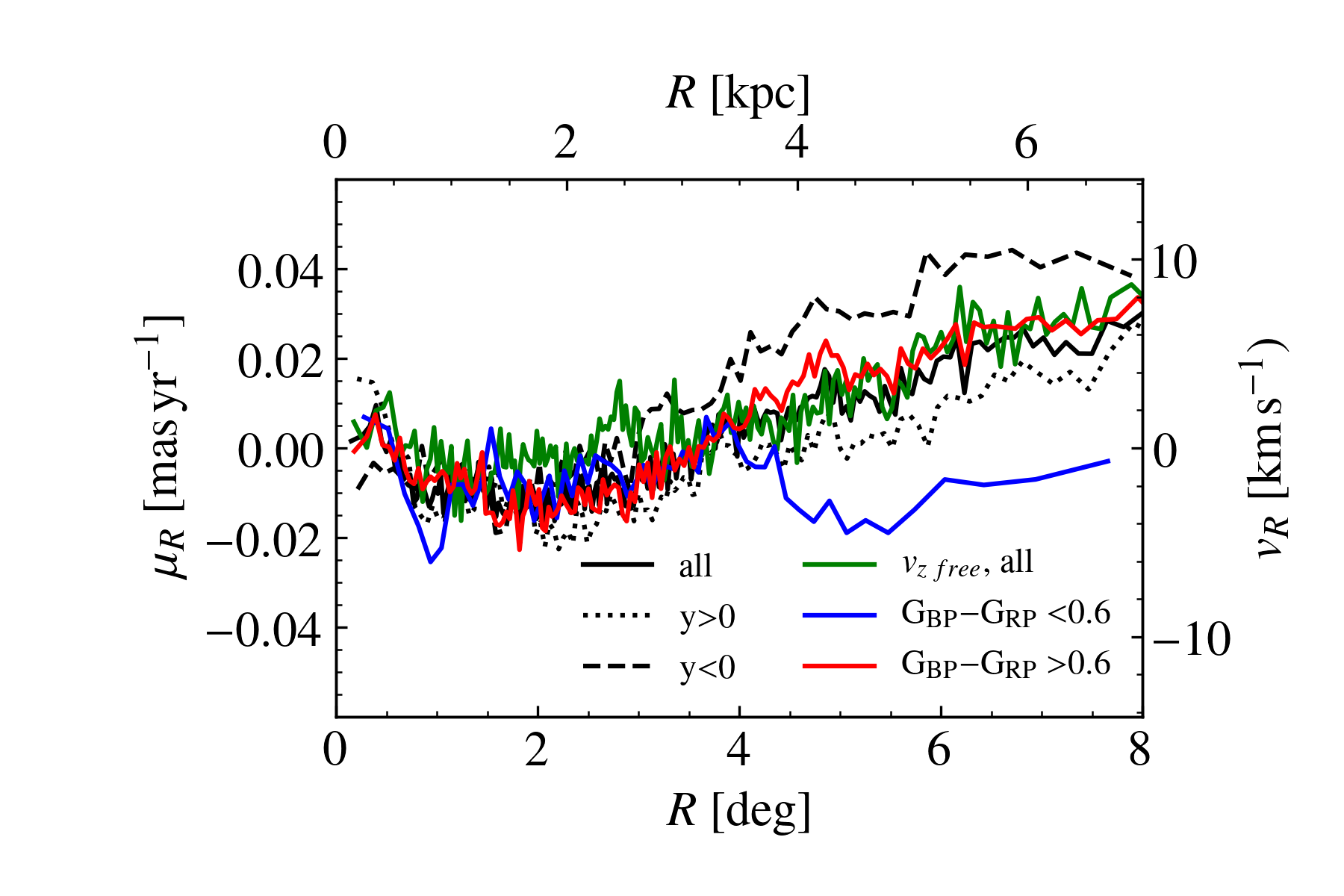}
\caption{Rotation curve (top) and median $v_R$ (bottom) of the LMC. The assumed values for the centre-of-mass velocity and orientation of the disc ($i$ and $\Omega$) are taken from a fit to all stars within angular radii $\rho<3\degr$ of the LMC centre. Angular distances and velocities given on the lower and left axes have been converted to real-space values  on the upper and right axes assuming a distance to the LMC of  $50.1{\rm kpc}$ \protect\citep{2001ApJ...553...47F}. The points shown in the upper panel are derived from observed line-of-sight velocities of old and young stars by \citet[][their Table 4.]{2014ApJ...781..121V} This figure, and its interpretation, has been changed from the originally published version following the erratum \cite{2020A&A...642..C1G}.}
\label{fig:LMCSMCRotCurve}
\end{figure}

\subsection{Rotation curve}

If we assume that we know the orientation of the LMC (or SMC), and that the motions we see are confined to the plane, it is possible to de-project the observed motions (minus the bulk motion) onto that plane. In Appendix~\ref{app:LMCSMC} we give the mathematical details of how this was performed. The SMC is less suitable for approximation as a simple flat rotating disc than the LMC, therefore we did not attempt this here. Tentative evidence that there is some sense of rotation of the SMC stars is provided by the consistent, but small, measurement of $\omega$ for different annuli of stars shown in Table~\ref{tab:SMCcircles}.

In Figure~\ref{fig:LMCSMCRotCurve} we show the resulting median tangential velocity, $v_T$ (the rotation curve) and median $v_R$ as a function of de-projected radius $R$ for the LMC (note that $R$, $v_R$ , and $v_T$ are de-projected position and velocity, i.e.~in the plane of the LMC), with
$v_x,v_y, i,\Omega$ as determined from a least-squares fit to the filtered data for angular radii $\rho<3\degr$, holding $v_z$ fixed. The
figures show the median value of $v_T$ (or $v_R$) as we increase $R$,
in non-overlapping bins of 40$\,$000 sources. In both cases we also
divide the sample into sources with $y>0$ and $y<0$ as a consistency
check. We also show the velocity curve derived from line-of-sight
velocities by \cite{2014ApJ...781..121V}, separately for young and old
stars. This shows that the precision of the PM rotation
curve from \Gaia is competitive with those in line-of-sight velocity
curves (which are derived from very many more
sources). The \Gaia data contain both old and young stars, therefore
it is expected that the \Gaia rotation curve lies between the two curves from
the old and the young populations.

The rotation curve for the LMC rises approximately linearly for $R\lesssim3\degr$ (which provides post hoc motivation for us to choose stars within a projected angular radius $\rho$ of $3\degr$ to determine the other parameters of the disc; the assumption of constant $\omega$ is reasonable over this radius). The dense coverage of the LMC provided by \Gaia allows us to resolve this rise in the rotation curve in a way that was not possible with the relatively sparse coverage provided by the HST \citep{2014ApJ...781..121V}. The scale over which the rise occurs is closer to that found for the old stellar population using line-of-sight velocities by \cite{2014ApJ...781..121V}.
We also show the rotation curve that we derived when we allowed $v_z$ to vary freely, which is broadly similar.

As a further sanity check, we divided the sources into blue and red bins, with the division at $G_{\rm BP}-G_{\rm RP} = 0.6$, and we plot the rotation curve in each case. As one would expect, the blue stars tend to have a higher rotation velocity out to about $4\degr$, by about $5-10\,\rm{km}\,\rm{s}^{-1}$, reflecting a lower asymmetric drift. 

The $v_R$ plot is rather harder to interpret. It is pleasingly close to $0$~km~s$^{-1}$ in the inner regions, and the variation we see at projected radii $< 2\degr$ may well be related to the effect of the bar. In the outer regions there is a small difference between the trends seen at positive and negative $y$-values, as in the rotation curve, and there is a small increase in $v_R$ for the redder stars. This might be due to non-equilibrium effects that are possibly caused by the past interaction of the LMC and SMC \citep[e.g.][]{2016ApJ...825...20B}, but further interpretation is beyond the scope of this paper.

\section{Orbital Integrations}
\label{sec:orbits}

In this section we present the results of the orbital integrations for
the globular clusters in our sample and for the dSph. We did not perform integrations for the Magellanic Clouds as this would require
consideration of dynamical friction, which introduces additional degrees of
freedom such as the total mass of each Cloud
\citep[e.g.][]{2013ApJ...764..161K}. For each object we integrated an
ensemble of 1000 orbits whose initial conditions were drawn via Monte
Carlo sampling the measurements and their uncertainties given in
Tables~\ref{tab:overview} and \ref{tab:PM_dwarfs} (i.e. we used the
full covariance matrix and assume Gaussian errors). We produced two sets of Monte Carlo samples,
considering only the random error on the observables, and considering
in addition a systematic error of amplitude 0.035 \masyr (as estimated
in Sec.~\ref{sec:dwarfs_sys}) for each PM component. We
then transformed these coordinates into Cartesian positions and
velocities. Because the \Gaia DR2 parallaxes (or distances)
have larger (systematic) uncertainties than the measurements
available in the literature (see Sec.~\ref{sec:gc} and \ref{sec:dw}),
we used published values from Harris10 and \cite{McConnachie12} for
the globular clusters and the dSph, respectively. The radial velocities were also taken from these databases. For the distance errors, we assumed an
uncertainty in the distance modulus of $\sim 0.05$~mag, which
corresponds to a relative distance error of $\sim 0.023$.

The initial conditions for the orbit integrations are listed in Tables
\ref{tab:gc_uvw} and \ref{tab:dw_uvw} for the globular clusters and
the dwarfs, respectively. In these Tables we give the
uncertainty of the positions and velocities derived from the 16th and
and 84th quantiles (which would be a 1$\sigma$ deviation if the
distributions were Gaussian), which were obtained by marginalising
over the other coordinates. However, as started earlier, we took into account the full covariance matrix
in the orbit integrations. We
considered three different Galactic potentials (labelled Model-1,
  -2, and -3) for the orbital integrations, and each of these
potentials corresponds to a model previously published in the
literature. Our goal was to understand how different their predictions
are, and also how robust the conclusions. In the near future, it
will be possible to use the data to constrain the model parameters and
to understand which model performs best, for example by imposing
self-consistency, but this exercise is beyond the scope of this paper.

Because each potential has its own set of characteristic parameters whose
values were derived by fitting different observables, they each
assume different values for the position of the Sun and its peculiar
velocity as well as for the motion of the local standard of rest. This
implies for example that the angular momentum of the dwarf galaxies
and globular clusters can be slightly different for the various potentials. 

\subsection{Description of the Galactic potentials}

The gravitational potential in Model-1 is axisymmetric, consisting of a
stellar bulge and discs (which have a combined stellar mass of
$5.4\times10^{10}\,{\rm M}_\odot$), two gas discs (with a combined
mass of $1.2\times10^{10}\,{\rm M}_\odot$), and a \citet[][NFW]{nfw1996} dark
matter halo. The virial mass is $1.37\times10^{12}\,{\rm
  M}_\odot$. The model was found using a Bayesian analysis of
kinematic tracers by \citet[cf. their Table 3]{McMillan2017}.

The Model-2 gravitational potential is axisymmetric, consists of a stellar
bulge and disc modelled as Miyamoto-Nagai potentials (which have a
combined stellar mass of $7.55\times10^{10}\,{\rm M}_\odot$), and a
spherical dark matter halo.  The mass at $R<200$ kpc is
$1.9\times10^{12}\,{\rm M}_\odot$. The model is the one of
\citet{Allen1991} but with revised parameters from \citet[][their Table~1]{Irrgang2013}.

The third potential we considered (Model-3) is the non-axisymmetric
mass model described in \citet{Robin2003,Robin2012}.  The properties of the
axisymmetric components (thin and thick stellar discs, stellar halo,
interstellar matter disc, dark matter halo) are described in
\citet{Robin2003}, while those of the rotating non-axisymmetric bar,
renormalised to have a total mass of $6.7\times 10^{9}\,{\rm
  M}_\odot$, are described in \citet{Robin2012}.  The mass of
the Galaxy for $R<100$ kpc is $ \sim 1.2 \times 10^{12}\,{\rm M}_\odot$. The
computation of the potential is described in \citet{Bienayme1987} for
the axisymmetric components and in Fernandez-Trincado et al. (2018, in prep) for
the bar and halo components.

The three potentials have very similar mass distributions between
$\sim 3$ and 40 kpc as measured by the circular velocity curves
(shown in Figure \ref{fig:circ_vel} of the appendix). However, the
potentials differ substantially both in the inner and in the
outer regions. This will lead to some differences in the orbits and
their characteristics, as we show below.

\begin{figure}[!h]
\centering
\includegraphics[width=8.5cm, trim={0 3cm 0 5cm},clip]{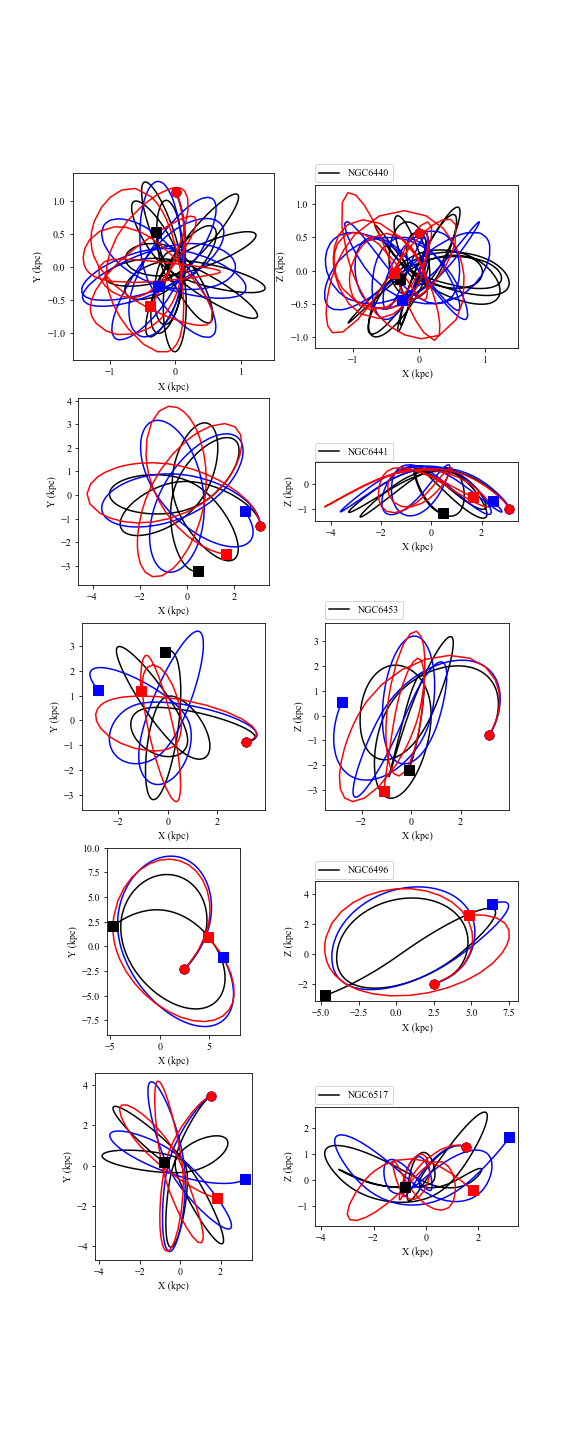}
\caption{Examples of the orbits of some of the globular clusters in
  our sample. The different colours correspond to the different
  potentials: Model-1 \citep[based on][in blue]{McMillan2017}, Model-2
    \citep[based on][in black]{Allen1991}, and Model-3 \citep[based
    on][in red]{Robin2003,Robin2012}. The orbits of clusters that remain
  in the inner few kpc are quite different for the various potentials,
  while as expected, the differences are much smaller for those that
  have pericentres greater than $\sim 2$ kpc (e.g. NGC6496, fourth
  row). In these cases, the location of streams, if present, can be
  predicted much more reliably.}
\label{fig:gc_orbit-examples}
\end{figure}
\begin{figure*}
\centering
\includegraphics[width=18cm, trim={0cm 3.5cm 0cm 6cm},clip]{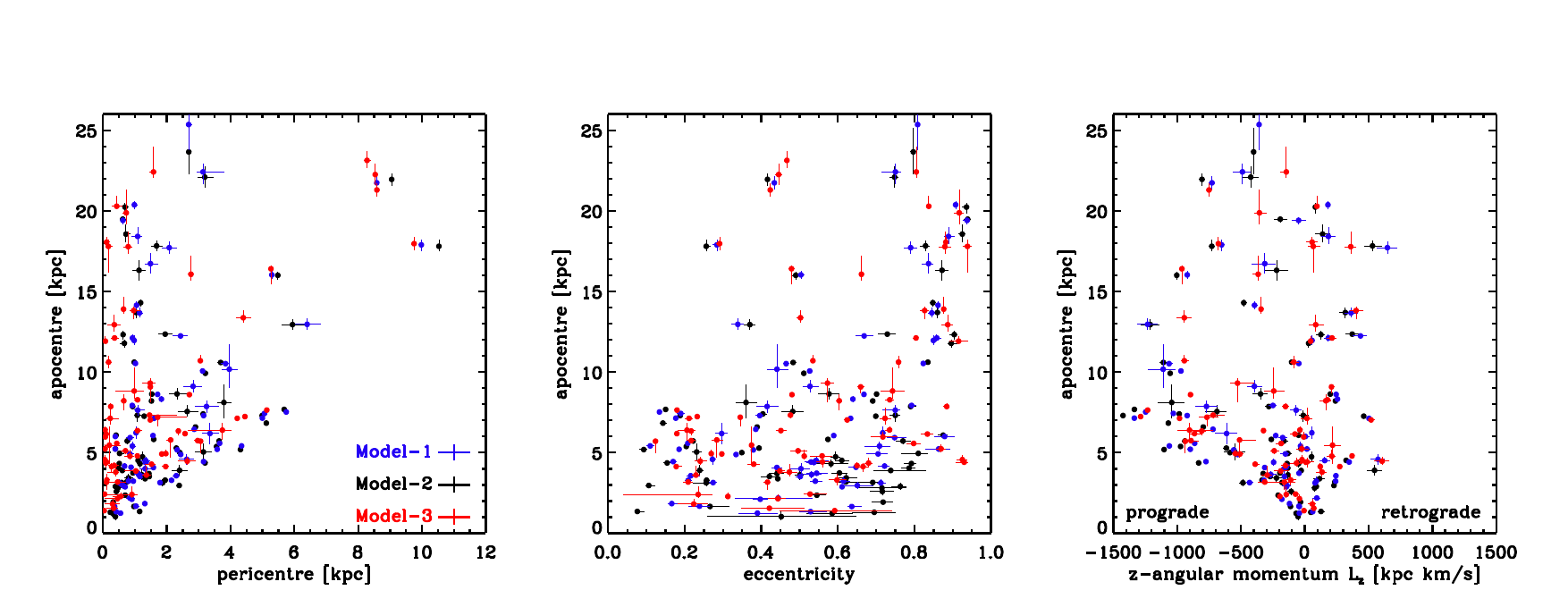}
\caption{Distribution of orbital parameters for the globular clusters
  in our sample. Three globular clusters are not shown in these plots:
  NGC~3201 and NGC~4590, which have apocentres $\sim 30$~kpc and large
  retrograde and prograde motions, respectively ($|L_z| >
  2000$~kpc~km~s$^{-1}$), and NGC~5466, which has an apocentre $\sim 50$~kpc
  \citep[given its orbital parameters, it was likely associated with the
  Sagittarius dwarf, as proposed by][]{2003AJ....125..188B}. We note the
  concentration of clusters with small pericentre and apocentre around
  $L_z \sim 0$ for Model-3. Such clusters are on box orbits
  in the barred potential of Model-3, and hence do not conserve $L_z$, whose
  time average over the 10 Gyr of integration is approximately
  zero. The error bars correspond to the 16th and 84th quantiles
  obtained from the orbit integration using the statistical
  errors. The effect of the systematic error on the orbits of the
  globular clusters is found to be negligible and is not shown here.}
\label{fig:gc_orbit_props}
\end{figure*}

\subsection{Results for the globular clusters}

Fig.~\ref{fig:gc_orbit-examples} shows some examples of orbits for the
globular clusters integrated backward in time for 0.25 Gyr. For all the
clusters, the orbits are very similar initially (at least for one
orbital period), but they then begin to diverge, reflecting the
differences in the Galactic potentials that were used (see also Fig.~\ref{fig:app_orbits_clusters}). The differences for clusters that penetrate the regions dominated by
the bar are particularly large; this is modelled as a non-axisymmetric component in Model-3 (in red),
but not in Models 1 and 2 (in blue and black, respectively). Furthermore, some clusters
appear to be on resonant orbits, and interestingly, as shown in the
second panel of Fig.~\ref{fig:gc_orbit-examples} for NGC 6441, this is
true for the three potentials. A quick exploration reveals that
several other clusters appear to be on similar types of resonant
orbits. Some clusters, on the other hand, seem to be on chaotic orbits,
but further analysis is required to establish this reliably.

The distribution of some of the orbital parameters for the globular
clusters in our sample is shown in Fig.~\ref{fig:gc_orbit_props} and
summarised in Table~\ref{tab:orb_prop_gc}. In these plots, the solid
circles show the median value of the time averages over 10 Gyr of
integration, while the error bars correspond to the 16th and 84th
percentiles derived using only the statistical (and not the
systematic) errors on the observables.  These figures show that the
orbits of globular clusters in our sample are very centrally
concentrated, as most have their orbital apocentres within 10 kpc. Most
of our clusters are on prograde orbits, and a small fraction are
retrograde. We recall that this sample is focused on the inner 20~kpc, so we
cannot establish with this dataset whether this is also the case for
the outer halo clusters. The clusters with the highest $L_z$ are
NGC~3201 (retrograde) and NGC~4590 (prograde) are not shown in this
figure, nor is NCG~5466 because of its large apocentre ($\sim
50$~kpc).

In Fig.~\ref{fig:gc_orbit_props} the eccentricity is defined as the
time-average of $(r_{apo} - r_{peri})/(r_{apo} + r_{peri})$. The
eccentricity distribution is extended and rather uniform with many
clusters on relatively radial orbits. The apparent trend that clusters
with larger apocentres are on more radial orbits may be due to our
sample selection (currently located within $\sim 20$~kpc from the Sun). A
cluster with a large apocentre will almost only be included in our sample
if it has a relatively radial orbit. It will be possible to draw more
reliable conclusions about the distribution of orbital parameters when
the whole globular cluster population has been analysed. Nonetheless,
Fig.~\ref{fig:gc_orbit_props} already highlights that the greater
differences in the orbital properties for the various potentials arise for clusters with apocentres in approximately the inner 5 kpc. These differences are in many cases larger than the
error bars. This implies that there is room for improvement by
performing a self-consistent dynamical model of the globular cluster
population and the mass distribution in our Galaxy \citep[e.g.][]{2017MNRAS.467.2446B}.

\subsection{Results for the dwarf galaxies}

\begin{figure*}
\centering
\includegraphics[width=19cm, trim={6cm 3cm 0cm 0cm},clip]{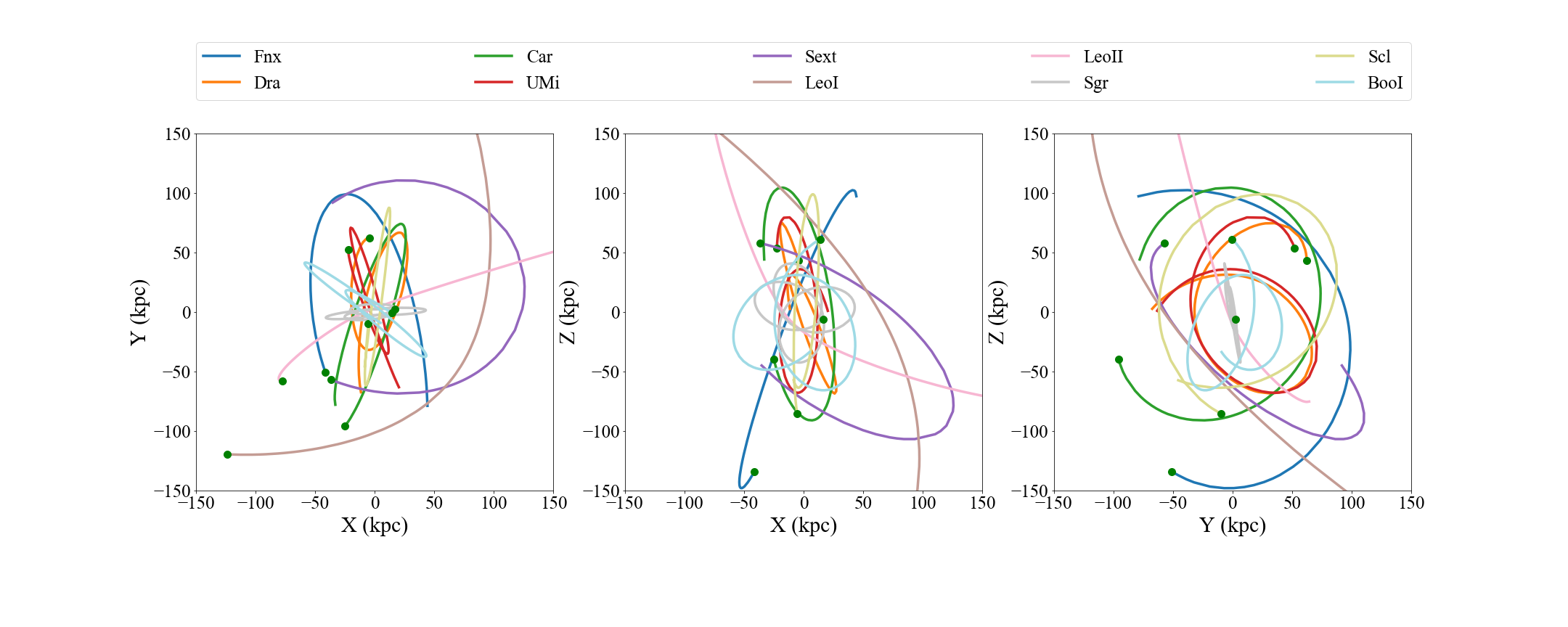}
\caption{Orbits integrated backward in time for 2.5 Gyr for the different dwarfs shown
  in different colours using the potential of Model-2.}
\label{fig:orbits-dw}
\end{figure*}

In Fig.~\ref{fig:orbits-dw} we plot the orbits of the different dwarf
(spheroidal) galaxies. This figure reveals for example that the orbits
of Draco and Ursa Minor look similar, and that the orbital planes of
most dSph are different, with the orbit of Sagittarius, for instance, being orthogonal to
those of Draco and Ursa Minor.

\begin{figure*}
\centering
\includegraphics[width=18cm, trim={0cm 3.6cm 0cm 7.1cm},clip]{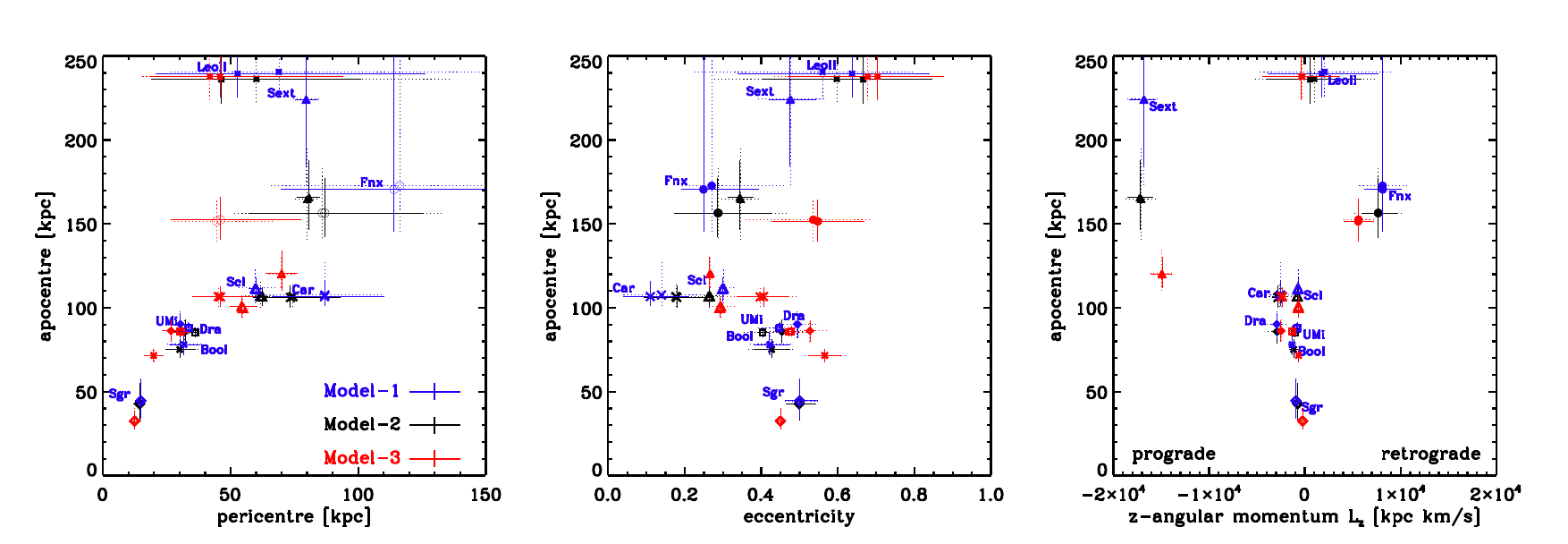}
\caption{Distribution of orbital parameters for the dSph. The
  different colours indicate computations with different
  potentials. The agreement is generally good, with Model-3
  systematically leading to smaller pericentres as a result of its
  higher mass at the radii probed by the systems. Larger differences
  are found for the more distant objects, revealing the sensitivity of
  their orbits to variations in the assumed mass distributions for the
  Galaxy. The dSph eccentricity distribution differs from that shown
  in Fig.~\ref{fig:gc_orbit_props} for the globular clusters in our
  sample. The symbols with solid error bars correspond to the median
  and uncertainties derived using the Monte Carlo realisations, and
  those with dotted error bars also take a systematic error of $0.035$
  \masyr on each of the PM components into account.}
\label{fig:dw_orbit_props}
\end{figure*}

Fig.~\ref{fig:dw_orbit_props} shows the distribution of orbital
parameters for the dSph (and listed in
Table~\ref{tab:orb_prop_dw}). We have plotted here the median values
(over 10 Gyr of integration) and the 16th and 84th percentile range
as the symbols with solid error bars (derived from the 1000 Monte
Carlo realisations of the observables and their uncertainties). Using
the same symbols, but now with dotted lines, we have plotted the
orbital parameters derived including the effect of the 0.035~\masyr
systematic uncertainty on each of the PM components. In
general, the effects of the systematic errors on the
characteristic properties of the orbits are relatively small. We note,
however, that in both cases these error bars do not always properly
reflect the uncertainties on the orbital parameters because of
degeneracies. We include examples of the Monte Carlo realisations in
the appendix (Figure~\ref{fig:app-orb-MCs}) to give examples
of these
degeneracies.

Fig.~\ref{fig:dw_orbit_props} reveals that most satellites are on (slightly) 
prograde orbits, while Fornax is retrograde (and possibly Leo~II
as well,
although it is also consistent with prograde at the 1$\sigma$ level),
as can be seen from the rightmost panel. This is qualitatively similar to what
we found for the globular clusters. However, the orbital
eccentricity distribution for the dSph (middle panel) is very
different from that of the clusters. Fewer dwarfs have very elongated
orbits as their eccentricity is typically lower than 0.6. Carina
even has a median eccentricity $\lesssim 0.2$ for Models 1 and 2.

This finding leads to two interesting preliminary conclusions. Firstly, there is a weak link at most between the globular clusters in our
sample and the dSph (although this is partly driven by our selection
of the globular cluster sample). Secondly, the eccentricity
distribution of the dwarfs is inconsistent with the predictions of
cosmological simulations, where satellites are expected to be on
rather radial orbits \citep[e.g.][]{2014MNRAS.437..959B}.

Fig.~\ref{fig:dw_orbit_props} also confirms that Draco and Ursa Minor
have very similar orbital properties and hence possibly constitute a
physically connected group. Bootes~I appears close to these objects in
all panels of this figure, but as we show below, the orientation of
its angular momentum differs by $\sim 140$~deg.

In Fig.~\ref{fig:dw_orbit_props} we have not plotted Leo~I. This is
because Leo~I has extreme orbital characteristics, and 
is unbound in roughly 20\% of the realisations for 
  Model-1\footnote{This is because this model has a lower dark matter
  halo mass. For example, the escape velocity from the location of the
  Sun for Model-2 is 812~km~s$^{-1}$, which may be compared to
  the value of $\sim 533$~\kms derived by \citet{2014A&A...562A..91P}
  from RAVE data and is consistent with that of Model-1.}, for example. In the
cases in which a bound orbit is found, the predicted median apocentres
are 819, 429, and 388 for Models 1, 2, and 3,
respectively. Evidently, the apocentre of Leo~I is beyond the likely
virial radius of the Milky Way \citep[estimated to be smaller than
300~kpc, see][and references therein]{2016ARA&A..54..529B}. The orbit
of Leo~I is quite eccentric, with medians in the range 0.6 to 0.8. The
radial period is greater than 5 Gyr and its estimate varies by a
factor of two for the various potentials, but a robust prediction is that
the last pericentric passage (at a distance of $\sim 100$~kpc) took
place approximately 1~Gyr ago.

\begin{figure}
\centering
\includegraphics[width=8.4cm, trim={0.9cm 0 1cm 8cm},clip]{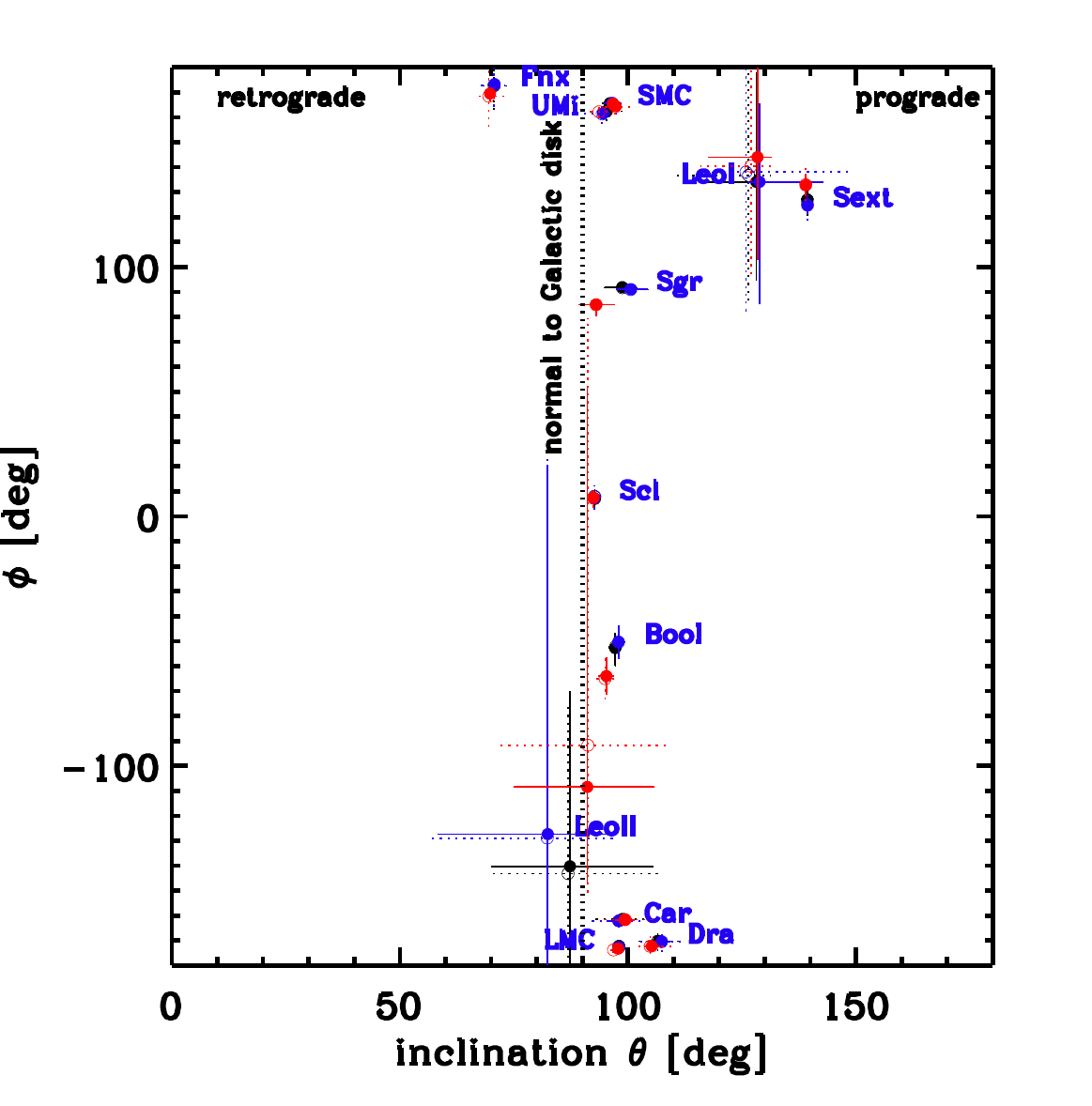}
\includegraphics[width=9.2cm, trim={0cm 0cm 0cm 18cm}, clip]{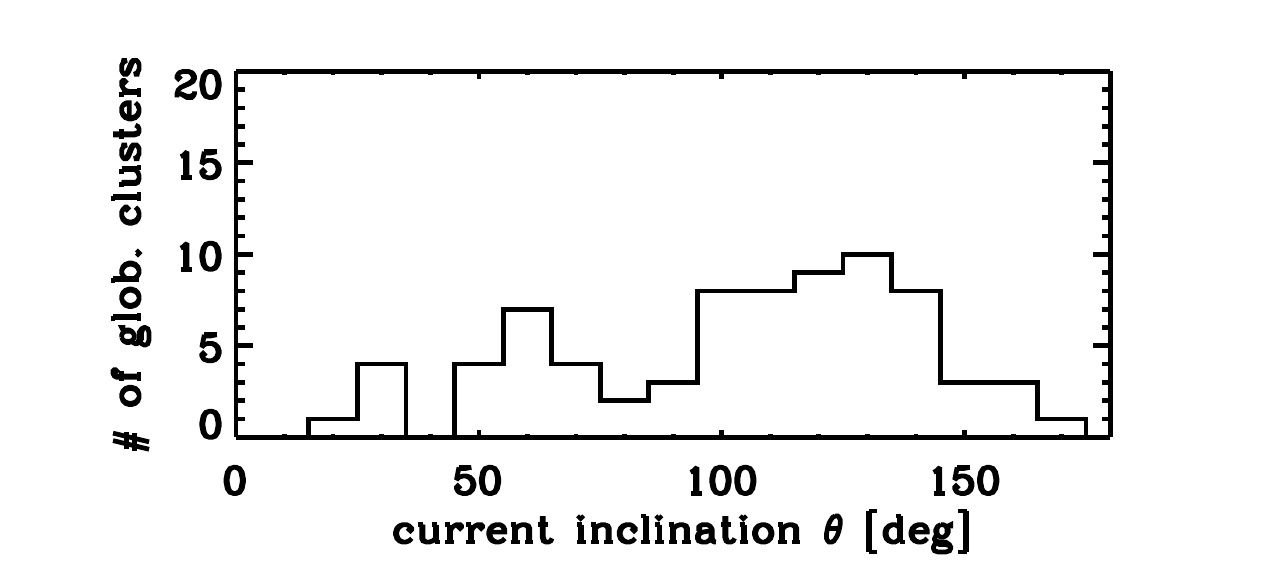}
\caption{{\it Top}: Orientation of the time-averaged orbital plane,
  defined by the angular momentum angles $(\phi,\theta)$ for the
  different dwarfs for the various potentials. The colours and error bars are the same as in
  Fig.~\ref{fig:dw_orbit_props}, and open and solid circles are the
  median values obtained by including or excluding our estimate of the systematic
  error on the PMs. Most dwarfs
  have highly inclined orbits with respect to the Galactic plane
  (i.e. $\theta \sim 90^\deg$). Their variation in orientation (angle
  $\phi$) over the 10~Gyr of integration is much smaller than the size
  of the error bars. {\it Bottom}: Histogram showing the present-day
  inclination of the orbits for the globular clusters.}
\label{fig:orientation-dw}
\end{figure}

The orientations of the orbital planes of the dwarfs, now including
the LMC and SMC\footnote{We include the Magellanic Clouds here because
  the orientation of their angular momenta has likely been less
  affected by dynamical friction than the other orbital parameters.}, averaged over the 10 Gyr of integration, are shown in the top
panel of Fig.~\ref{fig:orientation-dw}. It has been suggested that the
Milky Way dwarf galaxy satellites lie on a plane
\citep{2005A&A...431..517K}. We find that their orbits tend to be
perpendicular to the Galactic disc (the majority cluster at an
inclination of $\sim 90\pm20$~deg) but span a broad range of
orientations. This implies that even though the orientation of the
average plane of motion may be similar, they may rotate in the
opposite sense, such as Sculptor and Ursa Minor together with Draco, the LMC and SMC (in the YZ
plane). When we alternatively compare Sculptor and Sagittarius, they move in planes that are
nearly perpendicular to each other (and to the Galactic disc). This
ordered complexity might indicate (group) infall from a preferential
direction (from a cosmic web filament aligned with the $z$-axis), but it
seems  to disfavour one single event as the cause of these configurations.

For the globular clusters we plot only the distribution of the
current, instantaneous orbital plane inclination, defined as $\cos{\theta}
= L_z/|{\mathbf L}|$, since the angular momentum $|{\mathbf L}|$ is not
conserved for the majority of the clusters in our sample (for the 
barred potential of Model-3,  $L_z$ changes as well). The bottom panel of
Fig.~\ref{fig:orientation-dw} shows that the globular clusters in our sample
have a much broader distribution of inclinations than the dwarf
galaxies. This behaviour might again reflect a bias in the
selection of the sample of globular clusters, however, for example in the sense that there are fewer of the outer
clusters that might have been associated with dwarf galaxies.

\section{Discussion}
\label{sec:discussion}

\subsection{Brief summary of the systematics}

The analysis performed in this paper has served to highlight the
excellent quality of the \Gaia DR2, and also to pinpoint its
limitations. On this second aspect, we have in particular confirmed a
systematic offset in the parallax of sources
\citep[see][]{DR2-DPACP-39,DR2-DPACP-51}, which we find has an
amplitude of $\sim -0.049$~mas when averaged over all dSph, while
averaged over the globular clusters, it is $\sim -0.025$~mas. This
offset varies in amplitude between the locations on the sky, which
explains the difference found between our samples of dSph and globular
clusters. No such offset has been found in the PMs, for
example when studying open clusters \citep{DR2-DPACP-39}, from
which we conclude that if such a systematic is present, it is of small
amplitude. However, we do find that because of local variations
(driven by the non-uniform scanning of the sky), the PMs
might have an additional systematic uncertainty of $\sim
0.035$~mas~yr$^{-1}$ in each direction.

\subsection{Exemplifying the data quality}

\subsubsection{Galactic satellites}

We were able to determine the PMs of all 75
globular clusters in our sample reliably (with a significance far greater than
10$\sigma$). The effect of a systematic floor noise level
\citep[see][]{DR2-DPACP-39,DR2-DPACP-51} is negligible for the
clusters in our sample. This is also the case for the majority of the
dwarf galaxies we have analysed, possibly with the exception of Leo I
(because of its large distance), and Leo II (because it is small on
the sky). Nonetheless, even this systematic uncertainty is typically
smaller than the uncertainties of previously reported measurements of
the PM of globular clusters and dSph in the Milky Way halo.

In the case of the dSph, the possibility of selecting members now via
PM has revealed (and confirmed previous indications of) high
spatial asymmetries, and
possibly also tidal features in several systems. We did not analyse the significance of
these features, nor have we attempted to establish the presence of
tidal tails beyond the field of view of 2 deg that we have chosen to 
measure their mean PM. We expect the
community to explore this area with \Gaia DR2, especially now that the PMs are so accurate that reliable predictions can be made for the orbits
of these systems, and indications can be obtained on where streams, if present, might be expected. 
\begin{figure}
\centering
\includegraphics[width=8.5cm]{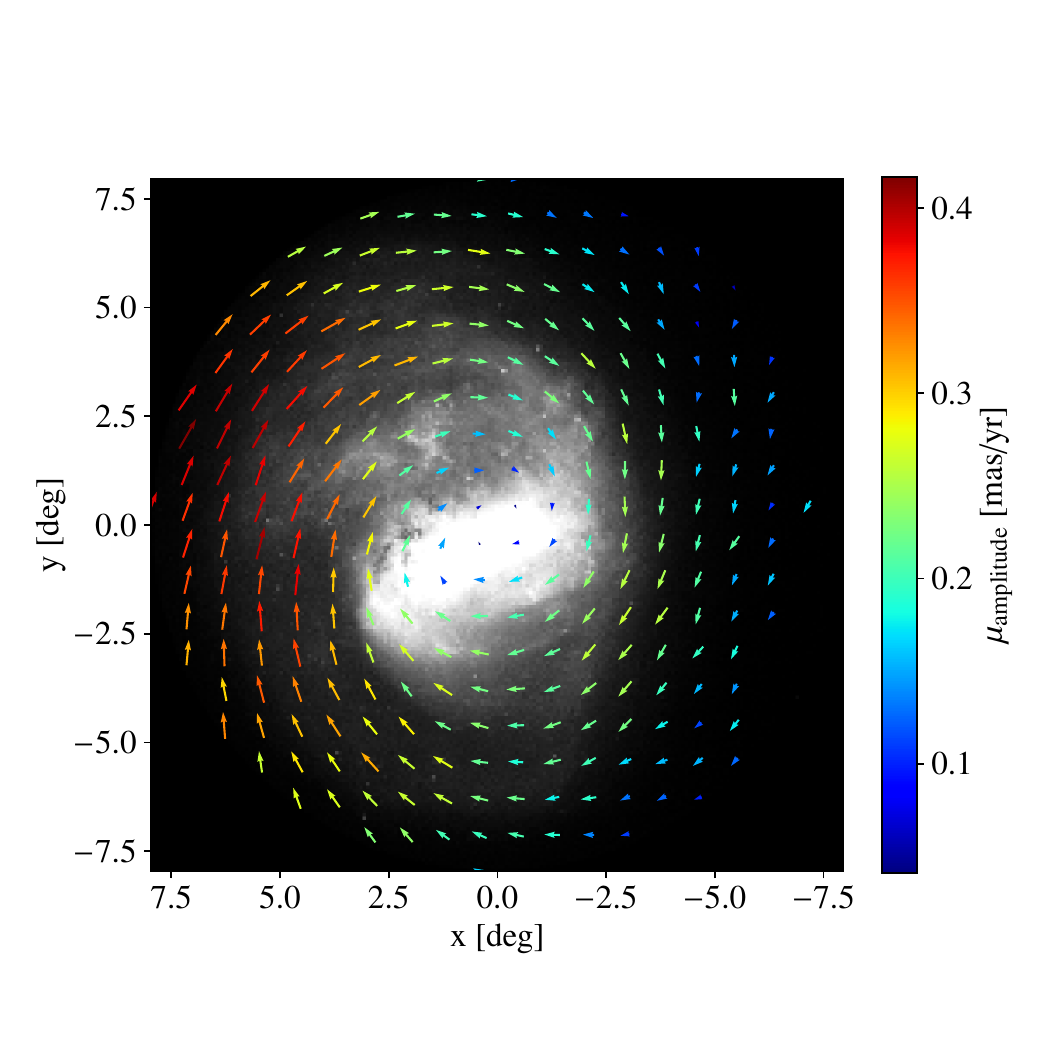}
\caption{PMs of stars in the LMC, represented as vectors,
  overlaid on a representation of the source density. This 
  figure shows the clear consistent rotation measured by \Gaia around
  the centre of the LMC.}
\label{fig:LMCPMarrows}
\end{figure}

The improved PMs for millions of stars in the Magellanic
Clouds offer a unique dataset for understanding the internal dynamics
of these systems.  For example, Fig.~\ref{fig:LMCPMarrows} shows the
velocity map obtained for the LMC, and reveals a high degree of order in
the rotational motion of this system. This has allowed us to derive a
rotation curve based on tangential velocities that is competitive to
that obtained using radial velocity information. However, we also find
indications in the PM residuals (after subtraction of a model of a rotating inclined disc), of streaming motion along the bar. Not only
do we learn about internal structure and mass distribution of the
Clouds, but this dataset will also help us in understanding how and
when the two galaxies interacted, and whether and how this is related
to the Magellanic stream \citep[e.g.][]{2006ApJ...652.1213K}. It might even be possible to find stars stripped from the Clouds
at much larger distances than previously attempted.

This brief discussion on the impressive astrometric quality of
  the \Gaia DR2 datasets for the Galactic satellites calls for high-precision radial velocity measurements and abundances
  follow-up for as many of the members identified by \Gaia
  as possible \citep[as planned e.g. by the WEAVE and 4MOST projects, see
  e.g.][]{2017arXiv170808884F}. Lists of possible members according to
  our analyses are given in Table~\ref{tab:members} for the globular
  clusters, dSph, and UFD galaxies, and for the Magellanic Clouds.
  
\subsubsection{Substructure and debris}
\label{sec:helmi-stream}

The methods we used thus far has relied on the objects
of interest being concentrated in a specific location on the sky.
Tidally torn satellites and streams, on the other hand, may extend
across great parts of the celestial sphere. Here we briefly
demonstrate the capacity of \Gaia DR2 to also investigate this type of
structure in the Milky Way using astrometric data alone. A further and
deeper analysis is left to the general users of \Gaia DR2.

One known substructure in the halo near the Sun that presumably
originated in a disrupted satellite was discovered by
\cite{1999Natur.402...53H}, defining a clump in the $L_z$ versus
$|L_\perp| = \sqrt{L_x^2+L_y^2}$ space. The physical reality of this
structure, sometimes called Helmi's stream, was later
independently confirmed by
e.g. \cite{2000AJ....119.2843C,2009MNRAS.399.1223S}, but no
significant additional members have been identified since then. \Gaia
DR2 will likely reveal a very large number of new members, but
measuring $L_z$ and $L_\perp$ without radial velocity can only be done
exactly in two small areas on the sky: in the directions of the
Galactic centre and anticentre. In this case, $\mu_l$ and $\mu_b$
translate directly into the space velocities $v_z$ and $v_\phi$ with
knowledge of the distance $D$.  The angular-momentum components are
then simply $L_z = x v_\phi$ and $L_\perp = L_y \sim -x v_z$, where $x
= D+R_\odot$, and $R_\odot$ is the Galactocentric distance of the Sun.

\begin{figure}[!h]
\centering
\includegraphics[width=8.5cm]{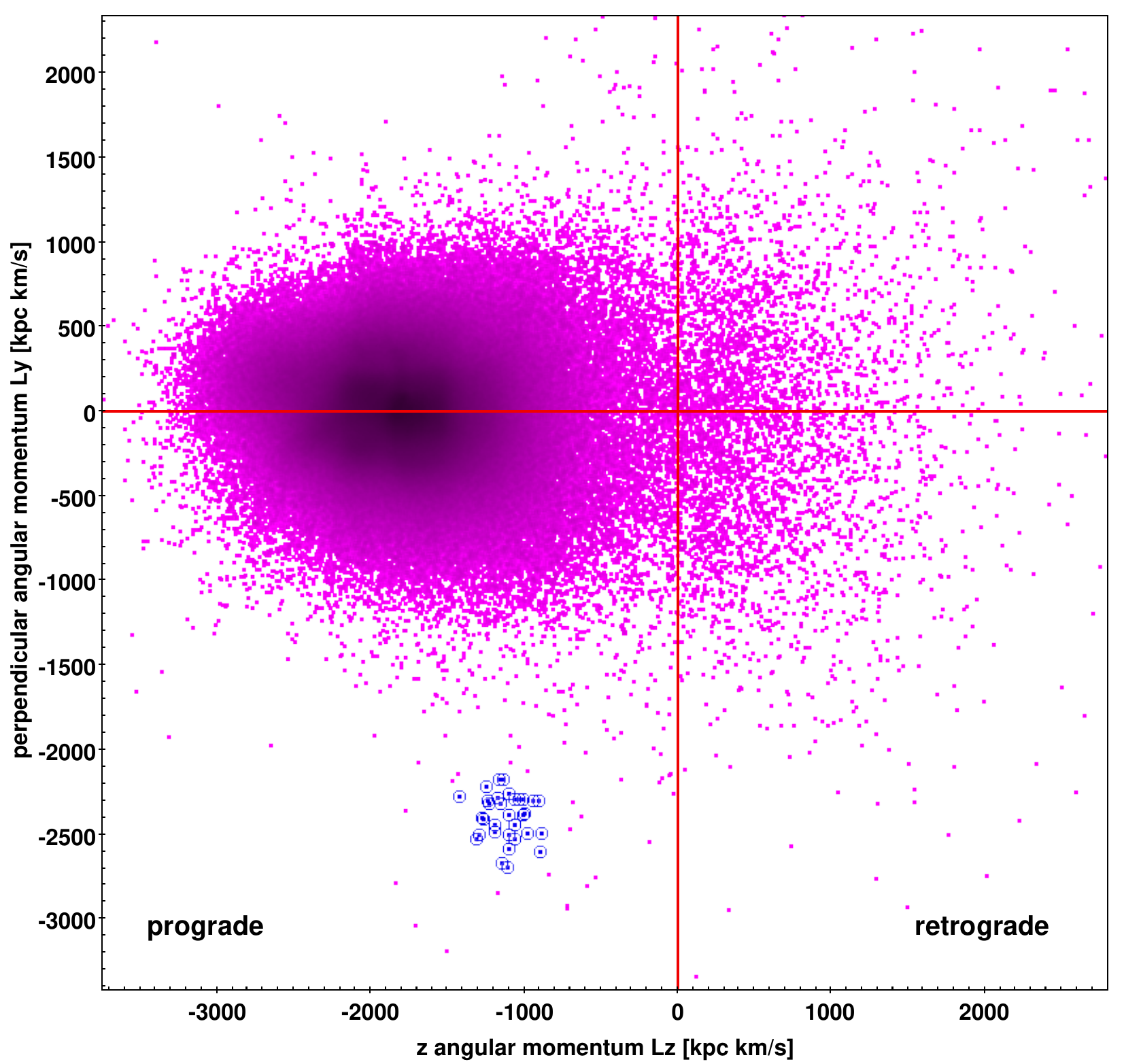}
\caption{Distribution of DR2 stars in a circle of 15\,degrees radius
  around the Galactic anticentre direction in the $L_\perp = L_y$ vs
  $L_z$ space. The Helmi stream is the distinct density enhancement
  near the bottom of the plot (in blue). We have assumed here $v_{rot}
  = 220$~\kms for the LSR velocity, $v_{z,\odot}$=5~km~s$^{-1}$, and
  $v_{\phi,\odot}$=7~\kms for the peculiar motion of the Sun, and
  $R_\odot = 8.1$~kpc. }
\label{fig:helmi-stream-Lz}
\end{figure}

\begin{figure}[!h]
\centering
\includegraphics[width=8.5cm]{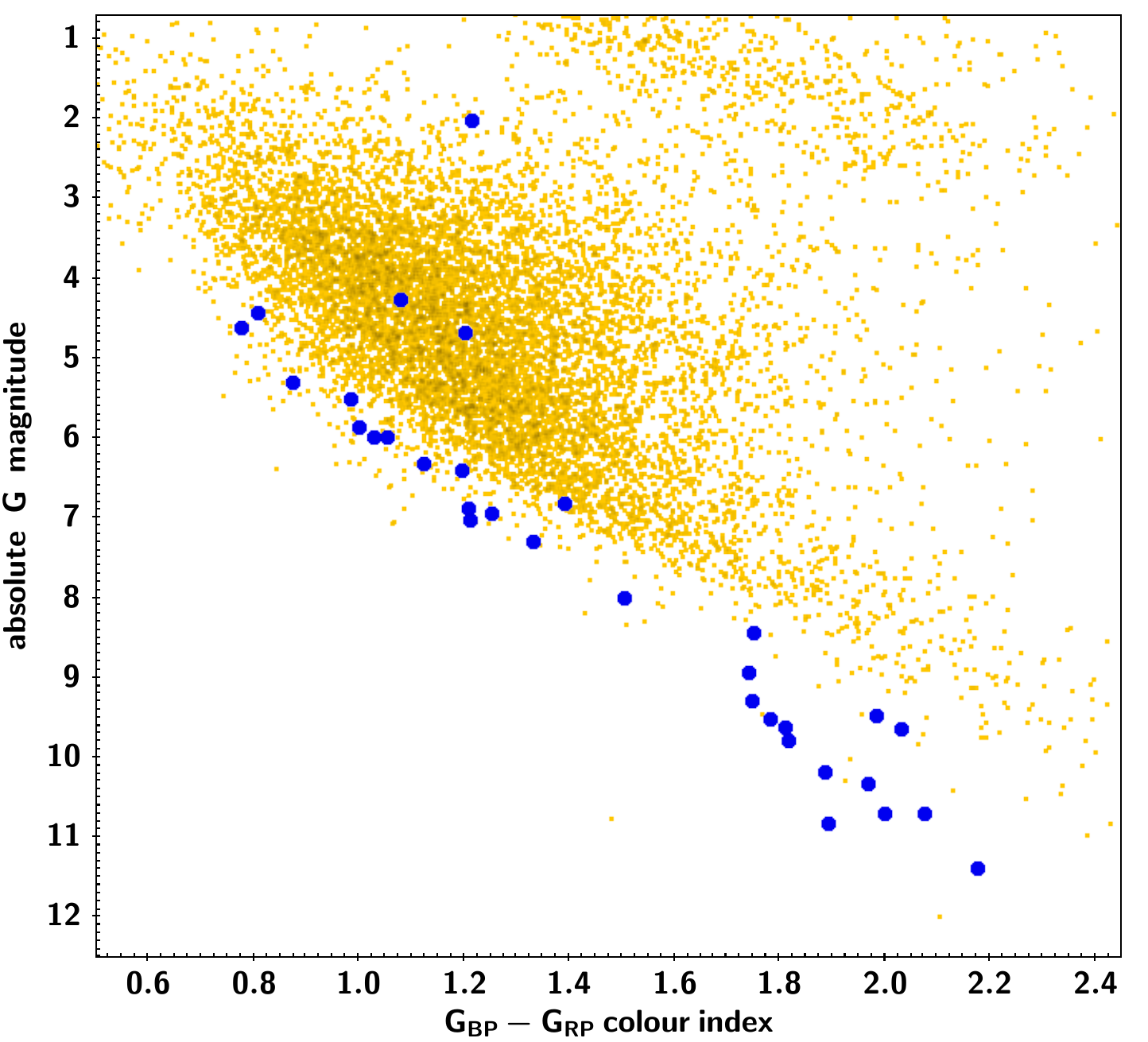}
\caption{Hertzprung-Russell diagram using \Gaia DR2 $G$\,magnitudes, colours, and parallaxes in the direction of the Galactic
  anticentre. The yellowish cloud represents a random sample of disc
  stars, selected on the basis of their disc-like $L_z$ and $L_\perp$,
  while the blue dots are the 32 new members of the Helmi stream.}
\label{fig:helmi-stream-HR}
\end{figure}

We show the distribution of stars within a circle of 15\,degrees
radius around the Galactic anticentre in the $L_z$ vs. $L_y$ space in
Fig.~\ref{fig:helmi-stream-Lz}. We considered here only stars with
$\varpi/\sigma_\varpi/ > 5$. In addition to the dominant disc centred on
$(-1800,0)$~kpc km~s$^{-1}$ and the more diffuse halo centred on
(0,0), stars are distinctly concentrated around
$(-1100,-2400)$~kpc km~s$^{-1}$, corresponding to the expected
location of the Helmi stream. A tight cut around the centre of this
clump gives 32 candidate members (with distances from 260\,pc to about
2\,kpc), more than tripling the original number reported in
\cite{1999Natur.402...53H}.  A\ Hertzsprung-Russel (HR) diagram of these 32 members using \Gaia DR2 $G$ magnitudes, parallaxes, and $G_{\rm BP}$--$G_{\rm RP}$ colours is shown in Fig.~\ref{fig:helmi-stream-HR}. Despite
the patchy extinction in this area of the sky, the HR diagram reveals
an old, metal-poor main sequence, offset from a corresponding disc
sequence by about 0.2 mag towards the blue. This
figure shows four probable binaries among the 32 new members,
and the three subgiants indicate a turnoff at an absolute magnitude
$M_G \sim4.5$. A straight extrapolation of the 32 new members from the
15-degree circle to the whole sky would give close to 2000 members in
all of DR2. These can in principle be identified using radial
velocity information as well.

Fig.~\ref{fig:helmi-stream-Lz} shows a clear over-density only for $L_y
\propto v_z < 0$, and not for $v_z > 0$ (at fixed $L_z$). This
stronger asymmetry in the direction of the anticentre than originally
reported by \cite{1999Natur.402...53H} (where the ratio was 3:1) may
be used to place constraints on the accretion time and suggests that the
merger may have taken place even more recently than argued by
e.g. \citet{2007AJ....134.1579K}.

\subsection{Implications for the formation and dynamics} 

The unprecedentedly accurate PMs derived for the Galactic satellites
from the \Gaia DR2 data will allow determining the mass distribution
of the Milky Way well into the realms of the dark matter halo. They
will enable breaking the degeneracy between the slope of the mass
density profile and the orbital anisotropy \citep{1999MNRAS.310..645W,
  2010MNRAS.406..264W}, the latter being the limiting factor thus far,
which \Gaia DR2 has turned into an observable.  Interestingly, our
measurements indicate that the orbits of the dSph are not very radial,
and this appears to challenge expectations derived from cosmological
simulations in the $\Lambda$CDM framework
\citep[e.g.][]{2010MNRAS.406.2312L,2017MNRAS.468L..41C}. Their
relatively low eccentricity ($r_{apo}/r_{peri} < 4$) and
non-penetrating orbits ($r_{peri} > 30$ kpc) may also disfavour models
for the transition of dwarf irregulars into dSph via a tidal-stirring
mechanism \citep{2017ApJ...836L..13K}.

We have found relatively small
differences in the orbits in three different realistic Galactic
potentials 
 when we integrated over short timescales \citep[the
potentials are based on][]{Allen1991,Robin2003,McMillan2017}. This is the case for
globular clusters that do not probe the inner few kpc of the Milky Way
and for the dwarf galaxies. This implies that the orbit-forecasting
power is high, and it might therefore be used to search for tidal
tails particularly for the globular clusters, both through their
predicted location on the sky but also using PM information
that is now available thanks to \Gaia DR2. This is a particularly
interesting avenue because of the very high constraining power of
streams on the mass distribution in our Galaxy \citep[see][and references therein]{2016ASSL..420..169J}.

Another interesting constraint comes from the most distant satellite
Leo~I. This object has most recently been used to derive a limit on
the mass of the dark matter halo of the Milky Way on the basis of the
so-called timing argument by \cite{2013ApJ...768..139S}. From our own
measurements of the PM of Leo~I, we find that it is barely
bound, with its orbit extending well beyond recent estimates of the virial
radius of our Galaxy. Although it is very unlikely that Leo~I is unbound
\citep{2013ApJ...768..140B}, we may use it to derive a
lower limit to the mass of the Milky Way if we assume it has the
escape velocity. Neglecting the contribution of the disc(s) and bulge,
for a (non-truncated) NFW halo,
\begin{equation}
M_{MW}(r_{Leo I}) = \frac{v_{Leo I}^2 r_{Leo I}}{2 G} \frac{\log(1+x_{LeoI}) - x_{LeoI}/(1+x_{LeoI})}{\log(1 + x_{LeoI})},
\end{equation}
with $x_{LeoI} = r_{LeoI}/r_s$. 
When we use our estimates of $v_{Leo I} \sim 217.3^{+62.6}_{ -48.9}$~\kms
and $r_{Leo I} \sim 257.8^{+16.8}_{-35.1}$~kpc, this implies a lower limit for the enclosed mass of the Milky Way of $M_{MW}(r_{Leo I}) =
9.1^{+6.2}_{-2.6} \times 10^{11}\,\sm$ assuming $r_s =
18.6$~kpc as in \cite{McMillan2017}, where the error bars indicate
the 16th and 84th percentiles. When we let this parameter vary in the
range $10 \le r_s \le 30$~kpc, the estimate of
the lower limit to this enclosed mass varies by $\sim 15\%$, that
is, within
the uncertainties bracketed by the measurement errors. This value is
in line with previous work \citep[see review by][and references
therein]{2016ARA&A..54..529B} and does not preclude a light dark
matter halo for our Galaxy \citep[of $\sim 10^{12}\sm$,
e.g.][]{2005MNRAS.364..433B,2014MNRAS.445.3788G}.

Leo~I is an intriguing object because it is very distant and its
velocity indicates that it is receding from us.  A possible explanation for
 a system such as Leo~I is that it has experienced a
three-body interaction with the Magellanic Clouds
\citep{2007MNRAS.379.1475S}. Although we do not explore this
possibility here, we find that the phase-space
distribution of dwarf galaxies is not homogeneous. These satellites tend to have orbits
with angular momenta perpendicular to the Galactic disc, meaning
that their
orbits take place in planes with varying orientations, but always with
high inclination. This supports the idea of filamentary infall
\citep[see][]{2005MNRAS.363..146L}, and might also imply that some of the
dSph in our sample have fallen in together as a group
\citep{1995MNRAS.275..429L,2008MNRAS.385.1365L}. For example, the
relative distances of Ursa Minor and Draco remain relatively small
when computed over our 10 Gyr long orbital integrations, thus favouring
some amount of group infall.

Our measurements rule out that the dwarf galaxies are on one single
narrow ``disc'', an idea that has led to an important debate in the
literature.  Furthermore, although some objects are on the same plane,
they rotate in a different sense around the Galaxy, making it less
likely that they formed in single event such as a major merger
\citep[][this scenario would also require that their eccentricities be
relatively high, in contrast to what we find from \Gaia
DR2 data]{2014MNRAS.442.2419Y}. On the other hand, our measurements will
finally enable exploring how the satellite population was
put in place and how this relates to the environment of the Milky Way
on a firm basis \citep{2015MNRAS.452.1052L}.

Future studies of the orbital properties of the more distant globular
clusters will allow us to establish their relation to the present-day
dwarf galaxy population \citep[and further test the ideas put
forward in][]{1995MNRAS.275..429L}. On the other hand, some of the
globular clusters in our sample may have been associated with
long-gone accreted galaxies, whose debris we might expect will be
discovered and characterised using \Gaia DR2 data. Together, this will
shed light on the build-up of the globular cluster population and its
link to former and current satellite galaxies.

\section{Conclusions}
\label{sec:concl}

The second data release from the \Gaia mission has delivered its
promise of new and accurate PM measurements for a billion
stars across the full Galaxy and its nearest neighbours. Our analysis
of the PMs of stars in roughly half of the Galactic
globular clusters, in all the known dSph galaxies, and in the
Magellanic Clouds has allowed us to derive their mean motions as they
orbit the Galaxy much more precisely than ever before, despite
systematics that are clearly present in this data release. The simple
analyses carried out in this paper have confirmed, and also revealed
for many globular clusters, previously reported internal dynamical
complexity (such as rotation and the presence of extended halos). The
PMs, and hence the orbits, of the dSph have finally been
pinned-down, and this has uncovered their relatively coherent
phase-space distribution, which is not consistent with a single
``disc of satellites'', however. The astounding dynamical maps of the Magellanic
clouds contain such richness that is has even been possible to derive for the first time a high quality
rotation curve for
the
Large Magellanic Cloud based on tangential velocities alone and to unveil the
dynamical imprint of the bar.

Much remains to be understood and discovered from
the PMs that have become available with \Gaia DR2. This
dataset will undoubtedly keep the Galactic astronomy community busy
for many years to come.

\begin{acknowledgements}

This work presents results from the European Space Agency (ESA) space mission {\it Gaia}. \Gaia data are being processed by the \Gaia Data Processing and Analysis Consortium (DPAC). Funding for the DPAC is provided by national institutions, in particular the institutions participating in the \Gaia\ MultiLateral Agreement (MLA). The \Gaia mission website is \url{https://www.cosmos.esa.int/gaia}. The \Gaia archive website is \url{https://archives.esac.esa.int/gaia}.

The \Gaia\ mission and data processing have financially been supported, in alphabetical order by country, by
the Algerian Centre de Recherche en Astronomie, Astrophysique et G\'{e}ophysique of Bouzareah Observatory;
the Austrian Fonds zur F\"{o}rderung der wissenschaftlichen Forschung (FWF) Hertha Firnberg Programme through grants T359, P20046, and P23737;
the BELgian federal Science Policy Office (BELSPO) through various PROgramme de D\'eveloppement d'Exp\'eriences scientifiques (PRODEX) grants and the Polish Academy of Sciences - Fonds Wetenschappelijk Onderzoek through grant VS.091.16N;
the Brazil-France exchange programmes Funda\c{c}\~{a}o de Amparo \`{a} Pesquisa do Estado de S\~{a}o Paulo (FAPESP) and Coordena\c{c}\~{a}o de Aperfeicoamento de Pessoal de N\'{\i}vel Superior (CAPES)-Comit\'{e} Fran\c{c}ais d'Evaluation de la Coop\'{e}ration Universitaire et Scientifique avec le Br\'{e}sil (COFECUB);
the Chilean Direcci\'{o}n de Gesti\'{o}n de la Investigaci\'{o}n (DGI) at the University of Antofagasta and the Comit\'e Mixto ESO-Chile;
the National Science Foundation of China (NSFC) through grants 11573054 and 11703065;  
the Czech Republic Ministry of Education, Youth, and Sports through grant LG 15010, the Czech Space Office through ESA PECS contract 98058, and Charles University Prague through grant PRIMUS/SCI/17;    
the Danish Ministry of Science;
the Estonian Ministry of Education and Research through grant IUT40-1;
the European Commission’s Sixth Framework Programme through the European Leadership in Space Astrometry (\url{https://www.cosmos.esa.int/web/gaia/elsa-rtn-programme}, ELSA) Marie Curie Research Training Network (MRTN-CT-2006-033481), through Marie Curie project PIOF-GA-2009-255267 (Space AsteroSeismology \& RR Lyrae stars, SAS-RRL), and through a Marie Curie Transfer-of-Knowledge (ToK) fellowship (MTKD-CT-2004-014188); the European Commission's Seventh Framework Programme through grant FP7-606740 (FP7-SPACE-2013-1) for the \Gaia\ European Network for Improved data User Services (\url{https://gaia.ub.edu/twiki/do/view/GENIUS/WebHome}, GENIUS) and through grant 264895 for the \Gaia\ Research for European Astronomy Training (\url{https://www.cosmos.esa.int/web/gaia/great-programme}, GREAT-ITN) network;
the European Research Council (ERC) through grants 320360 and 647208 and through the European Union’s Horizon 2020 research and innovation programme through grants 670519 (Mixing and Angular Momentum tranSport of massIvE stars -- MAMSIE) and 687378 (Small Bodies: Near and Far);
the European Science Foundation (ESF), in the framework of the \Gaia\ Research for European Astronomy Training Research Network Programme (\url{https://www.cosmos.esa.int/web/gaia/great-programme}, GREAT-ESF);
the European Space Agency (ESA) in the framework of the \Gaia\ project, through the Plan for European Cooperating States (PECS) programme through grants for Slovenia, through contracts C98090 and 4000106398/12/NL/KML for Hungary, and through contract 4000115263/15/NL/IB for Germany;
the European Union (EU) through a European Regional Development Fund (ERDF) for Galicia, Spain;    
the Academy of Finland and the Magnus Ehrnrooth Foundation;
the French Centre National de la Recherche Scientifique (CNRS) through action 'D\'efi MASTODONS', the Centre National d'Etudes Spatiales (CNES), the L'Agence Nationale de la Recherche (ANR) 'Investissements d'avenir' Initiatives D’EXcellence (IDEX) programme Paris Sciences et Lettres (PSL$\ast$) through grant ANR-10-IDEX-0001-02, the ANR 'D\'{e}fi de tous les savoirs' (DS10) programme through grant ANR-15-CE31-0007 for project 'Modelling the Milky Way in the Gaia era' (MOD4Gaia), the R\'egion Aquitaine, the Universit\'e de Bordeaux, and the Utinam Institute of the Universit\'e de Franche-Comt\'e, supported by the R\'egion de Franche-Comt\'e and the Institut des Sciences de l'Univers (INSU);
the German Aerospace Agency (Deutsches Zentrum f\"{u}r Luft- und Raumfahrt e.V., DLR) through grants 50QG0501, 50QG0601, 50QG0602, 50QG0701, 50QG0901, 50QG1001, 50QG1101, 50QG1401, 50QG1402, 50QG1403, and 50QG1404 and the Centre for Information Services and High Performance Computing (ZIH) at the Technische Universit\"{a}t (TU) Dresden for generous allocations of computer time;
the Hungarian Academy of Sciences through the Lend\"ulet Programme LP2014-17 and the J\'anos Bolyai Research Scholarship (L.~Moln\'ar and E.~Plachy) and the Hungarian National Research, Development, and Innovation Office through grants NKFIH K-115709, PD-116175, and PD-121203;
the Science Foundation Ireland (SFI) through a Royal Society - SFI University Research Fellowship (M.~Fraser);
the Israel Science Foundation (ISF) through grant 848/16;
the Agenzia Spaziale Italiana (ASI) through contracts I/037/08/0, I/058/10/0, 2014-025-R.0, and 2014-025-R.1.2015 to the Italian Istituto Nazionale di Astrofisica (INAF), contract 2014-049-R.0/1/2 to INAF dedicated to the Space Science Data Centre (SSDC, formerly known as the ASI Sciece Data Centre, ASDC), and contracts I/008/10/0, 2013/030/I.0, 2013-030-I.0.1-2015, and 2016-17-I.0 to the Aerospace Logistics Technology Engineering Company (ALTEC S.p.A.), and INAF;
the Netherlands Organisation for Scientific Research (NWO) through grant NWO-M-614.061.414 and through a VICI grant (A.~Helmi) and the Netherlands Research School for Astronomy (NOVA);
the Polish National Science Centre through HARMONIA grant 2015/18/M/ST9/00544 and ETIUDA grants 2016/20/S/ST9/00162 and 2016/20/T/ST9/00170;
the Portugese Funda\c{c}\~ao para a Ci\^{e}ncia e a Tecnologia (FCT) through grant SFRH/BPD/74697/2010; the Strategic Programmes UID/FIS/00099/2013 for CENTRA and UID/EEA/00066/2013 for UNINOVA;
the Slovenian Research Agency through grant P1-0188;
 the Spanish Ministry of Economy (MINECO/FEDER, UE) through grants ESP2014-55996-C2-1-R, ESP2014-55996-C2-2-R, ESP2016-80079-C2-1-R, and ESP2016-80079-C2-2-R, the Spanish Ministerio de Econom\'{\i}a, Industria y Competitividad through grant AyA2014-55216, the Spanish Ministerio de Educaci\'{o}n, Cultura y Deporte (MECD) through grant FPU16/03827, the Institute of Cosmos Sciences University of Barcelona (ICCUB, Unidad de Excelencia 'Mar\'{\i}a de Maeztu') through grant MDM-2014-0369, the Xunta de Galicia and the Centros Singulares de Investigaci\'{o}n de Galicia for the period 2016-2019 through the Centro de Investigaci\'{o}n en Tecnolog\'{\i}as de la Informaci\'{o}n y las Comunicaciones (CITIC), the Red Espa\~{n}ola de Supercomputaci\'{o}n (RES) computer resources at MareNostrum, and the Barcelona Supercomputing Centre - Centro Nacional de Supercomputaci\'{o}n (BSC-CNS) through activities AECT-2016-1-0006, AECT-2016-2-0013, AECT-2016-3-0011, and AECT-2017-1-0020;
the Swedish National Space Board (SNSB/Rymdstyrelsen);
the Swiss State Secretariat for Education, Research, and Innovation through the ESA PRODEX programme, the Mesures d’Accompagnement, the Swiss Activit\'es Nationales Compl\'ementaires, and the Swiss National Science Foundation;
the United Kingdom Rutherford Appleton Laboratory, the United Kingdom Science and Technology Facilities Council (STFC) through grant ST/L006553/1, the United Kingdom Space Agency (UKSA) through grant ST/N000641/1 and ST/N001117/1, as well as a Particle Physics and Astronomy Research Council Grant PP/C503703/1.

We are very grateful to the referee, Mario Mateo, for his
  prompt, very positive, and constructive report. 

We are grateful to Dana Casetti-Dinescu, Mike Boylan-Kolchin, Gary Mamon and Eero Vaher for suggesting various small corrections after this article went to print. 
\end{acknowledgements}

%
%

\bibliographystyle{aa} 

\appendix
\section{Additional descriptions}
\begin{figure*}[!ht]
\begin{center}
\includegraphics[totalheight=4.2cm]{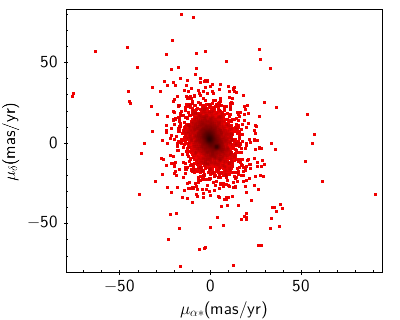}
\includegraphics[totalheight=4.2cm]{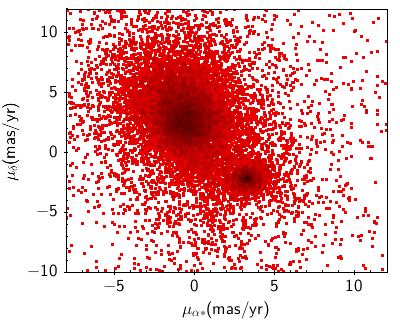}
\includegraphics[totalheight=4.2cm]{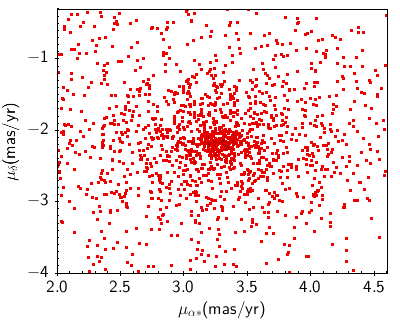}
\caption{PM of stars within 0.5 deg.\ from the centre of
  NGC2298 as extracted from the GACS archive. {\it Left:} All $\sim 259000$ 
  stars. {\it Middle:} Zoom-in on the PM field, which
  shows a concentration of the stars near position $(3.2, -2.2)$
  mas~yr$^{-1}$ corresponding to cluster members. {\it Right:} Another zoom-in on the PM field for the cluster.}
\label{fig:sequstep1}
\end{center}
\end{figure*}
\begin{figure*}
\begin{center}
\includegraphics[totalheight=4.2cm]{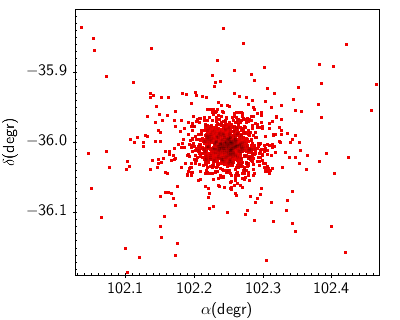}
\includegraphics[totalheight=4.2cm]{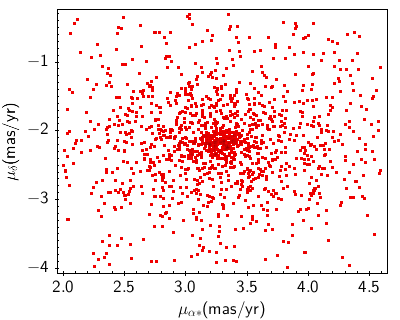}
\includegraphics[totalheight=4.2cm]{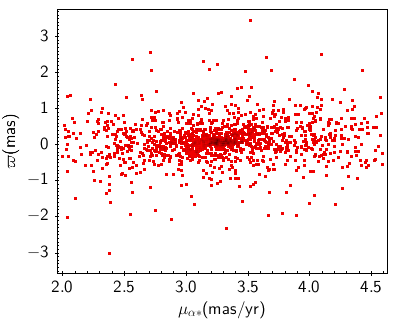}
\caption{From left to right: Distribution of positions on the sky for the first selection on cluster star PMs, showing the cluster in the centre, PM distribution for the cluster field, and distribution of parallax vs PM in RA.}
\label{fig:sequstep3a}
\end{center}
\end{figure*}

\subsection{Example of data extraction}
\label{app:sequence}

To illustrate the different stages of the data extraction, we show
here an example following the procedure used for globular cluster
NGC~2298.

\begin{figure}[!h]
\begin{center}
\includegraphics[width=4.3cm]{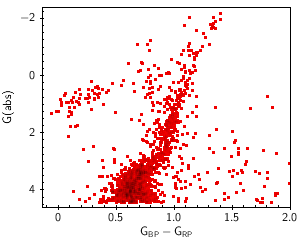}
\includegraphics[width=4.3cm]{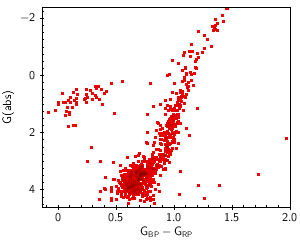}
\caption{HR diagram for the cluster field before (left) and after (right) the astrometric solution. }
\label{fig:sequstep4b}
\end{center}
\end{figure}

A first selection was made in a field with a radius of about 1~degree
around the cluster centre as given in Harris10. The PMs for $\sim 259000$
stars in that field are shown in the left panel of
Fig.~\ref{fig:sequstep1}.  A zoom-in on the diagram (middle panel)
shows a small concentration in the PM field near position
$(3.2, -2.2)$~mas~yr$^{-1}$, which is likely to be the cluster. No
other such concentrations are apparent in the same field. A further
zoom-in on the concentration (right panel of Fig.~\ref{fig:sequstep1})
shows it to be real. The data in this field are selected as a new
subset, for which the distribution of positions on the sky are shown
in the left panel of Fig.~\ref{fig:sequstep3a}, which reveals the
cluster more clearly. The PM distribution for this new
coverage is also improved, showing much reduced field star
contamination (middle panel). The distribution over parallaxes
now also clearly shows the cluster with little contamination (right
panel). Finally, for the pre-selection procedure, the colour-magnitude
diagram was created for the selection (Fig.~\ref{fig:sequstep4b}),
which now contains about 1500 stars. The selected data thus obtained
were saved as a CSV file for further analysis through the cluster
analysis software.

The analysis software iterates over an astrometric solution for the
cluster, in which the parallax and PM of the cluster are
solved for. The cluster centre is determined based on the mean
position of all selected members, without any weighting being
applied. First the cluster stars positions are projected on a
tangential plane with the assumed cluster centre as zero point. The
mean of the Cartesian positions is determined, which then is
de-projected to provide the new cluster centre on the sky. The
relevant equations have been presented in \cite{2017A&A...601A..19G} .

\begin{figure*}[ht]
\begin{center}
\includegraphics[width=10cm]{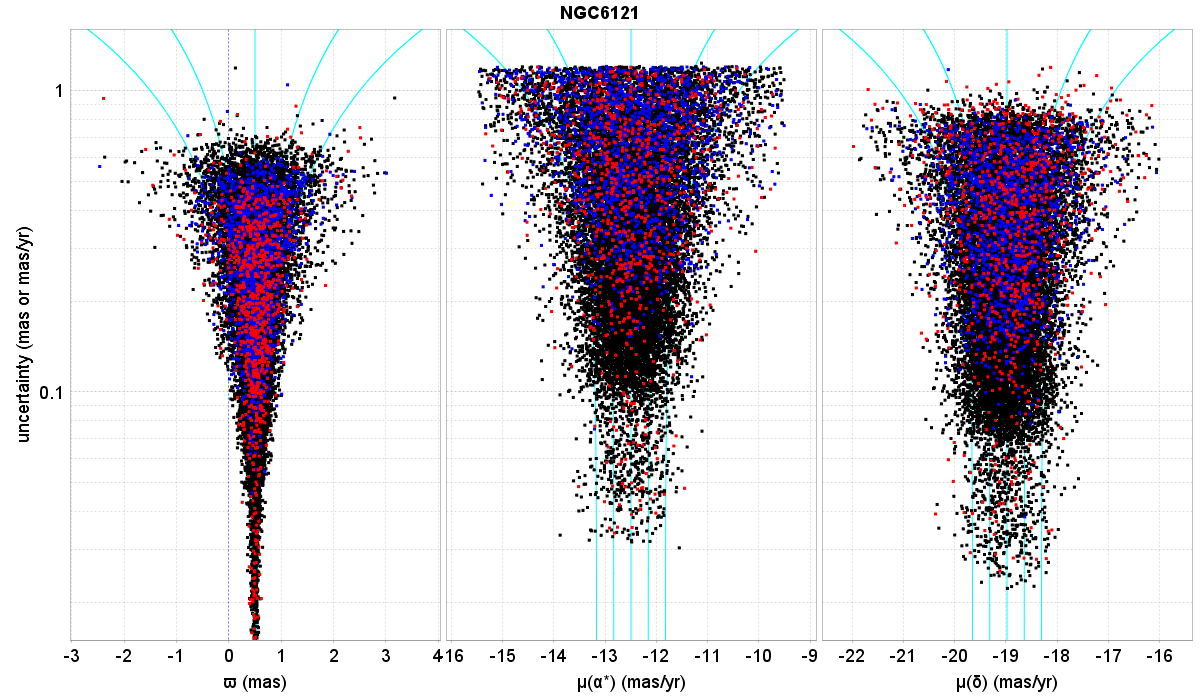}
\includegraphics[width=10cm]{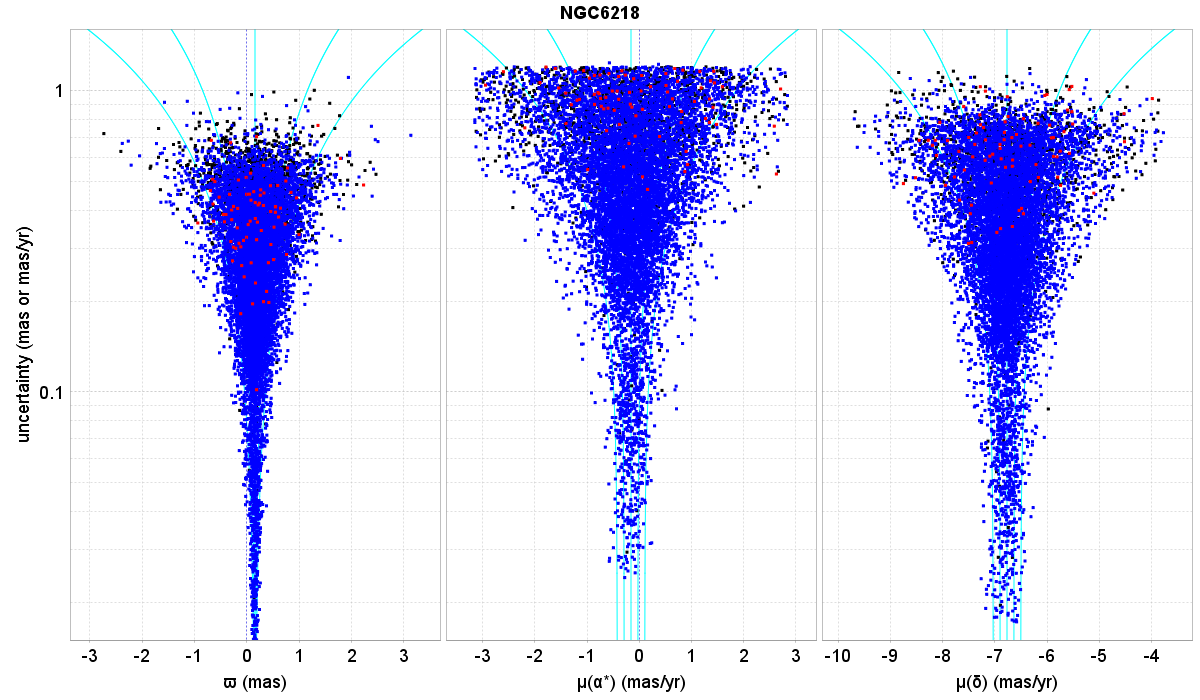}
\caption{Two examples of the parallax and PM dispersion diagrams for clusters at a distance of approximately 2 (top) and 5 (bottom) kpc. The parallax and PMs are plotted against the standard uncertainties. The blue lines show the expected uncertainty levels relative to the mean value at -2, -1 0 1 and 2 $\sigma$, including noise contributions from the internal velocity dispersion and the dispersion of the parallaxes from the depth of the cluster. At low levels of standard uncertainty, the contributions from the internal dispersion are easily detected. The colouring of the data points reflects the error correlations: red means strongly negative, blue means strongly positive, and black means low.}
\label{fig:parpm}.
\end{center}
\end{figure*}

The astrometric solution is based on the parallax and PM
determinations for the individual member stars. For each star the
contribution to the solution was normalised by the covariance matrix
\citep[for more details, see Appendix A.1
in][]{2017A&A...601A..19G}. This created for each star three
uncorrelated contributions with unit-weight error variance,
which contribute to the cluster astrometric parameter solution. Outliers
were rejected on the basis of the normalised residuals, except in
a few cases where field stars significantly disturbed
the astrometric parameters. In these cases, which were mainly found at
lower galactic latitudes, a 3$\sigma$ cut was applied to the
individual parallaxes and PMs as based on their standard
uncertainties and the estimate of the cluster PM and
parallax (see also Fig.~\ref{fig:parpm}). The astrometric solution
removes a large fraction, but not all, of the remaining field stars, as
can be seen from the HR diagram shown in the right panel of
Fig.~\ref{fig:sequstep4b}. When this residual population of field
stars still significantly disturbed the HR diagram, we
performed a final cleanup in TopCat on the basis of that diagram.

\begin{figure*}[ht]
\begin{center}
\includegraphics[totalheight=9cm]{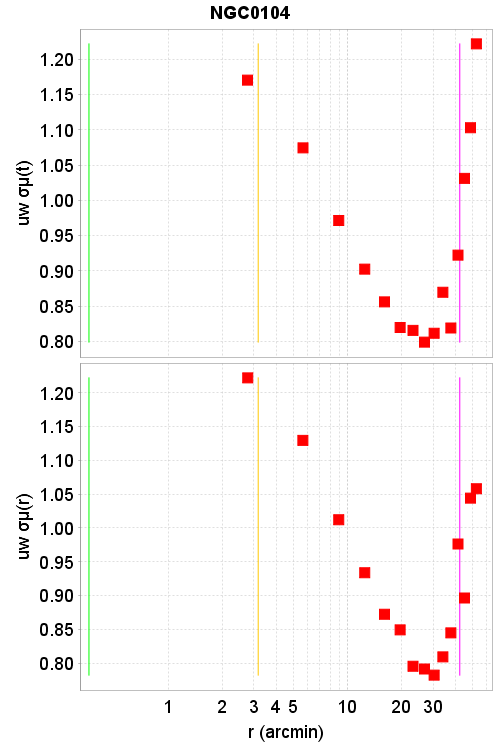}
\includegraphics[totalheight=9cm]{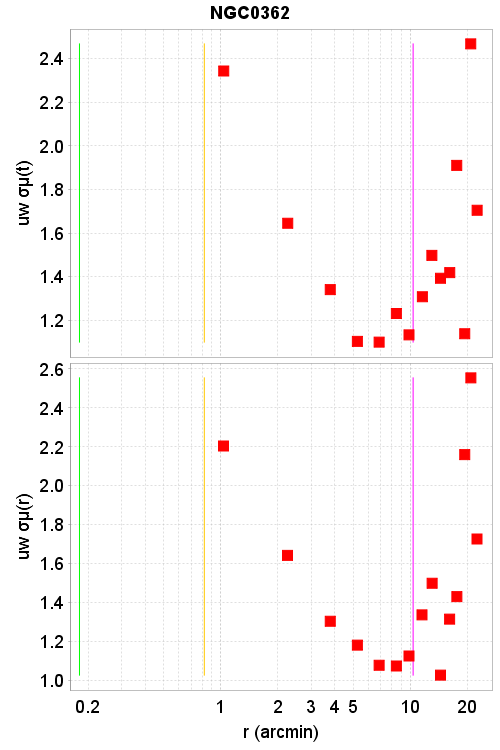}
\includegraphics[totalheight=9cm]{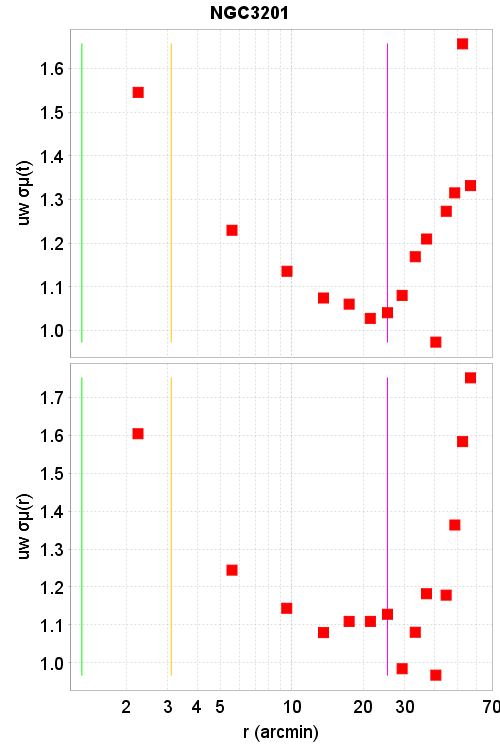}
\caption{Three examples of the PM dispersion in the radial and transverse directions as a function of distance to the cluster centre. The three vertical lines indicate, from left to right, the core radius, the half-light radius, and the tidal radius following Harris10.}
\label{fig:pmrtdisp}.
\end{center}
\end{figure*}

Two criteria are set in the iterations over the astrometric solution,
one of which could require an extended extract from the archive. The
first criterion concerns the internal PM dispersion. This
could be detected from the dispersion in the PMs of the
brightest cluster members. This is, however, not an unambiguous
process, as the observed internal velocity dispersion is often
observed to be strongly dependent on distance from the cluster
centre. The dispersions were examined in the radial and transverse
directions, and three examples are shown in
Fig.~\ref{fig:pmrtdisp}. The modelling of these dependencies is,
however, beyond the scope of the present paper, and an average
dispersion was used in all cases for the whole cluster. The values
found were generally between 2 and 8 km~s$^{-1}$. 
\begin{figure*}[ht]
\begin{center}
\includegraphics[totalheight=8cm]{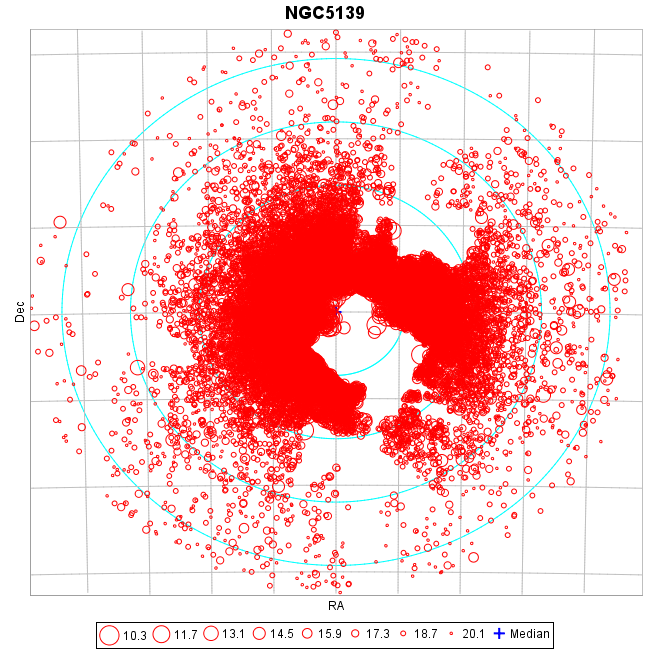}
\includegraphics[totalheight=8cm]{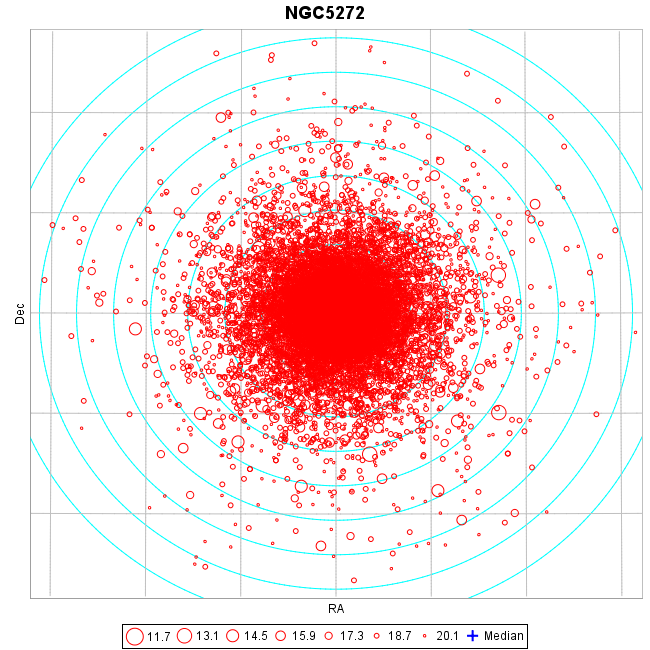}
\caption{Two examples of astrometric data coverage with five-parameter
  solutions. On the left, $\omega$~Cen, the worst case, on the right
  NGC~5272, a more average example of coverage. The gaps in the
  coverage for $\omega$~Cen are the result of the filters that have
  been applied to the astrometric data. The cyan circles are
  at intervals of 35 pc in $\omega$~Cen and 10~pc in NGC~5272.}
\label{fig:maps}
\end{center}
\end{figure*}

\begin{figure*}[ht]
\begin{center}
\includegraphics[totalheight=7cm]{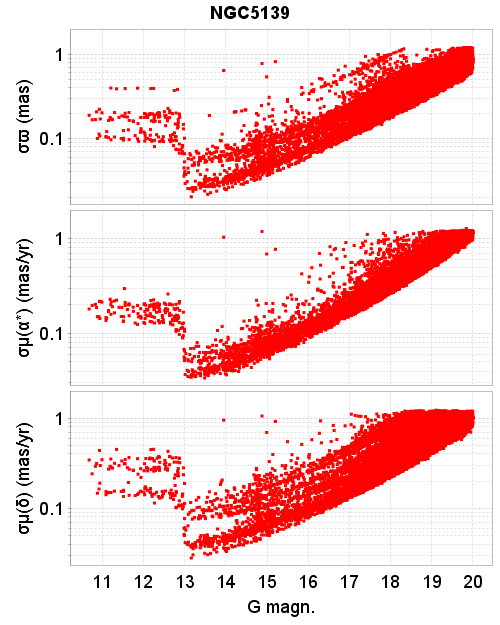}
\includegraphics[totalheight=7cm]{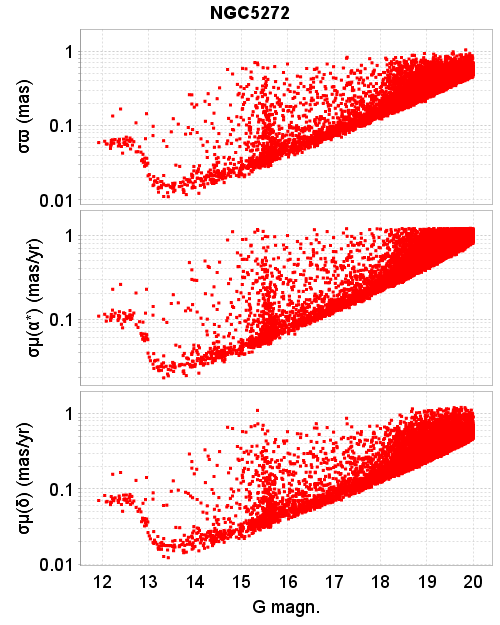}
\includegraphics[totalheight=7cm]{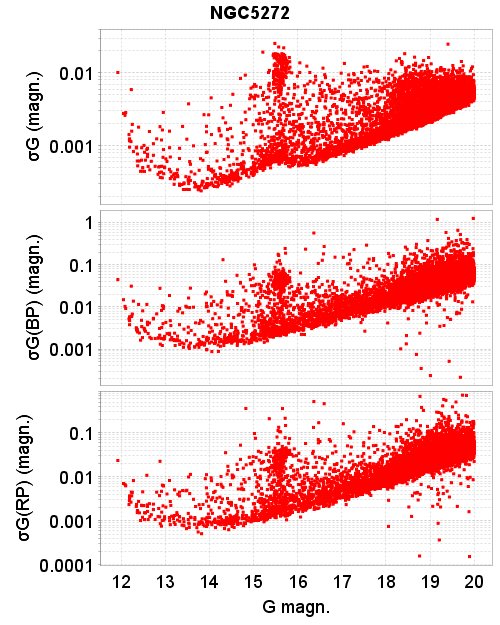}
\caption{From left to right: Standard uncertainties on the astrometric parameters of individual stars as a function of magnitude in $\omega$~Cen and NGC~5272, and standard uncertainties on the photometric data for NGC~5272. The photometric data clearly show the variability of the RR Lyrae stars (around $G=15.3$), while the astrometric data also show locally much increased uncertainties at the same brightness.}
\label{fig:dispersions}.
\end{center}
\end{figure*}

Strong variations in coverage of stars with five-parameter solutions in
the astrometry \citep{DR2-DPACP-51} also show the relation between standard
uncertainties on the astrometric parameters and the brightness of the
stars. With relatively homogeneous coverage, the
distribution is narrow, while, as is the case for $\omega$~Cen, a multi-layered
set of relations is observed for poorly covered objects (see
Fig.~\ref{fig:maps}). An additional feature appears to be the
increased noise on the astrometric data for variable stars with large
amplitudes, such as the RR Lyrae stars in globular clusters, as shown
in Figure~\ref{fig:dispersions}. The exact reason for this is still
unclear.

\subsection{Exemplifying some systematics in \Gaia DR2 using dwarf galaxies}
\label{app:app-ah}

\begin{figure}
\centering
\includegraphics[width=8cm, trim={0cm 0cm 0cm 0cm},clip]{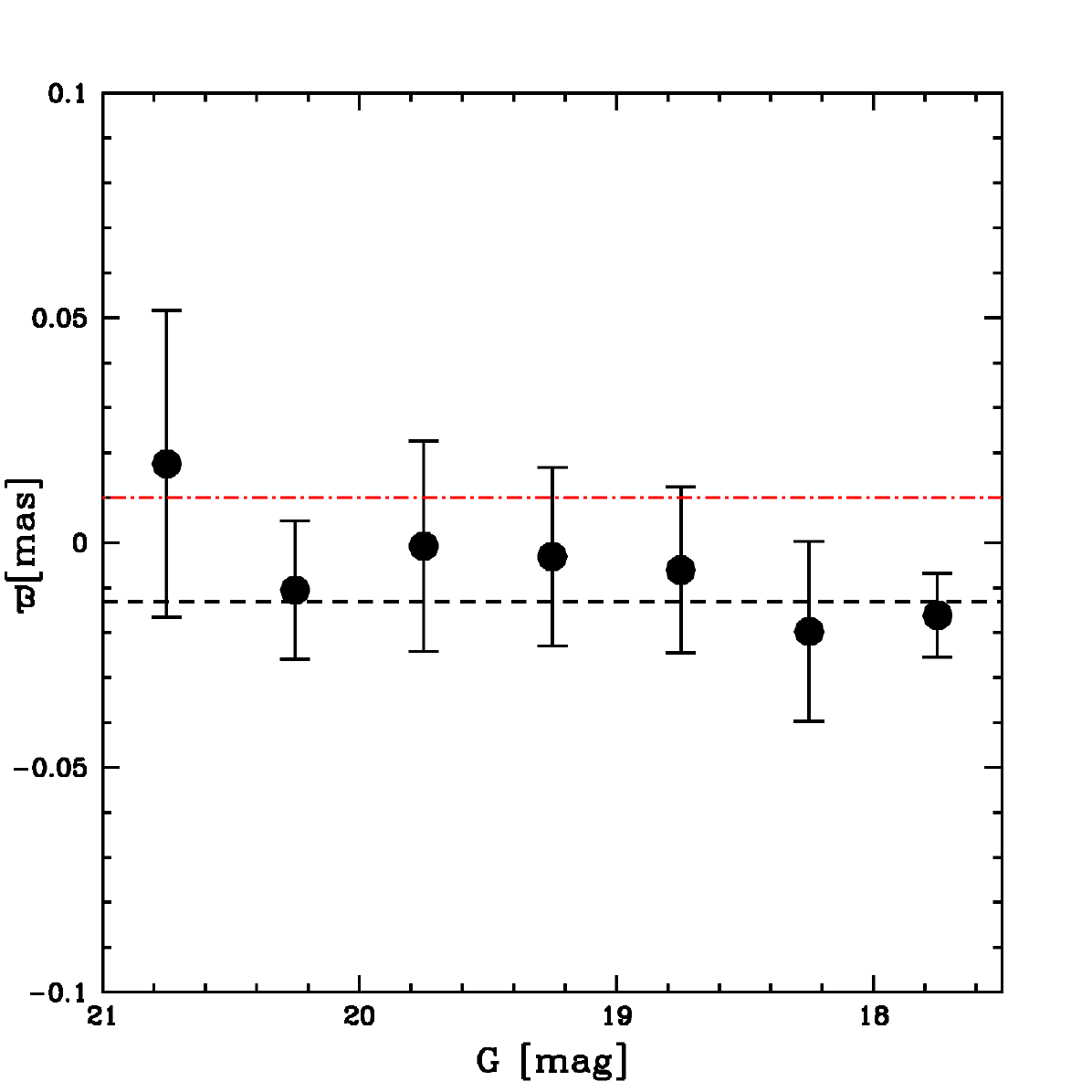}
\caption{Parallax of the stars in the Carina dSph as function of their
  $G$ magnitude. A clear systematic offset is apparent and is
significantly
  measured. The expected parallax (based on literature values) is indicated by the red dashed line.}
\label{fig:par-Gmag-Car}
\end{figure}

Figure~\ref{fig:par-Gmag-Car} shows the parallax zero-point offset for the stars
in Carina and demonstrates that it is very reliably measured and
is independent of the magnitude of the stars used. This indicates a
systematic effect present in the parallax measurements of the DR2
data, which for Carina is of the order of $\varpi_{DR2} - \varpi_{lit}
= -0.015 - 0.0095 \sim -0.024$~mas.

\begin{figure}
\centering
\includegraphics[width=8cm]{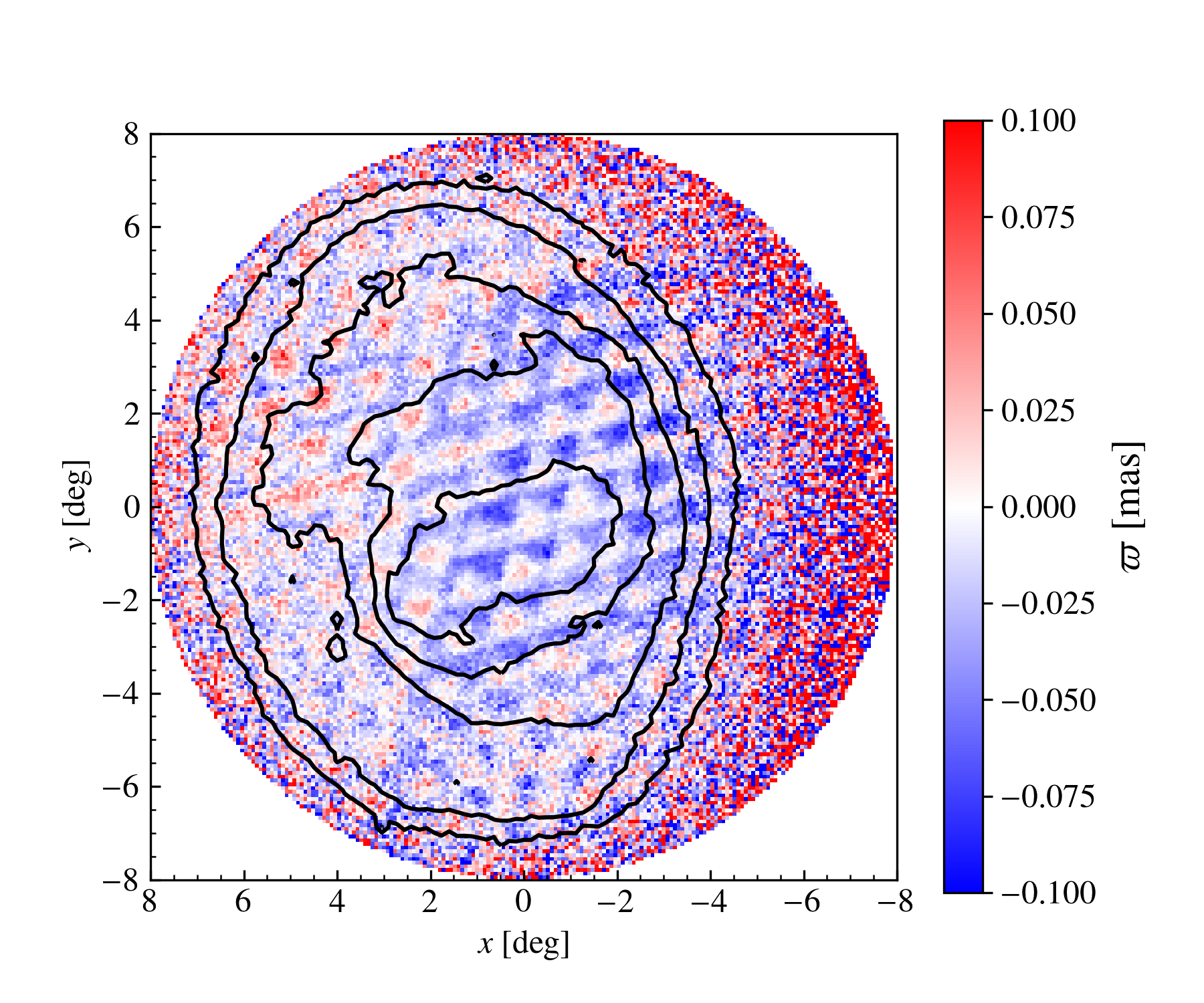}
\includegraphics[width=8cm]{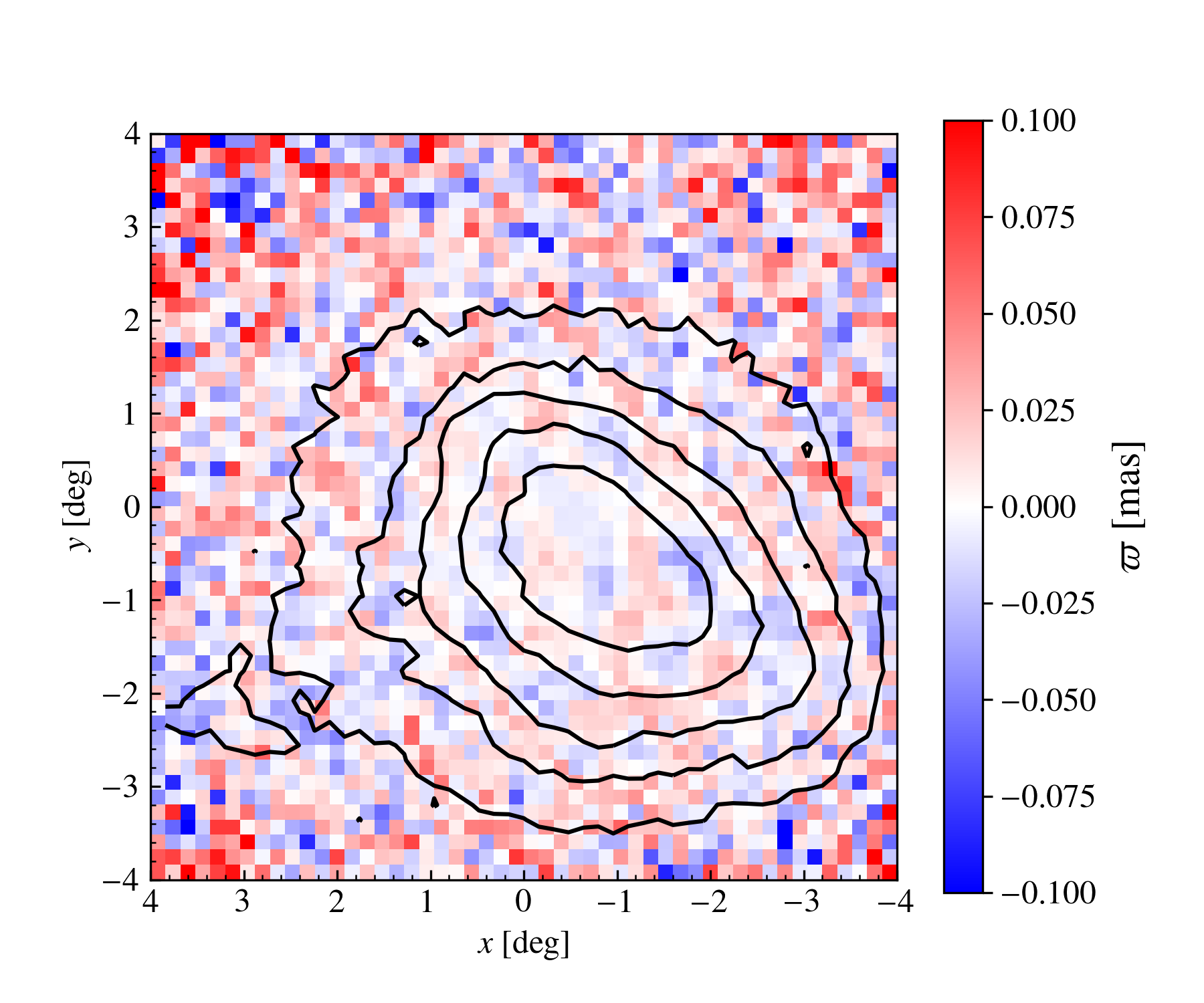}
\caption{Measured parallaxes of stars in the LMC (upper panel) and SMC (lower panel). The banding associated with the \textit{Gaia} scanning law is clearly visible.}
\label{fig:LMCSMCPar}
\end{figure}

\begin{figure}[t]
\centering
\includegraphics[width=8cm]{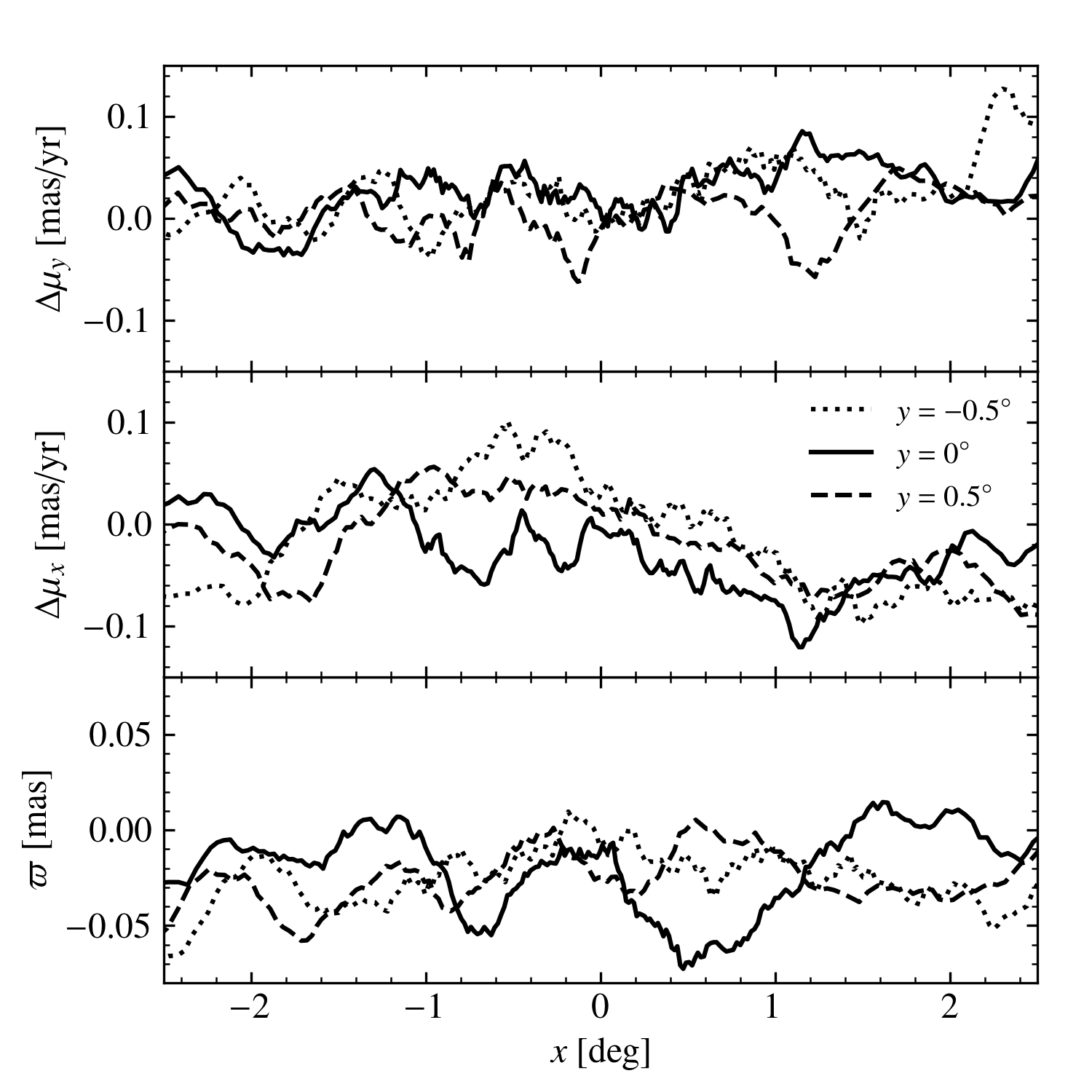}
\caption{Parallaxes and residual PMs of sources in the LMC. The lines show the median value as a function of coordinate position $x$ for all LMC sources within $0.2^\circ$ stripes centred at $y=-0.5,0$ or $0.5^\circ$. Residual PMs are calculated after subtraction of a model disc with parameters determined using only sources within $3^\circ$ of the centre. The non-zero values are due to both the systematic errors in the \textit{Gaia} data and the differences between the simple disc model and the true dynamics of the LMC.}
\label{fig:LMC1Dvar}
\end{figure}

Figure~\ref{fig:LMCSMCPar} shows the gridding pattern present in the
parallaxes, in this case, for stars in the LMC and the SMC \citep[see also][]{DR2-DPACP-51}. This
pattern has an amplitude of $\sim 0.03$~mas (see the bottom panel of
Fig.~\ref{fig:LMC1Dvar}), and is clearly apparent when analysing
sufficiently large objects on the sky, but it is likely to be present
throughout the full sky \citep[see e.g.][]{DR2-DPACP-39}.

Figure~\ref{fig:LMC1Dvar} shows the variation in PM and parallax along stripes of width $0.2^\circ$ in the LMC after a model for the PM field has been subtracted. The variation seen is produced both by the systematic errors in \textit{Gaia} DR2 and the shortcomings of the simple model.

\begin{figure}
\centering
\includegraphics[width=9cm,trim={0 1cm 0 1cm},clip]{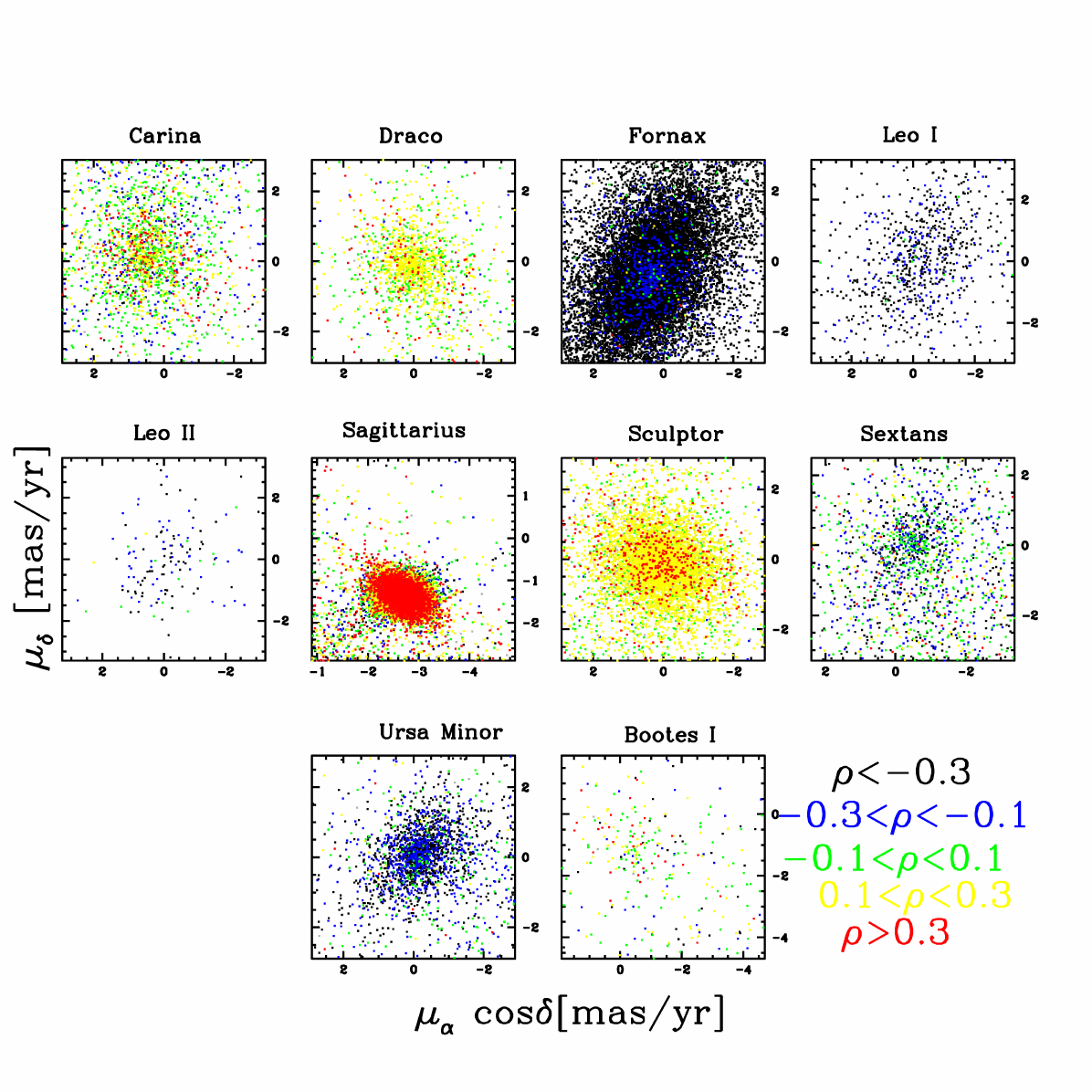}
\caption{Correlations in the PMs of the stars in the
  field of view towards the different dSph galaxies in our sample. The
  different colours indicate the amplitude of the correlations.}
\label{fig:corr_dwarfs}
\end{figure}

The correlation between the PMs of individual stars in the
field of view towards the dSph in our sample is is illustrated in
Fig.~\ref{fig:corr_dwarfs}, where we have plotted the individual stars
with different colours that indicate the amount of correlation (from
very negative, to none, to very positive) between $\mu_\alpha^*$ and
$\mu_\delta$. Fornax in particular shows a strong correlation in the
PM components of the individual stars, and these are of course then
reflected in the PM correlation coefficient given in the table. The
amplitude and orientation of the correlation differs among the dwarf galaxies, indicating that the correlations are localised on the
sky and do not have the same amplitude everywhere. Furthermore, there
are regions where these correlations are negligible when averaged
out, as in the case of Carina, Draco, Sculptor, and even the
Sagittarius dSph. We also note that all the stars in the field of view
towards these dwarfs are affected in a similar way, and not only the
members.

\section{Details of the LMC and SMC modelling} 
\label{app:LMCSMC}

In this appendix we provide a rigorous description of the coordinate
system used in Sections~\ref{sec:data-method-MCs} and \ref{sec:LMCSMC}, the modelling assumptions used to derive Eqs.~\ref{eq:xyderivs},
and the deprojection of the PMs shown in
Figure~\ref{fig:LMCSMCRotCurve}. We refer to the centre of the
Cloud as $C$, to the observer as $O,$ and to a source in the Cloud as $S$.

The vectors $[\vec{x}~\vec{y}~\vec{z}]$ form an inertially fixed right-handed orthogonal triad with $\vec{z}$ the unit vector from observer to  $C$ at the reference epoch, 
$\vec{x}$ the unit vector in the direction of increasing $\alpha$ at $C,$\ and 
$\vec{y}$ the unit vector in the direction of increasing $\delta$ at $C$.

\subsection{Position and PM in the $xyz$ system}

In terms of the celestial position $(\alpha,\delta)$ and PM components
$(\mu_{\alpha*},\mu_\delta)$, we have the unit vector from the observer to a source, $\vec{u}$, given by
\begin{equation}\label{e01}
\vec{u} = \vec{r}\, , \quad \vec{\dot{u}}=\vec{p}\mu_{\alpha*}+\vec{q}\mu_\delta \, ,
\end{equation}
where $[\vec{p}~\vec{q}~\vec{r}]$ is the local normal triad at $(\alpha,\delta)$
\begin{equation}\label{e02}
\vec{p}=\begin{bmatrix}-\sin\alpha\\ \cos\alpha\\ 0\end{bmatrix},\quad
\vec{q}=\begin{bmatrix}-\sin\delta\cos\alpha\\ -\sin\delta\sin\alpha\\ \cos\delta\end{bmatrix},\quad
\vec{r}=\begin{bmatrix} \cos\delta\cos\alpha\\ \cos\delta\sin\alpha\\ \sin\delta\end{bmatrix}.
\end{equation}

The $xyz$ system coincides with the local normal triad at 
$(\alpha_C,\delta_C)$
\begin{equation}\label{e04}
\vec{x}=\begin{bmatrix}-\sin\alpha_C\\ \cos\alpha_C\\ 0\end{bmatrix},\quad
\vec{y}=\begin{bmatrix}-\sin\delta_C\cos\alpha_C\\ -\sin\delta_C\sin\alpha_C\\ \cos\delta_C\end{bmatrix},\quad
\vec{z}=\begin{bmatrix} \cos\delta_C\cos\alpha_C\\ \cos\delta_C\sin\alpha_C\\ \sin\delta_C\end{bmatrix}.
\end{equation}
The components of $\vec{u}, \vec{\dot{u}}$ in the $xyz$ system are obtained as scalar products (e.g. $x = \vec{x}\cdot\vec{u}$, $\dot{x} = \vec{x}\cdot\vec{\dot{u}}$)
from which we can derive Eqs.~\ref{eq:xy}, where we refer to $\dot{x}, \dot{y}$ as $\mu_x, \mu_y$ (they are not strictly speaking PMs, but it is convenient to give them this notation). 

When $(\alpha_C,\delta_C)$ are used as a fixed reference point, the Cartesian coordinates 
$(x,y,z,\dot{x},\dot{y},\dot{z})$ provide a useful substitute for 
$(\alpha,\delta,\mu_{\alpha*},\mu_\delta)$. The six components are a redundant set, and when working in a limited area around $C$ (in principle
as long as $z>0$, i.e.\ within $90^\circ$ from $C$), it is possible to use the 
non-redundant set $(x,y,\dot{x},\dot{y})$, with
\begin{equation}\label{e09}
z = \sqrt{1-x^2-y^2}\, , \quad \dot{z}=-(x\dot{x}+y\dot{y})/z \, .
\end{equation}
$(x,y)$ is equivalent to the orthographic projection in cartography.%
\footnote{The gnomonic projection $(x/z,y/z)$ is more common in astrometry, where 
they are known as standard coordinates. For the current problem, they do not seem
to provide any particular advantage, and the expressions for the time derivatives become
much more complicated.} 

\subsection{Kinematic model}

Assuming a flat disc, the vector $\vec{R}$ from $C$ to $S$ must be in the plane
of the disc, which gives the condition
\begin{equation}\label{e13}
\vec{n}\cdot\vec{R} = 0 \, ,
\end{equation}
where we have reintroduced the vector $\vec{n}$, the normal to the disc plane (such that rotation about $\vec{n}$ is positive), and we also reintroduce the two normal unit vectors $\vec{l}$ and $\vec{m,}$ which form a right-handed triad with $\vec{n}$ (e.g. Eqs.~\ref{eq:lmn}). A source in the plane of the disc can be described in terms of rectangular coordinates $\xi,\eta,$ where 
\begin{equation}\label{e19}
\vec{R} =  \vec{l}\xi + \vec{m}\eta\, . 
\end{equation}

The motion of $S$ is the vectorial sum of the bulk motion (of $C$) and the peculiar 
motion of $S$ with respect to $C$. If we assume that the peculiar motion is circular with angular velocity $\omega(R),$ we have
\begin{equation}\label{e20}
\vec{v}_S=\vec{v}_C+(\vec{m}\xi-\vec{l}\eta)\,\omega(R) \, ,
\end{equation}
where $\vec{v}_C$ is the bulk motion and $R=\sqrt{\xi^2+\eta^2}$. 

At the reference epoch, we can write $\vec{R}$ in terms of the position of the source $\vec{s}$ and the centre \vec{c} as $\vec{R}=\vec{s}-\vec{c}=\vec{u}s-\vec{z}$, where s is the distance to the source.
Introducing the inverse distance factor $f=s^{-1}$ and inserting in Eq.~\ref{e13} gives
\begin{equation}\label{e21}
f = (\vec{n}\cdot\vec{u})/n_z = ax + by + z = (1-x^2-y^2)^{1/2} + ax + by \, ,
\end{equation}
where
\begin{equation}\label{e22}
a = n_x/n_z, \quad b = n_y/n_z,
\end{equation}
and $n_x=\vec{x}\cdot\vec{n}$ etc are the components of $\vec{n}$ in the $xyz$ system. These components, along with those of $\vec{l}$ and $\vec{m}$ and the values $a$ and $b$ are given as a function of the inclination $i$ and the line-of-nodes position angle $\Omega$ in the main text (Eqs.~\ref{eq:lmn} \& \ref{e42}).
We now find\footnote{This arXiv version has been altered to reflect the erratum to the published paper \citep{2020A&A...642..C1G}. This affects eqs.~\ref{e23} \& \ref{e46}. The other items noted in that erratum were always correct in the arXiv version of the paper.}
\begin{equation}\label{e23}
\begin{aligned}
\xi &= \vec{l}\cdot\vec{R}  =
\frac{l_xx+l_yy}{z+ax+by}\\[3pt]
\eta &= \vec{m}\cdot\vec{R}  =
\frac{(m_x-am_z)x+(m_y-bm_z)y}{z+ax+by} \end{aligned}
.\end{equation}

Turning now to the PMs, we seek the corresponding relationships 
between $(\dot{x},\dot{y})$ and $(\dot{\xi},\dot{\eta})$. We know that
\begin{equation}\label{e25}
\dot{\xi} =-\eta\,\omega(R), \quad
\dot{\eta} =\xi\,\omega(R).
\end{equation}
The PM vector is
\begin{equation}\label{e26}
\vec{\dot{u}} = \frac{\text{d}(\vec{s}f)}{\text{d}t} = \left(\vec{v}_S-\vec{u}(\vec{u}\cdot\vec{v}_S)\right)f\, ,
\end{equation}
which, using our previous results, we can rewrite as
\begin{multline}\label{e28}
\vec{\dot{u}} = \left(\vec{v}_C-\vec{u}(\vec{u}\cdot\vec{v}_C)\right)(ax+by+z)\;+ 
\bigl(( \vec{m}-\vec{u}(\vec{u}\cdot\vec{m}))(l_xx+l_yy) - \\
(\vec{l}-\vec{u}(\vec{u}\cdot\vec{l}))((m_x-am_z)x+(m_y-bm_z)y) \bigr)
\;\omega(R) \, . 
\end{multline}
Taking the scalar products with $\vec{x}$ and $\vec{y}$ gives explicit expressions
for $\dot{x}$ and $\dot{y}$ as functions of $x$ and $y$. If $\omega$ is
constant, the expressions contain terms up to the third power in $x$ and $y$.
At $C$ ($x=y=0$) we find, as expected,
\begin{equation}\label{e30}
\dot{x} = v_x, \quad \dot{y} = v_y,
\end{equation}
where $v_x,v_y,v_z$, etc. are the components of $\vec{v}_C$ in the $xyz$
system.

Retaining only first-order terms in $x$ and $y$ while assuming constant $\omega$, and then taking derivatives, we have
\begin{equation}\label{e31}
\begin{aligned}
{\partial\dot{x}}/{\partial x} &= av_x - v_z + al_xm_z\omega, \\[3pt]
{\partial\dot{x}}/{\partial y} &= bv_x - n_z\omega + bl_xm_z\omega, \\[3pt]
{\partial\dot{y}}/{\partial x} &= av_y + n_z\omega + al_ym_z\omega, \\[3pt]
{\partial\dot{y}}/{\partial y} &= bv_y - v_z + bl_ym_z\omega,
\end{aligned}
\end{equation}
which hold exactly at $C$ if $\text{d}\omega(r)/\text{d}R=0$ for $R=0$. These equations (writing $\dot{x},\dot{y}$ as $\mu_{x},\mu_{y}$) were used in this study to determine $v_z,i,\Omega,$ and $\omega$.

The orientation of the LMC plane is given by the unit vector $\vec{n}$, the direction of which is conventionally given by the two angles $i$ and $\Omega$:
\begin{equation}\label{e16}
\vec{n}=\vec{x}\sin i\cos\Omega - \vec{y}\sin i\sin\Omega+\vec{z}\cos i \, ,
\end{equation}
from which (with the definition of $\vec{l}$ and $\vec{m}$) we have the components of all three of these vectors in the $xyz$ system, as given in Eq.~\ref{eq:lmn}.

\subsection{Estimating the kinematic parameters for fixed $C$ and constant $\omega$}

From Eq.~\ref{e31} we see that $(\dot{x},\dot{y})$ should vary linearly
with $(x,y)$ for constant $\omega$. This is a reasonable approximation
in the inner few degrees of the LMC. Fitting the linear relation
\begin{equation}\label{e32}
\begin{aligned}
\dot{x} &= v_x + A_x x + A_y y \\[2pt]
\dot{y} &= v_y + B_x x + B_y y
\end{aligned} 
,\end{equation}
we immediately obtain estimates of $v_x$, $v_y$, and the four gradients 
$\partial\dot{x}/\partial x=A_x$, etc. We can use Eqs.~\ref{e31} to express these 
gradients as functions of the six kinematic parameters $v_x$, $v_y$, $v_z$, $i$, 
$\Omega$, and $\omega$ (assumed constant). In this study we usually then held $v_z$ constant, in which case we varied $i$, $\Omega$, and $\omega$ to minimise the sum of the square residuals
\begin{equation}
S = (\partial\dot{x}/\partial x-A_x)^2+ (\partial\dot{x}/\partial y-A_y)^2 +
 (\partial\dot{y}/\partial x-B_x)^2+ (\partial\dot{y}/\partial y-B_y)^2
,\end{equation}
where $A_x$ etc. are measured from the data, and $\partial\dot{x}/\partial x$ are predictions of the model. We can also leave $v_z$ free, in which case Eqs.~\ref{e31} can be directly solved to determine 
$v_z$, $i$, $\Omega$, and $\omega$.

\subsection{De-projection method}

The observed PMs, or $\dot{x}$, $\dot{y}$, are the projections of the
true space motions on the celestial sphere, or normal to $\vec{z}$. This projection
from 3D to 2D cannot be inverted, but if we assume that the true motions are
confined to the (known) plane of the LMC, it is possible to project the observed
motions back to that plane. We call this de-projection.
 
We assume that $\vec{v}_C$, $i$, and $\Omega$ are known. The starting point is Eq.~\ref{e26}, but with the velocity of $S$ written as
\begin{equation}\label{e44}
\vec{v}_S = \vec{v}_C + \vec{l}\dot{\xi} + \vec{m}\dot{\eta} \, ,
\end{equation}
which allows arbitrary motions in the plane. We have then
\begin{equation}\label{e45}
\vec{\dot{u}} = \left(\vec{v}_C-\vec{u}(\vec{u}\cdot\vec{v}_C)\right)f +
(\vec{l}-\vec{u}(\vec{u}\cdot\vec{l}))f\dot{\xi} + (\vec{m}-\vec{u}(\vec{u}\cdot\vec{m}))f\dot{\eta}\, . 
\end{equation}
Taking the scalar products with $\vec{x}$ and $\vec{y}$ gives two linear equations,
\begin{equation}\label{e46}
\begin{aligned}
(l_x-x(l_xx+l_yy))\dot{\xi} &+ (m_x-x(m_xx+m_yy+m_zz))\dot{\eta} \\
&=-v_x+x(v_xx+v_yy+v_zz) + \dot{x}/(ax+by+z)\\[6pt]
(l_y-y(l_xx+l_yy))\dot{\xi} &+ (m_y-y(m_xx+m_yy+m_zz))\dot{\eta} \\
&=-v_y+y(v_xx+v_yy+v_zz) + \dot{y}/(ax+by+z) ,
\end{aligned} 
\end{equation}
from which $\dot{\xi}$ and $\dot{\eta}$ can be solved.\footnote{These equations have been corrected reflecting  \cite{2020A&A...642..C1G}}
The corresponding position $(\xi,\eta)$ is obtained from (\ref{e23}).

To map the kinematics of the LMC, it is more convenient to transform the Cartesian 
$(\xi,\eta,\dot{\xi},\dot{\eta})$
into polar coordinates $R$, $\phi$ and the corresponding velocity components
$v_R$ (in the direction of increasing $R$) and $v_T$ (tangential velocity), as in Figure~\ref{fig:LMCSMCRotCurve}.

\begin{table*}
\small
\caption{Parameters of LMC based on sources with angular radii from the assumed LMC centre ($\rho$) in various ranges (with or without binning, see Section~\ref{sec:LMCSMCBasic}). We show the derived parameters when we left $v_z$ free, or when we held it fixed (the parameters were derived from the gradients, so
that the gradients are the same in either case). }
\label{tab:LMCcircles}
\begin{center}
\begin{tabular}{ccccccccccccc}
$\rho$ & $v_x$ & $v_y$ & 
$\partial \mu_x/\partial x$ & $\partial \mu_x/\partial y$ &$ \partial \mu_y/\partial x$ & 
$\partial \mu_y/\partial y $& N  &
$v_z$ & $v_{z,{\rm const}}$& $i$  & $\Omega$ & $\omega$   \\
 {[}deg] &  [mas/yr]  &  [mas/yr] 
  &[mas/yr & [mas/yr & [mas/yr & [mas/yr & 
 &  [mas/yr]  &  [mas/yr]  &[deg]  & [deg] &  [mas/yr]   \\
  &    &   & /rad] & /rad] & /rad] & /rad] & 
 &  &    & &  &     \vspace{2mm} \\ 
\hline 
 \noalign{\smallskip}
$<$ 2.0 & 1.848 & 0.236 & -2.256 & -5.441 & 4.838 & -0.225 & 2482275 & 1.930 & & 61.497 & -73.577 & 5.217 \\
$<$ 2.0 &   &   &   &   &   &   &   & & 1.104 & 40.008 & -68.576 & 5.643 \\
$<$ 3.0 & 1.850 & 0.234 & -1.577 & -4.565 & 4.765 & -0.284 & 4295125 & 1.493 & & 53.097 & -70.263 & 5.095 \\
$<$ 3.0 &   &   &   &   &   &   &   & & 1.104 & 33.982 & -61.916 & 5.095 \\
$<$ 4.0 & 1.849 & 0.234 & -1.330 & -3.936 & 4.543 & -0.420 & 5757649 & 1.407 & & 48.038 & -66.336 & 4.740 \\
$<$ 4.0 &   &   &   &   &   &   &   & & 1.104 & 30.795 & -55.251 & 4.604 \\
$<$ 5.0 & 1.849 & 0.235 & -1.183 & -3.493 & 4.337 & -0.482 & 6610213 & 1.379 & & 45.600 & -62.242 & 4.410 \\
$<$ 5.0 &   &   &   &   &   &   &   & & 1.104 & 30.018 & -49.346 & 4.236 \\
$<$ 6.0 & 1.849 & 0.237 & -1.046 & -3.134 & 4.092 & -0.511 & 7279181 & 1.328 & & 44.155 & -59.946 & 4.104 \\
$<$ 6.0 &   &   &   &   &   &   &   & & 1.104 & 30.108 & -45.378 & 3.909 \\
\hline
 \noalign{\smallskip}
\multicolumn{13}{c}{Annuli} \\ \hline
 \noalign{\smallskip}
 1.0--2.0 & 1.847 & 0.230 & -2.164 & -5.389 & 4.860 & -0.245 & 1679866 & 1.864 & & 60.448 & -73.439 & 5.259 \\
1.0--2.0 & & & & & & & & &  1.104 & 39.175 & -68.368 & 5.623 \\
2.0--3.0 & 1.847 & 0.231 & -1.278 & -4.255 & 4.751 & -0.302 & 1812850 & 1.308 & & 48.064 & -67.135 & 4.990 \\
2.0--3.0 & & & & & & & & &  1.104 & 31.471 & -55.591 & 4.870 \\
3.0--4.0 & 1.841 & 0.238 & -1.032 & -3.429 & 4.297 & -0.528 & 1462524 & 1.281 & & 41.805 & -60.172 & 4.336 \\
3.0--4.0 & & & & & & & & &  1.104 & 28.524 & -45.464 & 4.142 \\
4.0--5.0 & 1.833 & 0.251 & -0.941 & -2.741 & 3.794 & -0.521 & 852564 & 1.296 & & 43.761 & -58.173 & 3.750 \\
4.0--5.0 & & & & & & & & &  1.104 & 30.984 & -42.516 & 3.546 \\
5.0--6.0 & 1.835 & 0.264 & -0.701 & -2.300 & 3.215 & -0.477 & 668968 & 1.024 & & 39.870 & -64.539 & 3.285 \\
5.0--6.0 & & & & & & & & &  1.104 & 31.528 & -41.959 & 3.033 \\
\hline
 \noalign{\smallskip}
\multicolumn{13}{c}{Using binned data} \\ \hline
 \noalign{\smallskip}
$<$ 2.0 & 1.850 & 0.219 & -2.341 & -5.791 & 5.376 & -0.552 & & 2.041 & & 56.399 & -71.686 & 5.807 \\
$<$ 2.0 & & & & & & & & &  1.104 & 35.461 & -68.997 & 6.061 \\
$<$ 3.0 & 1.846 & 0.220 & -1.482 & -4.778 & 5.198 & -0.349 & & 1.449 & & 48.161 & -65.873 & 5.428 \\
$<$ 3.0 & & & & & & & & &  1.104 & 30.938 & -57.200 & 5.359 \\
$<$ 4.0 & 1.842 & 0.226 & -1.217 & -3.924 & 4.701 & -0.447 & & 1.367 & & 44.841 & -61.545 & 4.779 \\
$<$ 4.0 & & & & & & & & &  1.104 & 29.534 & -49.547 & 4.627 \\
$<$ 5.0 & 1.838 & 0.239 & -1.058 & -3.231 & 4.064 & -0.452 & & 1.270 & & 45.039 & -63.705 & 4.165 \\
$<$ 5.0 & & & & & & & & &  1.104 & 30.549 & -48.400 & 3.964 \\
$<$ 6.0 & 1.837 & 0.255 & -0.919 & -2.704 & 3.387 & -0.414 & & 1.058 & & 46.981 & -74.732 & 3.686 \\
$<$ 6.0 & & & & & & & & &  1.104 & 32.097 & -52.666 & 3.407 \\
\end{tabular}
\end{center}
\label{default1}
\end{table*}%

\begin{table*}
\small
\caption{Parameters of the SMC, with the assumed $v_z$ value.}
\label{tab:SMCcircles}
\begin{center}
\begin{tabular}{ccccccccccccc}
$r_{max}$ & $v_x$ & $v_y$ & 
$\partial \mu_x/\partial x$ & $\partial \mu_x/\partial y$ &$ \partial \mu_y/\partial x$ & 
$\partial \mu_y/\partial y $& N  &
$v_{z,{\rm const}}$& $i$  & $\Omega$ & $\omega$   \\
 {[}deg] &  [mas/yr]  &  [mas/yr] 
  &[mas/yr & [mas/yr & [mas/yr & [mas/yr & 
  &  [mas/yr]  &[deg]  & [deg] &  [mas/yr]   \\
  &    & 
  & /rad] & /rad] & /rad] & /rad] & 
 &  & &  &     \\ \hline 
 \noalign{\smallskip}
 $<$ 2.0 & 0.794 & -1.219 & 1.960 & 0.831 & -2.117 & 0.030 &  935265  & 0.489 & 73.995 & -6.613 & 0.595 \\
$<$ 3.0 & 0.797 & -1.220 & 2.247 & 0.383 & -1.932 & 0.067 &  1219082  & 0.489 & 74.199 & -1.682 & 0.643 \\
$<$ 4.0 & 0.799 & -1.221 & 2.295 & 0.253 & -1.948 & 0.176 &  1343140  & 0.489 & 73.678 & 1.257 & 0.610 \\
\hline
 \noalign{\smallskip}
\multicolumn{12}{c}{Annuli} \\ \hline
 \noalign{\smallskip}
1.0--2.0 & 0.797 & -1.219 & 2.014 & 0.715 & -2.269 & 0.094 &  543224  & 0.489 & 73.046 & -3.293 & 0.497 \\
2.0--3.0 & 0.810 & -1.234 & 2.632 & 0.128 & -1.929 & -0.090 &  283817  & 0.489 & 75.568 & -0.464 & 0.711 \\
3.0--4.0 & 0.824 & -1.235 & 2.455 & 0.150 & -2.019 & 0.299 &  124058  & 0.489 & 73.549 & 3.676 & 0.591 \\
\hline
 \noalign{\smallskip}
\multicolumn{12}{c}{Using binned data} \\
\hline
 \noalign{\smallskip}
 $<$ 2.0 & 0.796 & -1.225 & 2.709 & 0.300 & -2.585 & 0.242  & &  0.489 & 75.581 & 2.178 & 0.528 \\
$<$ 3.0 & 0.804 & -1.233 & 2.959 & -0.027 & -2.329 & 0.340  & &  0.489 & 75.895 & 5.386 & 0.609 \\
$<$ 4.0 & 0.817 & -1.237 & 2.671 & -0.054 & -2.215 & 0.417  & &  0.489 & 74.195 & 7.082 & 0.557 \\\end{tabular}
\end{center}
\label{default2}
\end{table*}%

\section{Globular clusters and dSph solution data}
\onecolumn
\begin{center}
\footnotesize{
\begin{longtable}{l|rrrrrrrrr}
  \caption{Overview of the results for globular clusters\label{tab:overview}. For each cluster we list the NGC name, the SIMBAD identifier, as well as the derived position on the sky ($\alpha$, $\delta$), parallax $\varpi$, PMs ($\mu_{\alpha*}$, $\mu_\delta$), and the elements of the covariance matrix $\epsilon_\varpi$, $\epsilon_{\mu\alpha *}$, $\epsilon_{\mu\delta}$ and correlation coefficients $C$. The entry r(max)\degr corresponds to the maximum radius at which PM members have been found, nMemb is the number of members used to derive the astrometric parameters, and uwsd$_{astr}$ is the unit-weight standard deviation of the astrometric solution. Lastly, Vrad, $\epsilon_{Vr}$, uwsd$_{Vr}$, and $N_{Vr}$ are the mean radial velocity derived from \Gaia DR2 data, its error, the unit-weight standard deviation for the radial velocity solution, and the number of stars used to derive these quantities, respectively.}  \\
  \hline\hline
  Name & $\alpha$  &$\varpi$ & $\mu_{\alpha *}$ & $\mu_\delta$ & C$_{\varpi,\mu_\alpha}$ & C$_{\mu_\alpha,\mu_\delta}$ & nMemb & Vrad & uwsd$_{Vr}$\\
  ClustId & $\delta$& $\epsilon_\varpi$ & $\epsilon_{\mu\alpha *}$ &
  $\epsilon_{\mu\delta}$ & C$_{\varpi,\mu_\delta}$  & r(max)\degr & uwsd$_{astr}$ & $\epsilon_{Vr}$ & $N_{Vr}$\\
  & [deg] & [mas] & [mas~yr$^{-1}$] & [mas~yr$^{-1}$] & & & & [km~s$^{-1}$] &  \\
  \hline
\endfirsthead

\multicolumn{10}{c}%
{{\bfseries \tablename\ \thetable{} -- continued from previous page}} \\
\hline 
  Name & $\alpha$  &$\varpi$ & $\mu_{\alpha *}$ & $\mu_\delta$ & C$_{\varpi,\mu_\alpha}$ & C$_{\mu_\alpha,\mu_\delta}$ & nMemb & Vrad & uwsd$_{Vr}$\\
  ClustId & $\delta$& $\epsilon_\varpi$ & $\epsilon_{\mu\alpha *}$ &
  $\epsilon_{\mu\delta}$ & C$_{\varpi,\mu_\delta}$  & r(max)\degr & uwsd$_{astr}$ & $\epsilon_{Vr}$ & $N_{Vr}$\\
  & [deg] & [mas] & [mas~yr$^{-1}$] & [mas~yr$^{-1}$] & & & & [km~s$^{-1}$] &  \\
\hline 
\endhead

\hline \multicolumn{10}{r}{{Continued on next page}} \\ \hline
\endfoot

\hline \hline
\endlastfoot
NGC0104 &   6.0194 &    0.1959 &   5.2477 &  -2.5189 & -0.01 & -0.06  & 60093 & -18.95 & 11.55  \\ 
C0021-723 & -72.0821 &   0.0002 &   0.0016 &   0.0015 &  -0.01 &  0.90 &  0.79 &  0.42 &  229 \\ 
NGC0288 &  13.1879 &    0.1401 &   4.2385 &  -5.6470 &  0.15 &  0.25  & 5897 & -49.06 &  2.87  \\ 
C0050-268 & -26.5858 &   0.0021 &   0.0035 &   0.0026 &  -0.13 &  0.33 &  0.99 &  0.32 &   11 \\ 
NGC0362 &  15.8099 &    0.0788 &   6.6954 &  -2.5184 & -0.04 & -0.09  & 6896 & 226.93 &  6.06  \\ 
C0100-711 & -70.8489 &   0.0012 &   0.0045 &   0.0034 &  -0.12 &  0.39 &  1.23 &  0.77 &   19 \\ 
NGC1851 &  78.5280 &    0.0298 &   2.1308 &  -0.6220 &  0.06 & -0.09  & 4044 & 323.36 &  3.74  \\ 
C0512-400 & -40.0456 &   0.0011 &   0.0037 &   0.0040 &  -0.07 &  0.28 &  1.07 &  1.04 &   17 \\ 
NGC1904 &  81.0463 &    0.0362 &   2.4702 &  -1.5603 &  0.05 & -0.03  & 2363 & 206.43 &  2.94  \\ 
C0522-245 & -24.5255 &   0.0017 &   0.0048 &   0.0054 &   0.04 &  0.14 &  1.14 &  0.87 &   14 \\ 
NGC2298 & 102.2464 &    0.0791 &   3.2762 &  -2.1913 &  0.08 &  0.07  & 1373 & 147.41 &  1.54  \\ 
C0647-359 & -36.0046 &   0.0019 &   0.0060 &   0.0061 &  -0.07 &  0.16 &  1.06 &  1.40 &    4 \\ 
NGC2808 & 138.0071 &    0.0560 &   1.0032 &   0.2785 &  0.05 & -0.08  & 6769 & 104.61 &  5.33  \\ 
C0911-646 & -64.8645 &   0.0006 &   0.0032 &   0.0032 &  -0.01 &  0.39 &  0.87 &  1.26 &   20 \\ 
NGC3201 & 154.3987 &    0.1724 &   8.3344 &  -1.9895 &  0.04 &  0.12  & 19921 & 494.62 &  5.09  \\ 
C1015-461 & -46.4125 &   0.0006 &   0.0021 &   0.0020 &  -0.02 &  0.98 &  0.97 &  0.37 &   64 \\ 
NGC4372 & 186.4587 &    0.1426 &  -6.3898 &   3.3266 &  0.03 &  0.01  & 10744 &  77.41 &  5.50  \\ 
C1223-724 & -72.6562 &   0.0006 &   0.0030 &   0.0025 &   0.03 &  0.46 &  0.83 &  0.58 &   42 \\ 
NGC4590 & 189.8651 &    0.0664 &  -2.7640 &   1.7916 & -0.00 & -0.29  & 3338 &      &      \\ 
C1236-264 & -26.7454 &   0.0025 &   0.0050 &   0.0039 &   0.13 &  0.24 &  0.99 &     &    0 \\ 
NGC4833 & 194.8978 &    0.1163 &  -8.3147 &  -0.9366 &  0.05 &  0.06  & 6269 & 207.86 &  5.97  \\ 
C1256-706 & -70.8718 &   0.0010 &   0.0036 &   0.0029 &   0.11 &  0.19 &  0.93 &  0.57 &   40 \\ 
NGC5024 & 198.2262 &    0.0143 &  -0.1466 &  -1.3514 & -0.12 & -0.28  & 2637 & -64.33 &      \\ 
C1310+184 &  18.1661 &   0.0018 &   0.0045 &   0.0032 &   0.08 &  0.26 &  1.14 &     &    1 \\ 
NGC5053 & 199.1124 &    0.0064 &  -0.3591 &  -1.2586 &  0.13 & -0.32  &  918 &      &      \\ 
C1313+179 &  17.7008 &   0.0040 &   0.0071 &   0.0048 &  -0.17 &  0.13 &  0.90 &     &    0 \\ 
NGC5139 & 201.7876 &    0.1237 &  -3.1925 &  -6.7445 & -0.04 & -0.03  & 32700 & 235.12 & 11.73  \\ 
C1323-472 & -47.4515 &   0.0011 &   0.0022 &   0.0019 &   0.17 &  1.09 &  0.91 &  0.59 &   88 \\ 
NGC5272 & 205.5486 &    0.0265 &  -0.1127 &  -2.6274 & -0.01 & -0.03  & 12057 & -146.48 &  5.54  \\ 
C1339+286 &  28.3760 &   0.0010 &   0.0029 &   0.0022 &  -0.05 &  0.47 &  1.01 &  0.66 &   35 \\ 
NGC5286 & 206.6136 &    0.0168 &   0.1836 &  -0.1477 & -0.02 & -0.01  & 1649 &  56.80 &  1.97  \\ 
C1343-511 & -51.3723 &   0.0025 &   0.0076 &   0.0068 &   0.08 &  0.16 &  1.21 &  1.66 &    7 \\ 
NGC5466 & 211.3614 &    0.0210 &  -5.4044 &  -0.7907 &  0.04 &  0.07  & 1772 & 109.41 &  0.41  \\ 
C1403+287 &  28.5331 &   0.0021 &   0.0042 &   0.0041 &   0.15 &  0.15 &  0.93 &  0.31 &    2 \\ 
NGC5634 & 217.4053 &    0.0039 &  -1.7309 &  -1.5283 & -0.06 & -0.02  &  602 &      &      \\ 
C1427-057 &  -5.9773 &   0.0047 &   0.0087 &   0.0074 &   0.06 &  0.09 &  1.04 &     &    0 \\ 
NGC5897 & 229.3515 &    0.0680 &  -5.4108 &  -3.4595 & -0.02 & -0.12  & 2613 &  99.92 &  1.77  \\ 
C1514-208 & -21.0115 &   0.0026 &   0.0053 &   0.0045 &   0.02 &  0.18 &  0.96 &  1.31 &    5 \\ 
NGC5904 & 229.6394 &    0.1135 &   4.0613 &  -9.8610 & -0.07 &  0.03  & 11741 &  54.54 &  7.56  \\ 
C1516+022 &   2.0766 &   0.0010 &   0.0032 &   0.0029 &   0.09 &  0.56 &  0.98 &  0.86 &   61 \\ 
NGC5927 & 232.0065 &    0.0996 &  -5.0470 &  -3.2325 & -0.00 & -0.08  & 2621 &      &      \\ 
C1524-505 & -50.6694 &   0.0021 &   0.0060 &   0.0055 &  -0.01 &  0.15 &  0.92 &     &    0 \\ 
NGC5946 & 233.8711 &    0.0444 &  -5.1909 &  -1.6522 & -0.04 & -0.10  &  757 & 131.88 &      \\ 
C1531-504 & -50.6617 &   0.0047 &   0.0124 &   0.0101 &  -0.06 &  0.09 &  0.94 &     &    1 \\ 
NGC5986 & 236.5211 &    0.0718 &  -4.2217 &  -4.5515 & -0.10 & -0.18  & 2477 &  98.90 &  2.38  \\ 
C1542-376 & -37.7826 &   0.0031 &   0.0084 &   0.0065 &  -0.01 &  0.16 &  1.38 &  1.06 &   11 \\ 
NGC6093 & 244.2564 &    0.0558 &  -2.9469 &  -5.5613 & -0.10 &  0.01  & 1927 &  12.01 &  8.34  \\ 
C1614-228 & -22.9723 &   0.0030 &   0.0090 &   0.0073 &   0.06 &  0.16 &  1.23 &  1.70 &   16 \\ 
NGC6121 & 245.8976 &    0.5001 & -12.4956 & -18.9789 & -0.08 &  0.08  & 19508 &  71.40 &  7.79  \\ 
C1620-264 & -26.5279 &   0.0007 &   0.0033 &   0.0030 &   0.03 &  1.13 &  1.02 &  0.30 &  182 \\ 
NGC6144 & 246.8061 &    0.0668 &  -1.7646 &  -2.6371 & -0.15 &  0.08  & 1882 & 195.85 &  1.38  \\ 
C1624-259 & -26.0301 &   0.0040 &   0.0085 &   0.0063 &   0.15 &  0.17 &  1.16 &  0.90 &    3 \\ 
NGC6171 & 248.1350 &    0.1480 &  -1.9359 &  -5.9487 & -0.13 &  0.03  & 4032 & -35.01 &  6.94  \\ 
C1629-129 & -13.0570 &   0.0026 &   0.0064 &   0.0048 &   0.17 &  0.33 &  1.08 &  0.89 &   15 \\ 
NGC6205 & 250.4217 &    0.0801 &  -3.1762 &  -2.5876 & -0.04 &  0.19  & 15634 & -245.62 &  9.79  \\ 
C1639+365 &  36.4596 &   0.0007 &   0.0027 &   0.0030 &   0.04 &  0.58 &  1.07 &  0.94 &   65 \\ 
NGC6218 & 251.8101 &    0.1563 &  -0.1577 &  -6.7683 & -0.06 &  0.29  & 10488 & -41.00 &  5.24  \\ 
C1644-018 &  -1.9510 &   0.0013 &   0.0040 &   0.0027 &   0.11 &  0.38 &  1.08 &  0.51 &   38 \\ 
NGC6235 & 253.3557 &    0.0618 &  -3.9442 &  -7.5615 & -0.18 &  0.25  &  882 &      &      \\ 
C1650-220 & -22.1798 &   0.0078 &   0.0130 &   0.0067 &   0.37 &  0.15 &  1.43 &     &    0 \\ 
NGC6254 & 254.2861 &    0.1511 &  -4.7031 &  -6.5285 & -0.07 &  0.21  & 13005 &  76.76 &  5.89  \\ 
C1654-040 &  -4.0981 &   0.0014 &   0.0039 &   0.0027 &   0.10 &  0.47 &  1.12 &  0.59 &   61 \\ 
NGC6266 & 255.2821 &    0.2187 &  -5.3269 &  -2.9818 & -0.11 &  0.23  & 3096 & -74.86 &  3.84  \\ 
C1658-300 & -30.0938 &   0.0036 &   0.0082 &   0.0052 &   0.16 &  0.17 &  1.01 &  0.79 &   14 \\ 
NGC6273 & 255.6561 &    0.0924 &  -3.2237 &   1.6059 & -0.07 &  0.15  & 4977 & 141.29 &  2.94  \\ 
C1659-262 & -26.2696 &   0.0022 &   0.0069 &   0.0050 &   0.11 &  0.21 &  1.44 &  1.00 &   16 \\ 
NGC6284 & 256.1187 &    0.0499 &  -3.1882 &  -2.0479 & -0.16 &  0.20  &  911 &  30.29 &  1.71  \\ 
C1701-246 & -24.7662 &   0.0056 &   0.0112 &   0.0075 &   0.19 &  0.10 &  1.06 &  1.80 &    6 \\ 
NGC6287 & 256.2882 &    0.1074 &  -4.8866 &  -1.9208 &  0.02 &  0.06  & 1518 & -292.45 &  2.62  \\ 
C1702-226 & -22.7183 &   0.0049 &   0.0117 &   0.0083 &   0.19 &  0.13 &  1.29 &  0.81 &    3 \\ 
NGC6293 & 257.5413 &    0.0696 &   0.8225 &  -4.3070 & -0.10 &  0.20  & 1036 & -143.65 &  1.70  \\ 
C1707-265 & -26.5799 &   0.0049 &   0.0093 &   0.0064 &   0.26 &  0.09 &  1.04 &  0.67 &    2 \\ 
NGC6304 & 258.6370 &    0.1077 &  -3.9478 &  -1.1248 & -0.04 &  0.16  & 1322 & -111.70 &  5.37  \\ 
C1711-294 & -29.4816 &   0.0034 &   0.0095 &   0.0069 &   0.13 &  0.16 &  1.05 &  1.33 &    6 \\ 
NGC6316 & 259.1534 &    0.0659 &  -4.8215 &  -4.6140 & -0.19 &  0.23  &  961 &      &      \\ 
C1713-280 & -28.1532 &   0.0094 &   0.0146 &   0.0095 &   0.34 &  0.10 &  1.22 &     &    0 \\ 
NGC6325 & 259.4962 &    0.1431 &  -8.3777 &  -9.0067 & -0.33 &  0.17  &  392 &      &      \\ 
C1714-237 & -23.7668 &   0.0160 &   0.0195 &   0.0135 &   0.34 &  0.09 &  1.29 &     &    0 \\ 
NGC6333 & 259.8021 &    0.0934 &  -2.2028 &  -3.2084 & -0.01 &  0.19  & 3478 &      &      \\ 
C1716-184 & -18.5146 &   0.0027 &   0.0075 &   0.0056 &   0.15 &  0.09 &  1.40 &     &    0 \\ 
NGC6341 & 259.2821 &    0.0564 &  -4.9367 &  -0.5559 & -0.04 &  0.11  & 7079 & -118.81 &  5.56  \\ 
C1715+432 &  43.1352 &   0.0008 &   0.0040 &   0.0040 &  -0.00 &  0.26 &  1.11 &  0.62 &   26 \\ 
NGC6342 & 260.2983 &    0.0973 &  -2.9475 &  -7.0059 & -0.01 &  0.14  & 1121 & 118.97 &    \\ 
C1718-195 & -19.6050 &   0.0057 &   0.0112 &   0.0083 &   0.22 &  0.13 &  1.05 &   &    1 \\ 
NGC6352 & 261.3739 &    0.1543 &  -2.1889 &  -4.4209 & -0.12 &  0.22  & 7255 & -123.25 &  2.80  \\ 
C1721-484 & -48.4270 &   0.0018 &   0.0046 &   0.0036 &   0.13 &  0.18 &  1.01 &  0.63 &   12 \\ 
NGC6356 & 260.8898 &    0.0791 &  -3.7683 &  -3.3746 & -0.24 &  0.27  & 2021 &      &      \\ 
C1720-177 & -17.8128 &   0.0066 &   0.0096 &   0.0063 &   0.31 &  0.09 &  1.59 &     &    0 \\ 
NGC6362 & 262.9772 &    0.0974 &  -5.5014 &  -4.7417 & -0.06 &  0.06  & 9169 &      &      \\ 
C1725-050 & -67.0492 &   0.0011 &   0.0028 &   0.0032 &   0.06 &  0.26 &  1.12 &     &    0 \\ 
NGC6366 & 261.9393 &    0.2292 &  -0.3835 &  -5.1309 & -0.08 &  0.27  & 7108 &      &      \\ 
C1726-670 &  -5.0752 &   0.0022 &   0.0054 &   0.0044 &   0.14 &  0.36 &  1.09 &     &    0 \\ 
NGC6380 & 263.6202 &    0.1014 &  -2.0984 &  -3.1922 & -0.25 & -0.05  &  988 &      &      \\ 
C1731-390 & -39.0694 &   0.0163 &   0.0183 &   0.0125 &   0.49 &  0.07 &  1.64 &     &    0 \\ 
NGC6388 & 264.0654 &    0.0482 &  -1.3548 &  -2.7144 & -0.12 &  0.11  & 3912 &  80.00 &  4.18  \\ 
C1732-447 & -44.7423 &   0.0034 &   0.0072 &   0.0061 &   0.18 &  0.11 &  1.51 &  2.49 &    9 \\ 
NGC6397 & 265.1697 &    0.3781 &   3.2908 & -17.5908 & -0.05 &  0.10  & 22116 &  19.18 &  8.00  \\ 
C1736-536 & -53.6773 &   0.0007 &   0.0026 &   0.0025 &   0.07 &  0.76 &  0.96 &  0.46 &   79 \\ 
NGC6401 & 264.6581 &    0.1156 &  -2.8193 &   1.4424 & -0.03 &  0.07  &  484 &      &      \\ 
C1735-238 & -23.9173 &   0.0055 &   0.0116 &   0.0095 &   0.15 &  0.11 &  0.94 &     &    0 \\ 
NGC6402 & 264.3984 &    0.0536 &  -3.6146 &  -5.0357 & -0.08 &  0.17  & 4203 &      &      \\ 
C1735-032 &  -3.2473 &   0.0027 &   0.0067 &   0.0059 &   0.15 &  0.20 &  1.33 &     &    0 \\ 
NGC6440 & 267.2028 &    0.0958 &  -1.2135 &  -3.8830 & -0.05 &  0.29  & 1033 & -72.58 &  1.75  \\ 
C1746-203 & -20.3521 &   0.0058 &   0.0124 &   0.0096 &   0.15 &  0.09 &  0.97 &  0.69 &    2 \\ 
NGC6441 & 267.5540 &    0.0403 &  -2.5394 &  -5.3010 & -0.18 &  0.17  & 2121 &      &      \\ 
C1746-370 & -37.0660 &   0.0037 &   0.0070 &   0.0057 &   0.17 &  0.12 &  1.03 &     &    0 \\ 
NGC6453 & 267.7197 &    0.0425 &   0.0699 &  -5.8521 & -0.11 &  0.14  &  710 & -94.94 &  0.52  \\ 
C1748-346 & -34.6002 &   0.0077 &   0.0164 &   0.0136 &   0.13 &  0.07 &  1.65 &  1.06 &    2 \\ 
NGC6496 & 269.7677 &    0.0803 &  -3.0290 &  -9.1971 & -0.15 &  0.07  & 1860 &      &      \\ 
C1755-442 & -44.2660 &   0.0031 &   0.0057 &   0.0050 &   0.17 &  0.13 &  0.87 &     &    0 \\ 
NGC6517 & 270.4528 &    0.0217 &  -1.5209 &  -4.2622 & -0.15 &  0.28  &  880 & -32.45 &      \\ 
C1759-089 &  -8.9568 &   0.0072 &   0.0139 &   0.0114 &   0.11 &  0.08 &  1.20 &     &    1 \\ 
NGC6522 & 270.8956 &    0.0697 &   2.5780 &  -6.3412 & -0.02 &  0.12  &  474 & -17.26 &  0.32  \\ 
C1800-300 & -30.0350 &   0.0050 &   0.0124 &   0.0109 &   0.09 &  0.09 &  1.00 &  0.38 &    2 \\ 
NGC6528 & 271.2039 &    0.0746 &  -2.1879 &  -5.5718 &  0.04 &  0.11  &  354 &      &      \\ 
C1801-300 & -30.0550 &   0.0074 &   0.0160 &   0.0137 &   0.08 &  0.05 &  0.94 &     &    0 \\ 
NGC6535 & 270.9590 &    0.1294 &  -4.2101 &  -2.9461 &  0.10 &  0.10  &  740 & -211.40 &  1.36  \\ 
C1801-003 &  -0.2953 &   0.0047 &   0.0115 &   0.0108 &  -0.02 &  0.11 &  0.87 &  0.30 &    2 \\ 
NGC6539 & 271.1924 &    0.0630 &  -6.8310 &  -3.4792 & -0.44 &  0.18  & 1149 &      &      \\ 
C1802-075 &  -7.5896 &   0.0084 &   0.0106 &   0.0085 &   0.22 &  0.26 &  1.38 &     &    0 \\ 
NGC6541 & 271.9827 &    0.1139 &   0.2762 &  -8.7659 & -0.04 & -0.03  & 2987 & -164.64 &  1.97  \\ 
C1804-437 & -43.7144 &   0.0025 &   0.0054 &   0.0048 &   0.08 &  0.27 &  0.91 &  1.09 &    8 \\ 
NGC6544 & 271.8438 &    0.3311 &  -2.3280 & -18.5574 & -0.04 &  0.11  & 3266 & -27.06 &  7.35  \\ 
C1804-250 & -25.0186 &   0.0022 &   0.0089 &   0.0083 &   0.04 &  0.25 &  1.03 &  0.76 &    5 \\ 
NGC6626 & 276.1349 &    0.1469 &  -0.4236 &  -8.8037 & -0.15 &  0.07  & 1969 &  14.55 &  4.41  \\ 
C1821-249 & -24.8430 &   0.0033 &   0.0087 &   0.0082 &   0.12 &  0.19 &  1.04 &  0.76 &   17 \\ 
NGC6637 & 277.8342 &    0.0746 &  -5.0669 &  -5.8017 & -0.01 &  0.23  &  773 &  47.19 &  2.61  \\ 
C1828-323 & -32.3565 &   0.0032 &   0.0104 &   0.0094 &  -0.05 &  0.10 &  0.93 &  1.71 &    3 \\ 
NGC6656 & 279.1048 &    0.2602 &   9.8019 &  -5.5643 & -0.05 &  0.15  & 16261 & -147.60 & 13.11  \\ 
C1833-239 & -23.9102 &   0.0009 &   0.0036 &   0.0034 &   0.02 &  0.85 &  1.01 &  0.57 &  116 \\ 
NGC6681 & 280.8020 &    0.1096 &   1.3853 &  -4.7174 & -0.24 &  0.26  & 1276 & 215.87 &  0.23  \\ 
C1840-323 & -32.2892 &   0.0038 &   0.0076 &   0.0065 &   0.10 &  0.08 &  0.99 &  0.35 &    2 \\ 
NGC6752 & 287.7175 &    0.2310 &  -3.1908 &  -4.0347 & -0.29 &  0.19  & 23684 & -26.12 &  7.82  \\ 
C1906-600 & -59.9833 &   0.0011 &   0.0018 &   0.0020 &   0.03 &  0.55 &  1.02 &  0.51 &   82 \\ 
NGC6779 & 289.1480 &    0.0702 &  -2.0092 &   1.6553 & -0.05 &  0.03  & 2379 & -136.67 &  2.56  \\ 
C1914+300 &  30.1840 &   0.0015 &   0.0051 &   0.0056 &  -0.03 &  0.12 &  1.15 &  1.00 &   11 \\ 
NGC6809 & 295.0046 &    0.1707 &  -3.4017 &  -9.2642 & -0.03 &  0.18  & 13046 & 176.46 &  4.64  \\ 
C1936-310 & -30.9621 &   0.0011 &   0.0031 &   0.0028 &   0.00 &  0.28 &  0.90 &  0.57 &   47 \\ 
NGC6838 & 298.4427 &    0.2252 &  -3.3842 &  -2.6528 & -0.11 &  0.11  & 6766 & -21.01 &  3.96  \\ 
C1951+186 &  18.7790 &   0.0010 &   0.0027 &   0.0028 &  -0.01 &  0.19 &  0.90 &  0.53 &   22 \\ 
NGC6864 & 301.5205 &    0.0208 &  -0.5869 &  -2.7839 & -0.35 &  0.21  &  946 & -185.33 &  0.72  \\ 
C2003-220 & -21.9213 &   0.0066 &   0.0088 &   0.0065 &  -0.22 &  0.05 &  1.46 &  1.51 &    3 \\ 
NGC6981 & 313.3662 &    0.0225 &  -1.2488 &  -3.3117 & -0.38 &  0.26  &  974 &      &      \\ 
C2050-127 & -12.5386 &   0.0063 &   0.0089 &   0.0068 &  -0.13 &  0.18 &  1.26 &     &    0 \\ 
NGC7078 & 322.4949 &    0.0568 &  -0.6238 &  -3.7960 & -0.02 & -0.04  & 4479 & -105.58 &  5.29  \\ 
C2127+119 &  12.1661 &   0.0014 &   0.0041 &   0.0039 &  -0.15 &  0.40 &  0.87 &  1.45 &   12 \\ 
NGC7089 & 323.3497 &    0.0591 &   3.4911 &  -2.1501 & -0.14 & -0.04  & 1259 &  -4.79 &  1.42  \\ 
C2130-010 &  -0.8177 &   0.0035 &   0.0077 &   0.0071 &  -0.14 &  0.20 &  1.03 &  0.27 &    5 \\ 
NGC7099 & 325.0888 &    0.0746 &  -0.7017 &  -7.2218 & -0.29 &  0.30  & 3554 & -186.48 &  3.80  \\ 
C2137-234 & -23.1792 &   0.0040 &   0.0063 &   0.0055 &  -0.27 &  0.26 &  1.17 &  0.93 &   13 \\ 
\\
\end{longtable}
}
\end{center}

\begin{table*}
\footnotesize
\caption{Overview of the astrometric parameters for dwarf spheroidal galaxies. For each dSph we include the derived position on the sky ($\alpha$, $\delta$), parallax $\varpi$, PMs ($\mu_{\alpha*}$, $\mu_\delta$), and the elements of the covariance matrix $\epsilon_\varpi$, $\epsilon_{\mu\alpha *}$, $\epsilon_{\mu\delta}$ and correlation coefficients $C$. The last two columns list the number of stars and the magnitude limit used for the determination of the astrometric parameters, respectively. The ($\alpha$, $\delta$) listed here are determined from stars with five-parameter solutions, and hence these coordinates might not provide the most accurate estimate of the centre of the dSph because of incompleteness in the spatial coverage of such solutions (see e.g. Fig.~\ref{fig:fov_dwarfs}). For the orbital integrations in Sec.~\ref{sec:orbits} we therefore
used ($\alpha$, $\delta$) sky coordinates from the literature.}\label{tab:PM_dwarfs} 
\begin{tabular}{lccccccccrrrcc}
\hline
Name & $\alpha$ & $\delta$ & $\varpi$ & $\epsilon_\varpi $  & $\mu_{\alpha^*}$ & $\epsilon_{\mu_{\alpha^*}}$ &  $\mu_\delta$ &  $\epsilon_{\mu_\delta}$  &  C$_{\varpi,\mu_\alpha}$ &  C$_{\varpi,\mu_\delta}$ &  C$_{\mu_\alpha,\mu_\delta}$ & nMemb & $G_{lim}$\\
 & [deg] & [deg] & [mas] & [mas] & [mas/yr] & [mas/yr]  & [mas/yr]  & [mas/yr] & & & &  & [mag] \\
\hline
Fnx  &  39.9971 & -34.4492   & -0.054  & 0.002           &           0.376        &             0.003        &       -0.413     &        0.003        &    0.16             &   -0.46       &            -0.09    &       7722  & 19.9\\
Dra   &  260.0517  & 57.9153  & -0.052  & 0.005          &           -0.019      &               0.009        &       -0.145     &       0.010       &    -0.18         &        0.12         &          -0.08        &     422 &  19.5 \\
Car   & 100.4029 & -50.9661  & -0.015  &  0.005      &                0.495        &             0.015         &       0.143     &        0.014      &     -0.00      &           0.02        &           -0.08    &         257  &  19.1 \\
U Min & 227.2854 & 67.2225  & -0.039  &  0.006             &       -0.182        &             0.010      &           0.074      &       0.008    &      -0.01      &          -0.31      &            -0.34       &      925 &   19.8 \\
Sext & 153.2625 & -1.6147  & -0.102 &   0.023         &            -0.496     &                0.025           &     0.077       &      0.020    &        0.28       &         -0.10     &              -0.45     &        205  & 19.7 \\
Leo~I      &  152.1171 & 12.3064 &  -0.214  &  0.065      &              -0.097       &              0.056      &          -0.091      &       0.047       &     0.29       &         -0.30          &         -0.51     &        174  & 19.9 \\
Leo~II     &  168.3700  & 22.1517 &  -0.001  &  0.037     &               -0.064          &           0.057      &          -0.210      &      0.054       &     -0.18      &          -0.24    &                0.05      &       116  & 20.0 \\
Sgr    & 283.8313 & -30.5453  &  0.003  &  0.001                 &    -2.692            &         0.001         &       -1.359   &         0.001     &       -0.17      &          0.21       &             0.09      &    23109  & 18.0 \\
Scl & 15.0392  & -33.7092  & -0.013  &  0.004    &                  0.082         &             0.005       &        -0.131     &       0.004      &       0.17     &            0.15      &            0.23       &    1592  & 19.5 \\
Boo~I  &  210.025  & 14.500  & -0.069  &  0.024    &                 -0.459        &              0.041      &         -1.064    &         0.029    &       0.01     &            0.11      &              0.16      &       115 &  19.7 \\
\hline
\end{tabular}
\end{table*}

\onecolumn
\begin{center}
\scriptsize{
\begin{longtable}{lcccccc}
\caption{Globular cluster position and velocity from the Sun, not corrected for the solar motion or the local standard of rest. Quoted values are medians, with errors that indicate uncertainties calculated from the 16th and 84th percentiles, and were obtained from Monte Carlo sampling the (statistical and our best estimates of the systematic) errors in the observables. X is towards $l=0$, Y is in the direction of Galactic rotation (towards $l=90^\circ$), and Z towards $b=90$ (the Galactic north pole). $U$, $V,$ and $W$ are the velocities in these directions.\label{tab:gc_uvw}}\\
\hline\hline 
Name & $X$  [kpc]& $Y$ [kpc]&$Z$ [kpc]&$U$ [km/s] & $V$ [km/s] & $W$ [km/s] \\
\hline 
\endfirsthead

\multicolumn{7}{c}%
{{\bfseries \tablename\ \thetable{} -- continued from previous page}} \\
\hline 
Name & $X$  [kpc]& $Y$ [kpc]&$Z$ [kpc]&$U$ [km/s] & $V$ [km/s] & $W$ [km/s] \\
\hline 
\endhead

\hline \multicolumn{7}{c}{{Continued on next page}} \\ \hline
\endfoot

\hline \hline
\endlastfoot
NGC0104 & $1.87^{+0.04}_{-0.04}$ & $-2.58^{+0.06}_{-0.06}$ & $-3.18^{+0.07}_{-0.07}$ & $-88.6^{+2.0}_{-2.0}$ & $-80.1^{+2.1}_{-2.2}$ & $38.5^{+0.8}_{-0.8}$ \\ \noalign{\smallskip}
NGC0288 & $-0.084^{+0.002}_{-0.002}$ & $0.046^{+0.001}_{-0.001}$ & $-8.89^{+0.21}_{-0.20}$ & $-19.3^{+1.6}_{-1.5}$ & $-297.3^{+7.0}_{-6.9}$ & $44.0^{+0.2}_{-0.2}$ \\ \noalign{\smallskip}
NGC0362 & $3.11^{+0.07}_{-0.07}$ & $-5.07^{+0.12}_{-0.11}$ & $-6.21^{+0.15}_{-0.14}$ & $-100.0^{+4.4}_{-4.4}$ & $-344.8^{+5.1}_{-5.0}$ & $-78.1^{+2.2}_{-2.2}$ \\ \noalign{\smallskip}
NGC1851 & $-4.26^{+0.10}_{-0.10}$ & $-8.94^{+0.20}_{-0.20}$ & $-6.95^{+0.16}_{-0.16}$ & $-94.5^{+1.9}_{-2.0}$ & $-319.5^{+2.4}_{-2.3}$ & $-88.9^{+2.7}_{-2.8}$ \\ \noalign{\smallskip}
NGC1904 & $-7.63^{+0.17}_{-0.17}$ & $-8.25^{+0.19}_{-0.19}$ & $-6.32^{+0.14}_{-0.14}$ & $-57.6^{+2.3}_{-2.3}$ & $-266.4^{+3.5}_{-3.6}$ & $-2.6^{+3.0}_{-3.0}$ \\ \noalign{\smallskip}
NGC2298 & $-4.29^{+0.10}_{-0.10}$ & $-9.46^{+0.22}_{-0.22}$ & $-2.98^{+0.07}_{-0.07}$ & $80.5^{+3.7}_{-3.7}$ & $-227.8^{+2.6}_{-2.7}$ & $67.5^{+3.1}_{-3.1}$ \\ \noalign{\smallskip}
NGC2808 & $1.99^{+0.05}_{-0.05}$ & $-9.20^{+0.21}_{-0.21}$ & $-1.87^{+0.04}_{-0.04}$ & $43.5^{+1.6}_{-1.7}$ & $-101.0^{+0.8}_{-0.8}$ & $21.7^{+1.9}_{-1.9}$ \\ \noalign{\smallskip}
NGC3201 & $0.61^{+0.01}_{-0.01}$ & $-4.80^{+0.11}_{-0.11}$ & $0.74^{+0.02}_{-0.02}$ & $244.8^{+4.3}_{-4.3}$ & $-450.6^{+0.8}_{-0.8}$ & $143.5^{+1.8}_{-1.8}$ \\ \noalign{\smallskip}
NGC4372 & $2.94^{+0.07}_{-0.07}$ & $-4.90^{+0.11}_{-0.11}$ & $-1.00^{+0.02}_{-0.02}$ & $-114.6^{+3.6}_{-3.7}$ & $-166.6^{+2.7}_{-2.7}$ & $59.7^{+1.9}_{-1.9}$ \\ \noalign{\smallskip}
NGC4590 & $4.12^{+0.10}_{-0.09}$ & $-7.24^{+0.17}_{-0.17}$ & $6.06^{+0.14}_{-0.14}$ & $-182.5^{+3.7}_{-3.7}$ & $38.2^{+1.4}_{-1.4}$ & $8.6^{+2.0}_{-2.1}$ \\ \noalign{\smallskip}
NGC4833 & $3.62^{+0.08}_{-0.08}$ & $-5.44^{+0.12}_{-0.12}$ & $-0.92^{+0.02}_{-0.02}$ & $-109.2^{+5.2}_{-5.1}$ & $-307.1^{+3.4}_{-3.4}$ & $-48.7^{+1.2}_{-1.2}$ \\ \noalign{\smallskip}
NGC5024 & $2.83^{+0.07}_{-0.07}$ & $-1.45^{+0.03}_{-0.03}$ & $17.61^{+0.41}_{-0.41}$ & $44.2^{+3.1}_{-3.2}$ & $-95.2^{+3.9}_{-3.7}$ & $-78.9^{+0.7}_{-0.7}$ \\ \noalign{\smallskip}
NGC5053 & $3.04^{+0.07}_{-0.07}$ & $-1.37^{+0.03}_{-0.03}$ & $17.08^{+0.40}_{-0.39}$ & $42.2^{+2.9}_{-3.1}$ & $-104.7^{+3.8}_{-3.6}$ & $28.9^{+0.8}_{-0.8}$ \\ \noalign{\smallskip}
NGC5139 & $3.17^{+0.07}_{-0.07}$ & $-3.90^{+0.09}_{-0.09}$ & $1.34^{+0.03}_{-0.03}$ & $87.7^{+1.4}_{-1.4}$ & $-268.7^{+2.3}_{-2.3}$ & $-88.3^{+3.5}_{-3.5}$ \\ \noalign{\smallskip}
NGC5272 & $1.48^{+0.03}_{-0.03}$ & $1.34^{+0.03}_{-0.03}$ & $10.00^{+0.23}_{-0.23}$ & $54.2^{+2.4}_{-2.5}$ & $-121.5^{+2.9}_{-2.9}$ & $-142.2^{+0.4}_{-0.4}$ \\ \noalign{\smallskip}
NGC5286 & $7.64^{+0.18}_{-0.18}$ & $-8.60^{+0.20}_{-0.20}$ & $2.15^{+0.05}_{-0.05}$ & $44.9^{+1.8}_{-1.8}$ & $-38.1^{+1.8}_{-1.7}$ & $0.5^{+2.0}_{-1.9}$ \\ \noalign{\smallskip}
NGC5466 & $3.35^{+0.08}_{-0.08}$ & $3.03^{+0.07}_{-0.07}$ & $15.34^{+0.35}_{-0.35}$ & $-230.9^{+6.4}_{-6.4}$ & $-284.7^{+7.7}_{-7.3}$ & $222.1^{+2.8}_{-2.8}$ \\ \noalign{\smallskip}
NGC5634 & $15.66^{+0.36}_{-0.36}$ & $-5.02^{+0.12}_{-0.12}$ & $19.09^{+0.44}_{-0.44}$ & $-87.6^{+5.6}_{-5.5}$ & $-259.3^{+7.8}_{-7.5}$ & $-55.9^{+5.8}_{-5.8}$ \\ \noalign{\smallskip}
NGC5897 & $10.32^{+0.24}_{-0.24}$ & $-3.17^{+0.07}_{-0.07}$ & $6.31^{+0.14}_{-0.15}$ & $-44.1^{+3.3}_{-3.3}$ & $-383.0^{+8.6}_{-8.4}$ & $81.1^{+2.0}_{-2.0}$ \\ \noalign{\smallskip}
NGC5904 & $5.12^{+0.12}_{-0.12}$ & $0.345^{+0.008}_{-0.008}$ & $5.47^{+0.13}_{-0.13}$ & $290.4^{+6.1}_{-6.1}$ & $-163.1^{+4.1}_{-4.1}$ & $-188.8^{+5.4}_{-5.4}$ \\ \noalign{\smallskip}
NGC5927 & $6.40^{+0.15}_{-0.15}$ & $-4.22^{+0.10}_{-0.10}$ & $0.65^{+0.02}_{-0.01}$ & $-210.2^{+3.0}_{-3.1}$ & $-123.2^{+4.4}_{-4.4}$ & $-2.7^{+1.3}_{-1.3}$ \\ \noalign{\smallskip}
NGC5946 & $8.92^{+0.20}_{-0.20}$ & $-5.67^{+0.13}_{-0.13}$ & $0.77^{+0.02}_{-0.02}$ & $-36.6^{+3.7}_{-3.8}$ & $-284.8^{+5.2}_{-5.1}$ & $94.6^{+2.6}_{-2.7}$ \\ \noalign{\smallskip}
NGC5986 & $9.32^{+0.22}_{-0.21}$ & $-3.95^{+0.09}_{-0.09}$ & $2.39^{+0.06}_{-0.05}$ & $-29.7^{+4.2}_{-4.2}$ & $-316.8^{+6.9}_{-7.0}$ & $-21.1^{+2.1}_{-2.1}$ \\ \noalign{\smallskip}
NGC6093 & $9.35^{+0.22}_{-0.22}$ & $-1.20^{+0.03}_{-0.03}$ & $3.33^{+0.08}_{-0.08}$ & $-6.3^{+1.6}_{-1.6}$ & $-290.3^{+7.0}_{-6.8}$ & $-68.6^{+2.4}_{-2.3}$ \\ \noalign{\smallskip}
NGC6121 & $2.09^{+0.05}_{-0.05}$ & $-0.332^{+0.007}_{-0.008}$ & $0.61^{+0.01}_{-0.01}$ & $40.4^{+0.7}_{-0.7}$ & $-243.6^{+5.2}_{-5.5}$ & $-15.9^{+0.9}_{-0.9}$ \\ \noalign{\smallskip}
NGC6144 & $8.48^{+0.19}_{-0.20}$ & $-1.20^{+0.03}_{-0.03}$ & $2.41^{+0.05}_{-0.06}$ & $171.0^{+0.8}_{-0.8}$ & $-158.1^{+3.4}_{-3.3}$ & $34.8^{+1.6}_{-1.5}$ \\ \noalign{\smallskip}
NGC6171 & $5.88^{+0.13}_{-0.14}$ & $0.347^{+0.008}_{-0.008}$ & $2.50^{+0.06}_{-0.06}$ & $3.7^{+0.9}_{-1.0}$ & $-179.2^{+4.2}_{-4.2}$ & $-71.2^{+1.7}_{-1.6}$ \\ \noalign{\smallskip}
NGC6205 & $2.76^{+0.06}_{-0.06}$ & $4.60^{+0.11}_{-0.10}$ & $4.65^{+0.11}_{-0.11}$ & $-43.4^{+1.6}_{-1.6}$ & $-263.1^{+2.6}_{-2.6}$ & $-86.9^{+1.9}_{-1.9}$ \\ \noalign{\smallskip}
NGC6218 & $4.14^{+0.09}_{-0.10}$ & $1.17^{+0.03}_{-0.03}$ & $2.13^{+0.05}_{-0.05}$ & $33.5^{+1.6}_{-1.7}$ & $-128.6^{+2.8}_{-2.7}$ & $-88.2^{+1.7}_{-1.8}$ \\ \noalign{\smallskip}
NGC6235 & $11.17^{+0.26}_{-0.26}$ & $-0.211^{+0.005}_{-0.005}$ & $2.69^{+0.06}_{-0.06}$ & $94.4^{+3.3}_{-3.3}$ & $-460.0^{+11.0}_{-10.8}$ & $-55.2^{+2.8}_{-2.7}$ \\ \noalign{\smallskip}
NGC6254 & $3.91^{+0.09}_{-0.09}$ & $1.06^{+0.02}_{-0.02}$ & $1.73^{+0.04}_{-0.04}$ & $106.2^{+1.2}_{-1.2}$ & $-144.8^{+3.8}_{-3.8}$ & $40.1^{+0.8}_{-0.8}$ \\ \noalign{\smallskip}
NGC6266 & $6.70^{+0.16}_{-0.15}$ & $-0.76^{+0.02}_{-0.02}$ & $0.87^{+0.02}_{-0.02}$ & $-99.2^{+1.5}_{-1.6}$ & $-170.5^{+4.3}_{-4.3}$ & $68.8^{+2.1}_{-2.1}$ \\ \noalign{\smallskip}
NGC6273 & $8.67^{+0.20}_{-0.20}$ & $-0.47^{+0.01}_{-0.01}$ & $1.43^{+0.03}_{-0.03}$ & $107.5^{+4.1}_{-4.0}$ & $-32.4^{+1.6}_{-1.6}$ & $167.9^{+3.8}_{-3.7}$ \\ \noalign{\smallskip}
NGC6284 & $15.07^{+0.35}_{-0.35}$ & $-0.44^{+0.01}_{-0.01}$ & $2.64^{+0.06}_{-0.06}$ & $2.6^{+1.8}_{-1.8}$ & $-256.7^{+6.4}_{-6.4}$ & $102.1^{+3.4}_{-3.6}$ \\ \noalign{\smallskip}
NGC6287 & $9.23^{+0.21}_{-0.21}$ & $0.021^{+0.000}_{-0.000}$ & $1.80^{+0.04}_{-0.04}$ & $-307.1^{+3.5}_{-3.5}$ & $-197.5^{+4.9}_{-4.8}$ & $69.0^{+3.4}_{-3.3}$ \\ \noalign{\smallskip}
NGC6293 & $9.40^{+0.22}_{-0.21}$ & $-0.391^{+0.009}_{-0.009}$ & $1.29^{+0.03}_{-0.03}$ & $-130.9^{+1.8}_{-1.7}$ & $-130.5^{+3.6}_{-3.5}$ & $-161.8^{+3.7}_{-3.6}$ \\ \noalign{\smallskip}
NGC6304 & $5.86^{+0.14}_{-0.14}$ & $-0.43^{+0.01}_{-0.01}$ & $0.55^{+0.01}_{-0.01}$ & $-119.8^{+3.6}_{-3.6}$ & $-81.1^{+2.3}_{-2.4}$ & $61.4^{+2.0}_{-2.0}$ \\ \noalign{\smallskip}
NGC6316 & $10.34^{+0.24}_{-0.24}$ & $-0.51^{+0.01}_{-0.01}$ & $1.04^{+0.02}_{-0.02}$ & $48.7^{+8.8}_{-9.0}$ & $-325.6^{+7.5}_{-7.8}$ & $71.0^{+2.5}_{-2.5}$ \\ \noalign{\smallskip}
NGC6325 & $7.72^{+0.17}_{-0.18}$ & $0.131^{+0.003}_{-0.003}$ & $1.09^{+0.02}_{-0.02}$ & $27.8^{+1.8}_{-1.8}$ & $-449.5^{+10.3}_{-10.3}$ & $70.9^{+2.1}_{-2.1}$ \\ \noalign{\smallskip}
NGC6333 & $7.73^{+0.18}_{-0.18}$ & $0.75^{+0.02}_{-0.02}$ & $1.47^{+0.03}_{-0.03}$ & $237.7^{+6.8}_{-7.0}$ & $-123.4^{+3.6}_{-3.6}$ & $44.8^{+1.8}_{-1.9}$ \\ \noalign{\smallskip}
NGC6341 & $2.51^{+0.06}_{-0.06}$ & $6.33^{+0.14}_{-0.14}$ & $4.74^{+0.11}_{-0.11}$ & $-36.7^{+1.4}_{-1.3}$ & $-208.6^{+2.8}_{-2.8}$ & $87.9^{+3.7}_{-3.7}$ \\ \noalign{\smallskip}
NGC6342 & $8.35^{+0.19}_{-0.19}$ & $0.72^{+0.02}_{-0.02}$ & $1.44^{+0.03}_{-0.03}$ & $148.9^{+1.7}_{-1.6}$ & $-289.3^{+7.0}_{-7.1}$ & $-36.4^{+2.0}_{-1.9}$ \\ \noalign{\smallskip}
NGC6352 & $5.27^{+0.12}_{-0.12}$ & $-1.77^{+0.04}_{-0.04}$ & $-0.70^{+0.02}_{-0.02}$ & $-172.2^{+1.4}_{-1.5}$ & $-79.3^{+2.9}_{-3.0}$ & $1.1^{+1.0}_{-1.0}$ \\ \noalign{\smallskip}
NGC6356 & $14.76^{+0.34}_{-0.34}$ & $1.74^{+0.04}_{-0.04}$ & $2.68^{+0.06}_{-0.06}$ & $50.6^{+4.3}_{-4.3}$ & $-345.8^{+8.6}_{-8.1}$ & $97.9^{+3.3}_{-3.5}$ \\ \noalign{\smallskip}
NGC6362 & $5.98^{+0.14}_{-0.13}$ & $-4.10^{+0.09}_{-0.10}$ & $-2.29^{+0.05}_{-0.05}$ & $-211.5^{+2.8}_{-2.8}$ & $-151.9^{+5.1}_{-5.1}$ & $124.7^{+2.4}_{-2.3}$ \\ \noalign{\smallskip}
NGC6366 & $3.19^{+0.07}_{-0.07}$ & $1.06^{+0.02}_{-0.02}$ & $0.97^{+0.02}_{-0.02}$ & $22.1^{+1.0}_{-0.9}$ & $-73.6^{+1.7}_{-1.7}$ & $-39.4^{+1.0}_{-1.0}$ \\ \noalign{\smallskip}
NGC6380 & $10.72^{+0.25}_{-0.26}$ & $-1.85^{+0.04}_{-0.04}$ & $-0.65^{+0.02}_{-0.02}$ & $-37.1^{+2.6}_{-2.6}$ & $-193.9^{+5.0}_{-4.9}$ & $2.5^{+2.1}_{-2.0}$ \\ \noalign{\smallskip}
NGC6388 & $9.52^{+0.22}_{-0.22}$ & $-2.45^{+0.06}_{-0.06}$ & $-1.16^{+0.03}_{-0.03}$ & $40.2^{+1.2}_{-1.2}$ & $-156.6^{+3.6}_{-3.5}$ & $-22.8^{+1.7}_{-1.7}$ \\ \noalign{\smallskip}
NGC6397 & $2.09^{+0.05}_{-0.05}$ & $-0.84^{+0.02}_{-0.02}$ & $-0.48^{+0.01}_{-0.01}$ & $-62.5^{+1.8}_{-1.9}$ & $-134.7^{+3.0}_{-3.0}$ & $-127.9^{+2.9}_{-2.9}$ \\ \noalign{\smallskip}
NGC6401 & $10.56^{+0.24}_{-0.24}$ & $0.64^{+0.01}_{-0.01}$ & $0.74^{+0.02}_{-0.02}$ & $-74.9^{+8.3}_{-8.7}$ & $-18.3^{+1.9}_{-1.9}$ & $153.6^{+4.2}_{-4.2}$ \\ \noalign{\smallskip}
NGC6402 & $8.37^{+0.19}_{-0.19}$ & $3.27^{+0.07}_{-0.08}$ & $2.38^{+0.05}_{-0.05}$ & $31.8^{+2.7}_{-2.8}$ & $-279.0^{+6.1}_{-6.1}$ & $13.0^{+1.7}_{-1.7}$ \\ \noalign{\smallskip}
NGC6440 & $8.41^{+0.19}_{-0.19}$ & $1.14^{+0.03}_{-0.03}$ & $0.56^{+0.01}_{-0.01}$ & $-51.8^{+2.7}_{-2.8}$ & $-167.9^{+4.0}_{-3.9}$ & $-43.1^{+1.7}_{-1.7}$ \\ \noalign{\smallskip}
NGC6441 & $11.48^{+0.27}_{-0.26}$ & $-1.30^{+0.03}_{-0.03}$ & $-1.01^{+0.02}_{-0.02}$ & $-22.3^{+1.4}_{-1.3}$ & $-321.7^{+7.4}_{-7.7}$ & $-28.9^{+2.1}_{-2.0}$ \\ \noalign{\smallskip}
NGC6453 & $11.54^{+0.26}_{-0.27}$ & $-0.86^{+0.02}_{-0.02}$ & $-0.78^{+0.02}_{-0.02}$ & $-115.1^{+8.2}_{-8.4}$ & $-267.2^{+6.7}_{-6.6}$ & $-160.9^{+4.4}_{-4.3}$ \\ \noalign{\smallskip}
NGC6496 & $10.89^{+0.25}_{-0.25}$ & $-2.31^{+0.05}_{-0.05}$ & $-1.97^{+0.05}_{-0.05}$ & $-229.9^{+6.2}_{-6.2}$ & $-473.4^{+11.5}_{-11.8}$ & $-70.0^{+2.9}_{-2.9}$ \\ \noalign{\smallskip}
NGC6517 & $9.94^{+0.23}_{-0.23}$ & $3.47^{+0.08}_{-0.08}$ & $1.25^{+0.03}_{-0.03}$ & $40.8^{+7.9}_{-7.7}$ & $-223.3^{+5.8}_{-5.8}$ & $-41.1^{+2.2}_{-2.3}$ \\ \noalign{\smallskip}
NGC6522 & $7.68^{+0.17}_{-0.18}$ & $0.137^{+0.003}_{-0.003}$ & $-0.53^{+0.01}_{-0.01}$ & $-31.7^{+3.4}_{-3.4}$ & $-156.7^{+3.9}_{-3.8}$ & $-193.3^{+4.7}_{-4.6}$ \\ \noalign{\smallskip}
NGC6528 & $7.88^{+0.18}_{-0.18}$ & $0.157^{+0.004}_{-0.004}$ & $-0.58^{+0.01}_{-0.01}$ & $208.3^{+1.4}_{-1.4}$ & $-218.1^{+5.4}_{-5.3}$ & $-44.7^{+1.6}_{-1.5}$ \\ \noalign{\smallskip}
NGC6535 & $5.95^{+0.14}_{-0.13}$ & $3.06^{+0.07}_{-0.07}$ & $1.23^{+0.03}_{-0.03}$ & $-133.2^{+1.5}_{-1.5}$ & $-233.8^{+3.3}_{-3.4}$ & $35.9^{+2.1}_{-2.1}$ \\ \noalign{\smallskip}
NGC6539 & $7.24^{+0.17}_{-0.16}$ & $2.75^{+0.06}_{-0.06}$ & $0.92^{+0.02}_{-0.02}$ & $94.2^{+2.2}_{-2.2}$ & $-214.5^{+5.2}_{-5.3}$ & $162.6^{+3.8}_{-3.8}$ \\ \noalign{\smallskip}
NGC6541 & $7.23^{+0.17}_{-0.17}$ & $-1.37^{+0.03}_{-0.03}$ & $-1.46^{+0.03}_{-0.03}$ & $-232.4^{+2.9}_{-2.8}$ & $-234.1^{+6.3}_{-6.2}$ & $-116.6^{+3.7}_{-3.6}$ \\ \noalign{\smallskip}
NGC6544 & $2.98^{+0.07}_{-0.07}$ & $0.305^{+0.007}_{-0.007}$ & $-0.115^{+0.003}_{-0.003}$ & $-5.8^{+3.8}_{-3.9}$ & $-248.7^{+5.6}_{-5.6}$ & $-98.1^{+2.3}_{-2.3}$ \\ \noalign{\smallskip}
NGC6626 & $5.42^{+0.12}_{-0.13}$ & $0.74^{+0.02}_{-0.02}$ & $-0.53^{+0.01}_{-0.01}$ & $35.9^{+1.1}_{-1.1}$ & $-205.8^{+5.0}_{-4.9}$ & $-96.8^{+2.5}_{-2.4}$ \\ \noalign{\smallskip}
NGC6637 & $8.65^{+0.20}_{-0.20}$ & $0.260^{+0.006}_{-0.006}$ & $-1.57^{+0.04}_{-0.04}$ & $63.8^{+2.8}_{-2.9}$ & $-308.0^{+7.1}_{-7.3}$ & $76.9^{+2.6}_{-2.5}$ \\ \noalign{\smallskip}
NGC6656 & $3.13^{+0.07}_{-0.07}$ & $0.55^{+0.01}_{-0.01}$ & $-0.421^{+0.010}_{-0.010}$ & $-163.3^{+0.5}_{-0.5}$ & $-38.2^{+0.6}_{-0.6}$ & $-150.0^{+3.9}_{-4.0}$ \\ \noalign{\smallskip}
NGC6681 & $8.78^{+0.20}_{-0.20}$ & $0.437^{+0.010}_{-0.010}$ & $-1.95^{+0.04}_{-0.04}$ & $193.1^{+1.1}_{-1.1}$ & $-149.6^{+3.9}_{-4.0}$ & $-181.3^{+3.5}_{-3.4}$ \\ \noalign{\smallskip}
NGC6752 & $3.31^{+0.08}_{-0.08}$ & $-1.44^{+0.03}_{-0.03}$ & $-1.73^{+0.04}_{-0.04}$ & $-38.3^{+0.6}_{-0.5}$ & $-77.3^{+2.1}_{-2.1}$ & $52.7^{+1.1}_{-1.2}$ \\ \noalign{\smallskip}
NGC6779 & $4.27^{+0.10}_{-0.10}$ & $8.26^{+0.20}_{-0.19}$ & $1.36^{+0.03}_{-0.03}$ & $-92.1^{+1.7}_{-1.6}$ & $-121.9^{+1.1}_{-1.1}$ & $92.2^{+3.0}_{-3.1}$ \\ \noalign{\smallskip}
NGC6809 & $4.90^{+0.11}_{-0.11}$ & $0.76^{+0.02}_{-0.02}$ & $-2.13^{+0.05}_{-0.05}$ & $199.7^{+1.0}_{-1.0}$ & $-224.6^{+5.8}_{-5.8}$ & $-63.2^{+0.8}_{-0.8}$ \\ \noalign{\smallskip}
NGC6838 & $2.19^{+0.05}_{-0.05}$ & $3.33^{+0.07}_{-0.08}$ & $-0.318^{+0.007}_{-0.007}$ & $52.4^{+1.6}_{-1.6}$ & $-58.7^{+1.0}_{-1.0}$ & $31.2^{+1.0}_{-1.0}$ \\ \noalign{\smallskip}
NGC6864 & $17.66^{+0.41}_{-0.40}$ & $6.53^{+0.15}_{-0.15}$ & $-9.08^{+0.21}_{-0.21}$ & $-80.5^{+4.1}_{-4.0}$ & $-326.9^{+7.2}_{-7.1}$ & $44.0^{+3.7}_{-3.6}$ \\ \noalign{\smallskip}
NGC6981 & $11.69^{+0.27}_{-0.27}$ & $8.24^{+0.19}_{-0.19}$ & $-9.17^{+0.21}_{-0.21}$ & $136.5^{+4.9}_{-4.8}$ & $-251.9^{+6.2}_{-6.3}$ & $-0.3^{+3.2}_{-3.2}$ \\ \noalign{\smallskip}
NGC7078 & $3.90^{+0.09}_{-0.09}$ & $8.38^{+0.19}_{-0.19}$ & $-4.77^{+0.11}_{-0.11}$ & $89.9^{+3.3}_{-3.4}$ & $-195.1^{+2.7}_{-2.7}$ & $-35.6^{+2.5}_{-2.4}$ \\ \noalign{\smallskip}
NGC7089 & $5.57^{+0.13}_{-0.13}$ & $7.49^{+0.17}_{-0.17}$ & $-6.72^{+0.15}_{-0.16}$ & $-86.3^{+2.8}_{-2.8}$ & $-103.9^{+3.1}_{-3.0}$ & $-178.2^{+4.7}_{-4.5}$ \\ \noalign{\smallskip}
NGC7099 & $4.93^{+0.11}_{-0.11}$ & $2.53^{+0.06}_{-0.06}$ & $-5.91^{+0.14}_{-0.13}$ & $-18.3^{+2.4}_{-2.5}$ & $-317.7^{+6.1}_{-6.0}$ & $101.2^{+1.2}_{-1.2}$ \\ \noalign{\smallskip}
\\
\end{longtable}
}
\end{center}

\begin{table*}
\small
\begin{center}
\begin{tabular}{crrrrrr}
\hline
Name & $X$  [kpc]& $Y$ [kpc]&$Z$ [kpc]&$U$ [km/s] & $V$ [km/s] & $W$ [km/s] \\
\hline 
 \noalign{\smallskip}
Fornax  & $-33.1^{+2.6}_{-2.7}$ & $-51.1^{+4.1}_{-4.2}$ & $-134.5^{+10.8}_{-11.0}$ & $34.2^{+22.5}_{-23.4}$ & $-386.0^{+38.0}_{-36.9}$ & $77.2^{+14.8}_{-14.3}$ \\ \noalign{\smallskip}
Draco  & $4.0^{+0.3}_{-0.3}$ & $62.6^{+5.2}_{-4.5}$ & $43.5^{+3.6}_{-3.1}$ & $35.9^{+13.9}_{-14.8}$ & $-247.6^{+7.1}_{-7.2}$ & $-157.7^{+10.1}_{-10.2}$ \\ \noalign{\smallskip}
Carina  & $-16.7^{+0.9}_{-0.9}$ & $-95.7^{+5.0}_{-5.3}$ & $-39.7^{+2.1}_{-2.2}$ & $-51.1^{+18.9}_{-18.1}$ & $-298.4^{+9.8}_{-8.9}$ & $151.4^{+21.1}_{-23.6}$ \\ \noalign{\smallskip}
Ursa Minor  & $-13.9^{+0.5}_{-0.6}$ & $52.1^{+2.1}_{-2.0}$ & $53.6^{+2.2}_{-2.0}$ & $-12.8^{+12.2}_{-12.5}$ & $-205.0^{+10.0}_{-10.3}$ & $-153.7^{+9.7}_{-8.8}$ \\ \noalign{\smallskip}
Sextans  & $-28.4^{+1.4}_{-1.3}$ & $-57.0^{+2.8}_{-2.5}$ & $57.9^{+2.6}_{-2.8}$ & $-253.0^{+17.8}_{-19.8}$ & $-161.1^{+13.0}_{-11.0}$ & $50.8^{+14.0}_{-12.9}$ \\ \noalign{\smallskip}
Leo~I  & $-115.5^{+7.6}_{-7.2}$ & $-119.6^{+7.9}_{-7.4}$ & $192.0^{+11.9}_{-12.6}$ & $-177.0^{+80.3}_{-75.9}$ & $-243.0^{+61.0}_{-55.4}$ & $113.2^{+44.8}_{-47.4}$ \\ \noalign{\smallskip}
Leo~II  & $-69.0^{+3.9}_{-3.8}$ & $-58.3^{+3.3}_{-3.2}$ & $215.2^{+11.9}_{-12.3}$ & $13.2^{+73.3}_{-69.1}$ & $-253.9^{+73.8}_{-66.9}$ & $18.9^{+27.5}_{-28.9}$ \\ \noalign{\smallskip}
Sagittarius  & $25.2^{+2.0}_{-1.8}$ & $2.5^{+0.2}_{-0.2}$ & $-6.4^{+0.5}_{-0.5}$ & $221.3^{+7.2}_{-6.2}$ & $-266.5^{+19.9}_{-22.5}$ & $197.4^{+18.6}_{-17.1}$ \\ \noalign{\smallskip}
Sculptor  & $3.1^{+0.2}_{-0.2}$ & $-9.8^{+0.7}_{-0.7}$ & $-85.4^{+5.7}_{-6.1}$ & $6.2^{+15.3}_{-14.1}$ & $-74.0^{+15.6}_{-14.0}$ & $-103.5^{+1.8}_{-1.8}$ \\ \noalign{\smallskip}
Bootes~I  & $22.7^{+1.1}_{-1.0}$ & $-0.76^{+0.03}_{-0.04}$ & $61.0^{+2.8}_{-2.7}$ & $124.9^{+14.1}_{-15.3}$ & $-344.6^{+22.3}_{-21.6}$ & $57.9^{+5.7}_{-5.1}$ \\ \noalign{\smallskip}
LMC & $7.1^{+0.3}_{-0.3}$ & $-41.0^{+2.0}_{-2.0}$ & $-27.8^{+1.4}_{-1.4}$ & $-68.6^{+10.2}_{-9.7}$ & $-468.4^{+13.8}_{-13.5}$ & $201.0^{+18.0}_{-18.8}$ \\ \noalign{\smallskip}
SMC & $23.3^{+0.9}_{-0.9}$ & $-38.1^{+1.5}_{-1.5}$ & $-44.1^{+1.7}_{-1.7}$ & $14.8^{+10.0}_{-10.0}$ & $-425.0^{+16.0}_{-15.2}$ & $167.5^{+13.0}_{-13.3}$ \\ \noalign{\smallskip}
\hline
\end{tabular}
\caption{Dwarf positions and velocities from the Sun, not corrected for the solar motion or the local standard of rest. Quoted values are medians, with errors that indicate uncertainties calculated from the 16th and 84th percentiles, and were obtained from Monte Carlo sampling the (statistical and our best estimates of the systematic) errors in the observables. The conventions are the same as for the globular clusters in Table~\ref{tab:gc_uvw}.}
\label{tab:dw_uvw} 
\end{center}
\label{default}
\end{table*}%

\section{Orbital integrations}
\label{sec:app_orbits}

\begin{figure}
\centering
\includegraphics[width=9.5cm, trim={0 0 0 10cm},clip]{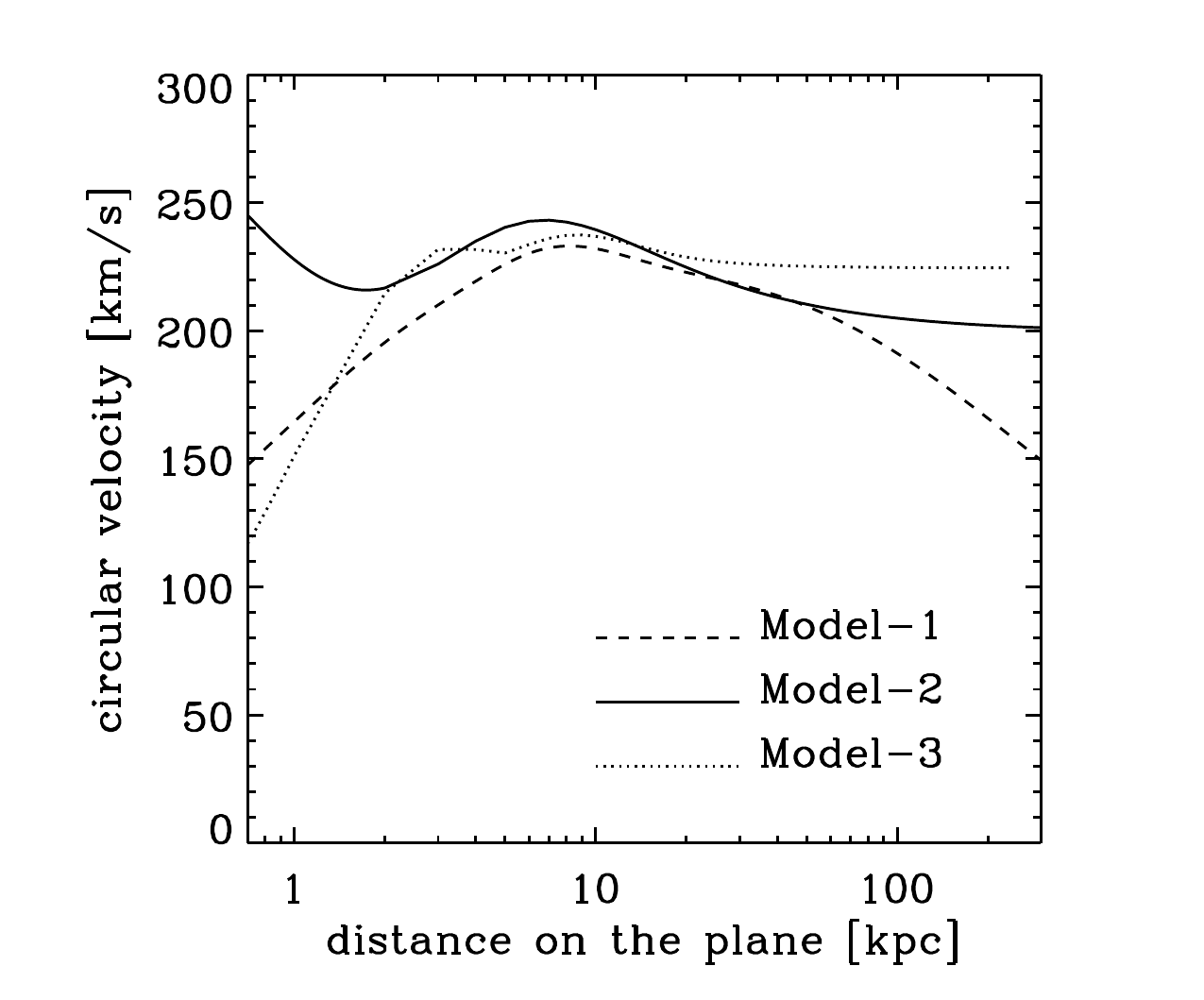}
\caption{Circular velocity curves for the three Galactic potentials considered for the orbit integrations in Sec.~\ref{sec:orbits}.}
\label{fig:circ_vel}
\end{figure}

Figure \ref{fig:circ_vel} compares the circular velocity curves for
the three different Galactic potentials. The curves are
quite similar between $\sim 3$ and $\sim 40$ kpc, but they differ
substantially in the inner as well as in the outer Galaxy. This
will lead to some of the differences in the orbits, as discussed in Sec.~\ref{sec:orbits}.

\begin{figure}
\centering
\includegraphics[width=18cm, trim={0 3cm 0 5cm},clip]{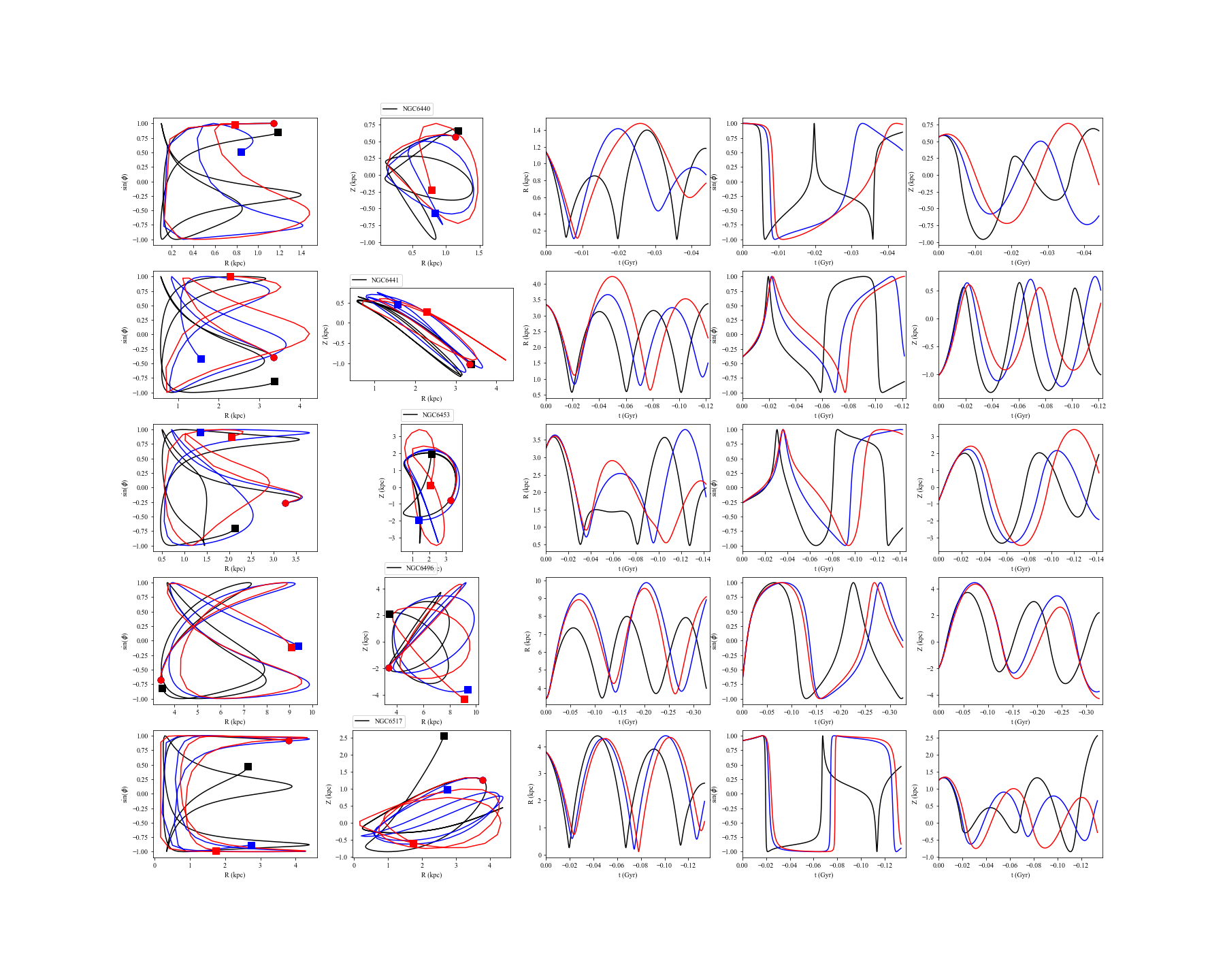}
\caption{Orbits of the globular clusters shown in
  Fig.~\ref{fig:gc_orbit-examples}, now plotted in cylindrical
  coordinates and as a function of time (right panels) for shorter integration times
  (approximately three radial oscillations). The different colours
  correspond to the axisymmetric potentials of Models 1 (blue) and 2
  (black) and to the barred Model-3 (red). The solid circles in the
  leftmost panels denote the present-day positions, while the squares
  are the positions for each of the models at the end of the chosen
  integration time.}
\label{fig:app_orbits_clusters}
\end{figure}

Figure \ref{fig:app_orbits_clusters} shows some of these
  differences for the subset of globular clusters shown in
  Fig.~\ref{fig:gc_orbit-examples}. We plot here the orbits in
  cylindrical coordinates and for a shorter period of time to show
  more clearly how the orbits diverge from each other due to the
  differences in the gravitational potentials of the various models.

\begin{figure*}
\centering
\includegraphics[width=18cm, trim={0cm 0cm 0cm 0cm},clip]{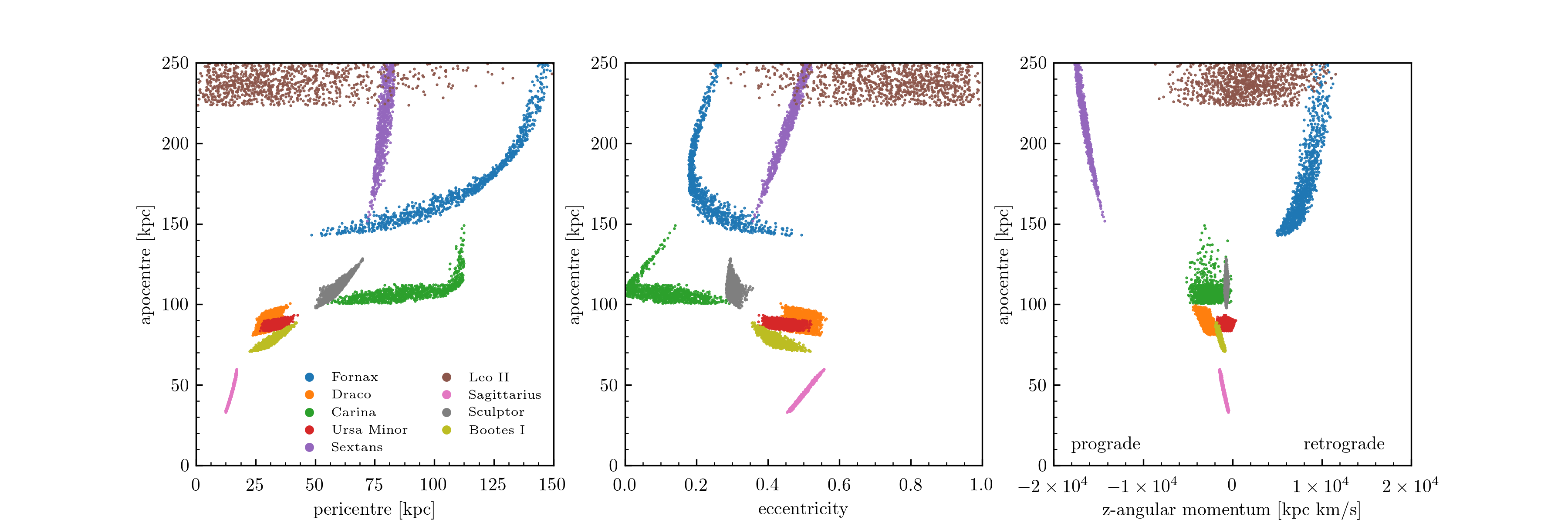}
\caption{Distribution of orbital parameters for the dwarfs (in
  different colours) showing the results obtained from the different
  Monte Carlo realisations (only for those within the 1$\sigma$
  uncertainties on the observables), computed using the potential of Model-1.}
\label{fig:app-orb-MCs}
\end{figure*}

Figure \ref{fig:app-orb-MCs} shows the distribution of orbital parameters for the dSph in
our sample, derived by drawing 1000 Monte Carlo realisations of the observables
taking into account their uncertainties. The correlations seen are not
all due to correlations in the errors themselves, but also reflect
that orbital parameters are not really fully independent. It is
important to bear this in mind when interpreting
Figs.~\ref{fig:gc_orbit_props} and \ref{fig:dw_orbit_props}, where we
have plotted uncorrelated error bars to facilitate visual inspection.

\onecolumn
\begin{center}
\scriptsize{
\begin{longtable}{lcccccc}
  \caption{Globular cluster orbital properties. We list here the
    median apocentre, pericentre, eccentricity, inclination of the
    orbital plane $\theta$, and radial period $T_r$. The median and
    errors (defined by the 16th and 84th percentiles) were
    obtained from Monte Carlo realisations sampling the errors in the
    observables using the full covariance matrix. For each globular
    cluster we quote values derived from orbits integrated for 10 Gyr in the three Galactic potentials in
    Sec.~\ref{sec:orbits}: Models 1, 2, and 3 in the first, second, and
    third rows, respectively.
\label{tab:orb_prop_gc}}\\
  \hline\hline
  Name & apocentre & pericentre & eccentr. & $L_z$ & inclin. $\theta$ & $T_r$ \\
 & [kpc] & [kpc] &  & [km/s kpc] & [deg]  & [Myr]  \\
  \hline
 \endfirsthead

\multicolumn{7}{c}%
{{\bfseries \tablename\ \thetable{} -- continued from previous page}} \\
\hline 
  Name & apocentre & pericentre & eccentr. & $L_z$ & inclin. $\theta$ & $T_r$ \\
 & [kpc] & [kpc] &  & [km/s kpc] & [deg]  & [Myr]  \\
 \hline
\endhead

\hline \multicolumn{7}{c}{{Continued on next page}} \\ \hline
\endfoot

\hline \hline
\endlastfoot
NGC0104 &  7.52$^{+ 0.01}_{- 0.01}$ &  5.74$^{+ 0.02}_{- 0.02}$ &  0.13$^{+ 0.01}_{- 0.01}$ &  -1247.5$^{+    10.7}_{-    11.2}$ & 151.6$^{+  0.6}_{-  0.6}$ &  126.9$^{+   0.1}_{-   0.1}$ \\ \noalign{\smallskip}
 &  7.68$^{+ 0.01}_{- 0.01}$ &  5.68$^{+ 0.03}_{- 0.03}$ &  0.15$^{+ 0.01}_{- 0.01}$ &  -1335.8$^{+    10.6}_{-    12.2}$ & 147.6$^{+  0.6}_{-  0.6}$ &  123.5$^{+   0.1}_{-   0.1}$ \\ \noalign{\smallskip}
 &  7.64$^{+ 0.02}_{- 0.02}$ &  5.13$^{+ 0.02}_{- 0.02}$ &  0.18$^{+ 0.01}_{- 0.01}$ &  -1231.7$^{+     6.5}_{-     8.8}$ & 148.6$^{+  0.7}_{-  0.8}$ &  122.0$^{+   1.5}_{-   1.9}$ \vspace*{0.1cm} \\ \noalign{\smallskip}
NGC0288 & 12.24$^{+ 0.15}_{- 0.16}$ &  2.43$^{+ 0.22}_{- 0.24}$ &  0.67$^{+ 0.02}_{- 0.02}$ &    436.0$^{+    52.5}_{-    60.3}$ &  57.5$^{+  1.7}_{-  1.1}$ &  151.1$^{+   3.2}_{-   3.4}$ \\ \noalign{\smallskip}
 & 12.35$^{+ 0.15}_{- 0.14}$ &  1.95$^{+ 0.23}_{- 0.21}$ &  0.73$^{+ 0.02}_{- 0.02}$ &    371.4$^{+    58.8}_{-    55.9}$ &  64.7$^{+  1.6}_{-  1.4}$ &  144.5$^{+   3.0}_{-   3.0}$ \\ \noalign{\smallskip}
 & 12.93$^{+ 0.64}_{- 0.42}$ &  0.36$^{+ 0.21}_{- 0.21}$ &  0.89$^{+ 0.01}_{- 0.02}$ &     85.3$^{+    65.4}_{-    56.6}$ &  77.2$^{+  8.3}_{-  6.7}$ &  139.2$^{+   3.6}_{-   2.8}$ \vspace*{0.1cm} \\ \noalign{\smallskip}
NGC0362 & 11.96$^{+ 0.31}_{- 0.27}$ &  0.98$^{+ 0.01}_{- 0.03}$ &  0.85$^{+ 0.01}_{- 0.01}$ &     56.6$^{+    13.3}_{-    13.4}$ &  80.1$^{+  3.6}_{-  5.3}$ &  138.9$^{+   3.5}_{-   3.6}$ \\ \noalign{\smallskip}
 & 11.78$^{+ 0.21}_{- 0.26}$ &  0.67$^{+ 0.06}_{- 0.13}$ &  0.90$^{+ 0.02}_{- 0.01}$ &     29.3$^{+    11.8}_{-    14.5}$ &  80.6$^{+  5.2}_{-  4.4}$ &  129.5$^{+   3.0}_{-   3.0}$ \\ \noalign{\smallskip}
 & 11.91$^{+ 0.29}_{- 0.13}$ &  0.08$^{+ 0.05}_{- 0.03}$ &  0.92$^{+ 0.03}_{- 0.02}$ &     45.5$^{+    20.4}_{-    15.2}$ &  80.5$^{+  3.3}_{-  3.0}$ &  127.8$^{+   3.1}_{-   3.9}$ \vspace*{0.1cm} \\ \noalign{\smallskip}
NGC1851 & 20.37$^{+ 0.22}_{- 0.29}$ &  0.98$^{+ 0.10}_{- 0.07}$ &  0.91$^{+ 0.01}_{- 0.01}$ &    181.0$^{+    16.9}_{-    20.1}$ &  69.8$^{+  3.3}_{-  3.3}$ &  230.9$^{+   3.3}_{-   3.3}$ \\ \noalign{\smallskip}
 & 20.23$^{+ 0.26}_{- 0.41}$ &  0.69$^{+ 0.11}_{- 0.07}$ &  0.94$^{+ 0.01}_{- 0.01}$ &     82.4$^{+    19.0}_{-    17.9}$ &  71.2$^{+  4.1}_{-  4.1}$ &  224.5$^{+   3.5}_{-   3.0}$ \\ \noalign{\smallskip}
 & 20.28$^{+ 0.63}_{- 0.18}$ &  0.43$^{+ 0.19}_{- 0.15}$ &  0.84$^{+ 0.01}_{- 0.01}$ &     94.9$^{+    17.7}_{-    22.4}$ &  78.6$^{+  3.4}_{-  3.4}$ &  227.3$^{+   4.0}_{-   4.1}$ \vspace*{0.1cm} \\ \noalign{\smallskip}
NGC1904 & 19.41$^{+ 0.22}_{- 0.27}$ &  0.63$^{+ 0.14}_{- 0.02}$ &  0.94$^{+ 0.01}_{- 0.01}$ &    -48.7$^{+    59.0}_{-    55.0}$ &  96.8$^{+  9.1}_{-  8.1}$ &  217.3$^{+   4.1}_{-   3.2}$ \\ \noalign{\smallskip}
 & 19.48$^{+ 0.18}_{- 0.22}$ &  0.62$^{+ 0.09}_{- 0.09}$ &  0.94$^{+ 0.01}_{- 0.01}$ &   -192.4$^{+    56.5}_{-    48.9}$ & 128.1$^{+  8.0}_{- 11.4}$ &  214.0$^{+   3.5}_{-   2.5}$ \\ \noalign{\smallskip}
 & 19.88$^{+ 1.45}_{- 0.69}$ &  0.73$^{+ 0.08}_{- 0.11}$ &  0.92$^{+ 0.01}_{- 0.01}$ &   -355.9$^{+    55.1}_{-    37.9}$ & 134.6$^{+  4.4}_{-  6.1}$ &  208.5$^{+   4.7}_{-   4.0}$ \vspace*{0.1cm} \\ \noalign{\smallskip}
NGC2298 & 17.71$^{+ 0.38}_{- 0.36}$ &  2.08$^{+ 0.24}_{- 0.23}$ &  0.79$^{+ 0.02}_{- 0.02}$ &    647.0$^{+    79.9}_{-    82.5}$ &  55.2$^{+  2.1}_{-  1.8}$ &  207.0$^{+   5.9}_{-   5.5}$ \\ \noalign{\smallskip}
 & 17.82$^{+ 0.35}_{- 0.30}$ &  1.68$^{+ 0.19}_{- 0.16}$ &  0.83$^{+ 0.01}_{- 0.01}$ &    530.2$^{+    73.0}_{-    67.9}$ &  51.3$^{+  2.0}_{-  1.5}$ &  203.0$^{+   5.5}_{-   4.5}$ \\ \noalign{\smallskip}
 & 17.78$^{+ 0.93}_{- 0.44}$ &  0.79$^{+ 0.13}_{- 0.15}$ &  0.88$^{+ 0.02}_{- 0.02}$ &    360.0$^{+    44.6}_{-    52.1}$ &  49.5$^{+  4.6}_{-  4.2}$ &  190.7$^{+   2.5}_{-   7.6}$ \vspace*{0.1cm} \\ \noalign{\smallskip}
NGC2808 & 14.14$^{+ 0.23}_{- 0.22}$ &  1.05$^{+ 0.04}_{- 0.03}$ &  0.86$^{+ 0.01}_{- 0.01}$ &   -394.3$^{+    22.8}_{-    23.5}$ & 168.8$^{+  0.7}_{-  0.9}$ &  157.9$^{+   2.4}_{-   2.2}$ \\ \noalign{\smallskip}
 & 14.29$^{+ 0.23}_{- 0.24}$ &  1.18$^{+ 0.04}_{- 0.05}$ &  0.85$^{+ 0.01}_{- 0.01}$ &   -478.7$^{+    23.9}_{-    25.6}$ & 142.8$^{+  1.5}_{-  1.0}$ &  157.0$^{+   2.5}_{-   2.5}$ \\ \noalign{\smallskip}
 & 13.90$^{+ 0.78}_{- 0.28}$ &  0.64$^{+ 0.05}_{- 0.04}$ &  0.88$^{+ 0.01}_{- 0.01}$ &   -343.1$^{+    18.0}_{-    19.9}$ & 147.3$^{+  3.1}_{-  4.3}$ &  148.9$^{+   3.6}_{-   3.7}$ \vspace*{0.1cm} \\ \noalign{\smallskip}
NGC3201 & 29.25$^{+ 1.04}_{- 1.06}$ &  8.60$^{+ 0.02}_{- 0.07}$ &  0.55$^{+ 0.01}_{- 0.01}$ &   2793.7$^{+    37.7}_{-    40.8}$ &  27.5$^{+  0.1}_{-  0.1}$ &  388.8$^{+  11.8}_{-  12.5}$ \\ \noalign{\smallskip}
 & 26.82$^{+ 0.97}_{- 0.95}$ &  8.53$^{+ 0.06}_{- 0.07}$ &  0.52$^{+ 0.01}_{- 0.01}$ &   2762.7$^{+    37.1}_{-    39.6}$ &  27.6$^{+  0.1}_{-  0.1}$ &  365.0$^{+  12.0}_{-  11.5}$ \\ \noalign{\smallskip}
 & 22.25$^{+ 0.69}_{- 0.64}$ &  8.53$^{+ 0.07}_{- 0.07}$ &  0.45$^{+ 0.01}_{- 0.01}$ &   2707.1$^{+    37.1}_{-    35.7}$ &  24.4$^{+  0.1}_{-  0.0}$ &  307.1$^{+   6.8}_{-   6.8}$ \vspace*{0.1cm} \\ \noalign{\smallskip}
NGC4372 &  7.30$^{+ 0.03}_{- 0.03}$ &  3.15$^{+ 0.10}_{- 0.08}$ &  0.40$^{+ 0.01}_{- 0.01}$ &   -922.3$^{+    13.0}_{-    13.1}$ & 150.6$^{+  0.6}_{-  0.5}$ &  100.4$^{+   1.0}_{-   1.1}$ \\ \noalign{\smallskip}
 &  7.40$^{+ 0.03}_{- 0.03}$ &  3.14$^{+ 0.08}_{- 0.09}$ &  0.40$^{+ 0.01}_{- 0.01}$ &   -984.4$^{+    12.3}_{-    13.5}$ & 154.5$^{+  0.8}_{-  0.5}$ &   97.0$^{+   0.1}_{-   1.0}$ \\ \noalign{\smallskip}
 &  7.33$^{+ 0.11}_{- 0.19}$ &  1.46$^{+ 0.86}_{- 0.10}$ &  0.48$^{+ 0.01}_{- 0.01}$ &   -718.6$^{+    87.5}_{-    58.8}$ & 151.3$^{+  2.9}_{-  3.7}$ &   90.9$^{+   1.1}_{-   1.4}$ \vspace*{0.1cm} \\ \noalign{\smallskip}
NGC4590 & 28.76$^{+ 0.76}_{- 0.79}$ &  9.13$^{+ 0.13}_{- 0.13}$ &  0.52$^{+ 0.01}_{- 0.01}$ &  -2402.1$^{+    22.2}_{-    22.8}$ & 139.1$^{+  0.6}_{-  0.5}$ &  388.8$^{+   9.8}_{-  10.0}$ \\ \noalign{\smallskip}
 & 29.57$^{+ 0.92}_{- 0.82}$ &  9.10$^{+ 0.13}_{- 0.11}$ &  0.53$^{+ 0.01}_{- 0.01}$ &  -2492.4$^{+    20.7}_{-    24.4}$ & 138.5$^{+  0.5}_{-  0.5}$ &  403.0$^{+  12.0}_{-  10.5}$ \\ \noalign{\smallskip}
 & 23.13$^{+ 0.57}_{- 0.49}$ &  8.28$^{+ 0.12}_{- 0.10}$ &  0.47$^{+ 0.01}_{- 0.01}$ &  -2156.2$^{+    15.8}_{-    15.5}$ & 137.2$^{+  0.5}_{-  0.6}$ &  316.5$^{+   5.7}_{-   7.7}$ \vspace*{0.1cm} \\ \noalign{\smallskip}
NGC4833 &  7.93$^{+ 0.15}_{- 0.16}$ &  0.92$^{+ 0.05}_{- 0.08}$ &  0.79$^{+ 0.01}_{- 0.01}$ &   -249.8$^{+    30.9}_{-    31.4}$ & 141.8$^{+  3.4}_{-  3.9}$ &   90.7$^{+   1.9}_{-   1.9}$ \\ \noalign{\smallskip}
 &  7.86$^{+ 0.15}_{- 0.12}$ &  0.93$^{+ 0.02}_{- 0.09}$ &  0.79$^{+ 0.01}_{- 0.01}$ &   -282.2$^{+    26.2}_{-    30.0}$ & 133.2$^{+  3.4}_{-  1.6}$ &   86.0$^{+   1.5}_{-   1.5}$ \\ \noalign{\smallskip}
 &  7.86$^{+ 0.18}_{- 0.07}$ &  0.24$^{+ 0.07}_{- 0.08}$ &  0.88$^{+ 0.01}_{- 0.01}$ &   -151.4$^{+    25.9}_{-    28.6}$ & 127.4$^{+  4.2}_{-  3.8}$ &   83.8$^{+   3.0}_{-   2.0}$ \vspace*{0.1cm} \\ \noalign{\smallskip}
NGC5024 & 21.73$^{+ 0.39}_{- 0.39}$ &  8.58$^{+ 0.02}_{- 0.02}$ &  0.43$^{+ 0.01}_{- 0.01}$ &   -727.5$^{+    26.1}_{-    25.0}$ & 105.6$^{+  0.7}_{-  0.7}$ &  307.3$^{+   4.3}_{-   4.3}$ \\ \noalign{\smallskip}
 & 21.95$^{+ 0.39}_{- 0.38}$ &  9.05$^{+ 0.04}_{- 0.04}$ &  0.42$^{+ 0.01}_{- 0.01}$ &   -805.9$^{+    25.2}_{-    26.2}$ & 105.5$^{+  0.6}_{-  0.6}$ &  316.5$^{+   5.0}_{-   5.0}$ \\ \noalign{\smallskip}
 & 21.29$^{+ 0.44}_{- 0.42}$ &  8.59$^{+ 0.05}_{- 0.04}$ &  0.42$^{+ 0.01}_{- 0.01}$ &   -751.0$^{+    23.4}_{-    25.5}$ & 105.0$^{+  0.7}_{-  0.6}$ &  298.1$^{+   5.8}_{-   6.1}$ \vspace*{0.1cm} \\ \noalign{\smallskip}
NGC5053 & 17.89$^{+ 0.34}_{- 0.39}$ &  9.98$^{+ 0.05}_{- 0.05}$ &  0.28$^{+ 0.01}_{- 0.01}$ &   -652.1$^{+    23.5}_{-    27.1}$ & 103.7$^{+  0.7}_{-  0.6}$ &  279.3$^{+   3.6}_{-   3.8}$ \\ \noalign{\smallskip}
 & 17.80$^{+ 0.39}_{- 0.33}$ & 10.53$^{+ 0.06}_{- 0.06}$ &  0.26$^{+ 0.01}_{- 0.01}$ &   -729.1$^{+    27.5}_{-    24.1}$ & 103.7$^{+  0.6}_{-  0.6}$ &  285.0$^{+   4.5}_{-   3.5}$ \\ \noalign{\smallskip}
 & 17.96$^{+ 0.45}_{- 0.42}$ &  9.75$^{+ 0.06}_{- 0.05}$ &  0.29$^{+ 0.01}_{- 0.01}$ &   -677.5$^{+    20.9}_{-    21.7}$ & 103.3$^{+  0.6}_{-  0.5}$ &  272.3$^{+   3.9}_{-   3.7}$ \vspace*{0.1cm} \\ \noalign{\smallskip}
NGC5139 &  7.12$^{+ 0.02}_{- 0.02}$ &  1.60$^{+ 0.01}_{- 0.01}$ &  0.63$^{+ 0.01}_{- 0.01}$ &    502.8$^{+    12.6}_{-    13.2}$ &  41.4$^{+  0.8}_{-  0.9}$ &   86.7$^{+   0.6}_{-   0.5}$ \\ \noalign{\smallskip}
 &  7.26$^{+ 0.03}_{- 0.04}$ &  1.29$^{+ 0.01}_{- 0.02}$ &  0.70$^{+ 0.01}_{- 0.01}$ &    461.6$^{+    13.7}_{-    14.4}$ &  35.3$^{+  2.0}_{-  1.3}$ &   82.5$^{+   0.1}_{-   0.5}$ \\ \noalign{\smallskip}
 &  7.03$^{+ 0.13}_{- 0.05}$ &  1.48$^{+ 0.03}_{- 0.08}$ &  0.62$^{+ 0.01}_{- 0.01}$ &    519.0$^{+    27.3}_{-    26.9}$ &  26.4$^{+  0.9}_{-  1.8}$ &   81.8$^{+   1.8}_{-   2.1}$ \vspace*{0.1cm} \\ \noalign{\smallskip}
NGC5272 & 16.03$^{+ 0.26}_{- 0.24}$ &  5.29$^{+ 0.05}_{- 0.06}$ &  0.50$^{+ 0.01}_{- 0.01}$ &   -921.7$^{+    16.9}_{-    15.6}$ & 122.6$^{+  0.5}_{-  0.5}$ &  215.4$^{+   2.3}_{-   2.1}$ \\ \noalign{\smallskip}
 & 16.00$^{+ 0.24}_{- 0.25}$ &  5.48$^{+ 0.05}_{- 0.05}$ &  0.49$^{+ 0.01}_{- 0.01}$ &  -1003.6$^{+    15.8}_{-    17.1}$ & 119.9$^{+  0.5}_{-  0.4}$ &  214.5$^{+   2.0}_{-   2.5}$ \\ \noalign{\smallskip}
 & 16.40$^{+ 0.18}_{- 0.98}$ &  5.27$^{+ 0.03}_{- 0.05}$ &  0.48$^{+ 0.01}_{- 0.01}$ &   -963.3$^{+    13.0}_{-    11.7}$ & 120.2$^{+  0.2}_{-  0.5}$ &  204.0$^{+   4.5}_{-   5.0}$ \vspace*{0.1cm} \\ \noalign{\smallskip}
NGC5286 & 13.67$^{+ 0.34}_{- 0.30}$ &  1.15$^{+ 0.10}_{- 0.09}$ &  0.85$^{+ 0.01}_{- 0.01}$ &    362.3$^{+    52.2}_{-    49.9}$ &  55.4$^{+  3.4}_{-  3.1}$ &  155.3$^{+   4.3}_{-   3.6}$ \\ \noalign{\smallskip}
 & 13.69$^{+ 0.31}_{- 0.34}$ &  1.04$^{+ 0.09}_{- 0.07}$ &  0.86$^{+ 0.01}_{- 0.01}$ &    317.2$^{+    51.1}_{-    50.0}$ &  53.4$^{+  2.7}_{-  2.7}$ &  151.5$^{+   4.0}_{-   4.0}$ \\ \noalign{\smallskip}
 & 13.81$^{+ 0.25}_{- 0.54}$ &  0.96$^{+ 0.20}_{- 0.12}$ &  0.83$^{+ 0.01}_{- 0.01}$ &    401.7$^{+    58.3}_{-    46.5}$ &  46.0$^{+  3.6}_{-  4.0}$ &  145.9$^{+   4.2}_{-   4.0}$ \vspace*{0.1cm} \\ \noalign{\smallskip}
NGC5466 & 48.10$^{+ 3.80}_{- 2.80}$ &  6.17$^{+ 0.58}_{- 0.46}$ &  0.77$^{+ 0.01}_{- 0.01}$ &    854.1$^{+    67.6}_{-    58.6}$ &  71.8$^{+  0.2}_{-  0.1}$ &  590.1$^{+  51.1}_{-  37.1}$ \\ \noalign{\smallskip}
 & 48.16$^{+ 3.71}_{- 3.04}$ &  6.04$^{+ 0.55}_{- 0.51}$ &  0.78$^{+ 0.01}_{- 0.01}$ &    822.5$^{+    67.4}_{-    64.8}$ &  72.8$^{+  0.0}_{-  0.0}$ &  596.0$^{+  47.5}_{-  40.0}$ \\ \noalign{\smallskip}
 & 33.96$^{+ 1.31}_{- 1.47}$ &  4.24$^{+ 0.45}_{- 0.43}$ &  0.78$^{+ 0.01}_{- 0.01}$ &    526.8$^{+    60.8}_{-    50.8}$ &  74.4$^{+  0.6}_{-  0.2}$ &  393.5$^{+  22.9}_{-  18.0}$ \vspace*{0.1cm} \\ \noalign{\smallskip}
NGC5634 & 22.41$^{+ 0.52}_{- 0.78}$ &  3.14$^{+ 0.66}_{- 0.19}$ &  0.75$^{+ 0.02}_{- 0.04}$ &   -489.0$^{+    66.1}_{-    74.8}$ & 112.0$^{+  1.2}_{-  1.4}$ &  267.7$^{+  10.6}_{-   7.7}$ \\ \noalign{\smallskip}
 & 22.09$^{+ 0.65}_{- 0.63}$ &  3.19$^{+ 0.27}_{- 0.22}$ &  0.75$^{+ 0.01}_{- 0.01}$ &   -424.0$^{+    60.7}_{-    70.2}$ & 106.9$^{+  1.5}_{-  1.6}$ &  265.0$^{+  10.0}_{-   8.5}$ \\ \noalign{\smallskip}
 & 22.41$^{+ 1.56}_{- 0.34}$ &  1.58$^{+ 0.06}_{- 0.14}$ &  0.80$^{+ 0.01}_{- 0.01}$ &   -145.0$^{+    20.5}_{-    48.0}$ &  98.1$^{+  3.3}_{-  1.2}$ &  256.2$^{+   5.4}_{-   4.9}$ \vspace*{0.1cm} \\ \noalign{\smallskip}
NGC5897 &  9.11$^{+ 0.41}_{- 0.40}$ &  2.82$^{+ 0.28}_{- 0.30}$ &  0.53$^{+ 0.02}_{- 0.02}$ &   -395.5$^{+    60.8}_{-    60.2}$ & 119.2$^{+  1.0}_{-  1.2}$ &  120.9$^{+   6.5}_{-   6.5}$ \\ \noalign{\smallskip}
 &  8.64$^{+ 0.38}_{- 0.35}$ &  2.32$^{+ 0.31}_{- 0.25}$ &  0.58$^{+ 0.03}_{- 0.03}$ &   -347.4$^{+    53.4}_{-    63.7}$ & 112.3$^{+  1.5}_{-  1.5}$ &  107.0$^{+   6.5}_{-   5.5}$ \\ \noalign{\smallskip}
 &  8.82$^{+ 1.48}_{- 0.46}$ &  0.98$^{+ 0.30}_{- 0.58}$ &  0.74$^{+ 0.03}_{- 0.03}$ &   -243.3$^{+    88.0}_{-    75.9}$ & 113.6$^{+  5.2}_{-  4.2}$ &  105.2$^{+   4.5}_{-   5.8}$ \vspace*{0.1cm} \\ \noalign{\smallskip}
NGC5904 & 25.35$^{+ 1.48}_{- 1.58}$ &  2.69$^{+ 0.11}_{- 0.08}$ &  0.81$^{+ 0.01}_{- 0.01}$ &   -358.5$^{+    15.0}_{-    18.3}$ & 106.6$^{+  1.7}_{-  1.4}$ &  298.8$^{+  17.5}_{-  18.5}$ \\ \noalign{\smallskip}
 & 23.66$^{+ 1.50}_{- 1.39}$ &  2.69$^{+ 0.06}_{- 0.04}$ &  0.80$^{+ 0.01}_{- 0.01}$ &   -401.4$^{+    17.5}_{-    19.0}$ & 108.2$^{+  1.3}_{-  1.4}$ &  279.5$^{+  18.5}_{-  16.0}$ \\ \noalign{\smallskip}
 & 16.06$^{+ 1.16}_{- 0.39}$ &  2.76$^{+ 0.09}_{- 0.11}$ &  0.66$^{+ 0.01}_{- 0.01}$ &   -365.7$^{+    25.7}_{-    42.0}$ & 106.7$^{+  3.1}_{-  1.3}$ &  191.0$^{+  10.4}_{-   7.0}$ \vspace*{0.1cm} \\ \noalign{\smallskip}
NGC5927 &  5.42$^{+ 0.10}_{- 0.09}$ &  4.34$^{+ 0.02}_{- 0.01}$ &  0.11$^{+ 0.01}_{- 0.01}$ &  -1061.8$^{+     6.8}_{-     8.0}$ & 171.6$^{+  0.2}_{-  0.2}$ &   90.0$^{+   0.8}_{-   0.7}$ \\ \noalign{\smallskip}
 &  5.19$^{+ 0.09}_{- 0.07}$ &  4.31$^{+ 0.04}_{- 0.03}$ &  0.09$^{+ 0.01}_{- 0.01}$ &  -1102.7$^{+     5.0}_{-     5.5}$ & 167.2$^{+  0.2}_{-  0.2}$ &   84.0$^{+   0.5}_{-   0.5}$ \\ \noalign{\smallskip}
 &  5.75$^{+ 0.08}_{- 0.06}$ &  2.97$^{+ 0.06}_{- 0.10}$ &  0.22$^{+ 0.01}_{- 0.01}$ &   -892.5$^{+     7.6}_{-     7.8}$ & 166.6$^{+  0.3}_{-  0.3}$ &   84.8$^{+   1.8}_{-   2.9}$ \vspace*{0.1cm} \\ \noalign{\smallskip}
NGC5946 &  5.93$^{+ 0.17}_{- 0.16}$ &  0.85$^{+ 0.11}_{- 0.13}$ &  0.75$^{+ 0.03}_{- 0.02}$ &   -173.5$^{+    32.7}_{-    36.4}$ & 107.4$^{+  2.5}_{-  2.5}$ &   71.6$^{+   2.6}_{-   2.4}$ \\ \noalign{\smallskip}
 &  5.72$^{+ 0.19}_{- 0.20}$ &  0.76$^{+ 0.09}_{- 0.01}$ &  0.77$^{+ 0.02}_{- 0.02}$ &   -163.6$^{+    33.1}_{-    30.3}$ & 119.5$^{+  1.9}_{-  2.3}$ &   65.5$^{+   2.0}_{-   3.0}$ \\ \noalign{\smallskip}
 &  6.41$^{+ 0.23}_{- 0.12}$ &  0.07$^{+ 0.03}_{- 0.02}$ &  0.73$^{+ 0.01}_{- 0.01}$ &    -38.2$^{+    21.5}_{-    25.7}$ & 100.7$^{+  6.4}_{-  6.2}$ &   68.8$^{+   3.1}_{-   2.7}$ \vspace*{0.1cm} \\ \noalign{\smallskip}
NGC5986 &  4.92$^{+ 0.14}_{- 0.16}$ &  0.85$^{+ 0.14}_{- 0.10}$ &  0.70$^{+ 0.02}_{- 0.03}$ &   -152.2$^{+    32.4}_{-    38.4}$ & 124.1$^{+  3.3}_{-  3.8}$ &   61.7$^{+   2.4}_{-   2.1}$ \\ \noalign{\smallskip}
 &  4.95$^{+ 0.10}_{- 0.16}$ &  0.52$^{+ 0.16}_{- 0.08}$ &  0.81$^{+ 0.03}_{- 0.05}$ &   -133.1$^{+    34.1}_{-    32.4}$ & 124.3$^{+  4.1}_{-  2.3}$ &   55.5$^{+   2.0}_{-   2.0}$ \\ \noalign{\smallskip}
 &  5.24$^{+ 0.28}_{- 0.24}$ &  0.07$^{+ 0.03}_{- 0.02}$ &  0.87$^{+ 0.01}_{- 0.01}$ &    -24.0$^{+    53.3}_{-    45.6}$ &  96.6$^{+ 10.6}_{- 15.0}$ &   56.6$^{+   2.0}_{-   2.1}$ \vspace*{0.1cm} \\ \noalign{\smallskip}
NGC6093 &  3.72$^{+ 0.13}_{- 0.13}$ &  1.10$^{+ 0.14}_{- 0.12}$ &  0.54$^{+ 0.03}_{- 0.03}$ &    -45.5$^{+    15.9}_{-    18.6}$ & 100.7$^{+  2.6}_{-  2.6}$ &   51.8$^{+   2.4}_{-   2.3}$ \\ \noalign{\smallskip}
 &  3.89$^{+ 0.09}_{- 0.07}$ &  0.60$^{+ 0.14}_{- 0.23}$ &  0.74$^{+ 0.09}_{- 0.05}$ &    -28.5$^{+    13.3}_{-    15.9}$ &  97.8$^{+  2.0}_{-  1.3}$ &   45.5$^{+   2.5}_{-   1.5}$ \\ \noalign{\smallskip}
 &  4.41$^{+ 0.16}_{- 0.19}$ &  0.10$^{+ 0.04}_{- 0.04}$ &  0.93$^{+ 0.01}_{- 0.02}$ &      7.9$^{+    35.8}_{-    50.0}$ &  88.5$^{+  9.9}_{-  8.0}$ &   50.1$^{+   1.0}_{-   1.9}$ \vspace*{0.1cm} \\ \noalign{\smallskip}
NGC6121 &  6.00$^{+ 0.12}_{- 0.13}$ &  0.39$^{+ 0.09}_{- 0.09}$ &  0.88$^{+ 0.03}_{- 0.03}$ &      5.0$^{+    33.7}_{-    33.1}$ &  77.2$^{+ 70.3}_{- 42.7}$ &   69.2$^{+   0.6}_{-   0.5}$ \\ \noalign{\smallskip}
 &  6.08$^{+ 0.01}_{- 0.11}$ &  0.41$^{+ 0.05}_{- 0.04}$ &  0.88$^{+ 0.01}_{- 0.01}$ &    -49.4$^{+    31.9}_{-    36.1}$ & 110.8$^{+  9.1}_{- 10.5}$ &   65.5$^{+   1.0}_{-   0.5}$ \\ \noalign{\smallskip}
 &  6.15$^{+ 0.09}_{- 0.06}$ &  0.10$^{+ 0.09}_{- 0.05}$ &  0.83$^{+ 0.01}_{- 0.02}$ &    -77.5$^{+    48.6}_{-    40.3}$ & 118.9$^{+ 13.6}_{- 19.7}$ &   68.7$^{+   2.2}_{-   1.1}$ \vspace*{0.1cm} \\ \noalign{\smallskip}
NGC6144 &  3.57$^{+ 0.04}_{- 0.02}$ &  2.30$^{+ 0.09}_{- 0.12}$ &  0.22$^{+ 0.02}_{- 0.01}$ &    243.6$^{+    21.1}_{-    20.4}$ &  64.0$^{+  1.9}_{-  1.9}$ &   57.5$^{+   1.2}_{-   0.6}$ \\ \noalign{\smallskip}
 &  2.96$^{+ 0.05}_{- 0.02}$ &  2.39$^{+ 0.07}_{- 0.09}$ &  0.11$^{+ 0.02}_{- 0.01}$ &    228.5$^{+    21.0}_{-    23.7}$ &  68.2$^{+  1.8}_{-  1.5}$ &   36.0$^{+   1.0}_{-   0.5}$ \\ \noalign{\smallskip}
 &  4.14$^{+ 0.07}_{- 0.04}$ &  1.97$^{+ 0.12}_{- 0.11}$ &  0.18$^{+ 0.02}_{- 0.01}$ &    273.8$^{+    25.7}_{-    20.1}$ &  62.8$^{+  1.4}_{-  1.6}$ &   55.6$^{+   2.1}_{-   1.9}$ \vspace*{0.1cm} \\ \noalign{\smallskip}
NGC6171 &  3.52$^{+ 0.03}_{- 0.05}$ &  1.16$^{+ 0.06}_{- 0.03}$ &  0.50$^{+ 0.01}_{- 0.01}$ &   -160.1$^{+    17.3}_{-    18.4}$ & 126.2$^{+  1.9}_{-  2.0}$ &   50.2$^{+   0.6}_{-   0.8}$ \\ \noalign{\smallskip}
 &  3.74$^{+ 0.03}_{- 0.02}$ &  0.94$^{+ 0.06}_{- 0.07}$ &  0.60$^{+ 0.02}_{- 0.02}$ &   -195.0$^{+    19.2}_{-    19.5}$ & 121.8$^{+  1.5}_{-  1.6}$ &   47.0$^{+   1.0}_{-   0.5}$ \\ \noalign{\smallskip}
 &  3.86$^{+ 0.16}_{- 0.16}$ &  1.03$^{+ 0.10}_{- 0.38}$ &  0.45$^{+ 0.01}_{- 0.01}$ &   -189.0$^{+    67.0}_{-    50.8}$ & 117.5$^{+  5.1}_{-  8.3}$ &   46.1$^{+   2.2}_{-   0.7}$ \vspace*{0.1cm} \\ \noalign{\smallskip}
NGC6205 &  8.61$^{+ 0.09}_{- 0.08}$ &  1.71$^{+ 0.01}_{- 0.01}$ &  0.67$^{+ 0.01}_{- 0.01}$ &    245.3$^{+     9.8}_{-     9.9}$ &  69.3$^{+  1.3}_{-  1.4}$ &  107.2$^{+   0.7}_{-   0.8}$ \\ \noalign{\smallskip}
 &  8.63$^{+ 0.09}_{- 0.07}$ &  1.53$^{+ 0.02}_{- 0.02}$ &  0.70$^{+ 0.01}_{- 0.01}$ &    199.4$^{+    10.5}_{-    10.1}$ &  72.1$^{+  1.1}_{-  1.2}$ &  101.5$^{+   0.5}_{-   0.5}$ \\ \noalign{\smallskip}
 &  9.08$^{+ 0.08}_{- 0.37}$ &  1.49$^{+ 0.01}_{- 0.03}$ &  0.66$^{+ 0.01}_{- 0.01}$ &    208.8$^{+    24.8}_{-    15.5}$ &  73.0$^{+  1.3}_{-  2.6}$ &  105.2$^{+   0.9}_{-   4.7}$ \vspace*{0.1cm} \\ \noalign{\smallskip}
NGC6218 &  4.88$^{+ 0.04}_{- 0.04}$ &  2.44$^{+ 0.05}_{- 0.05}$ &  0.33$^{+ 0.01}_{- 0.01}$ &   -526.3$^{+    19.5}_{-    19.4}$ & 141.7$^{+  1.0}_{-  1.0}$ &   73.0$^{+   0.8}_{-   0.7}$ \\ \noalign{\smallskip}
 &  5.00$^{+ 0.04}_{- 0.04}$ &  2.41$^{+ 0.06}_{- 0.05}$ &  0.35$^{+ 0.01}_{- 0.01}$ &   -586.1$^{+    21.1}_{-    19.8}$ & 138.1$^{+  1.0}_{-  1.1}$ &   69.5$^{+   0.5}_{-   1.0}$ \\ \noalign{\smallskip}
 &  4.92$^{+ 0.10}_{- 0.11}$ &  1.84$^{+ 0.04}_{- 0.06}$ &  0.30$^{+ 0.01}_{- 0.01}$ &   -508.3$^{+    26.8}_{-    16.7}$ & 138.2$^{+  1.7}_{-  1.3}$ &   70.6$^{+   1.2}_{-   5.6}$ \vspace*{0.1cm} \\ \noalign{\smallskip}
NGC6235 &  6.19$^{+ 0.68}_{- 0.65}$ &  3.35$^{+ 0.28}_{- 0.26}$ &  0.30$^{+ 0.02}_{- 0.02}$ &   -613.0$^{+    85.2}_{-    91.1}$ & 130.2$^{+  2.0}_{-  2.2}$ &   94.1$^{+   9.0}_{-   8.6}$ \\ \noalign{\smallskip}
 &  5.04$^{+ 0.55}_{- 0.50}$ &  3.15$^{+ 0.27}_{- 0.26}$ &  0.23$^{+ 0.01}_{- 0.01}$ &   -549.0$^{+    80.0}_{-    85.4}$ & 125.9$^{+  2.2}_{-  2.4}$ &   76.5$^{+   7.5}_{-   6.5}$ \\ \noalign{\smallskip}
 &  5.79$^{+ 0.09}_{- 1.12}$ &  2.12$^{+ 0.05}_{- 0.13}$ &  0.28$^{+ 0.02}_{- 0.01}$ &   -511.5$^{+   129.6}_{-    11.2}$ & 127.9$^{+  0.5}_{-  4.5}$ &   66.7$^{+   5.1}_{-   5.3}$ \vspace*{0.1cm} \\ \noalign{\smallskip}
NGC6254 &  5.17$^{+ 0.07}_{- 0.07}$ &  2.33$^{+ 0.05}_{- 0.05}$ &  0.38$^{+ 0.01}_{- 0.01}$ &   -556.8$^{+    20.0}_{-    19.4}$ & 142.7$^{+  1.1}_{-  1.1}$ &   74.4$^{+   1.0}_{-   1.0}$ \\ \noalign{\smallskip}
 &  5.28$^{+ 0.08}_{- 0.07}$ &  2.30$^{+ 0.06}_{- 0.06}$ &  0.39$^{+ 0.01}_{- 0.01}$ &   -614.6$^{+    21.8}_{-    24.1}$ & 142.1$^{+  1.2}_{-  1.1}$ &   71.0$^{+   0.1}_{-   1.0}$ \\ \noalign{\smallskip}
 &  4.94$^{+ 0.26}_{- 0.11}$ &  1.97$^{+ 0.02}_{- 0.05}$ &  0.27$^{+ 0.01}_{- 0.01}$ &   -544.5$^{+     8.1}_{-    25.2}$ & 141.1$^{+  2.2}_{-  1.0}$ &   73.7$^{+   3.4}_{-   2.2}$ \vspace*{0.1cm} \\ \noalign{\smallskip}
NGC6266 &  2.11$^{+ 0.13}_{- 0.13}$ &  0.91$^{+ 0.01}_{- 0.02}$ &  0.40$^{+ 0.02}_{- 0.02}$ &   -180.2$^{+    16.3}_{-    15.3}$ & 144.5$^{+  1.6}_{-  2.0}$ &   33.4$^{+   1.4}_{-   1.4}$ \\ \noalign{\smallskip}
 &  2.36$^{+ 0.10}_{- 0.01}$ &  0.69$^{+ 0.07}_{- 0.07}$ &  0.55$^{+ 0.02}_{- 0.02}$ &   -209.3$^{+    16.9}_{-    18.5}$ & 145.9$^{+  0.4}_{-  0.2}$ &   31.0$^{+   1.5}_{-   1.5}$ \\ \noalign{\smallskip}
 &  2.28$^{+ 0.27}_{- 0.08}$ &  0.55$^{+ 0.12}_{- 0.25}$ &  0.31$^{+ 0.01}_{- 0.01}$ &   -195.6$^{+    18.2}_{-    18.2}$ & 138.2$^{+  6.1}_{-  2.5}$ &   42.9$^{+   1.8}_{-  20.1}$ \vspace*{0.1cm} \\ \noalign{\smallskip}
NGC6273 &  4.51$^{+ 0.21}_{- 0.18}$ &  1.34$^{+ 0.14}_{- 0.16}$ &  0.54$^{+ 0.03}_{- 0.02}$ &    155.8$^{+    41.0}_{-    43.5}$ &  68.7$^{+  5.5}_{-  4.8}$ &   61.7$^{+   3.1}_{-   3.0}$ \\ \noalign{\smallskip}
 &  3.38$^{+ 0.20}_{- 0.16}$ &  1.32$^{+ 0.09}_{- 0.07}$ &  0.44$^{+ 0.01}_{- 0.01}$ &    115.4$^{+    44.1}_{-    47.5}$ &  75.3$^{+  5.6}_{-  4.5}$ &   47.0$^{+   2.5}_{-   2.5}$ \\ \noalign{\smallskip}
 &  4.78$^{+ 0.43}_{- 0.23}$ &  0.85$^{+ 0.14}_{- 0.17}$ &  0.51$^{+ 0.01}_{- 0.01}$ &    211.5$^{+    32.1}_{-    39.9}$ &  60.3$^{+  4.2}_{-  3.2}$ &   58.6$^{+   4.4}_{-   4.2}$ \vspace*{0.1cm} \\ \noalign{\smallskip}
NGC6284 &  7.63$^{+ 0.33}_{- 0.36}$ &  1.09$^{+ 0.22}_{- 0.11}$ &  0.75$^{+ 0.01}_{- 0.03}$ &    -68.0$^{+    39.4}_{-    46.4}$ &  95.5$^{+  3.1}_{-  3.0}$ &   92.9$^{+   5.1}_{-   4.5}$ \\ \noalign{\smallskip}
 &  7.31$^{+ 0.32}_{- 0.39}$ &  1.07$^{+ 0.12}_{- 0.10}$ &  0.75$^{+ 0.01}_{- 0.01}$ &    -12.0$^{+    40.6}_{-    36.6}$ &  91.6$^{+  4.1}_{-  5.8}$ &   84.5$^{+   4.5}_{-   4.5}$ \\ \noalign{\smallskip}
 &  8.21$^{+ 0.38}_{- 0.57}$ &  0.65$^{+ 0.05}_{- 0.11}$ &  0.60$^{+ 0.01}_{- 0.01}$ &    164.5$^{+    42.3}_{-    50.0}$ &  67.4$^{+  6.3}_{-  6.4}$ &   91.8$^{+   5.9}_{-   5.4}$ \vspace*{0.1cm} \\ \noalign{\smallskip}
NGC6287 &  6.23$^{+ 0.38}_{- 0.36}$ &  1.04$^{+ 0.02}_{- 0.04}$ &  0.72$^{+ 0.01}_{- 0.01}$ &     54.8$^{+     4.8}_{-     7.5}$ &  84.9$^{+  0.4}_{-  0.2}$ &   77.6$^{+   4.1}_{-   4.0}$ \\ \noalign{\smallskip}
 &  4.74$^{+ 0.33}_{- 0.27}$ &  1.24$^{+ 0.05}_{- 0.05}$ &  0.59$^{+ 0.01}_{- 0.01}$ &     52.5$^{+     8.0}_{-     8.4}$ &  83.6$^{+  0.8}_{-  0.6}$ &   59.5$^{+   3.5}_{-   3.0}$ \\ \noalign{\smallskip}
 &  7.13$^{+ 0.65}_{- 0.40}$ &  0.23$^{+ 0.29}_{- 0.14}$ &  0.72$^{+ 0.02}_{- 0.01}$ &     16.2$^{+    40.9}_{-    50.7}$ &  86.5$^{+  7.4}_{-  5.6}$ &   76.7$^{+   4.9}_{-   3.3}$ \vspace*{0.1cm} \\ \noalign{\smallskip}
NGC6293 &  3.23$^{+ 0.31}_{- 0.16}$ &  0.80$^{+ 0.01}_{- 0.03}$ &  0.61$^{+ 0.02}_{- 0.02}$ &     90.4$^{+    19.1}_{-    19.2}$ &  46.0$^{+  5.9}_{-  1.7}$ &   44.4$^{+   3.3}_{-   2.2}$ \\ \noalign{\smallskip}
 &  2.81$^{+ 0.47}_{- 0.23}$ &  0.47$^{+ 0.11}_{- 0.07}$ &  0.71$^{+ 0.07}_{- 0.06}$ &     76.6$^{+    23.6}_{-    22.9}$ &  64.5$^{+  5.0}_{- 11.0}$ &   34.5$^{+   4.0}_{-   3.0}$ \\ \noalign{\smallskip}
 &  3.80$^{+ 0.48}_{- 0.29}$ &  0.40$^{+ 0.09}_{- 0.26}$ &  0.47$^{+ 0.02}_{- 0.02}$ &    134.8$^{+    35.1}_{-    50.1}$ &  60.6$^{+  9.9}_{-  3.5}$ &   49.6$^{+   2.7}_{-   2.4}$ \vspace*{0.1cm} \\ \noalign{\smallskip}
NGC6304 &  3.14$^{+ 0.16}_{- 0.17}$ &  1.79$^{+ 0.08}_{- 0.08}$ &  0.27$^{+ 0.01}_{- 0.01}$ &   -433.0$^{+    26.3}_{-    27.1}$ & 160.1$^{+  0.8}_{-  0.9}$ &   49.7$^{+   2.1}_{-   2.1}$ \\ \noalign{\smallskip}
 &  3.13$^{+ 0.20}_{- 0.20}$ &  1.87$^{+ 0.05}_{- 0.04}$ &  0.26$^{+ 0.02}_{- 0.04}$ &   -485.0$^{+    25.0}_{-    27.8}$ & 155.5$^{+  1.5}_{-  2.1}$ &   48.0$^{+   1.5}_{-   2.5}$ \\ \noalign{\smallskip}
 &  3.18$^{+ 0.14}_{- 0.09}$ &  0.46$^{+ 0.26}_{- 0.22}$ &  0.21$^{+ 0.01}_{- 0.01}$ &   -314.2$^{+    74.0}_{-    51.0}$ & 152.8$^{+  4.1}_{-  6.8}$ &   46.0$^{+   2.9}_{-   3.4}$ \vspace*{0.1cm} \\ \noalign{\smallskip}
NGC6316 &  2.96$^{+ 0.25}_{- 0.23}$ &  0.63$^{+ 0.13}_{- 0.13}$ &  0.65$^{+ 0.04}_{- 0.04}$ &   -139.9$^{+    33.1}_{-    36.4}$ & 141.2$^{+  1.5}_{-  1.7}$ &   39.1$^{+   3.2}_{-   2.7}$ \\ \noalign{\smallskip}
 &  2.59$^{+ 0.22}_{- 0.26}$ &  0.42$^{+ 0.11}_{- 0.08}$ &  0.72$^{+ 0.03}_{- 0.03}$ &   -107.0$^{+    28.6}_{-    32.0}$ & 128.3$^{+  3.5}_{-  3.4}$ &   30.0$^{+   3.5}_{-   3.0}$ \\ \noalign{\smallskip}
 &  3.30$^{+ 0.39}_{- 0.31}$ &  0.12$^{+ 0.10}_{- 0.07}$ &  0.60$^{+ 0.02}_{- 0.02}$ &   -110.7$^{+    46.7}_{-    39.0}$ & 124.7$^{+ 11.3}_{- 12.7}$ &   40.5$^{+   3.1}_{-   3.4}$ \vspace*{0.1cm} \\ \noalign{\smallskip}
NGC6325 &  2.18$^{+ 0.17}_{- 0.16}$ &  0.84$^{+ 0.17}_{- 0.13}$ &  0.44$^{+ 0.09}_{- 0.11}$ &     93.8$^{+    30.2}_{-    35.7}$ &  69.1$^{+  8.2}_{-  7.1}$ &   33.9$^{+   0.8}_{-   0.5}$ \\ \noalign{\smallskip}
 &  1.35$^{+ 0.03}_{- 0.00}$ &  1.15$^{+ 0.04}_{- 0.02}$ &  0.08$^{+ 0.02}_{- 0.01}$ &    126.7$^{+    26.8}_{-    28.9}$ &  64.2$^{+  6.0}_{-  5.3}$ &   16.5$^{+   0.1}_{-   0.5}$ \\ \noalign{\smallskip}
 &  1.81$^{+ 0.33}_{- 0.17}$ &  0.30$^{+ 0.22}_{- 0.24}$ &  0.22$^{+ 0.04}_{- 0.04}$ &     59.0$^{+    26.1}_{-    38.3}$ &  66.1$^{+ 14.3}_{-  9.7}$ &   25.2$^{+   1.4}_{-   1.3}$ \vspace*{0.1cm} \\ \noalign{\smallskip}
NGC6333 &  4.40$^{+ 0.23}_{- 0.22}$ &  1.33$^{+ 0.03}_{- 0.03}$ &  0.54$^{+ 0.02}_{- 0.02}$ &   -244.8$^{+    17.5}_{-    18.9}$ & 120.6$^{+  1.7}_{-  1.7}$ &   60.0$^{+   2.3}_{-   2.1}$ \\ \noalign{\smallskip}
 &  3.52$^{+ 0.19}_{- 0.18}$ &  1.45$^{+ 0.02}_{- 0.03}$ &  0.42$^{+ 0.02}_{- 0.02}$ &   -272.4$^{+    20.7}_{-    22.8}$ & 123.8$^{+  2.3}_{-  2.2}$ &   49.0$^{+   1.5}_{-   1.5}$ \\ \noalign{\smallskip}
 &  5.11$^{+ 0.21}_{- 0.23}$ &  0.71$^{+ 0.09}_{- 0.14}$ &  0.50$^{+ 0.02}_{- 0.02}$ &   -240.0$^{+    36.5}_{-    31.5}$ & 124.8$^{+  3.0}_{-  3.2}$ &   62.6$^{+   2.7}_{-   4.2}$ \vspace*{0.1cm} \\ \noalign{\smallskip}
NGC6341 & 10.53$^{+ 0.15}_{- 0.13}$ &  1.02$^{+ 0.01}_{- 0.02}$ &  0.82$^{+ 0.01}_{- 0.01}$ &    -47.1$^{+    21.7}_{-    20.7}$ &  94.5$^{+  2.4}_{-  2.2}$ &  124.0$^{+   1.9}_{-   1.9}$ \\ \noalign{\smallskip}
 & 10.61$^{+ 0.13}_{- 0.13}$ &  0.98$^{+ 0.03}_{- 0.00}$ &  0.83$^{+ 0.01}_{- 0.01}$ &   -103.5$^{+    21.9}_{-    21.8}$ & 104.6$^{+  3.4}_{-  3.5}$ &  119.0$^{+   2.0}_{-   1.5}$ \\ \noalign{\smallskip}
 & 10.62$^{+ 0.41}_{- 0.36}$ &  0.18$^{+ 0.10}_{- 0.08}$ &  0.76$^{+ 0.01}_{- 0.01}$ &    -86.3$^{+    37.3}_{-    32.5}$ & 107.4$^{+  5.2}_{-  6.8}$ &  114.6$^{+   1.9}_{-   3.2}$ \vspace*{0.1cm} \\ \noalign{\smallskip}
NGC6342 &  1.84$^{+ 0.06}_{- 0.04}$ &  1.31$^{+ 0.08}_{- 0.07}$ &  0.17$^{+ 0.02}_{- 0.01}$ &   -120.6$^{+    11.3}_{-    13.6}$ & 116.9$^{+  1.2}_{-  0.9}$ &   23.7$^{+   0.8}_{-   0.6}$ \\ \noalign{\smallskip}
 &  1.66$^{+ 0.00}_{- 0.00}$ &  0.97$^{+ 0.11}_{- 0.09}$ &  0.27$^{+ 0.05}_{- 0.05}$ &   -112.4$^{+     8.5}_{-     9.9}$ & 112.7$^{+  0.8}_{-  0.6}$ &   27.0$^{+   0.1}_{-   1.0}$ \\ \noalign{\smallskip}
 &  2.40$^{+ 0.53}_{- 0.19}$ &  0.91$^{+ 0.17}_{- 0.76}$ &  0.23$^{+ 0.04}_{- 0.20}$ &   -145.7$^{+     6.0}_{-    12.5}$ & 119.2$^{+  2.9}_{-  1.0}$ &   30.2$^{+   2.2}_{-   3.6}$ \vspace*{0.1cm} \\ \noalign{\smallskip}
NGC6352 &  4.44$^{+ 0.14}_{- 0.14}$ &  3.15$^{+ 0.04}_{- 0.03}$ &  0.17$^{+ 0.01}_{- 0.01}$ &   -774.8$^{+    19.4}_{-    20.6}$ & 167.6$^{+  0.4}_{-  0.4}$ &   72.1$^{+   1.5}_{-   1.4}$ \\ \noalign{\smallskip}
 &  4.36$^{+ 0.15}_{- 0.13}$ &  3.20$^{+ 0.04}_{- 0.03}$ &  0.15$^{+ 0.01}_{- 0.01}$ &   -832.1$^{+    19.8}_{-    21.7}$ & 163.1$^{+  0.6}_{-  0.5}$ &   67.5$^{+   1.5}_{-   1.0}$ \\ \noalign{\smallskip}
 &  5.71$^{+ 0.04}_{- 0.72}$ &  3.06$^{+ 0.03}_{- 0.07}$ &  0.12$^{+ 0.01}_{- 0.01}$ &   -936.6$^{+   103.7}_{-     5.6}$ & 164.7$^{+  0.2}_{-  0.8}$ &   69.9$^{+   2.3}_{-   3.9}$ \vspace*{0.1cm} \\ \noalign{\smallskip}
NGC6356 &  7.86$^{+ 0.39}_{- 0.38}$ &  3.25$^{+ 0.39}_{- 0.37}$ &  0.41$^{+ 0.03}_{- 0.03}$ &   -770.8$^{+    85.7}_{-    90.9}$ & 138.0$^{+  1.0}_{-  1.1}$ &  109.2$^{+   7.2}_{-   6.8}$ \\ \noalign{\smallskip}
 &  7.55$^{+ 0.40}_{- 0.34}$ &  2.64$^{+ 0.35}_{- 0.30}$ &  0.48$^{+ 0.03}_{- 0.03}$ &   -688.7$^{+    76.6}_{-    91.9}$ & 134.9$^{+  1.2}_{-  1.1}$ &   96.5$^{+   7.0}_{-   5.5}$ \\ \noalign{\smallskip}
 &  9.31$^{+ 0.29}_{- 1.17}$ &  1.48$^{+ 0.17}_{- 0.24}$ &  0.57$^{+ 0.02}_{- 0.02}$ &   -527.7$^{+   129.8}_{-    56.6}$ & 138.4$^{+  2.6}_{-  6.5}$ &   97.8$^{+   4.9}_{-   6.0}$ \vspace*{0.1cm} \\ \noalign{\smallskip}
NGC6362 &  7.43$^{+ 0.16}_{- 0.16}$ &  5.06$^{+ 0.01}_{- 0.01}$ &  0.19$^{+ 0.01}_{- 0.01}$ &  -1030.3$^{+     5.5}_{-     6.2}$ & 141.4$^{+  0.3}_{-  0.3}$ &  120.8$^{+   1.8}_{-   1.7}$ \\ \noalign{\smallskip}
 &  6.83$^{+ 0.15}_{- 0.13}$ &  5.12$^{+ 0.01}_{- 0.00}$ &  0.14$^{+ 0.01}_{- 0.01}$ &  -1069.7$^{+     4.1}_{-     4.9}$ & 138.8$^{+  0.4}_{-  0.3}$ &  111.0$^{+   1.5}_{-   1.5}$ \\ \noalign{\smallskip}
 &  7.13$^{+ 0.17}_{- 0.13}$ &  4.21$^{+ 0.03}_{- 0.02}$ &  0.21$^{+ 0.01}_{- 0.01}$ &   -981.9$^{+     3.7}_{-     4.0}$ & 142.4$^{+  0.3}_{-  0.2}$ &  108.1$^{+   2.8}_{-   1.4}$ \vspace*{0.1cm} \\ \noalign{\smallskip}
NGC6366 &  5.21$^{+ 0.07}_{- 0.06}$ &  3.58$^{+ 0.08}_{- 0.07}$ &  0.19$^{+ 0.01}_{- 0.01}$ &   -897.4$^{+    18.8}_{-    21.4}$ & 165.2$^{+  0.5}_{-  0.5}$ &   82.7$^{+   1.2}_{-   1.0}$ \\ \noalign{\smallskip}
 &  5.38$^{+ 0.07}_{- 0.06}$ &  3.56$^{+ 0.07}_{- 0.07}$ &  0.20$^{+ 0.01}_{- 0.01}$ &   -973.7$^{+    19.7}_{-    20.3}$ & 161.2$^{+  0.6}_{-  0.6}$ &   80.0$^{+   1.0}_{-   0.1}$ \\ \noalign{\smallskip}
 &  6.18$^{+ 0.05}_{- 0.12}$ &  2.57$^{+ 0.09}_{- 0.04}$ &  0.18$^{+ 0.01}_{- 0.01}$ &   -864.4$^{+    52.0}_{-    39.6}$ & 162.5$^{+  1.8}_{-  1.2}$ &   77.0$^{+   2.4}_{-   3.3}$ \vspace*{0.1cm} \\ \noalign{\smallskip}
NGC6380 &  3.12$^{+ 0.22}_{- 0.18}$ &  0.53$^{+ 0.01}_{- 0.01}$ &  0.71$^{+ 0.01}_{- 0.01}$ &     80.2$^{+     4.2}_{-     4.5}$ &  12.1$^{+  1.4}_{-  0.7}$ &   39.8$^{+   2.3}_{-   1.9}$ \\ \noalign{\smallskip}
 &  2.90$^{+ 0.21}_{- 0.22}$ &  0.39$^{+ 0.05}_{- 0.03}$ &  0.76$^{+ 0.01}_{- 0.02}$ &     90.1$^{+     4.4}_{-     5.4}$ &  55.8$^{+  3.1}_{-  2.7}$ &   34.0$^{+   2.0}_{-   2.0}$ \\ \noalign{\smallskip}
 &  4.13$^{+ 0.24}_{- 0.31}$ &  0.27$^{+ 0.15}_{- 0.06}$ &  0.67$^{+ 0.01}_{- 0.01}$ &    120.8$^{+    11.4}_{-    14.4}$ &  49.6$^{+  2.3}_{-  2.4}$ &   43.5$^{+   4.7}_{-   2.1}$ \vspace*{0.1cm} \\ \noalign{\smallskip}
NGC6388 &  3.23$^{+ 0.12}_{- 0.09}$ &  0.95$^{+ 0.08}_{- 0.06}$ &  0.54$^{+ 0.02}_{- 0.01}$ &    241.2$^{+    16.5}_{-    16.4}$ &  31.3$^{+  1.9}_{-  1.8}$ &   44.2$^{+   1.5}_{-   1.2}$ \\ \noalign{\smallskip}
 &  3.18$^{+ 0.10}_{- 0.10}$ &  0.74$^{+ 0.07}_{- 0.10}$ &  0.62$^{+ 0.04}_{- 0.02}$ &    235.5$^{+    16.7}_{-    20.5}$ &  36.5$^{+  0.6}_{-  2.5}$ &   39.5$^{+   1.5}_{-   1.0}$ \\ \noalign{\smallskip}
 &  5.46$^{+ 1.17}_{- 1.17}$ &  0.20$^{+ 0.35}_{- 0.11}$ &  0.37$^{+ 0.01}_{- 0.01}$ &    216.9$^{+    69.2}_{-    66.3}$ &  43.2$^{+ 18.8}_{- 12.8}$ &   48.6$^{+   3.5}_{-   1.6}$ \vspace*{0.1cm} \\ \noalign{\smallskip}
NGC6397 &  6.45$^{+ 0.02}_{- 0.02}$ &  2.86$^{+ 0.05}_{- 0.05}$ &  0.39$^{+ 0.01}_{- 0.01}$ &   -719.9$^{+    20.1}_{-    20.5}$ & 134.6$^{+  1.2}_{-  1.2}$ &   91.0$^{+   0.5}_{-   0.5}$ \\ \noalign{\smallskip}
 &  6.59$^{+ 0.03}_{- 0.02}$ &  2.90$^{+ 0.05}_{- 0.05}$ &  0.39$^{+ 0.01}_{- 0.01}$ &   -795.8$^{+    23.1}_{-    22.2}$ & 143.5$^{+  1.2}_{-  1.2}$ &   88.0$^{+   0.5}_{-   0.5}$ \\ \noalign{\smallskip}
 &  7.20$^{+ 0.11}_{- 0.57}$ &  1.72$^{+ 0.92}_{- 0.10}$ &  0.34$^{+ 0.01}_{- 0.01}$ &   -767.0$^{+    92.2}_{-    61.2}$ & 150.8$^{+  2.8}_{-  3.8}$ &   82.0$^{+   3.4}_{-   3.1}$ \vspace*{0.1cm} \\ \noalign{\smallskip}
NGC6401 &  4.58$^{+ 0.35}_{- 0.35}$ &  2.63$^{+ 0.23}_{- 0.25}$ &  0.27$^{+ 0.01}_{- 0.01}$ &    573.6$^{+    52.8}_{-    56.4}$ &  36.8$^{+  0.5}_{-  0.4}$ &   71.4$^{+   5.1}_{-   5.4}$ \\ \noalign{\smallskip}
 &  3.90$^{+ 0.35}_{- 0.33}$ &  2.40$^{+ 0.26}_{- 0.25}$ &  0.24$^{+ 0.01}_{- 0.01}$ &    546.2$^{+    58.8}_{-    58.2}$ &  39.1$^{+  0.3}_{-  0.2}$ &   59.0$^{+   5.5}_{-   4.5}$ \\ \noalign{\smallskip}
 &  4.47$^{+ 0.26}_{- 0.26}$ &  2.64$^{+ 0.27}_{- 0.30}$ &  0.24$^{+ 0.02}_{- 0.02}$ &    605.6$^{+    56.5}_{-    63.0}$ &  35.0$^{+  1.4}_{-  1.0}$ &   72.8$^{+   5.5}_{-   4.9}$ \vspace*{0.1cm} \\ \noalign{\smallskip}
NGC6402 &  4.09$^{+ 0.10}_{- 0.09}$ &  0.88$^{+ 0.09}_{- 0.08}$ &  0.65$^{+ 0.02}_{- 0.02}$ &   -147.1$^{+    18.9}_{-    19.6}$ & 129.7$^{+  0.8}_{-  0.9}$ &   53.6$^{+   1.6}_{-   1.5}$ \\ \noalign{\smallskip}
 &  4.32$^{+ 0.13}_{- 0.12}$ &  0.51$^{+ 0.05}_{- 0.04}$ &  0.79$^{+ 0.01}_{- 0.02}$ &   -139.4$^{+    13.2}_{-    16.3}$ & 130.3$^{+  3.6}_{-  3.1}$ &   49.0$^{+   1.5}_{-   1.5}$ \\ \noalign{\smallskip}
 &  4.35$^{+ 0.31}_{- 0.24}$ &  0.09$^{+ 0.12}_{- 0.04}$ &  0.68$^{+ 0.01}_{- 0.01}$ &    -65.4$^{+    42.0}_{-    50.8}$ & 110.9$^{+  9.1}_{- 12.9}$ &   51.8$^{+   1.4}_{-   2.7}$ \vspace*{0.1cm} \\ \noalign{\smallskip}
NGC6440 &  1.34$^{+ 0.05}_{- 0.02}$ &  0.41$^{+ 0.05}_{- 0.05}$ &  0.53$^{+ 0.04}_{- 0.03}$ &     61.7$^{+    14.3}_{-    15.9}$ &  54.4$^{+  7.8}_{-  6.0}$ &   22.0$^{+   0.9}_{-   0.6}$ \\ \noalign{\smallskip}
 &  1.30$^{+ 0.04}_{- 0.05}$ &  0.23$^{+ 0.06}_{- 0.04}$ &  0.69$^{+ 0.06}_{- 0.06}$ &     47.1$^{+    15.8}_{-    20.0}$ &  57.6$^{+ 13.4}_{-  4.2}$ &   15.0$^{+   1.0}_{-   0.1}$ \\ \noalign{\smallskip}
 &  1.55$^{+ 0.09}_{- 0.09}$ &  0.34$^{+ 0.14}_{- 0.23}$ &  0.42$^{+ 0.09}_{- 0.07}$ &     70.3$^{+    20.1}_{-    22.7}$ &  58.5$^{+  8.0}_{-  5.1}$ &   31.2$^{+   1.5}_{-   1.5}$ \vspace*{0.1cm} \\ \noalign{\smallskip}
NGC6441 &  3.64$^{+ 0.22}_{- 0.25}$ &  1.12$^{+ 0.16}_{- 0.15}$ &  0.53$^{+ 0.03}_{- 0.03}$ &   -266.9$^{+    46.7}_{-    47.2}$ & 160.1$^{+  0.9}_{-  1.1}$ &   48.2$^{+   3.3}_{-   3.5}$ \\ \noalign{\smallskip}
 &  3.43$^{+ 0.26}_{- 0.25}$ &  0.80$^{+ 0.14}_{- 0.12}$ &  0.62$^{+ 0.03}_{- 0.03}$ &   -222.0$^{+    39.8}_{-    45.8}$ & 137.6$^{+  2.8}_{-  2.9}$ &   41.0$^{+   4.0}_{-   3.0}$ \\ \noalign{\smallskip}
 &  4.19$^{+ 0.24}_{- 0.13}$ &  0.35$^{+ 0.14}_{- 0.19}$ &  0.65$^{+ 0.01}_{- 0.01}$ &   -152.7$^{+    36.0}_{-    32.0}$ & 134.3$^{+  3.8}_{-  4.9}$ &   51.2$^{+   2.5}_{-   3.4}$ \vspace*{0.1cm} \\ \noalign{\smallskip}
NGC6453 &  4.00$^{+ 0.33}_{- 0.33}$ &  1.32$^{+ 0.21}_{- 0.18}$ &  0.50$^{+ 0.02}_{- 0.03}$ &   -163.4$^{+    28.7}_{-    32.8}$ & 105.1$^{+  1.4}_{-  1.4}$ &   56.4$^{+   5.0}_{-   4.7}$ \\ \noalign{\smallskip}
 &  3.62$^{+ 0.32}_{- 0.32}$ &  1.22$^{+ 0.14}_{- 0.12}$ &  0.50$^{+ 0.01}_{- 0.01}$ &   -131.8$^{+    25.8}_{-    27.3}$ & 107.5$^{+  1.7}_{-  2.0}$ &   48.5$^{+   4.0}_{-   4.5}$ \\ \noalign{\smallskip}
 &  4.43$^{+ 0.28}_{- 0.32}$ &  0.10$^{+ 0.11}_{- 0.04}$ &  0.56$^{+ 0.02}_{- 0.02}$ &     17.3$^{+    52.1}_{-    47.8}$ &  86.1$^{+  9.8}_{-  9.4}$ &   55.9$^{+   3.2}_{-   4.3}$ \vspace*{0.1cm} \\ \noalign{\smallskip}
NGC6496 & 10.17$^{+ 1.53}_{- 1.18}$ &  3.96$^{+ 0.26}_{- 0.23}$ &  0.44$^{+ 0.03}_{- 0.03}$ &  -1109.9$^{+   100.2}_{-   119.7}$ & 148.9$^{+  0.8}_{-  0.8}$ &  137.7$^{+  18.8}_{-  14.0}$ \\ \noalign{\smallskip}
 &  8.11$^{+ 1.10}_{- 0.97}$ &  3.79$^{+ 0.25}_{- 0.21}$ &  0.36$^{+ 0.03}_{- 0.03}$ &  -1043.0$^{+   101.5}_{-   107.5}$ & 144.9$^{+  1.0}_{-  1.0}$ &  111.0$^{+  13.5}_{-  12.0}$ \\ \noalign{\smallskip}
 &  6.39$^{+ 0.46}_{- 0.60}$ &  3.74$^{+ 0.29}_{- 1.07}$ &  0.21$^{+ 0.02}_{- 0.02}$ &   -903.4$^{+   182.7}_{-    94.5}$ & 145.8$^{+  1.2}_{-  3.0}$ &   91.2$^{+   8.4}_{-   5.0}$ \vspace*{0.1cm} \\ \noalign{\smallskip}
NGC6517 &  4.16$^{+ 0.16}_{- 0.16}$ &  0.68$^{+ 0.06}_{- 0.05}$ &  0.72$^{+ 0.01}_{- 0.02}$ &   -143.4$^{+    26.5}_{-    25.4}$ & 127.0$^{+  3.4}_{-  4.3}$ &   51.5$^{+   1.9}_{-   1.9}$ \\ \noalign{\smallskip}
 &  4.04$^{+ 0.19}_{- 0.20}$ &  0.49$^{+ 0.05}_{- 0.05}$ &  0.79$^{+ 0.02}_{- 0.02}$ &   -134.0$^{+    24.2}_{-    23.5}$ & 130.8$^{+  4.6}_{-  5.0}$ &   46.0$^{+   2.0}_{-   2.0}$ \\ \noalign{\smallskip}
 &  4.56$^{+ 0.31}_{- 0.28}$ &  0.05$^{+ 0.04}_{- 0.02}$ &  0.93$^{+ 0.01}_{- 0.02}$ &    -22.0$^{+    39.8}_{-    37.5}$ &  98.9$^{+ 11.9}_{- 16.2}$ &   51.6$^{+   2.9}_{-   2.3}$ \vspace*{0.1cm} \\ \noalign{\smallskip}
NGC6522 &  1.25$^{+ 0.08}_{- 0.04}$ &  0.55$^{+ 0.10}_{- 0.08}$ &  0.39$^{+ 0.05}_{- 0.05}$ &    -42.7$^{+    15.9}_{-    19.5}$ & 116.0$^{+  1.6}_{-  1.4}$ &   21.0$^{+   1.6}_{-   1.1}$ \\ \noalign{\smallskip}
 &  1.23$^{+ 0.11}_{- 0.21}$ &  0.31$^{+ 0.11}_{- 0.07}$ &  0.58$^{+ 0.05}_{- 0.07}$ &    -68.4$^{+    20.5}_{-    19.5}$ & 121.9$^{+  1.3}_{-  7.7}$ &   15.0$^{+   2.5}_{-   2.5}$ \\ \noalign{\smallskip}
 &  1.40$^{+ 0.04}_{- 0.07}$ &  0.04$^{+ 0.02}_{- 0.02}$ &  0.59$^{+ 0.15}_{- 0.14}$ &     -6.7$^{+    21.7}_{-    16.7}$ &  94.3$^{+  7.9}_{- 14.5}$ &   20.6$^{+   1.3}_{-   1.2}$ \vspace*{0.1cm} \\ \noalign{\smallskip}
NGC6528 &  1.65$^{+ 0.13}_{- 0.07}$ &  0.36$^{+ 0.00}_{- 0.00}$ &  0.64$^{+ 0.03}_{- 0.01}$ &    -43.7$^{+     4.8}_{-     6.5}$ & 107.4$^{+  1.9}_{-  1.4}$ &   25.0$^{+   1.3}_{-   0.7}$ \\ \noalign{\smallskip}
 &  1.03$^{+ 0.31}_{- 0.24}$ &  0.39$^{+ 0.08}_{- 0.11}$ &  0.45$^{+ 0.20}_{- 0.19}$ &    -53.3$^{+     7.3}_{-     8.5}$ & 110.9$^{+ 10.8}_{-  3.6}$ &   14.0$^{+   2.5}_{-   2.0}$ \\ \noalign{\smallskip}
 &  2.43$^{+ 0.14}_{- 0.11}$ &  0.05$^{+ 0.03}_{- 0.02}$ &  0.53$^{+ 0.04}_{- 0.02}$ &    -72.6$^{+    25.5}_{-    23.8}$ & 123.7$^{+  9.0}_{- 12.9}$ &   33.6$^{+   1.0}_{-   1.1}$ \vspace*{0.1cm} \\ \noalign{\smallskip}
NGC6535 &  4.42$^{+ 0.05}_{- 0.04}$ &  1.36$^{+ 0.06}_{- 0.06}$ &  0.53$^{+ 0.02}_{- 0.02}$ &    347.6$^{+    13.3}_{-    14.3}$ &  20.4$^{+  1.1}_{-  1.1}$ &   57.6$^{+   0.1}_{-   0.1}$ \\ \noalign{\smallskip}
 &  4.52$^{+ 0.05}_{- 0.04}$ &  1.10$^{+ 0.04}_{- 0.06}$ &  0.61$^{+ 0.02}_{- 0.02}$ &    323.6$^{+    14.7}_{-    16.4}$ &  38.8$^{+  1.2}_{-  1.1}$ &   53.5$^{+   0.5}_{-   0.1}$ \\ \noalign{\smallskip}
 &  4.80$^{+ 0.06}_{- 0.06}$ &  1.08$^{+ 0.06}_{- 0.07}$ &  0.56$^{+ 0.03}_{- 0.03}$ &    370.0$^{+    12.8}_{-    12.6}$ &  26.5$^{+  0.9}_{-  1.0}$ &   57.3$^{+   1.1}_{-   1.6}$ \vspace*{0.1cm} \\ \noalign{\smallskip}
NGC6539 &  3.28$^{+ 0.05}_{- 0.03}$ &  2.15$^{+ 0.07}_{- 0.06}$ &  0.21$^{+ 0.01}_{- 0.01}$ &   -320.4$^{+     5.2}_{-     5.0}$ & 122.0$^{+  0.6}_{-  0.6}$ &   54.8$^{+   0.7}_{-   0.6}$ \\ \noalign{\smallskip}
 &  3.29$^{+ 0.02}_{- 0.02}$ &  1.95$^{+ 0.05}_{- 0.04}$ &  0.26$^{+ 0.01}_{- 0.01}$ &   -335.3$^{+     5.0}_{-     5.7}$ & 125.4$^{+  0.9}_{-  0.8}$ &   50.5$^{+   0.1}_{-   0.5}$ \\ \noalign{\smallskip}
 &  3.61$^{+ 0.60}_{- 0.16}$ &  1.36$^{+ 0.14}_{- 0.12}$ &  0.23$^{+ 0.01}_{- 0.01}$ &   -294.9$^{+    39.2}_{-    47.6}$ & 126.8$^{+  4.2}_{-  3.3}$ &   48.8$^{+   2.3}_{-   1.9}$ \vspace*{0.1cm} \\ \noalign{\smallskip}
NGC6541 &  4.08$^{+ 0.05}_{- 0.05}$ &  1.58$^{+ 0.06}_{- 0.05}$ &  0.44$^{+ 0.01}_{- 0.01}$ &   -314.6$^{+     3.7}_{-     4.1}$ & 139.2$^{+  2.0}_{-  1.9}$ &   58.5$^{+   0.8}_{-   0.6}$ \\ \noalign{\smallskip}
 &  3.70$^{+ 0.07}_{- 0.05}$ &  1.44$^{+ 0.07}_{- 0.05}$ &  0.44$^{+ 0.02}_{- 0.02}$ &   -326.8$^{+     3.5}_{-     4.0}$ & 130.8$^{+  1.1}_{-  1.1}$ &   50.5$^{+   0.5}_{-   0.5}$ \\ \noalign{\smallskip}
 &  4.28$^{+ 0.04}_{- 0.05}$ &  1.52$^{+ 0.13}_{- 0.22}$ &  0.38$^{+ 0.01}_{- 0.02}$ &   -389.3$^{+    34.6}_{-     8.7}$ & 135.1$^{+  1.0}_{-  1.8}$ &   58.8$^{+   2.2}_{-   1.7}$ \vspace*{0.1cm} \\ \noalign{\smallskip}
NGC6544 &  5.22$^{+ 0.12}_{- 0.14}$ &  0.38$^{+ 0.09}_{- 0.08}$ &  0.86$^{+ 0.03}_{- 0.03}$ &     16.7$^{+    27.0}_{-    29.2}$ &  88.0$^{+  3.5}_{-  3.2}$ &   61.5$^{+   0.6}_{-   0.6}$ \\ \noalign{\smallskip}
 &  5.23$^{+ 0.09}_{- 0.12}$ &  0.39$^{+ 0.05}_{- 0.03}$ &  0.86$^{+ 0.01}_{- 0.01}$ &    -29.8$^{+    32.7}_{-    34.5}$ & 105.8$^{+  8.3}_{- 18.6}$ &   57.5$^{+   0.1}_{-   1.0}$ \\ \noalign{\smallskip}
 &  5.57$^{+ 0.14}_{- 0.15}$ &  0.43$^{+ 0.06}_{- 0.06}$ &  0.80$^{+ 0.02}_{- 0.02}$ &   -185.7$^{+    30.6}_{-    43.7}$ & 133.3$^{+  3.2}_{-  2.7}$ &   60.9$^{+   3.8}_{-   2.1}$ \vspace*{0.1cm} \\ \noalign{\smallskip}
NGC6626 &  2.88$^{+ 0.12}_{- 0.10}$ &  0.70$^{+ 0.04}_{- 0.04}$ &  0.61$^{+ 0.01}_{- 0.01}$ &   -144.0$^{+    16.6}_{-    19.2}$ & 117.3$^{+  2.7}_{-  2.5}$ &   39.7$^{+   1.5}_{-   1.3}$ \\ \noalign{\smallskip}
 &  3.15$^{+ 0.09}_{- 0.16}$ &  0.58$^{+ 0.12}_{- 0.06}$ &  0.69$^{+ 0.03}_{- 0.06}$ &   -178.3$^{+    20.0}_{-    19.8}$ & 136.7$^{+  5.1}_{-  3.5}$ &   38.5$^{+   1.0}_{-   2.0}$ \\ \noalign{\smallskip}
 &  3.17$^{+ 0.16}_{- 0.46}$ &  0.11$^{+ 0.16}_{- 0.07}$ &  0.42$^{+ 0.01}_{- 0.01}$ &   -160.0$^{+    88.1}_{-    79.8}$ & 132.8$^{+ 13.2}_{- 20.8}$ &   40.3$^{+   1.5}_{-   1.8}$ \vspace*{0.1cm} \\ \noalign{\smallskip}
NGC6637 &  1.68$^{+ 0.09}_{- 0.06}$ &  1.03$^{+ 0.12}_{- 0.09}$ &  0.24$^{+ 0.03}_{- 0.03}$ &    -45.9$^{+    13.4}_{-    19.9}$ & 105.1$^{+  3.2}_{-  2.8}$ &   22.6$^{+   1.1}_{-   0.8}$ \\ \noalign{\smallskip}
 &  1.93$^{+ 0.12}_{- 0.01}$ &  0.32$^{+ 0.04}_{- 0.03}$ &  0.72$^{+ 0.03}_{- 0.03}$ &    -33.8$^{+    11.8}_{-    13.3}$ & 109.5$^{+  4.2}_{-  5.1}$ &   23.5$^{+   1.5}_{-   1.5}$ \\ \noalign{\smallskip}
 &  2.14$^{+ 0.28}_{- 0.20}$ &  0.45$^{+ 0.16}_{- 0.40}$ &  0.44$^{+ 0.02}_{- 0.03}$ &    -42.5$^{+    20.4}_{-    17.0}$ & 102.3$^{+  3.9}_{-  6.0}$ &   30.3$^{+   1.4}_{-   2.1}$ \vspace*{0.1cm} \\ \noalign{\smallskip}
NGC6656 & 10.06$^{+ 0.01}_{- 0.01}$ &  3.12$^{+ 0.04}_{- 0.04}$ &  0.53$^{+ 0.01}_{- 0.01}$ &   -970.8$^{+    18.3}_{-    17.5}$ & 145.7$^{+  0.8}_{-  0.8}$ &  129.2$^{+   0.2}_{-   0.2}$ \\ \noalign{\smallskip}
 &  9.92$^{+ 0.02}_{- 0.02}$ &  3.21$^{+ 0.04}_{- 0.04}$ &  0.51$^{+ 0.01}_{- 0.01}$ &  -1055.2$^{+    19.2}_{-    19.6}$ & 149.7$^{+  0.8}_{-  0.9}$ &  124.0$^{+   0.5}_{-   0.5}$ \\ \noalign{\smallskip}
 &  8.60$^{+ 0.15}_{- 0.10}$ &  2.71$^{+ 0.04}_{- 0.06}$ &  0.48$^{+ 0.01}_{- 0.01}$ &   -895.9$^{+    11.3}_{-    19.2}$ & 154.6$^{+  1.1}_{-  0.7}$ &  105.7$^{+   2.1}_{-   2.6}$ \vspace*{0.1cm} \\ \noalign{\smallskip}
NGC6681 &  5.41$^{+ 0.30}_{- 0.26}$ &  0.92$^{+ 0.10}_{- 0.06}$ &  0.71$^{+ 0.03}_{- 0.04}$ &    -33.9$^{+    14.0}_{-    16.8}$ &  96.1$^{+  2.2}_{-  2.1}$ &   68.6$^{+   2.7}_{-   2.1}$ \\ \noalign{\smallskip}
 &  4.31$^{+ 0.26}_{- 0.23}$ &  1.16$^{+ 0.10}_{- 0.09}$ &  0.58$^{+ 0.05}_{- 0.05}$ &    -49.5$^{+    16.7}_{-    19.2}$ &  96.5$^{+  2.1}_{-  1.9}$ &   54.5$^{+   2.0}_{-   1.5}$ \\ \noalign{\smallskip}
 &  5.97$^{+ 0.29}_{- 0.19}$ &  0.07$^{+ 0.03}_{- 0.02}$ &  0.72$^{+ 0.04}_{- 0.03}$ &     -9.1$^{+    20.3}_{-    18.9}$ &  92.2$^{+  5.0}_{-  5.4}$ &   65.8$^{+   3.2}_{-   3.3}$ \vspace*{0.1cm} \\ \noalign{\smallskip}
NGC6752 &  5.58$^{+ 0.04}_{- 0.04}$ &  3.66$^{+ 0.08}_{- 0.08}$ &  0.21$^{+ 0.01}_{- 0.01}$ &   -863.3$^{+    20.9}_{-    20.2}$ & 155.3$^{+  0.6}_{-  0.6}$ &   88.4$^{+   0.9}_{-   0.9}$ \\ \noalign{\smallskip}
 &  5.72$^{+ 0.04}_{- 0.05}$ &  3.64$^{+ 0.08}_{- 0.08}$ &  0.22$^{+ 0.01}_{- 0.01}$ &   -939.8$^{+    22.6}_{-    22.3}$ & 151.0$^{+  0.7}_{-  0.8}$ &   85.0$^{+   1.0}_{-   1.0}$ \\ \noalign{\smallskip}
 &  6.34$^{+ 0.15}_{- 0.28}$ &  2.36$^{+ 0.07}_{- 0.07}$ &  0.22$^{+ 0.01}_{- 0.01}$ &   -808.0$^{+    33.5}_{-    58.8}$ & 151.5$^{+  1.7}_{-  1.0}$ &   85.0$^{+   1.1}_{-   0.8}$ \vspace*{0.1cm} \\ \noalign{\smallskip}
NGC6779 & 12.11$^{+ 0.22}_{- 0.23}$ &  0.92$^{+ 0.10}_{- 0.10}$ &  0.86$^{+ 0.01}_{- 0.01}$ &    184.1$^{+    33.5}_{-    33.6}$ &  75.5$^{+  2.0}_{-  1.9}$ &  138.0$^{+   3.1}_{-   3.1}$ \\ \noalign{\smallskip}
 & 12.32$^{+ 0.24}_{- 0.29}$ &  0.63$^{+ 0.08}_{- 0.06}$ &  0.90$^{+ 0.01}_{- 0.01}$ &    125.9$^{+    33.7}_{-    36.8}$ &  62.6$^{+  5.6}_{-  5.4}$ &  133.5$^{+   3.0}_{-   3.0}$ \\ \noalign{\smallskip}
 & 12.11$^{+ 0.14}_{- 0.10}$ &  0.36$^{+ 0.13}_{- 0.10}$ &  0.80$^{+ 0.01}_{- 0.01}$ &    213.3$^{+    33.1}_{-    45.8}$ &  56.3$^{+  6.2}_{-  3.8}$ &  120.7$^{+   3.5}_{-   4.5}$ \vspace*{0.1cm} \\ \noalign{\smallskip}
NGC6809 &  6.06$^{+ 0.08}_{- 0.07}$ &  1.57$^{+ 0.00}_{- 0.00}$ &  0.59$^{+ 0.01}_{- 0.01}$ &   -228.0$^{+    14.5}_{-    17.9}$ & 109.8$^{+  1.6}_{-  1.3}$ &   79.1$^{+   0.8}_{-   0.6}$ \\ \noalign{\smallskip}
 &  5.82$^{+ 0.09}_{- 0.11}$ &  1.59$^{+ 0.00}_{- 0.00}$ &  0.57$^{+ 0.01}_{- 0.01}$ &   -262.5$^{+    19.5}_{-    18.9}$ & 115.3$^{+  1.8}_{-  1.9}$ &   72.5$^{+   0.5}_{-   0.1}$ \\ \noalign{\smallskip}
 &  6.36$^{+ 0.13}_{- 0.09}$ &  1.08$^{+ 0.14}_{- 0.12}$ &  0.45$^{+ 0.02}_{- 0.01}$ &   -324.0$^{+     3.9}_{-     3.0}$ & 121.2$^{+  1.0}_{-  1.4}$ &   77.2$^{+   3.9}_{-   1.4}$ \vspace*{0.1cm} \\ \noalign{\smallskip}
NGC6838 &  7.14$^{+ 0.00}_{- 0.00}$ &  5.00$^{+ 0.03}_{- 0.03}$ &  0.18$^{+ 0.01}_{- 0.01}$ &  -1336.3$^{+     5.2}_{-     5.4}$ & 168.1$^{+  0.3}_{-  0.2}$ &  111.6$^{+   0.2}_{-   0.2}$ \\ \noalign{\smallskip}
 &  7.30$^{+ 0.00}_{- 0.00}$ &  4.99$^{+ 0.03}_{- 0.03}$ &  0.19$^{+ 0.01}_{- 0.01}$ &  -1424.7$^{+     5.7}_{-     5.8}$ & 170.2$^{+  0.2}_{-  0.2}$ &  110.0$^{+   0.1}_{-   0.5}$ \\ \noalign{\smallskip}
 &  7.24$^{+ 0.01}_{- 0.01}$ &  4.45$^{+ 0.03}_{- 0.03}$ &  0.23$^{+ 0.01}_{- 0.01}$ &  -1286.8$^{+     5.5}_{-     6.0}$ & 170.9$^{+  0.2}_{-  0.2}$ &  106.0$^{+   2.6}_{-   2.6}$ \vspace*{0.1cm} \\ \noalign{\smallskip}
NGC6864 & 16.72$^{+ 0.64}_{- 0.61}$ &  1.49$^{+ 0.23}_{- 0.18}$ &  0.84$^{+ 0.01}_{- 0.02}$ &   -313.9$^{+    84.4}_{-   101.6}$ & 126.1$^{+  3.7}_{-  5.0}$ &  193.6$^{+   8.2}_{-   7.6}$ \\ \noalign{\smallskip}
 & 16.31$^{+ 0.62}_{- 0.62}$ &  1.13$^{+ 0.22}_{- 0.20}$ &  0.87$^{+ 0.02}_{- 0.02}$ &   -221.2$^{+    92.3}_{-    93.8}$ & 114.3$^{+  4.1}_{-  5.2}$ &  183.5$^{+   8.0}_{-   8.0}$ \\ \noalign{\smallskip}
 & 18.05$^{+ 0.32}_{- 0.29}$ &  0.11$^{+ 0.10}_{- 0.06}$ &  0.88$^{+ 0.01}_{- 0.01}$ &     57.0$^{+    43.5}_{-    44.7}$ &  78.9$^{+  8.5}_{-  6.7}$ &  183.5$^{+   5.1}_{-   9.5}$ \vspace*{0.1cm} \\ \noalign{\smallskip}
NGC6981 & 12.96$^{+ 0.39}_{- 0.34}$ &  6.40$^{+ 0.44}_{- 0.39}$ &  0.34$^{+ 0.02}_{- 0.02}$ &  -1231.9$^{+    75.9}_{-    85.3}$ & 131.9$^{+  0.4}_{-  0.5}$ &  192.1$^{+   8.2}_{-   7.2}$ \\ \noalign{\smallskip}
 & 12.94$^{+ 0.35}_{- 0.33}$ &  5.94$^{+ 0.40}_{- 0.33}$ &  0.37$^{+ 0.01}_{- 0.02}$ &  -1211.1$^{+    69.8}_{-    75.1}$ & 127.7$^{+  0.6}_{-  0.6}$ &  184.5$^{+   7.5}_{-   6.5}$ \\ \noalign{\smallskip}
 & 13.37$^{+ 0.47}_{- 0.32}$ &  4.40$^{+ 0.26}_{- 0.24}$ &  0.50$^{+ 0.01}_{- 0.01}$ &   -944.6$^{+    55.9}_{-    62.7}$ & 125.7$^{+  0.8}_{-  0.8}$ &  175.3$^{+   7.5}_{-   7.5}$ \vspace*{0.1cm} \\ \noalign{\smallskip}
NGC7078 & 10.52$^{+ 0.16}_{- 0.17}$ &  3.84$^{+ 0.14}_{- 0.14}$ &  0.46$^{+ 0.01}_{- 0.01}$ &  -1063.8$^{+    29.3}_{-    29.3}$ & 150.4$^{+  0.6}_{-  0.6}$ &  141.1$^{+   3.0}_{-   2.9}$ \\ \noalign{\smallskip}
 & 10.59$^{+ 0.16}_{- 0.16}$ &  3.69$^{+ 0.13}_{- 0.12}$ &  0.48$^{+ 0.01}_{- 0.01}$ &  -1108.7$^{+    27.2}_{-    27.5}$ & 143.9$^{+  0.2}_{-  0.2}$ &  136.5$^{+   2.5}_{-   3.0}$ \\ \noalign{\smallskip}
 & 10.70$^{+ 0.37}_{- 0.16}$ &  3.05$^{+ 0.06}_{- 0.10}$ &  0.53$^{+ 0.01}_{- 0.01}$ &   -944.1$^{+    37.6}_{-    30.2}$ & 142.7$^{+  0.4}_{-  0.7}$ &  127.5$^{+   2.6}_{-   2.5}$ \vspace*{0.1cm} \\ \noalign{\smallskip}
NGC7089 & 18.41$^{+ 0.60}_{- 0.48}$ &  1.09$^{+ 0.14}_{- 0.20}$ &  0.89$^{+ 0.02}_{- 0.01}$ &    184.4$^{+    57.4}_{-    52.6}$ &  60.6$^{+  3.5}_{-  2.2}$ &  210.1$^{+   7.3}_{-   6.8}$ \\ \noalign{\smallskip}
 & 18.55$^{+ 0.59}_{- 0.53}$ &  0.72$^{+ 0.13}_{- 0.09}$ &  0.93$^{+ 0.01}_{- 0.01}$ &    139.2$^{+    56.2}_{-    56.7}$ &  69.7$^{+  3.9}_{-  4.5}$ &  206.5$^{+   7.5}_{-   7.0}$ \\ \noalign{\smallskip}
 & 17.79$^{+ 0.42}_{- 1.63}$ &  0.17$^{+ 0.12}_{- 0.08}$ &  0.94$^{+ 0.01}_{- 0.01}$ &     66.0$^{+    53.9}_{-    42.3}$ &  78.8$^{+  6.9}_{-  6.9}$ &  176.7$^{+   3.5}_{-   5.2}$ \vspace*{0.1cm} \\ \noalign{\smallskip}
NGC7099 &  8.32$^{+ 0.17}_{- 0.18}$ &  1.83$^{+ 0.01}_{- 0.01}$ &  0.64$^{+ 0.01}_{- 0.01}$ &    255.7$^{+     5.3}_{-     7.3}$ &  59.0$^{+  1.5}_{-  0.9}$ &  104.9$^{+   1.9}_{-   1.8}$ \\ \noalign{\smallskip}
 &  8.21$^{+ 0.15}_{- 0.15}$ &  1.52$^{+ 0.00}_{- 0.00}$ &  0.69$^{+ 0.01}_{- 0.01}$ &    239.8$^{+     7.4}_{-     9.5}$ &  68.0$^{+  0.8}_{-  0.5}$ &   96.5$^{+   1.5}_{-   1.5}$ \\ \noalign{\smallskip}
 &  8.28$^{+ 0.16}_{- 0.12}$ &  1.08$^{+ 0.08}_{- 0.10}$ &  0.74$^{+ 0.01}_{- 0.01}$ &    177.8$^{+    24.8}_{-    37.2}$ &  73.5$^{+  3.1}_{-  2.9}$ &   91.7$^{+   3.0}_{-   1.7}$ \vspace*{0.1cm} \\ \noalign{\smallskip}
\\
\end{longtable}
}
\end{center}

\begin{table*}
\footnotesize
\caption{Dwarf spheroidal orbital properties. In addition to the parameters listed in Table~\ref{tab:orb_prop_gc}, we also include the angles $(\theta,\phi)$ of the angular momentum vector, and the time elapsed since the last apocentre and pericentre, $T_a$ and $T_p$ respectively. The errors indicate the 16th and 84th percentiles, which were obtained from Monte Carlo realisations sampling the (statistical and systematic) errors in the observables. For each dSph we quote values derived from orbits integrated for 10 Gyr in the Galactic potentials of Models 1, 2, and 3 in the first, second, and third row, respectively.  Since for Model-1 there is a significant fraction of the realisations for which Leo~I does not complete one radial oscillation in the 10~Gyr of integration, the values quoted here for this dSph for this model were derived for an integration time of 100~Gyr.} 
\label{tab:orb_prop_dw}
\begin{tabular}{lccccccccc}
\hline
  Name & apocentre & pericentre & eccentr. & $L_z$ & inclin. $\theta$ & $\phi$ & $T_r$ & $T_a$ & $T_p$ \\
  & [kpc] & [kpc] &  & [km/s kpc] & [deg] & [deg]  & [Gyr]  & [Gyr] & [Gyr] \\
\hline
Fnx & 172.6$^{+114.9}_{- 27.3}$ & 116.3$^{+ 34.5}_{- 50.4}$ &  0.27$^{+ 0.19}_{- 0.08}$ &   8123.6$^{+  2696.9}_{-  2444.9}$ &  70.7$^{+  2.9}_{-  2.4}$ &  172.9$^{+   8.7}_{-   8.5}$ &  3.95$^{+ 2.91}_{- 1.23}$ &  0.98$^{+ 2.01}_{- 0.64}$ &  2.97$^{+ 3.44}_{- 1.25}$ \\ \noalign{\smallskip}
&156.4$^{+ 26.9}_{- 15.1}$ &  85.9$^{+ 47.7}_{- 34.4}$ &  0.29$^{+ 0.18}_{- 0.12}$ &   7659.9$^{+  2586.6}_{-  2516.9}$ &  70.7$^{+  2.9}_{-  2.6}$ &  172.6$^{+   9.4}_{-   9.8}$ &  2.65$^{+ 0.77}_{- 0.52}$ &  0.31$^{+ 0.04}_{- 0.60}$ &  1.67$^{+ 0.40}_{- 0.96}$  \\ \noalign{\smallskip}
 & 152.4$^{+ 13.9}_{- 11.9}$ &  46.0$^{+ 31.6}_{- 19.6}$ &  0.54$^{+ 0.15}_{- 0.18}$ &   5610.0$^{+  2524.4}_{-  1818.0}$ &  69.5$^{+  3.4}_{-  2.8}$ &  168.4$^{+  11.0}_{-  12.4}$ &  2.06$^{+ 0.44}_{- 0.28}$ &  0.16$^{+ 0.08}_{- 0.03}$ &  1.18$^{+ 0.30}_{- 0.17}$ \vspace*{0.1cm} \\ \noalign{\smallskip}
Dra &  90.3$^{+  7.7}_{-  8.2}$ &  30.4$^{+  6.1}_{-  5.2}$ &  0.49$^{+ 0.05}_{- 0.05}$ &  -2919.2$^{+   954.0}_{-  1030.3}$ & 107.4$^{+  4.7}_{-  5.1}$ & -170.4$^{+   4.2}_{-   4.4}$ &  1.38$^{+ 0.18}_{- 0.16}$ &  0.30$^{+ 0.06}_{- 0.05}$ &  0.99$^{+ 0.15}_{- 0.12}$ \\ \noalign{\smallskip}
& 85.7$^{+  7.3}_{-  6.9}$ &  32.0$^{+  6.1}_{-  5.3}$ &  0.45$^{+ 0.05}_{- 0.05}$ &  -2879.5$^{+   866.1}_{-   977.0}$ & 106.5$^{+  4.6}_{-  4.2}$ & -170.3$^{+   4.4}_{-   3.9}$ &  1.29$^{+ 0.15}_{- 0.12}$ &  0.18$^{+ 0.04}_{- 0.02}$ &  0.89$^{+ 0.09}_{- 0.11}$  \\ \noalign{\smallskip}
 &  86.2$^{+  7.0}_{-  6.2}$ &  26.6$^{+  4.5}_{-  3.5}$ &  0.53$^{+ 0.04}_{- 0.04}$ &  -2433.7$^{+   722.1}_{-   912.5}$ & 104.9$^{+  4.6}_{-  3.8}$ & -172.3$^{+   4.1}_{-   3.0}$ &  1.17$^{+ 0.11}_{- 0.10}$ &  0.22$^{+ 0.03}_{- 0.02}$ &  0.80$^{+ 0.08}_{- 0.07}$ \vspace*{0.1cm} \\ \noalign{\smallskip}
Car & 107.5$^{+ 19.4}_{-  6.7}$ &  87.0$^{+ 23.0}_{- 24.8}$ &  0.14$^{+ 0.11}_{- 0.10}$ &  -2537.8$^{+  1837.6}_{-  1882.9}$ &  98.0$^{+  6.0}_{-  5.9}$ & -162.1$^{+   2.6}_{-   2.5}$ &  2.39$^{+ 0.68}_{- 0.42}$ &  0.05$^{+ 1.55}_{- 0.03}$ &  1.21$^{+ 1.64}_{- 0.24}$ \\ \noalign{\smallskip}
&106.7$^{+  7.8}_{-  6.3}$ &  74.5$^{+ 23.7}_{- 19.5}$ &  0.18$^{+ 0.12}_{- 0.10}$ &  -2663.3$^{+  1725.2}_{-  1851.0}$ &  98.7$^{+  5.8}_{-  5.8}$ & -161.5$^{+   2.8}_{-   2.6}$ &  1.96$^{+ 0.30}_{- 0.27}$ &  0.08$^{+ 0.01}_{- 0.17}$ &  1.06$^{+ 0.16}_{- 0.29}$  \\ \noalign{\smallskip}
 & 106.4$^{+  6.2}_{-  6.1}$ &  46.0$^{+ 13.2}_{- 11.2}$ &  0.40$^{+ 0.10}_{- 0.10}$ &  -2426.0$^{+  1538.0}_{-  1451.5}$ &  99.4$^{+  6.1}_{-  5.8}$ & -161.7$^{+   2.8}_{-   2.4}$ &  1.55$^{+ 0.19}_{- 0.14}$ &  1.54$^{+ 0.19}_{- 0.14}$ &  0.76$^{+ 0.09}_{- 0.07}$ \vspace*{0.1cm} \\ \noalign{\smallskip}
UMi &  88.0$^{+  3.7}_{-  3.9}$ &  33.5$^{+  5.6}_{-  4.9}$ &  0.45$^{+ 0.05}_{- 0.06}$ &   -786.2$^{+   708.9}_{-   749.9}$ &  94.4$^{+  4.0}_{-  4.0}$ &  161.3$^{+   4.0}_{-   3.7}$ &  1.39$^{+ 0.11}_{- 0.09}$ &  0.26$^{+ 0.04}_{- 0.03}$ &  0.96$^{+ 0.10}_{- 0.08}$ \\ \noalign{\smallskip}
& 85.3$^{+  3.1}_{-  3.1}$ &  36.1$^{+  5.2}_{-  5.1}$ &  0.40$^{+ 0.06}_{- 0.05}$ &  -1008.4$^{+   690.8}_{-   698.2}$ &  95.3$^{+  3.7}_{-  3.7}$ &  162.3$^{+   3.9}_{-   3.9}$ &  1.32$^{+ 0.08}_{- 0.09}$ &  0.19$^{+ 0.01}_{- 0.02}$ &  0.88$^{+ 0.07}_{- 0.07}$  \\ \noalign{\smallskip}
 &  85.6$^{+  3.8}_{-  2.9}$ &  30.5$^{+  3.9}_{-  3.8}$ &  0.47$^{+ 0.05}_{- 0.04}$ &  -1314.4$^{+   574.2}_{-   583.4}$ &  97.2$^{+  3.4}_{-  3.2}$ &  164.2$^{+   3.6}_{-   3.1}$ &  1.19$^{+ 0.06}_{- 0.06}$ &  0.20$^{+ 0.02}_{- 0.02}$ &  0.80$^{+ 0.05}_{- 0.04}$ \vspace*{0.1cm} \\ \noalign{\smallskip}
Sext & 224.3$^{+ 79.5}_{- 51.5}$ &  79.6$^{+  5.0}_{-  5.3}$ &  0.48$^{+ 0.09}_{- 0.09}$ & -16800.2$^{+  1592.0}_{-  1720.0}$ & 139.4$^{+  0.1}_{-  0.3}$ &  124.8$^{+   6.2}_{-   6.8}$ &  4.29$^{+ 1.58}_{- 0.99}$ &  2.34$^{+ 0.77}_{- 0.44}$ &  0.20$^{+ 0.05}_{- 0.04}$ \\ \noalign{\smallskip}
&164.8$^{+ 30.8}_{- 24.6}$ &  80.1$^{+  4.6}_{-  4.8}$ &  0.35$^{+ 0.05}_{- 0.05}$ & -17094.0$^{+  1565.6}_{-  1627.7}$ & 139.3$^{+  0.1}_{-  0.3}$ &  126.8$^{+   6.2}_{-   6.4}$ &  2.67$^{+ 0.43}_{- 0.34}$ &  1.54$^{+ 0.19}_{- 0.21}$ &  0.18$^{+ 0.02}_{- 0.10}$  \\ \noalign{\smallskip}
 & 120.5$^{+ 13.6}_{- 10.3}$ &  70.0$^{+  6.2}_{-  6.4}$ &  0.27$^{+ 0.01}_{- 0.00}$ & -14938.4$^{+  1333.5}_{-  1492.2}$ & 139.0$^{+  0.3}_{-  0.7}$ &  133.0$^{+   6.8}_{-   6.5}$ &  1.94$^{+ 0.21}_{- 0.17}$ &  1.31$^{+ 0.08}_{- 0.06}$ &  0.34$^{+ 0.05}_{- 0.05}$ \vspace*{0.1cm} \\ \noalign{\smallskip}
Leo I & 820.0$^{+920.3}_{-243.4}$ &  89.5$^{+ 55.9}_{- 47.5}$ &  0.83$^{+ 0.07}_{- 0.03}$ & -16313.6$^{+ 12245.8}_{- 10347.6}$ & 125.9$^{+  5.4}_{- 23.4}$ &  138.4$^{+  55.9}_{-  48.5}$ & 18.35$^{+28.13}_{- 7.19}$ & 10.30$^{+13.83}_{- 3.69}$ &  1.01$^{+ 0.08}_{- 0.11}$ \\ \noalign{\smallskip}
&429.0$^{+126.1}_{- 60.6}$ & 112.6$^{+ 58.4}_{- 60.6}$ &  0.61$^{+ 0.16}_{- 0.05}$ & -21186.6$^{+ 13321.0}_{- 12256.5}$ & 126.5$^{+  4.9}_{- 15.7}$ &  136.7$^{+  47.2}_{-  50.2}$ &  5.04$^{+ 0.63}_{- 0.54}$ &  4.01$^{+ 0.70}_{- 1.10}$ &  0.92$^{+ 0.16}_{- 0.08}$  \\ \noalign{\smallskip}
 & 388.0$^{+ 86.9}_{- 37.0}$ &  86.9$^{+ 59.2}_{- 44.4}$ &  0.63$^{+ 0.16}_{- 0.10}$ & -18831.8$^{+  9856.3}_{- 11928.2}$ & 127.0$^{+  4.5}_{- 11.0}$ &  140.4$^{+  46.0}_{-  43.3}$ &  4.92$^{+ 1.34}_{- 0.71}$ &  3.43$^{+ 0.67}_{- 0.40}$ &  0.91$^{+ 0.06}_{- 0.09}$ \vspace*{0.1cm} \\ \noalign{\smallskip}
Leo~II & 240.5$^{+ 20.2}_{- 14.5}$ &  68.9$^{+115.9}_{- 42.5}$ &  0.56$^{+ 0.24}_{- 0.33}$ &   2055.2$^{+  7219.4}_{-  6753.6}$ &  82.3$^{+ 25.3}_{- 14.6}$ & -128.9$^{+  80.9}_{- 152.5}$ &  4.38$^{+ 2.21}_{- 0.73}$ &  4.13$^{+ 1.35}_{- 0.68}$ &  1.82$^{+ 0.42}_{- 0.29}$ \\ \noalign{\smallskip}
&236.4$^{+ 14.9}_{- 14.0}$ &  60.1$^{+ 75.7}_{- 36.8}$ &  0.60$^{+ 0.22}_{- 0.32}$ &    972.6$^{+  6352.4}_{-  6214.4}$ &  86.9$^{+ 19.9}_{- 16.5}$ & -143.2$^{+  68.9}_{- 160.3}$ &  3.28$^{+ 0.76}_{- 0.34}$ &  3.21$^{+ 0.37}_{- 0.57}$ &  1.51$^{+ 0.17}_{- 0.25}$  \\ \noalign{\smallskip}
 & 237.7$^{+ 15.3}_{- 12.5}$ &  45.8$^{+ 48.4}_{- 26.9}$ &  0.68$^{+ 0.17}_{- 0.24}$ &   -280.3$^{+  5319.4}_{-  4976.1}$ &  91.2$^{+ 17.1}_{- 19.5}$ &  -91.7$^{+ 171.3}_{-  59.0}$ &  3.01$^{+ 0.45}_{- 0.29}$ &  2.90$^{+ 0.40}_{- 0.28}$ &  1.39$^{+ 0.18}_{- 0.13}$ \vspace*{0.1cm} \\ \noalign{\smallskip}
Sgr &  44.5$^{+ 13.3}_{- 10.4}$ &  14.8$^{+  2.1}_{-  2.3}$ &  0.50$^{+ 0.05}_{- 0.04}$ &   -927.1$^{+   437.9}_{-   472.7}$ & 100.6$^{+  3.1}_{-  3.8}$ &   90.9$^{+   0.9}_{-   1.0}$ &  0.62$^{+ 0.18}_{- 0.14}$ &  0.35$^{+ 0.09}_{- 0.07}$ &  0.04$^{+ 0.00}_{- 0.00}$ \\ \noalign{\smallskip}
& 42.6$^{+ 14.0}_{-  9.7}$ &  14.3$^{+  2.3}_{-  2.3}$ &  0.50$^{+ 0.05}_{- 0.03}$ &   -748.7$^{+   399.4}_{-   469.1}$ &  98.8$^{+  3.3}_{-  4.1}$ &   91.6$^{+   1.1}_{-   2.0}$ &  0.61$^{+ 0.18}_{- 0.14}$ &  0.34$^{+ 0.07}_{- 0.09}$ &  0.05$^{+ 0.01}_{- 0.00}$  \\ \noalign{\smallskip}
 &  32.3$^{+  6.5}_{-  5.0}$ &  12.3$^{+  2.1}_{-  2.0}$ &  0.45$^{+ 0.01}_{- 0.01}$ &   -222.5$^{+   290.1}_{-   364.5}$ &  93.0$^{+  3.6}_{-  4.1}$ &   84.9$^{+   2.1}_{-   4.8}$ &  0.45$^{+ 0.09}_{- 0.07}$ &  0.28$^{+ 0.05}_{- 0.04}$ &  0.06$^{+ 0.00}_{- 0.01}$ \vspace*{0.1cm} \\ \noalign{\smallskip}
Scl & 111.8$^{+ 12.4}_{-  9.1}$ &  59.7$^{+  7.7}_{-  6.7}$ &  0.30$^{+ 0.03}_{- 0.01}$ &   -713.6$^{+   188.9}_{-   174.0}$ &  92.6$^{+  0.7}_{-  0.7}$ &    8.1$^{+   5.3}_{-   4.5}$ &  2.07$^{+ 0.31}_{- 0.21}$ &  1.47$^{+ 0.14}_{- 0.12}$ &  0.43$^{+ 0.06}_{- 0.06}$ \\ \noalign{\smallskip}
&107.1$^{+  8.4}_{-  7.3}$ &  61.7$^{+  6.4}_{-  6.2}$ &  0.27$^{+ 0.03}_{- 0.01}$ &   -787.4$^{+   190.5}_{-   173.5}$ &  92.8$^{+  0.7}_{-  0.7}$ &    7.8$^{+   4.8}_{-   4.1}$ &  1.82$^{+ 0.16}_{- 0.14}$ &  1.31$^{+ 0.13}_{- 0.14}$ &  0.41$^{+ 0.06}_{- 0.05}$  \\ \noalign{\smallskip}
 & 100.9$^{+  6.5}_{-  6.8}$ &  54.4$^{+  6.1}_{-  4.8}$ &  0.29$^{+ 0.04}_{- 0.03}$ &   -675.1$^{+   167.8}_{-   150.6}$ &  92.4$^{+  0.7}_{-  0.6}$ &    7.4$^{+   3.7}_{-   3.8}$ &  1.57$^{+ 0.12}_{- 0.11}$ &  1.23$^{+ 0.08}_{- 0.09}$ &  0.44$^{+ 0.04}_{- 0.04}$ \vspace*{0.1cm} \\ \noalign{\smallskip}
Boo~I &  78.2$^{+  7.1}_{-  6.3}$ &  31.7$^{+  7.1}_{-  6.4}$ &  0.42$^{+ 0.06}_{- 0.05}$ &  -1334.0$^{+   372.0}_{-   405.2}$ &  97.9$^{+  1.3}_{-  1.6}$ &  -50.4$^{+   6.6}_{-   7.6}$ &  1.23$^{+ 0.18}_{- 0.16}$ &  0.92$^{+ 0.09}_{- 0.09}$ &  0.30$^{+ 0.01}_{- 0.01}$ \\ \noalign{\smallskip}
& 75.2$^{+  5.9}_{-  5.1}$ &  30.1$^{+  7.5}_{-  5.8}$ &  0.43$^{+ 0.06}_{- 0.06}$ &  -1177.7$^{+   357.3}_{-   420.6}$ &  97.3$^{+  1.4}_{-  1.7}$ &  -52.2$^{+   7.1}_{-   8.1}$ &  1.14$^{+ 0.15}_{- 0.11}$ &  0.88$^{+ 0.09}_{- 0.09}$ &  0.30$^{+ 0.01}_{- 0.01}$  \\ \noalign{\smallskip}
 &  71.6$^{+  3.9}_{-  3.9}$ &  19.8$^{+  4.0}_{-  3.8}$ &  0.57$^{+ 0.05}_{- 0.05}$ &   -635.7$^{+   273.1}_{-   351.1}$ &  95.0$^{+  2.0}_{-  1.9}$ &  -65.1$^{+   9.2}_{-   8.3}$ &  0.94$^{+ 0.08}_{- 0.06}$ &  0.76$^{+ 0.06}_{- 0.05}$ &  0.29$^{+ 0.02}_{- 0.02}$ \vspace*{0.1cm} \\ \noalign{\smallskip}

\end{tabular}
\end{table*}

\onecolumn

\begin{table*}
\small
\begin{center}
\begin{tabular}{cccc}
\hline
 {\tt source\_id} & {\tt ra} & {\tt dec} &$G$ ({\tt phot\_g\_mean\_mag})\\
\hline 
 \noalign{\smallskip}
  4689633627542489728 &    6.77396 &  -71.99619 &   19.378 \\
  4689634074221208320 &    6.48790 &  -72.01599 &   17.090\\
  4689638128697225216 &    5.61038 &  -72.13231 &   19.695\\
  4689637514491268480 &    5.75409 &  -72.11833 &   17.506\\
\hline
\end{tabular}
\caption{Example of part of a list of possible cluster members. Here we list the first four entries of the compilation of members of the globular cluster NGC 104. The full table for this cluster is available in electronic format, along with separate tables for each of the rest of the globular clusters in our sample, each of the 9 dSph, the Bootes I UFD, and the LMC and SMC. The stars in these lists were selected and used to determine the astrometric parameters of the corresponding objects following either the procedures described in Sec.~\ref{sec:data-method-gcdw} (for the clusters and dwarfs) or in Sec.~\ref{sec:data-method-MCs} (for the LMC and SMC).}
\label{tab:members}
\end{center}
\end{table*}%

\end{document}